\documentclass[preprint,12pt,authoryear]{elsarticle}
\bibpunct[]{(}{)}{;}{a}{}{,}

\usepackage{amssymb}
\usepackage{xcolor}

\journal{XXXXX}

\begin{document}

\begin{frontmatter}




%\title{Part 2-B. Observations of the gas phase: Complex organic molecules including mapping}

\title{Part 2-B. Observations of complex organic molecules in the gas phase of the interstellar medium}


\author[inst1]{Izaskun Jim\'enez-Serra}

\affiliation[inst1]{organization={Centro de Astrobiologia (CAB), CSIC-INTA},%Department and Organization
            addressline={Ctra. de Ajalvir km 4}, 
            city={Torrejon de Ardoz},
            postcode={E-28806}, 
            %state={State One},
            country={Spain}}

\author[inst2,inst3]{Claudio Codella}
%\author[inst1,inst2]{Author Three}
\affiliation[inst2]{organization={INAF, Osservatorio Astrofisico di Arcetri},%Department and Organization
            addressline={Largo E. Fermi 5}, 
            city={Firenze},
            postcode={50125}, 
            %state={State Two},
            country={Italy}}
\affiliation[inst3]{organization={Universite Grenoble Alpes, CNRS, IPAG},%Department and Organization
            addressline={}, 
            city={Grenoble},
            postcode={38000}, 
            %state={State Two},
            country={France}}

\author[inst4]{Arnaud Belloche}
%\author[inst1,inst2]{Author Three}
\affiliation[inst4]{organization={Max-Planck-Institut fur Radioastronomie},%Department and Organization
            addressline={Auf dem Hugel 69}, 
            city={Bonn},
            postcode={53121}, 
            %state={State Two},
            country={Germany}}

\begin{abstract}
Thanks to the advent of sensitive and broad bandwidth instrumentation, complex organic molecules (COMs) have been found in a wide variety of interstellar environments, not only in our Galaxy but also in external galaxies up to a redshift of 0.89. The detection of COMs in cold environments such as starless or prestellar cores has challenged our understanding of COM formation and new ideas are being implemented in chemical models and explored in laboratory experiments. At the protostellar stage, the advent of new interferometers such as the Atacama Large Millimeter/submillimeter Array (ALMA) has allowed the mapping of the weak emission of COMs in the protostellar envelopes and protoplanetary disks around both low-mass and high-mass protostars, pinpointing their location and revealing differentiation between the different families of molecules. In this way, thermal and non-thermal desorption mechanisms can be probed, constraining the efficiency of formation of COMs in the gas phase versus on grain surfaces. Some degree of continuity in the COM composition is found from the early to late stages of star formation, suggesting that a significant fraction of COMs are formed at the initial conditions of star formation. For extreme environments such as the Galactic Center, cosmic rays and low-velocity shocks seem to influence the COM composition of low and high-density gas components. The spectral confusion limit will be a major challenge for the detection of new COMs in future spectroscopic surveys. However, low-frequency interferometers targeting sources with low-excitation temperatures may help to overcome this limit. 
\end{abstract}



\begin{keyword}
astrochemistry \sep complex organic molecules \sep star formation \sep interstellar medium \sep hot cores \sep hot corinos \sep prestellar cores \sep outflows \sep protostellar and protoplanetary disks
\end{keyword}

\end{frontmatter}



%%       \citet{<label>} ==> Jones et al. (2015)
%%       \citep{<label>} ==> (Jones et al., 2015)
%Test of citation \citep[see, e.g.,][]{Jorgensen20}. Another test with \citet{Jorgensen20}.

\section{The importance of complex organic molecules}
\label{s:importance}

In previous chapters, we have seen that molecules can form in the interstellar medium (ISM), either in the gas phase or on grain surfaces, despite its extreme conditions of low temperature, low density, and energetic processing by UV radiation and cosmic rays. Some of these molecules are large and present a complex structure. One of the largest molecules detected so far in the ISM is cyanopyrene (C$_{16}$H$_9$CN), a four-ring polycyclic aromatic hydrocarbon (PAH) with several isomers discovered \citep[][]{wenzel2024a,wenzel2024b}. In the field of astrochemistry, this type of molecules are called complex organic molecules (COMs), and are defined as carbon-bearing molecules made of six or more atoms \citep[][]{Herbst09}. 
%COMs are important in Astrochemistry because 
They are believed to be the precursors of the building blocks of life. Important biomolecules such as amino acids, nucleobases and sugars have been found in meteorites, comets, and asteroids \citep{cronin1983,altwegg2016,koga2017,furukawa2019,oba2023,glavin2025}. Although aqueous alteration in the parent body is likely responsible for the formation of biomolecules in some of these objects \citep[as the amino acids reported in some meteorites; see, e.g.,][]{glavin2009}, for others like the amino acid glycine in comet 67P/Churyumov-Gerasimenko \citep[][]{altwegg2016} an interstellar origin has been invoked. Therefore, studying COMs along all the stages of star formation is essential to understand not only the chemical inheritance of COMs throughout the star and planet formation process, but ultimately, our cosmic origins \citep[e.g.,][]{Ceccarelli2023}. 

\section{Census of complex organic molecules in the Galactic and extragalactic interstellar medium}
\label{s:census}

\begin{figure}[!t]
    \centering
    \includegraphics[angle=90,width=0.52\textwidth]{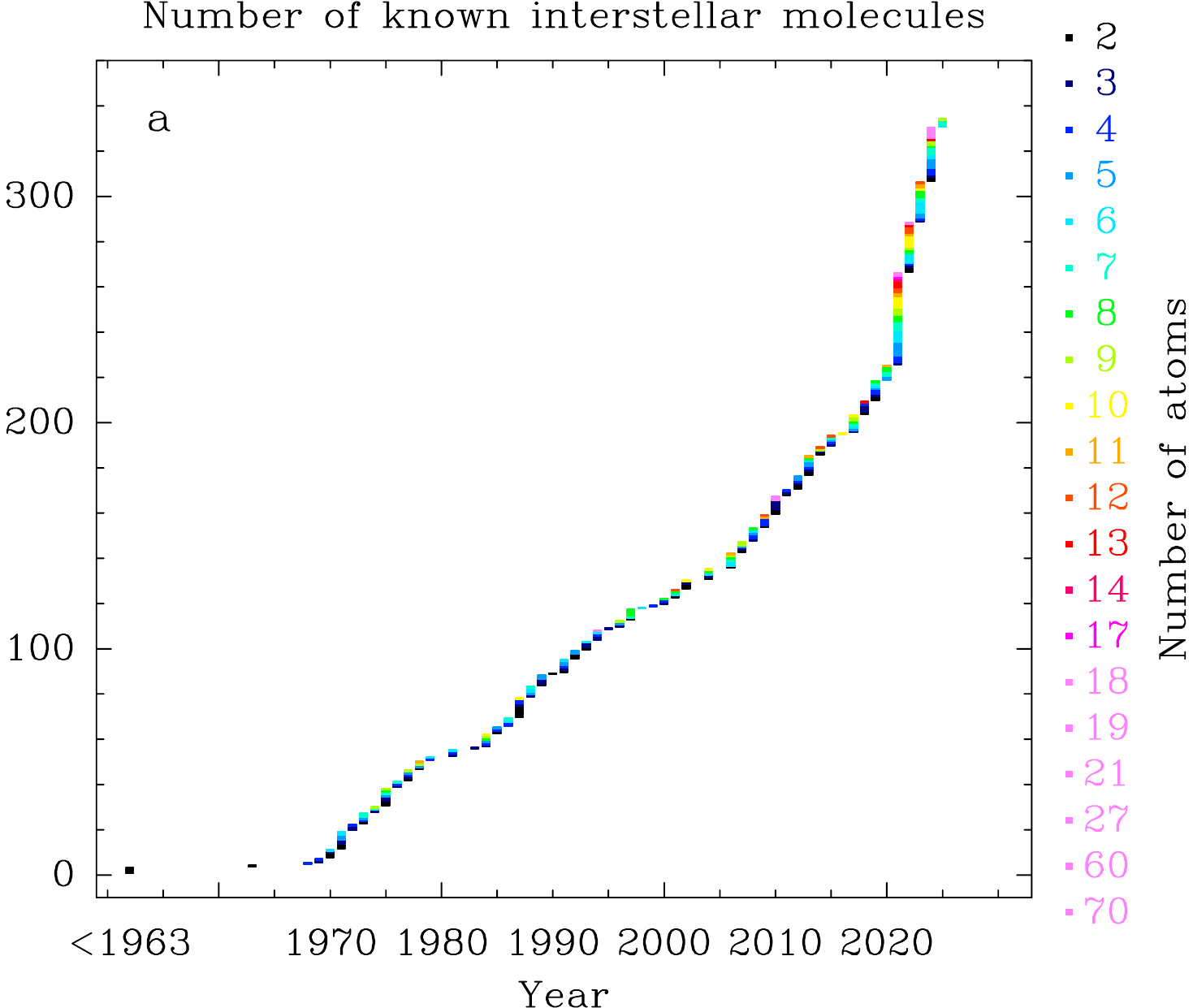}\hspace*{0.01\textwidth}\includegraphics[angle=90,width=0.52\textwidth]{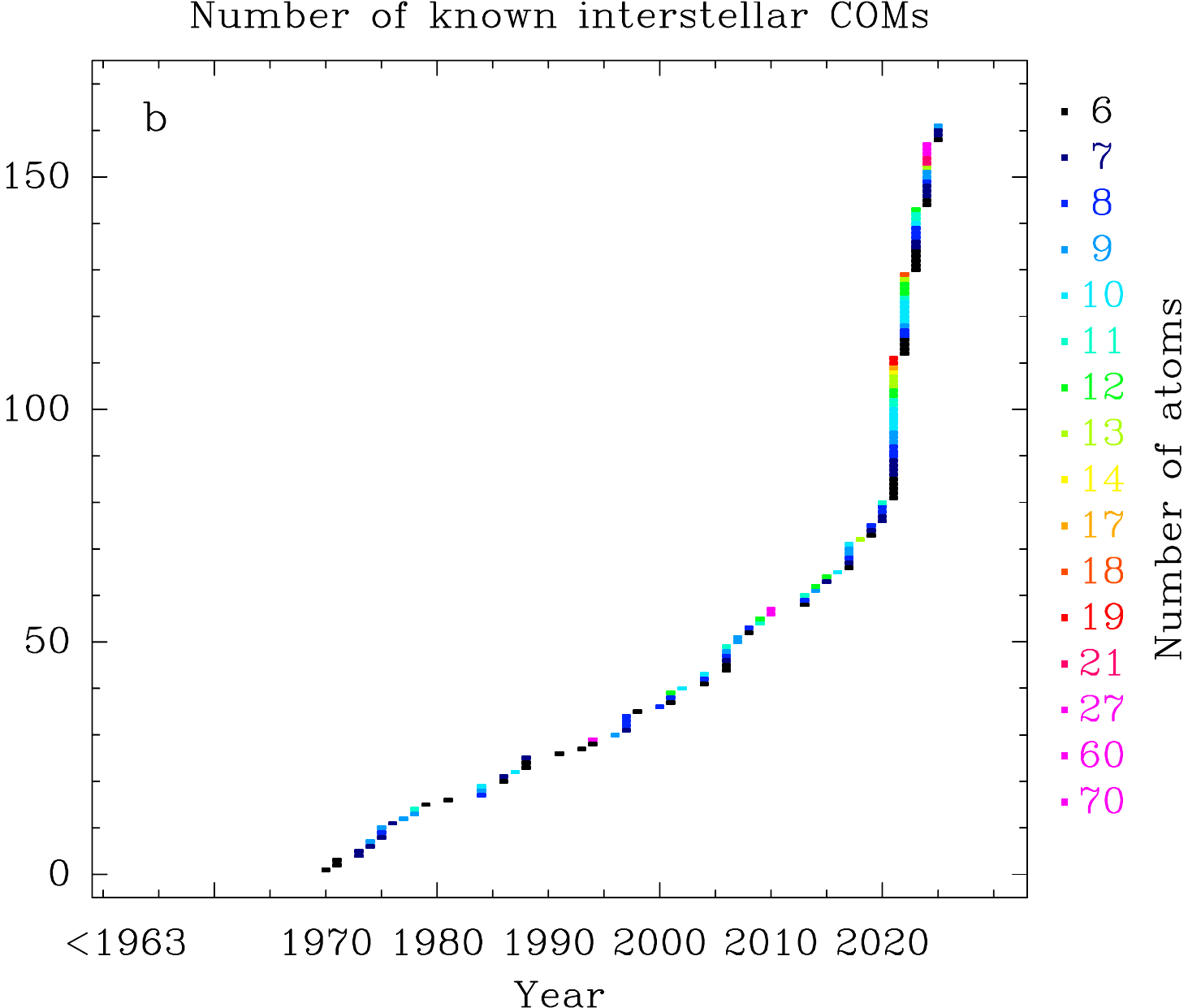}
    \caption{Cumulative number of known interstellar molecules (left) and COMs (right) as a function of time. The molecules are color-coded according to the number of their constituent atoms. The color-coding differs between the panels. This figure is based on the census of molecules in space maintained by H. M\"uller at the Cologne Database for Molecular Spectroscopy (CDMS, https://cdms.astro.uni-koeln.de/) as well as our own compilation of results found in the literature. See also \citet{McGuire22}.}
    \label{f:spacemol}
\end{figure}

\begin{figure}[!t]
    \centering
    \includegraphics[width=0.7\textwidth,angle=90]{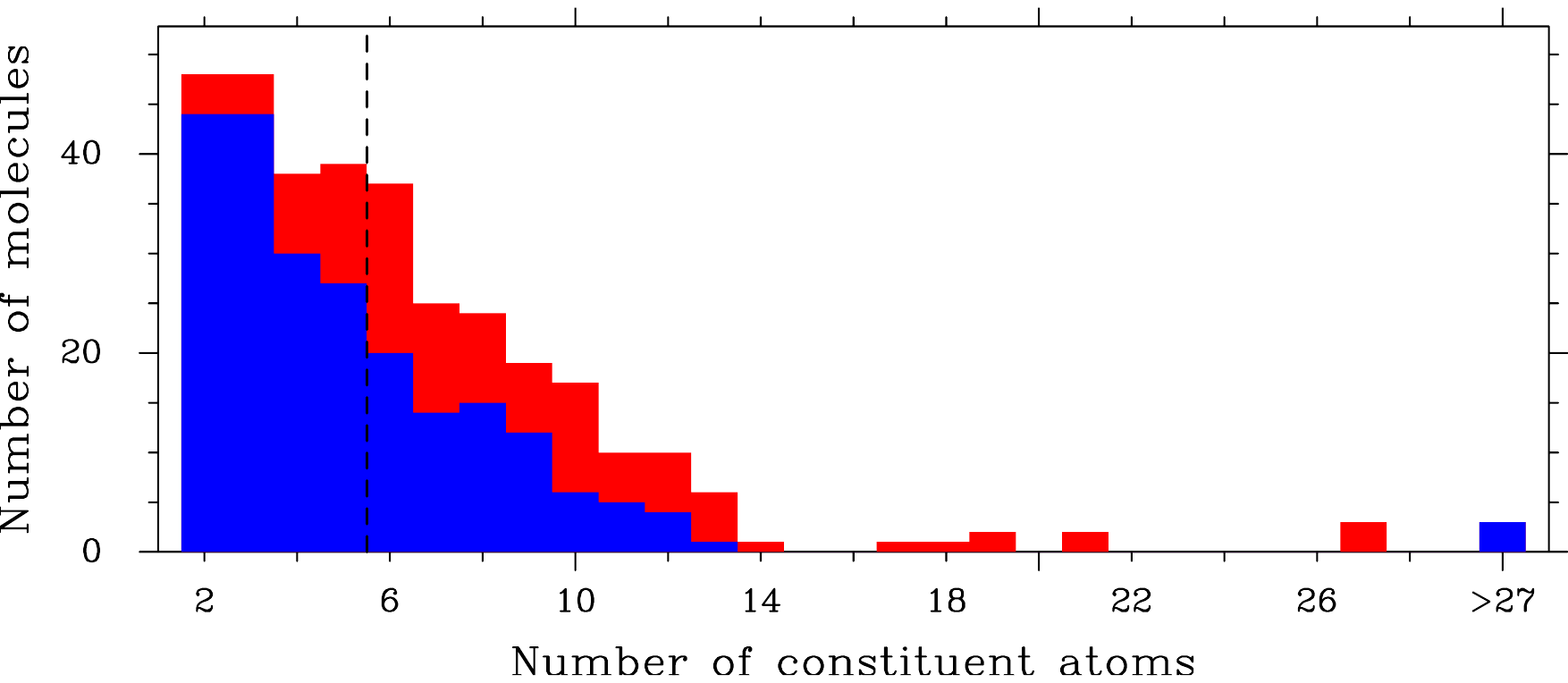}
    \vspace*{-1ex}
    \caption{Distribution of known interstellar molecules as a function of the number of their constituent atoms. The red boxes highlight the molecules that were discovered in 2021--25. COMs are located to the right of the dashed line. This figure is based on the CDMS census of molecules as well as our own compilation of the literature. See also \citet{McGuire22}.}
    \label{f:histomol}
\end{figure}

While the first seven molecules discovered in the ISM in the 1930s, 1940s, and 1960s were smaller ones (CH, CH$^+$, CN, OH, H$_2$O, NH$_3$, H$_2$CO), the first interstellar COMs were identified as early as 1970 \citep[CH$_3$OH;][]{Ball70} thanks to the progress made with instrumentation in radio astronomy and molecular spectroscopy in the laboratory. As of February 2025, 334 molecules have been identified in the ISM (Fig.~\ref{f:spacemol}a), among which 161 are COMs (Fig.~\ref{f:spacemol}b). COMs thus represent 48\% of all known interstellar molecules (Fig.~\ref{f:histomol}). 

The rate of discoveries, for both molecules in general and COMs in particular, has been roughly constant for about four decades since the 1970s but it experienced an important increase in the mid 2000s thanks to the advent of broadband receivers and backends at various radio astronomical facilities. Figure~\ref{f:spacemol} also reveals an explosion of discoveries in 2021--25, with 81 COMs out of 109 molecules newly identified, revealing in particular a new territory of molecules made of two dozen atoms. We may be witnessing the opening of a golden age of COM astrochemistry.

While the Atacama Large Millimeter/submillimeter Array (ALMA), with its high sensitivity, high angular resolution, and large spectral coverage, was expected to contribute much to discoveries of COMs in the ISM when it started operating in the early 2010s, it turns out that the huge increase in new detections recorded in 2021--25 results almost entirely from observations carried out with single-dish radio telescopes: the Yebes 40~m, IRAM 30~m, and Green Bank 100~m telescopes. Three main reasons explain this breakthrough: hundreds of hours have been invested in spectral line surveys with these telescopes over the past few years; recent progress with their instrumentation has increased their instantaneous spectral coverage and sensitivity, thereby boosting the efficiency of these surveys; and the targeted astronomical sources have, as we will see later, physical conditions that lead to low excitation temperatures of COMs, implying in turn less spectral confusion than in the spectra of compact, hot sources that ALMA excels in imaging.

All COMs detected in the ISM were first identified in our Galaxy. Many of them were discovered in star forming regions, but several have also been detected in other types of environments, as we will describe in the next sections. COMs have even been observed in external galaxies. The census of extragalactic molecules on the website of the Cologne Database for Molecular Spectroscopy (CDMS, https://cdms.astro.uni-koeln.de/classic/molecules) indicates that nearly 20\% of the approximately 74 molecules detected as of March 2024 in external galaxies are COMs \citep[see also][]{McGuire22}. The broad linewidths that lead to spectral confusion and the large distances that result in faint signals make the detection and identification of COMs in external galaxies challenging. Here, sensitive and broadband interferometers such as ALMA and the Northern Extended Millimeter Array (NOEMA) have played (and will continue playing) a major role.

In this chapter, we discuss the presence and distribution of COMs in interstellar regions forming low-mass stars (Sect.~\ref{s:lowmass}) and in regions forming high-mass stars (Sect.~\ref{s:highmass}). COMs are also present in other Galactic environments, which we address in Sect.~\ref{s:galenv}. As mentioned above, COMs have also been detected in external galaxies, which is the topic of Sect.~\ref{s:extragal}. Section~\ref{s:comparison} compares the COM chemical composition across different environments. Finally, our conclusions and an outlook are presented in Sect.~\ref{s:conclusions}.

\section{Regions forming low-mass stars}
\label{s:lowmass}

\subsection{Starless and prestellar cores}
\label{ss:prestellar}

One of the main results of the {\it Herschel} satellite mission is that the interstellar medium is arranged in filamentary dust and molecular structures, where the future generation of stars will be born \citep{andre10,andre19}. In molecular clouds forming low-mass stars such as the Taurus molecular complex, star formation is initiated within cold ($T\leq$10$\,$K) and dense molecular condensations \citep[H$_2$ gas densities $\geq$10$^5$$\,$cm$^{-3}$,][]{caselli02c,tafalla04,belloche11} called {\it starless cores} because they do not host any protostellar object yet. Not all starless cores will form stars but a subset of them that show signatures of gravitational collapse \citep[in the form of asymmetric blue-skewed self-absorbed profiles in optically thick lines of dense gas tracers such as N$_2$H$^+$;][]{caselli02a,devries05,keto15}. These cores are called {\it prestellar} and represent the initial conditions of Solar-system formation. Due to their cold gas and dust temperatures, the chemistry of starless/prestellar cores was initially believed to be dominated by simple molecular species whose formation occurs mainly in the gas phase. One good example are simple deuterated molecules (e.g., DCO$^+$ or N$_2$D$^+$), whose formation is favoured by the low temperatures and strong depletion of CO toward the densest, innermost regions of the cores, especially at the prestellar stage \citep[][]{bergin97,caselli02b,crapsi05,redaelli19}. 

This view has however changed in the past decade thanks to the advent of more sensitive and broad bandwidth instrumentation, which has revealed the presence of COMs in starless/prestellar cores \citep[][]{marcelino07,oberg10,bacmann12,vastel14,jimenez16,jimenez21,taquet17,soma18,Scibelli20,scibelli21,agundez19,agundez21, agundez23}. COMs such as propylene (CH$_2$CHCH$_3$), acetaldehyde (CH$_3$CHO), methyl formate (CH$_3$OCHO), dimethyl ether (CH$_3$OCH$_3$), propynal (HCCCHO), propenal (C$_2$H$_3$CHO), vinyl alcohol (C$_2$H$_3$OH), or ethanol (C$_2$H$_5$OH), are present in starless/prestellar cores with abundances typically in the range between 10$^{-12}$ and a few 10$^{-10}$ with respect to H$_2$ \citep[see, e.g.,][]{bacmann12,jimenez16,agundez21,agundez23}. Pure hydrocarbon cycles such as  cyclopentadiene (c-C$_5$H$_6$) and indene (c-C$_9$H$_8$) \citep{cernicharo21}, or polycyclic molecules such as 1- and 2-cyanonaphtalene \citep{mcguire21} and 1-, 2-, and 4-cyanopyrene \citep{wenzel2024a,wenzel2024b}, have also been reported toward the cold dark cloud TMC-1, which enables the study of the intermediate species likely responsible for the bottom-up synthesis of PAHs \citep[][]{Burkhardt21,garcia-concepcion2023}.

All this observational work has triggered a great interest in the astrochemistry community to investigate, from both a theoretical and experimental point of view, the processes that lead to the formation of COMs at the coldest conditions found in the ISM \citep[e.g.,][]{rawlings13,vasyunin13,balucani15,ivlev15,chuang16,holdship19,dartois19,jin20,wakelam21,Ceccarelli2023}. To constrain their formation mechanisms, maps of the distribution of methanol have been obtained toward a sample of cores \citep[][]{bizzocchi14,spezzano16,Scibelli20,spezzano20,punanova22,taillard23}.

Mapping the emission of larger COMs is, however, challenging because of the high sensitivity needed to detect the emission of these species. A recent example is given by \citet{cernicharo23}, who mapped the weak emission of benzonitrile (C$_6$H$_5$CN) toward the cold dark cloud TMC-1 using 100 hours of observing time. This map allowed the authors to draw conclusions about the bottom-up formation of aromatic molecules in this cold environment. An alternative method consists in carrying out deep observations of COMs toward two positions in a core: 1) the dust continuum  peak, which traces its central position; and 2) a second location at a distance ranging from some 1000 au to $\sim$10000 au, where the emission of CH$_3$OH peaks and that is representative of the intermediate-density, external layers of the core \citep[][]{jimenez16,jimenez21,megias23}. The prestellar core L1544 \citep{jimenez16} and the starless cores L1517B and L1498 \citep{jimenez21,megias23} were observed using this method. These high-sensitivity observations revealed that COMs are enhanced at the intermediate-density, outer layers of the cores where CO is catastrophically frozen onto dust grains (the CO-snow line in starless cores).  

In order to explain this COM enhancement, \citet[][]{vasyunin17} modelled the chemistry of COMs (including large O-bearing COMs such as methyl formate, CH$_3$OCHO, and dimethylether, CH$_3$OCH$_3$) as a function of the radial distance within L1544. In this model, precursors of COMs are non-thermally desorbed from the mantles of dust grains via chemical reactive desorption \citep[e.g.,][]{minissale16} and undergo gas-phase neutral-neutral reactions yielding COMs. The formation of large O-bearing COMs is enhanced at the location where CO severely freezes out onto dust grains \citep[i.e. at the location of the CO-snow line;][]{caselli99}, which activates the chemistry of O-bearing COM precursors on grain surfaces. This model successfully reproduces the COM enhancements observed toward the outer shells of L1544 \citep{vasyunin17}, although it overproduces the abundance of methanol measured in the ice toward this prestellar core \citep[see][]{goto21}. The model has also been applied to the L1498 and L1517B starless cores, providing an overall good match to the observed COM abundances, but larger discrepancies are found for L1517B \citep[][]{jimenez21,megias23}.

An interesting aspect to consider is that these cores are believed to be at different stages of evolution, where L1517B is the youngest followed by L1498 and L1544. The comparison among these cores of the chemical inventory of COMs (O-bearing vs. N-bearing), and of their abundances, reveals that COM chemistry evolves with the dynamical age of the core. Indeed, while N-bearing COMs are more abundant at the starless core stage, O-bearing COMs are enhanced in the prestellar core phase, likely due to the catastrophic depletion of CO onto dust grains, which activates a rich grain surface chemistry in O-bearing COMs \citep[see][]{jimenez21,megias23}. This trend is also consistent with the results obtained toward the young starless core L1521E \citep{nagy19,scibelli21}, but it still needs to be confirmed with a much larger sample of cores. 

In the modelling of \citet{vasyunin17}, the non-thermal desorption of COMs from grains occurs via chemical reactive desorption. However, the debate about the dominant mechanism(s) responsible for the non-thermal desorption of COMs in starless/prestellar cores remains unsettled. UV photodesorption has been proposed as a possible mechanism \citep{vastel14}. However, laboratory experiments have found that the amount of methanol (the simplest COM) desorbed by UV photons is negligible \citep{Cruz-Diaz16,Bertin16}. In contrast, IR photons seem to be efficient in the desorption of this molecule from ices \citep{santos23}. Alternatively, \citet{dartois19} and \citet{wakelam21} have proposed that COMs could be desorbed from ices in cold cores by cosmic-ray sputtering. By comparing gas-phase CH$_3$OH maps of the LDN 429-C core with ice measurements obtained with {\it Spitzer}, \citet{taillard23} inferred that the non-thermal desorption efficiency for CH$_3$OH increases with density from values of 0.002\% to 0.09\% toward the densest regions of the core. If the non-thermal desorption of CH$_3$OH in LDN 429-C is caused by cosmic-ray sputtering, its varying efficiency is likely due to the ice composition changing as a function of density, from a water-dominated ice to a mixed composition ice regime. Future comparisons between gas-phase and ice COM observations obtained from the ground and with {\it JWST} will help us to constrain what chemical/physical mechanisms govern the formation and/or non-thermal desorption of COMs under the coldest conditions of the ISM in starless/prestellar cores. 

\subsection{Hot corinos and Warm Carbon-Chain Cores (WCCCs)}
\label{ss:hotcorinos}

After the prestellar phase, the gravitational collapse of the core leads to the formation of a protostar that heats the surrounding envelope producing changes in its chemical composition. Depending on such composition, we can distinguish two types of objects: hot corinos and warm carbon-chain chemistry cores (or WCCCs). Hot corinos represent those dusty cocoons that present bright emission in COMs within their innermost ($\leq$100 au) and warm ($T$$\geq$100 K) regions. Examples of hot corinos are IRAS~16293--2422B and NGC 1333-IRAS4A \citep[see, e.g.,][]{vanDishoeck95,cazaux03,bottinelli04a,bottinelli04b,Jorgensen16}. WCCC sources, instead, present bright emission from unsaturated carbon-chain species such as C$_2$H or C$_4$H within the central 1000 au of these objects \citep[][]{sakai08,sakai09}. Examples of WCCC sources are L1527 and IRAS 15398-3359 \citep[][]{aikawa08,sakai13}.

The origin of such chemical differences has been the topic of numerous theoretical and observational studies of hot corinos and WCCCs. It is well known that COMs thermally desorb from dust grains at temperatures $\geq$100~K at the same time as water \citep[see, e.g.,][]{Viti04,Garrod06,Lee22,Busch22,Ceccarelli2023}. In contrast, unsaturated carbon-chain molecules are formed in the gas phase once CH$_4$ is sublimated at temperatures $\geq$25--30~K. If the central protostar were not bright enough, this would explain why WCCCs are deficient in COMs; but this idea cannot explain why hot corinos are deficient in carbon-chain molecules. Alternatively, with the discovery of the first WCCCs, \citet{sakai08,sakai09} proposed that the radical chemical variations between hot corinos and WCCCs could be due to differences in the duration of the prestellar phase. More recent chemical modelling \citep[][]{aikawa20} has revealed that other parameters such as the temperature and extinction during the prestellar (static) phase also play an essential role in the chemical diversity between hot corinos, WCCCs, and hybrid sources, where both COM and warm carbon-chain chemistries co-exist \citep[see][]{Oya17}.  

As a result of their high level of chemical complexity, hot corinos have been prime objects for the search for new COMs in low-mass protostars. The complete chemical inventory of COMs in hot corinos, however, has been studied only toward a few sources. The spectral survey performed at 0.8~mm with ALMA in the context of the PILS project \citep[Protostellar Interferometric Line Survey:][]{Jorgensen16,Jorgensen18} can be considered a breakthrough in obtaining the census of COMs in hot corinos. PILS used as target the prototypical IRAS~16293--2422 protostellar region, hosting the so-called A and B Class 0 sources. PILS detected and imaged more than twenty COMs \citep[see][]{Jorgensen16,Jorgensen18,Calcutt18,Manigand20,Manigand21}, which include tentative detections \citep[e.g., 3-hydroxypropenal, HOCHCHCHO;][]{coutens22} and first detections toward a low-mass protostar \citep[e.g., methyl isocyanate, CH$_3$NCO;][]{Ligterink17,MartinDomenech17}. Using ALMA data toward IRAS~16293--2422 at lower frequencies, \citet[][]{Zeng19} also reported the discovery of glycolonitrile (HOCH$_2$CN) in the ISM, an isomer of CH$_3$NCO and a molecule that has also been found recently toward the intermediate-mass hot corino Serpens SMM1-a \citep[][]{Ligterink21}. Other large programs on astrochemistry toward hot corinos have been performed or are currently in progress: IRAM 30-m ASAI \citep[][]{Lefloch18}, IRAM-PdBI CALYPSO \citep[][]{Belloche20}, ALMA FAUST \citep[][]{Codella21}, and ALMA PEACHES \citep[][]{Yang21}. 

Noticeably, PILS reported O-bearing, S-bearing (CH$_3$SH), as well as N-bearing COMs such as CH$_3$CN. \citet{Droz18,Droz19} compared the PILS results with the composition of comet 67P Churyumov-Gerasimenko \citep[see][]{Altwegg19}, as determined with the in situ monitoring carried out by {\it Rosetta} using the {\it Rosina} spectrometer. The authors found reasonable correlations for CHO-, N-, and S-bearing molecules \citep[see also][]{Altwegg19}. A similar result was reported by \citet{Bianchi19a,Bianchi19b} using a survey of COMs associated with the Class I SVS13-A protostellar system. These findings suggest that Solar System relics such as comets have partially inherited the volatile compositions from the earliest phases of the star formation process.

What do we see if we image the inner 100 au of a protostellar region associated with a hot corino? Besides the thermal heating induced by the protostar, fast jets and slower disk winds can alter the chemical content of the gas phase (see Sect. \ref{ss:outflows}). In addition, a complex chemistry can also be triggered by accretion streamers and/or infalling envelope material that crashes against the accretion disk. In both cases, low-velocity ($\sim$~1~km~s$^{-1}$) shocks are expected.
The study of accretion streamers feeding the protostellar system has been boosted by recent results using relatively simple species such as CO \citep{Akiyama19}, HCO$^{+}$ \citep{yen19}, and HC$_3$N \citep{Pineda20}. In addition, \citet{Garufi22} detected SO and SO$_2$ in portions of the HL Tau and DG Tau disks associated with shocked accretion streamers. In the context of COMs, \citet{Bianchi22} imaged CH$_3$OH emission associated with slow shocks occurring where the accretion streamers are feeding both disks of the SVS13-A binary.

Moreover, shocks occurring at the interface between the infalling material and the disk \citep{Stahler94}, close to the centrifugal barrier, also enrich the gas. As a consequence, chemically enriched rotating rings were predicted and recently observed in several targets, such as the prototypical L1527 protostar \citep{Sakai14b,Sakai14a}. Later on, COMs (CH$_3$OH, CH$_3$OCHO) were also observed at the centrifugal barriers around IRAS 16293--2422A and IRAS~04368+2557 by \citet{Oya16} and \citet{Sakai17}, while a large number of COMs (including, e.g., CH$_3$OCH$_3$ and NH$_2$CHO) have been observed in the edge-on HH212 disk, possibly associated with the shocked rings at the centrifugal barrier (see Sect. \ref{ss:disks}). These recent findings on COMs imaging at 20-50 au scales clearly call for 10 au--scale surveys to properly identify the disk components that are chemically enriched.
   
The study of the inner 100 au of protostellar regions is key to understand whether hot-corino and WCCC chemistries co-exist. This is one of the goals of the ALMA FAUST large program \citep{Codella21}, which will provide the chemical content of a large sample of protostars with a spatial resolution of 50 au. Enlightening results have been obtained by \citet{Oya17}, who studied with ALMA the Class 0 protostar L483, a WCCC source, using CCH and COMs. \citet{Oya17} imaged CH$_3$OCHO and NH$_2$CHO close to the protostar, and concluded that a WCCC chemistry is at work in the molecular envelope ($\sim$ 1000 au), while the inner 100 au region hosts a hot corino.
    
Future directions for the investigation of hot corinos has been drawn by  \citet{Desimone20b}, who studied the NGC1333-IRAS4 A1 and A2 binary system in Perseus, using both mm- (NOEMA) and cm-wavelength (Very Large Array, VLA) imaging. The goal was to understand why only one of them, A2, was associated with hot-corino activity \citep[e.g.,][]{Taquet15,DeSimone17,Belloche20,Yang21}. \citet{Desimone20b} found that as a matter of fact the lack of COM emission toward A1 is not due to a different chemistry but, instead, to the optically thick dust emission at higher frequencies. Methanol lines in the cm spectral range reveal the hot-corino toward A1. The IRAS4A binary case calls for further high-sensitivity and high-resolution surveys in the cm wavelength regime and at the same time shows that molecular abundances derived using (sub-)mm data could be underestimated by $\sim$ 30\%  \citep{Desimone20b}.

\subsection{Outflows}
\label{ss:outflows}

Outflowing gas driven by protostars is one of the most spectacular events occurring during the earliest stages leading to the formation of both low- and high-mass stars \citep[e.g.,][, and references therein]{Arce07,Bally07,Frank14,Motte18}. Angular momentum has to be extracted from the protostellar system to allow disk accretion onto the protostar itself. A major role is played by fast ($\sim$ 100 km s$^{-1}$) collimated jets, plausibly originated by magneto-centrifugal winds emitted from the rotating disk \citep[e.g.,][, and references therein]{Tabone20}. In turn, jets sweep up the envelope material creating high-velocity ($\geq$ 10 km s$^{-1}$) shocks \citep[e.g.,][]{Frank14} as well as slower outflows, observed on large spatial scales (up to pc scales). Shocks heat the gas and trigger several processes such as endothermic chemical reactions and ice grain mantle sublimation or sputtering/shattering \citep[e.g.,][, and references therein]{JimenezSerra08,Gusdorf08a,Gusdorf08b,Guillet11}. As a consequence, several molecular species (e.g., Si-bearing and S-bearing molecules) undergo significant enhancements in their abundances, as observed at millimeter wavelengths toward a number of outflows driven by low- and high-mass young stellar objects  \citep[YSOs; e.g.,][]{Bachiller01,Codella05,jimenezserra05,Herbst09,Lopez11,Sanchez13,Leurini13}. 

Shocked regions are characterized by different atomic and molecular components with different physical properties. In order to carefully image these regions, high spatial resolution (at least down to about 1000 au) is needed. Therefore, a detailed picture is provided by nearby Sun-like star forming regions, such as Taurus, Perseus, and Ophiuchus, located at distances shorter than 400~pc. A good example is provided by the protostellar shocked region L1157-B1 at a distance of 250~pc in Cepheus, as investigated with both IRAM mm-telescopes and {\it Herschel} Far Infrared (FIR) instruments \citep{Benedettini12,Busquet14}. Several components were revealed, from small (2$''$--5$''$), hot ($T_{\rm kin}$ $\sim$~1000~K), and relatively low-density ($n_{\rm H_2}$ down to 10$^3$~cm$^{-3}$) gas to cold ($\sim$ 20~K), dense (up to 10$^8$~cm$^{-3}$), extended (larger than 40$''$) cavities  \citep[see also][]{Lefloch12}.

What about COMs in outflows? The first COM observed in protostellar shocks was methanol (CH$_3$OH), which has been routinely observed in low-mass star forming regions such as IRAS~16293--2422, NGC1333-IRAS4A, L1448-mm, BHR71, and L1157 \citep[e.g.,][]{vanDishoeck95,Blake95,Bachiller98,Jorgensen04,jimenezserra05}. \citet{Arce08}, observing L1157-B1 with the IRAM 30-m telescope, revealed for the first time emission due to methyl formate (CH$_3$OCHO), ethanol (C$_2$H$_5$OH), and methyl cyanide (CH$_3$CN). Detections of acethaldehyde (CH$_3$CHO) and formamide (NH$_2$CHO) were reported, again toward L1157-B1, by \citet{Sugimura11} and \citet{Yamaguchi12} using the NRO 45m antenna. Finally, the IRAM 30-m Large Program ASAI \citep{Lefloch17,Lefloch18} did a significant step ahead with an unbiased spectral survey at 1, 2, and 3~mm toward L1157-B1, adding new species such as dimethyl ether (CH$_3$OCH$_3$), ketene (H$_2$CCO), and glycolaldehyde (HCOCH$_2$OH). The COM column densities and, in turn, their abundances are usually derived assuming LTE, given the lack of knowledge on collisional coefficients of many COMs. Interestingly, the COM abundances are comparable to or higher (up to a factor 10) than those in hot corinos \citep{Lefloch17}.

\begin{figure}[!t]
    \centering
    \includegraphics[angle=0,width=0.7\textwidth]{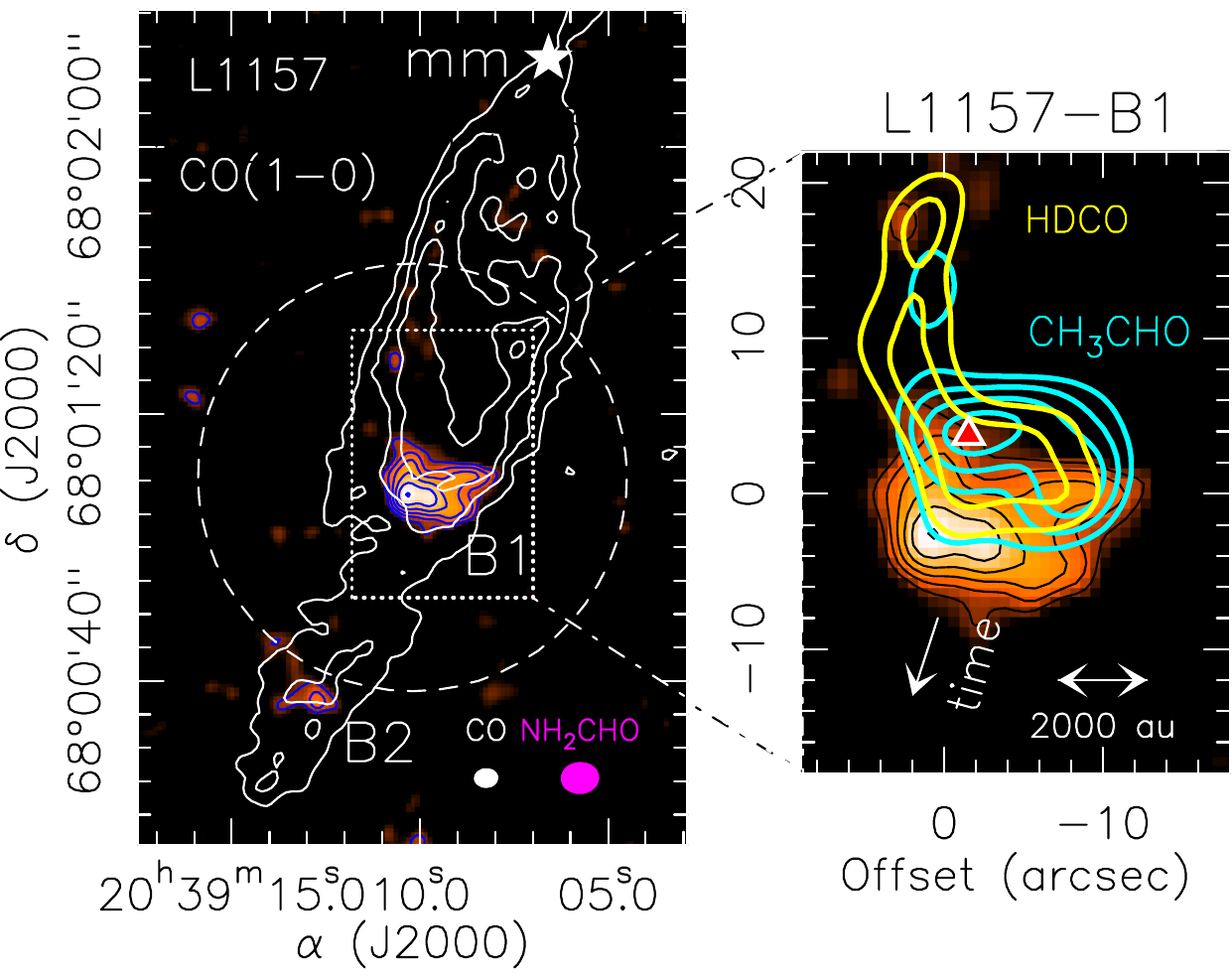}
    \vspace*{-4ex}
    \caption{{\it Left panel:} southern blueshifted lobe of the outflow of L1157 in CO \citep[white contours;][]{Gueth96}. The jet driven by L1157-mm (white star) produced two main cavities, the brightest being named B1. The emission map of the NH$_2$CHO(4$_{\rm 1,4}$–-3$_{\rm 1,3}$) line is shown in colour \citep{Codella17}. The dashed circle indicates the primary beam of the NH$_2$CHO image. The magenta and white ellipses depict the synthesised beams of the IRAM-NOEMA observations. {\it Right panel:} Chemical segregation in L1157-B1. The spatial distribution of formamide (in color and black contours) is compared with those of HDCO \citep[yellow contours,][]{Fontani14} and CH$_3$CHO \citep[cyan contours,][]{Codella15,Codella17}, once smoothed to the same angular resolution. The red triangle marks the peak position of the high-velocity SiO emission \citep{Gueth98}, indicating the youngest shocked region within L1157-B1.}
    \label{fig:l1157}
\end{figure}

These findings opened a rush in interferometric imaging of COM emission in L1157-B1. As a matter of fact, imaging the spatial distribution helps to constrain the formation of COMs in shocked regions, where dust ice mantles and refractory cores are sputtered
\citep[][, and references therein]{Ceccarelli2023}. An illustrative example is shown in Fig.~\ref{fig:l1157}, reporting in the left panel the southern blueshifted lobe of the L1157 outflow, with NH$_2$CHO emission tracing the apex of the B1 cavity opened by the precessing jet of L1157 \citep{Codella17}. The right panel of Fig. \ref{fig:l1157} shows the chemical differentiation revealed using the IRAM NOEMA interferometer in the context of the IRAM SOLIS large program \citep{Ceccarelli17}. The keys to interpret the complex scenario are:
(i) the kinematical time of the shocked regions increases from North to South \citep[white arrow; see the precessing model by][]{Podio16}, and 
(ii) the red triangle stands for the peak position of the high-velocity SiO emission \citep{Gueth98}, i.e. the region where the youngest ($\sim$ 1000--2000~yr) shock event (impact jet-cavity) in B1 occurs. A spatial dichotomy is clearly revealed: (1) HDCO is at the interface between the fast jet and the slower ambient medium, indicating that it was formed on grain mantles and then released into the gas due to shock sputtering/shattering \citep{Fontani14}. CH$_3$CHO shows a spatial distribution consistent with that of HDCO \citep{Codella15}; (2) NH$_2$CHO is detected in the post-shocked gas \citep{Codella17}, downstream of the sputtered region. There is then evidence that the formation time of formamide is delayed with respect to the initial shock event (SiO peak), allowing us to assess that, at least in L1157-B1, gas-phase chemistry could play a role in the formation of formamide \citep[see][]{Skouteris17}. However, note that the formation pathways of formamide both in the gas phase and on dust grains are still hotly debated \citep[e.g.,][]{dulieu19,Douglas22}.

The next step is to image other shocked regions located in different star forming regions and with different kinematical ages to confirm on a statistical basis what was found so far in L1157-B1. In this vein, a second interstellar laboratory has been recently opened by \citet{Desimone20a}, who used NOEMA to detect and image the emission of methanol, acetaldehyde, dimethyl ether, and formamide in two outflows driven by the NGC1333-4A1+A2 system in Perseus. Finally, we note that \citet{Maureira2022} recently obtained with ALMA a dust temperature distribution of IRAS~16293--2422A on 10 au scales. Their findings support that mechanical processes such as shocks play an important role in heating the protostellar disk. This in turn calls for further interferometric studies to verify if and how COMs abundances are affected by shocks in Class 0 disks.

\subsection{Protostellar and protoplanetary disks}
\label{ss:disks}

\begin{figure}[!t]
    \centering
    \includegraphics[width=0.7\textwidth]{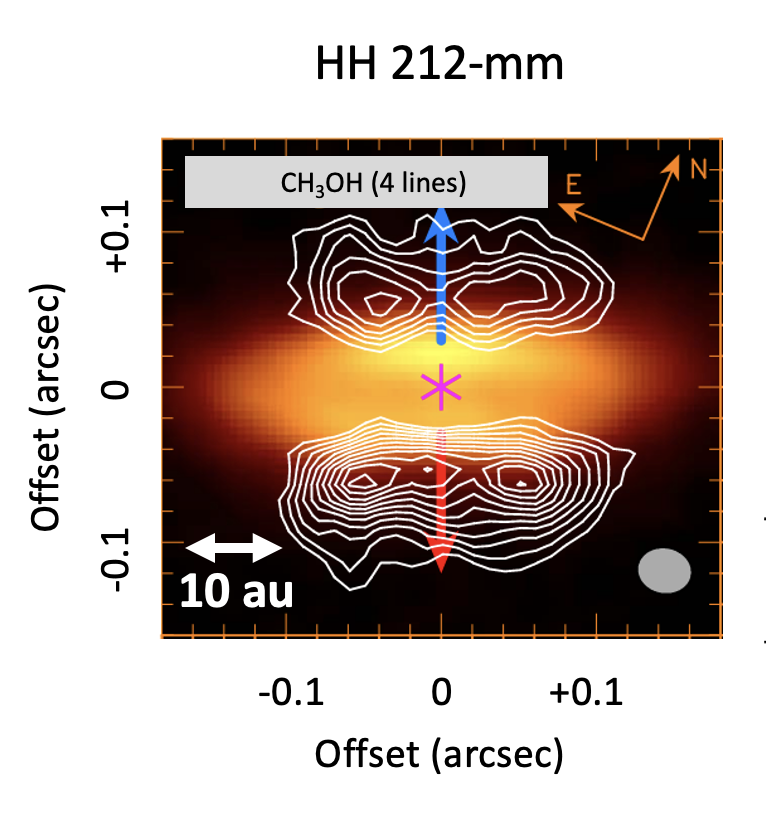}
    \vspace*{-3ex}
    \caption{The protostellar disk associated with the HH 212-mm Class 0 protostar. Methanol emission (contours) as imaged with ALMA overlaid on the continuum map (color image). The line images have been obtained by stacking four lines of methanol. The blue and red arrows indicate the axis of the blue- and red-shifted jet components, respectively. Adapted from \citet{Lee19}. 
    %{\it Right panel:} CH$_3$OH emission (two lines stacked) as imaged in the disk around the Class II (A-type) HD~100546 object \citep[][]{Booth21}.
    }
    \label{fig:HH212}
\end{figure}

The earliest phases of the accretion disks occur when the protostar is embedded in its dense  envelope. Disentangling signatures of the disk from those of the other physical components around the protostar is then really challenging. In particular, this applies to usually weak COM emission lines. Up to now, the unique protostellar disk  that has been successfully spatially resolved and chemically characterized at spatial scales down to 10 au is the one surrounding the HH~212-mm Class 0 protostar. Figure~\ref{fig:HH212} shows the HH~212 disk observed by \citet{Lee17a,Lee17b} at extremely high spatial resolution ($\sim$10 au). The dust continuum emission obscures the equatorial plane due to high optical depth. Methanol emission (white contours) is detected only in the outer layers, at about $\pm$40~au from the equatorial plane. Several other COMs were detected by \citet{Lee17b}, e.g., CH$_3$CHO, CH$_3$OCHO, and NH$_2$CHO, and found to have a spatial distribution consistent with that of methanol. The analysis of the kinematics reveals that the COM emission in HH~212 occurs in rings rotating in the same way as both the inner envelope \citep[e.g.,][]{Codella18} and the SiO jet \citep{Lee17c}. Such images of COM emission at extremely high spatial resolution indicate that either the rings are those associated with accretion shocks close to the centrifugal barrier or, alternatively, the COM emission reveals disk outer layers illuminated by the protostar \citep[e.g.,][]{Codella19,Lee19,Lee22}. Continuum modelling allowed \citet{Lee17c} to estimate, for the first time on this spatial scale, COM abundances (with respect to H$_2$): CH$_3$OH (10$^{-7}$), CH$_3$CHO (10$^{-9}$), CH$_3$OCHO (10$^{-9}$), NH$_2$CHO (10$^{-10}$). Observations at lower frequencies using, e.g., the cm spectral window, would be instructive in order to reveal the chemical composition of the gas in the equatorial plane which cannot be probed at mm wavelengths because of the high dust opacity.

If COM emission in Class 0  disks is difficult to reveal because it is deeply hidden in the dense  surrounding envelopes, observing COMs in more evolved, Class I or Class II disks (also called protoplanetary disks) is also challenging. More precisely, a protoplanetary disk can be subdivided into three main regions \citep[e.g.,][, and references therein]{Walsh14}: (1) the hot surface layer, where molecules are photodissociated, (2) the warm molecular layer, where molecules are in the gas phase, and (3) the cold outer midplane where molecules freeze out onto dust grains. As a matter of fact, the regions expected to be chemically rich in the gas phase are those few astronomical units close to the protostar, where we can invoke a sort of hot-corino chemistry regime. This is a region where line emission, again, is expected to be hugely affected by the high optical depth of the dust emission. 

The very close (60 pc), Solar-type, young T-Tauri star TW Hya has been used as laboratory to detect methyl cyanide \citep{Oberg15a} and methanol \citep{Walsh16}. Even in TW Hya the methanol emission is so weak that only staking different lines allowed \citet{Walsh16} to reveal the molecule. More recently, \citet{Podio20} detected methanol toward the Class I disk IRAS~04302+2247 by integrating the emission over the whole disk, while \citet{Booth21} imaged CH$_3$OH in the disk around the Class II (A-type) HD~100546 YSO. Methanol abundances (with respect to H$_2$) are estimated to fall in the 10$^{-12}$--10$^{-11}$ range.

\begin{figure}[!t]
    \centering
    \includegraphics[angle=90,width=1.0\textwidth]{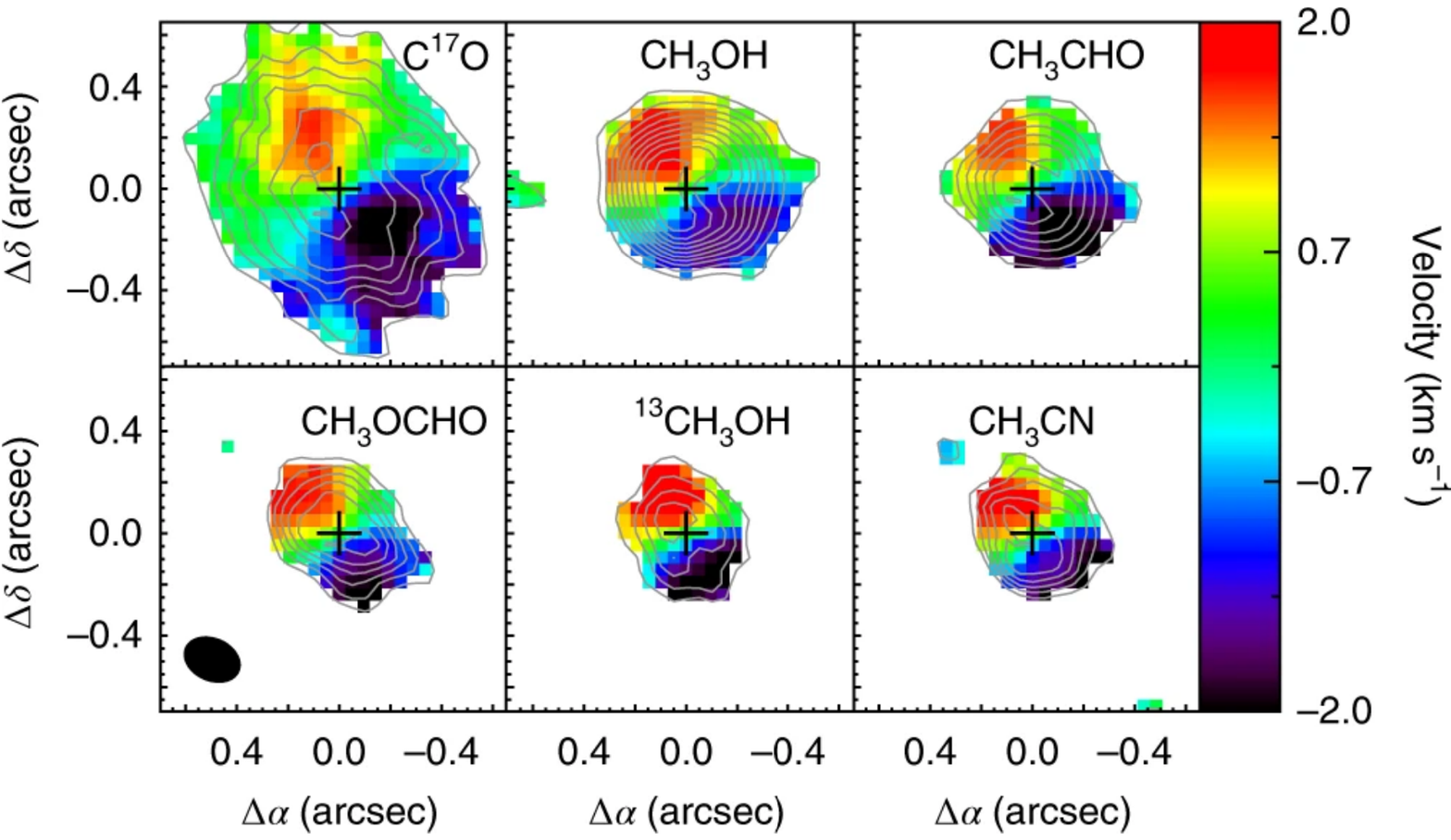}
    \caption{From \citet{LeeJ19}: 
    the rotating disk around the FU Ori object V833 Ori.
    The imaged species is indicated in each panel.}
    \label{fig:FuOri}
\end{figure}

On the other hand, methyl cyanide has been imaged toward a dozen of protoplanetary disks \citep{Oberg15a,Bergner18,Loomis18}. The recent ALMA large program MAPS \citep[Molecules with ALMA at Planet-forming Scales;][]{Oberg21} has reported the detection of CH$_3$CN in almost all targets \citep{Ilee21}. Typical abundances (with respect to H$_2$) are 10$^{-13}$--10$^{-12}$. The methyl cyanide emission is located close to the protostar or in rings, where the gas temperature is below 50 K. Interestingly, the fact that CH$_3$CN is easier to detect than methanol, the most common COM at the protostellar stage, suggests that the disk COM chemical inventory is not simply inherited from the protostellar stage.

Finally, a particular class of disks is represented by protoplanetary disks around FU Orionis objects (hereafter FU Ori), which can induce illumination changes on the surrounding material by up to three orders of magnitude. The consequence is the sublimation of ices in the outer portions of the disk. In other words, the snowline moves outwards each time there is a luminosity increase, releasing fresh material from dust ices into the gas phase. In this vein, the ALMA observations of the V883~Ori protoplanetary disk are instructive (see Fig. \ref{fig:FuOri}): Four COMs (CH$_3$OH, CH$_3$CN, CH$_3$CHO, CH$_3$OCHO) were imaged \citep{vantHoff18,LeeJ19}, tracing  portions of the rotating protoplanetary disk. The abundances of acetaldehyde (CH$_3$CHO) and methyl formate (CH$_3$OCHO) with respect to methanol in the V883~Ori disk are similar to those found in Class 0 and I hot corinos (Fig. \ref{fig:comparison}). The case of CH$_3$CN is different, with a much lower CH$_3$CN to CH$_3$OH abundance ratio compared to protostars. Obviously, a statistical approach is needed before supporting or ruling out the inheritance scenario. The next breakthrough will be to draw a thread to follow COM abundances from young (10$^4$ yr) to old (10$^6$ yr) disks. In this context, a key detection is that reported by \citet{Brunken22}, who imaged dimethyl ether (CH$_3$OCH$_3$) and, tentatively, methyl formate (CH$_3$OCHO) toward the Herbig IRS 48 Class II transition disk. This opens the search for COMs different from methanol in larger samples of objects.

\section{High-mass regime}
\label{s:highmass}

\subsection{Galactic hot cores}
\label{ss:hotcores}

\subsubsection{Definition and historical perspective}

Like hot corinos, their lower-mass counterparts (see Sect.~\ref{ss:hotcorinos}), hot cores are molecular regions around newly born high-mass (proto)stars that are heated up above 100~K by their central YSOs. They are also compact and dense \citep[$<0.1$~pc, $n_{{\rm H}_2} > 10^6$~cm$^{-3}$, e.g.,][]{Garay99,Osorio99}. High-mass star forming regions have historically been the places of choice to detect COMs in the interstellar medium because of their high H$_2$ column densities, and in turn high COM column densities, which makes the rotational emission of COMs easier to detect despite their larger distances compared to nearby low-mass star forming regions. In the early days of radio astronomy, many COMs were first detected toward the well-known high-mass star forming regions Sagittarius (Sgr) B2 \citep[e.g., ethanol, C$_2$H$_5$OH;][]{Zuckerman75} and Orion KL \citep[e.g., dimethyl ether, CH$_3$OCH$_3$;][]{Snyder74} thanks to single-dish spectroscopic surveys carried out at millimeter and sub-millimeter wavelengths \citep{blake86,blake87,Cummins86,Turner89,Turner91,tercero10,tercero12,neill12,comito05,belloche08,belloche13}. Interferometric follow-ups of these spectroscopic surveys have enabled not only the detection of many other COMs \citep[e.g.,][]{beltran09,belloche14,Belloche19}, but also the study of the spatial distribution and physical properties of the gas traced by COMs in these high-mass star-forming regions \citep{calcutt14,favre11,Tercero18}. 

COMs, such as CH$_3$OH and CH$_3$CN, are indeed excellent probes of the physical properties of the dense gas in hot molecular cores. The spatial distribution of the emission of COMs provides information about the innermost physical structure of hot cores, unveiling the presence of circumstellar disks, internal holes and energetic shocks associated with high-velocity jets \citep[see the examples of Cepheus A HW2, AFGL2591, and IRAS 20126+4104, respectively;][]{patel05,jimenez12,cesaroni14,palau17}. At larger scales, COMs also efficiently probe the regions recently processed by shocks, as seen for the Orion KL high-mass star-forming region \citep{favre11,Tercero18}. 

ALMA has represented a real game-changer in our understanding of COM spatial distribution, COM properties, and COM chemistry in hot molecular cores. Thanks to ALMA's exquisite angular resolution, sensitivity, and imaging fidelity, COMs have allowed us to probe the dynamics and physics of the regions very close to the central high-mass protostars, revealing their multiplicity, infall motions, spin-up rotation, and disk dynamics \citep{Beltran18,williams22}. All this is described in detail in the following sections.

\subsubsection{Spatial distribution of COMs}

\begin{figure}[!t]
    \centering
    \includegraphics[width=0.44\textwidth,angle=0]{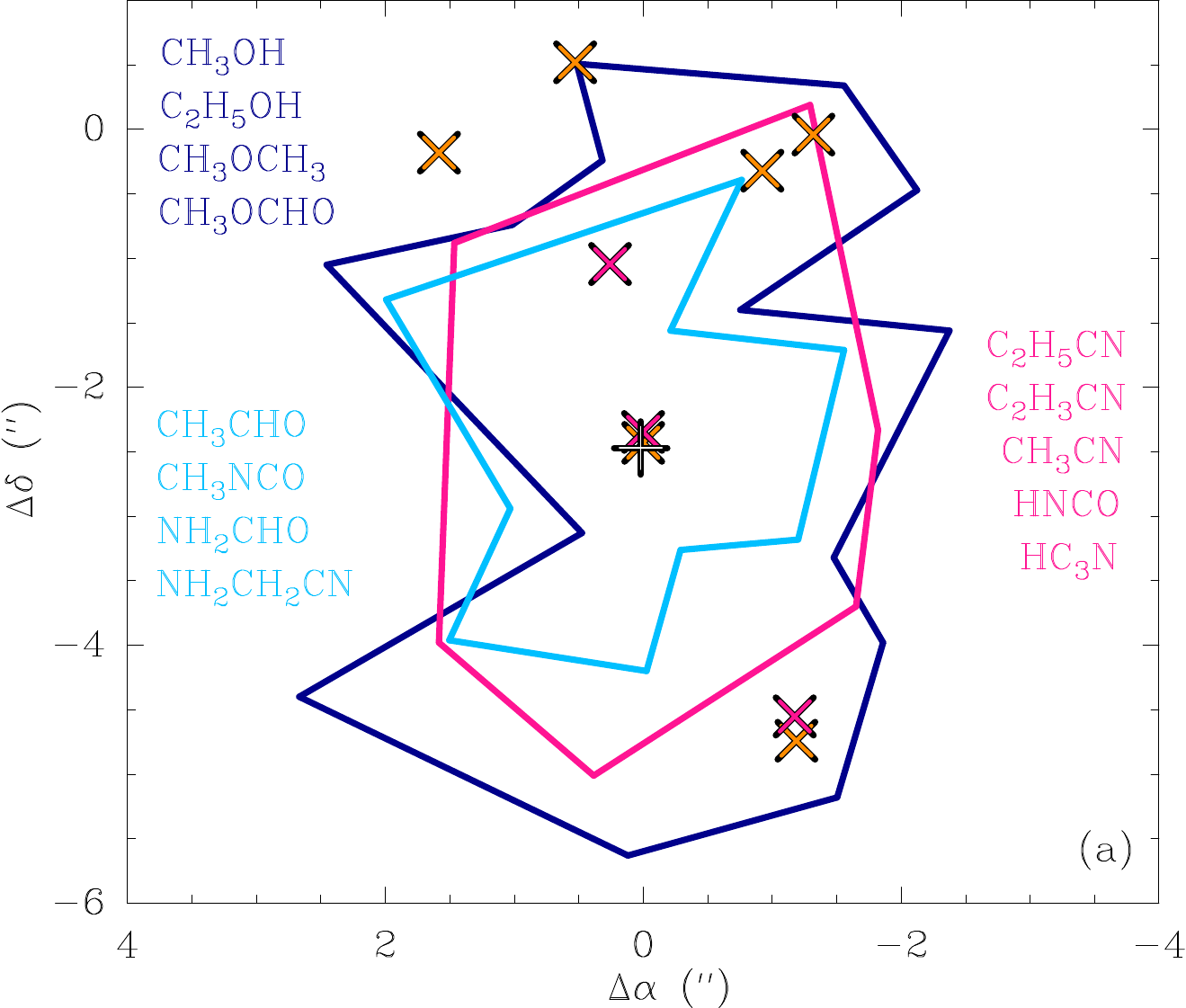}\includegraphics[width=0.56\textwidth,angle=0]{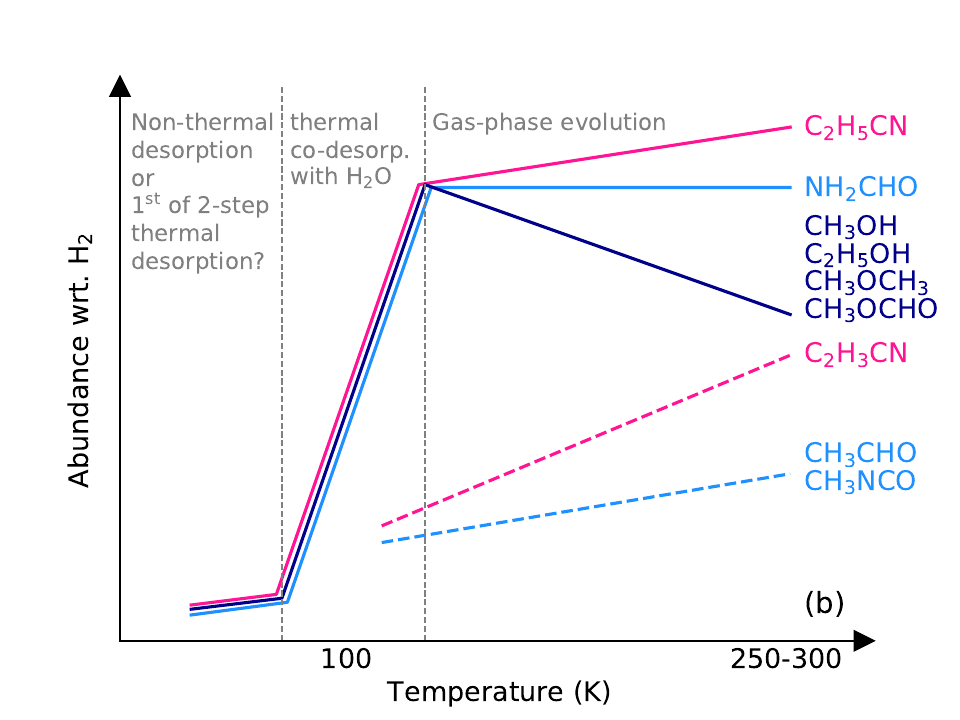}
    \caption{Overview sketch of the different emission morphologies (a) and abundance profiles (b) of a dozen of complex and more simple organic molecules in Sgr~B2(N1) (white cross), the main hot core of the protocluster Sgr~B2(N), as derived from the ReMoCA survey performed with ALMA. In panel b, the temperature axis points toward the central object. The abundance profiles reveal two regimes of COM desorption. Figure extracted from \citet{Busch22}.}
    \label{f:sgrb2n1}
\end{figure}

Thanks to the high angular resolution of the ReMoCA survey performed with ALMA toward the protocluster Sgr~B2(N) \citep[][]{Belloche19}, the spatial distribution of COMs in its main hot core, Sgr~B2(N1), could be probed in detail. \citet{Busch22} studied the morphology of the emission of a dozen of complex and more simple organic molecules (Fig.~\ref{f:sgrb2n1}a) and derived their abundance and temperature profiles as a function of distance to the central object. The COM abundance profiles as a function of temperature (Fig.~\ref{f:sgrb2n1}b) show a sharp increase at a temperature of about 100~K, revealing the transition between two regimes of COM desorption, spatially resolved for the first time in a hot core: thermal desorption at a temperature of about 100~K, interpreted as a co-desorption of COMs with water rather than desorption depending on their binding energies, and another regime of desorption below 100~K, revealed by lower but non-zero abundances at lower temperatures. Two interpretations were proposed by \citeauthor{Busch22} for the latter regime: either non-thermal desorption (e.g., chemical desorption or desorption induced by shocks or secondary UV photons produced by cosmic rays, or sputtering by cosmic rays) or partial thermal desorption related to the outer, CO-rich (and thus less polar) ice layers of dust grains. In contrast, using nine of the most line-rich hot cores of the ALMAGAL large program performed with ALMA (PI: S. Molinari), \citet{Nazari22} found a trend for the rotational temperature of the investigated molecules (five COMs and HNCO) to increase with their binding energy, potentially suggesting that the binding energy does have an impact on the thermal desorption behaviour of these molecules. Measuring spatially-resolved abundance profiles of these molecules in these sources is needed to investigate if this is in tension with the Sgr~B2(N1) result.

Like in Sgr~B2(N1), an abundance increase at 80--100~K was also reported by \citet{Mottram20} for CH$_3$OH and CH$_3$CN around the hot core W3 IRS4 in the frame of the NOEMA large program CORE \citep[][]{Beuther18}, but the evidence is less clear because the abundance profiles as a function of temperature were constructed using a few specific positions (continuum peaks and outflow) within the region rather than a continuous cut across the hot core envelope.

The spatial distribution of COM emission was mapped with the Submillimeter Array (SMA), ALMA, or NOEMA at high angular resolution in several other high-mass star forming regions. In G10.6--0.4, \citet{Law21} found that methyl formate and dimethyl ether correlate well over a wide range of temperatures across the region, confirming a correlation found in a large sample of sources \citep[e.g.,][]{Coletta20} but this time resolving the emission. However, \citet{Law21} noted that the correlation brakes down in the hottest ($>$275~K) region of the main hot core where the abundance ratio of methyl formate to dimethyl ether is much higher, suggesting an additional, high-temperature formation route of methyl formate in the gas phase.
Strong correlations were also reported for CH$_3$CN and CH$_3$OH,
CH$_3$CN and C$_2$H$_5$CN, CH$_3$OH and CH$_3$OCH$_3$, as well as HNCO and NH$_2$CHO. However, CH$_3$CHO poorly correlates with other COMs, possibly pointing to multiple formation pathways across different temperature regimes. \citeauthor{Law21} presented resolved profiles of COM column densities as a function of temperature. A natural next step would be to check if the COM abundances relative to H$_2$ in G10.6--0.4 show the same desorption pattern as in Sgr~B2(N). 

The COM chemistry of G31.41+0.31 was investigated in the frame of the GUAPOS project with ALMA \citep[][]{Mininni20,Colzi21}. In particular, the relative abundances of organic molecules bearing a NCO bond (HNCO, NH$_2$CHO, CH$_3$NCO, CH$_3$NHCHO, CH$_3$CONH$_2$) were found to be similar to those of Sgr~B2(N2), except for formamide, an exception that would be interesting to understand. Three of these molecules (HNCO, NH$_2$CHO, CH$_3$NCO) in G31.41+0.31 have remarkably similar emission morphologies at a resolution of 0.2'' (750~au), while significant morphological differences exist between HNCO and the two amides NH$_2$CHO and CH$_3$C(O)NH$_2$ in other hot cores such as the ones in NGC~6334I \citep[][]{Ligterink20}.

The dominant hot core of NGC~6334I was resolved with ALMA into seven hot components within a radius of 1000~au, with at least four showing evidence for internal heating \citep[][]{Brogan16}. A detailed investigation of the COM composition of this cluster of YSOs in various states of formation and activity would be extremely interesting, in particular given that one of them experienced a strong luminosity outburst in 2015 \citep[][]{Hunter17}, which certainly had an impact on the chemistry of the region. NGC6334I is also the region where a bimodal abundance pattern of CH$_2$(OH)CHO with respect to its isomers CH$_3$OCHO and CH$_3$COOH was discovered by \citet{ElAbd19}, a pattern that the authors also found in a sample of star forming regions collected from the literature. The fact that the two modes are found in NGC~6334I suggests that the bimodality does not originate in differences in the large-scale environment of the sources but rather in their stage of evolution.

The COM chemical composition of the high-mass protostar AFGL 4176 relative to methanol was found by \citet{Bogelund19} to resemble more closely that of the Class 0 protostar IRAS~16293--2422 than that of Galactic center sources, and this despite their difference in luminosity. This led the authors to conclude that AFGL 4176 is a young source and that the COM composition is already set in the ice during the cold prestellar phase. In the same source, \citet{Johnston20} classified the emission of 25 species in four morphological types with the O-bearing and N-bearing molecules belonging predominantly to different types, the former potentially tracing shocks at the disk's centrifugal barrier and the latter a region closer to the protostar (the disk?), like in G328.2551–0.5321 \citep[][]{Csengeri18}.

\subsubsection{COM properties in large samples of hot cores}

Complementing the detailed studies of individual objects, surveys of large samples of hot cores provide further insight into the complex organic chemistry of high-mass star forming regions. The ALMA Three-millimeter Observations of Massive Star-forming regions survey (ATOMS) targeted 146 active high-mass star forming regions at moderately high angular resolution \citep[][]{Liu20}. Out of the 453 compact dense cores identified in continuum emission, 91 were found to enclose hyper- or ultracompact HII (H/UC-HII) regions, 82 showed COM emission (infered from a high spectral line density) but no sign of H/UC-HII region, and 53 belonged to both categories \citep[][]{LiuHL21,LiuHL22}. The authors concluded from these results that the lifetime of hot cores is at least comparable to the lifetime of H/UC-HII regions. In addition, the overlap between both categories suggests that hot cores persist during more than half of the lifetime of H/UC-HII regions even though the central objects have started to ionize their surroundings. However, higher angular resolution observations will be needed to check if the (residual) COM emission surrounds the ionized region or if multiple sources in different evolutionary stages are present in the ATOMS beam. 

A more detailed investigation of the COM emission in the ATOMS data set identified 60 hot cores traced by compact ($< 0.1$~pc) and hot ($> 100$~K) emission of C$_2$H$_5$CN, CH$_3$OCHO, and CH$_3$OH, 75\% out of which are new sources \citep[][]{Qin22}. The authors found in about half of the sample spatial offsets between the emission of C$_2$H$_5$CN and CH$_3$OCHO. Such a differentiation between N-bearing and O-bearing species was seen in individual objects in the past \citep[see discussion in][]{Jorgensen20}. The ATOMS results establish it as a widespread phenomenon, the origin of which still needs to be understood.

\begin{figure}[!t]
    \centering
    \includegraphics[width=\textwidth,angle=0]{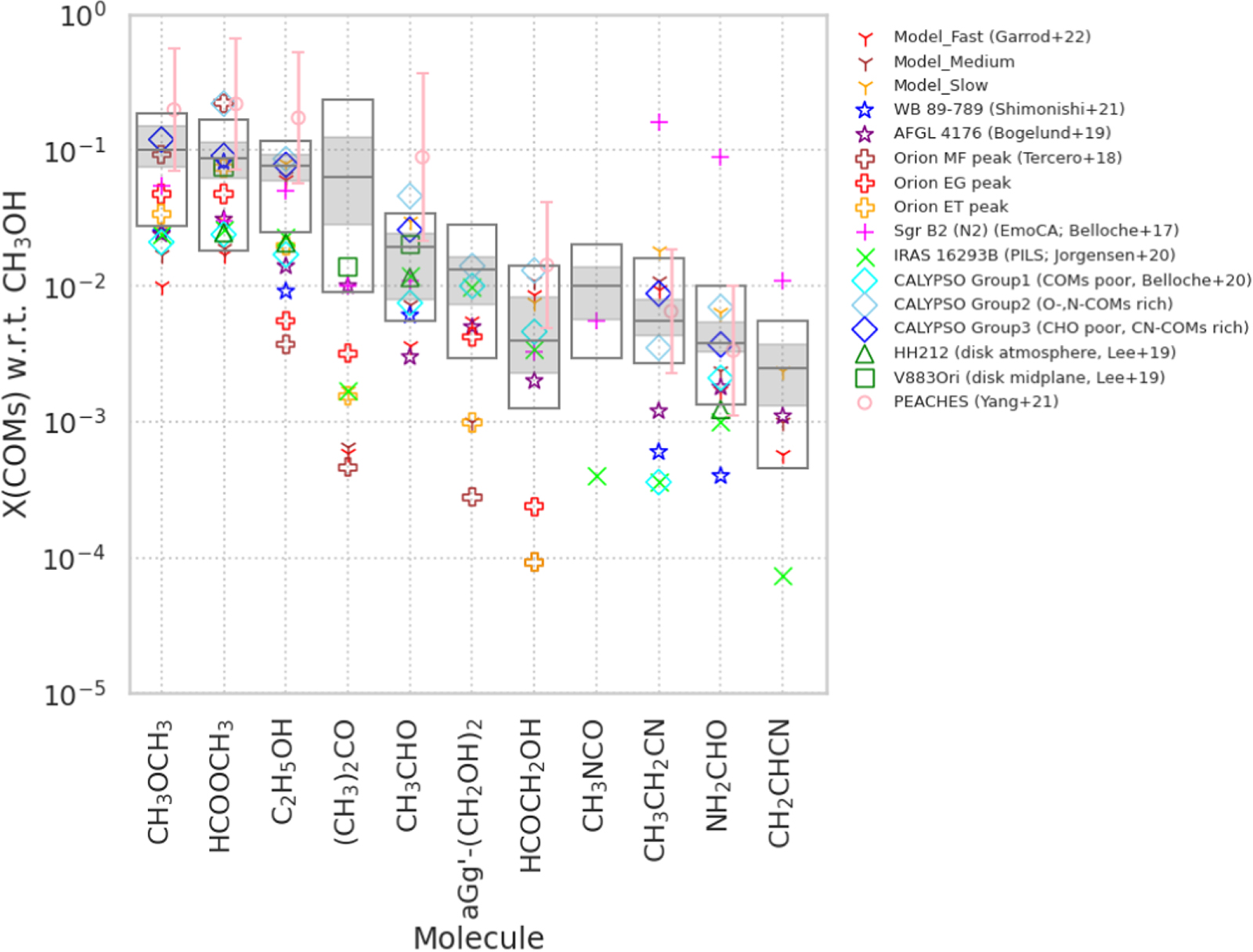}
    \vspace*{-4ex}
    \caption{Abundances of COMs relative to methanol for a large sample of hot cores and hot corinos. Predictions of chemical models of \citet{Garrod22} are shown as well. The references for HH212 and V883 Ori are \citet{Lee19} and \citet{LeeJ19}, respectively. One reference, \citet{Belloche17}, is not cited elsewhere in this chapter. Figure extracted from \citet{Baek22}.
    }
    \label{f:baek22_fig10}
\end{figure}

Another ALMA survey of 12 high-mass star forming regions harboring Class I or II methanol maser emission revealed 28 hot cores out of 68 detected continuum sources \citep[][]{Baek22}. The cores associated with 6.7~GHz Class II methanol maser emission showed on average a higher number of detected COMs, making this maser a good tracer of COM-rich hot cores \citep[see also][ for the case of Sgr~B2(N)]{Bonfand17}. This survey revealed a flattening of the methanol column densities above a peak H$_2$ column density $N_{{\rm H}_2}$ of the continuum source of $10^{24.7}$~cm$^{-2}$, likely due to the dust becoming optically thick, thus preventing the detection of COM emission in the innermost parts (see Sect.~\ref{ss:hotcorinos} for similar optical depth issues in the low-mass case). The COM abundances relative to methanol do not show a dependence on $N_{{\rm H}_2}$ for CH$_3$OCH$_3$, CH$_3$OCHO, CH$_3$CHO, and C$_2$H$_5$CN in this hot core sample while they tend to increase with $N_{{\rm H}_2}$ for eight other O-bearing and/or N-bearing COMs. This might indicate that molecules in the latter group thermally desorb from the grain mantles at higher temperatures or have additional gas-phase formation routes after thermal desorption. Overall, the COM abundances derived by \citeauthor{Baek22} relative to methanol show a good agreement with the predictions of the non-diffusive chemical models of \citet{Garrod22}, except for CH$_3$OCH$_3$ and CH$_3$C(O)CH$_3$ which the models underestimate (Fig.~\ref{f:baek22_fig10}). Acetone is also one of the COMs with the largest dispersion of its abundance relative to methanol in the sample of sources shown in Fig.~\ref{f:baek22_fig10} (more than two orders of magnitude). Acetone thus seems to be particularly sensitive to the local physical conditions.

The methanol emission of an unprecedently large sample of 184 low-mass and high-mass protostars, most of them observed with ALMA, was investigated by \citet{vanGelder22a}. The study revealed a correlation between the mass of methanol normalized to the continuum mass of the inner envelope (plus disk) and the source luminosity. This correlation matches the simple expectation that the radius of the methanol snow line correlates with the luminosity. However, the dispersion of the data points around the correlation is large (several orders of magnitude). The authors proposed that it could be due in part to the presence of disks affecting the structure of the region traced by methanol in some sources and to the high optical depth of the continuum emission that can hide the COM emission in some others.

\subsubsection{COMs in hot cores as tracers of physical properties}

In addition to chemistry, COMs can be used as tracers of physical properties. A nice example can be found in the work of \citet{Beltran18} toward the hot molecular core G31.41+0.31. Taking advantage of many vibrationally excited lines of CH$_3$CN and CH$_3$OCHO covering a broad range of energies, \citet{Beltran18} found accelerating infall motions and rotational spin-up toward the center of the core down to spatial scales of $\sim$1700 au, consistent with gravitational collapse. This core is also fragmenting into at least four massive sources as shown by even higher-angular resolution ALMA observations at spatial scales of $\sim$340 au \citep{beltran22}. This ALMA data set at 1.4~mm also reveals the difficulty to peer into the inner regions of hot cores because of high dust opacity at this wavelength.

\citet{Gieser21} also used methyl cyanide (and formaldehyde) as thermometers to trace the temperature profiles of 22 continuum cores in 18 high-mass star forming regions observed in the frame of the NOEMA large program CORE. They derived an average power-law index of $-0.4\pm0.1$, consistent with internal heating through an optically thin envelope.

The deuteration level of methanol in a large sample of 99 high-mass protostars observed with ALMA was studied by \citet{vanGelder22b} who found that it is on average more than one order of magnitude lower than in low-mass protostars. This low deuteration level was interpreted as resulting either from a higher temperature during the prestellar phase of high-mass protostars or from a shorter duration of their prestellar phase.

\subsubsection{Hot cores in the outer Galaxy}
Finally, a hot core was recently detected with ALMA in the source WB89–789 that is located in the extreme outer Galaxy at a galactocentric radius of 19~kpc, where the metallicity is a factor $\sim$4 lower than in the inner Galaxy \citep[][]{Shimonishi21}. Nine COMs were detected with abundances relative to H$_2$ that globally appear to scale with the metallicity of the source. The COM abundances relative to methanol are remarkably similar within a factor of two to those measured by \citet{Fuente14} in the intermediate-mass protostar NGC~7192~FIRS2 which is located only slightly beyond the solar circle. Except for an overall scaling, it thus seems that metallicity has a minor effect on the COM chemical production and composition of star forming regions. However, CH$_3$CN and C$_2$H$_5$OH deviate from this correlation by a factor of $\sim$4, which calls for further investigations of a broader sample of sources in order to probe the effect of metallicity on COM chemistry on a statistically more robust basis. Surprisingly, the molecular composition of WB89–789 was found by \citeauthor{Shimonishi21} to differ significantly from that of hot cores recently identified in the Large Magellanic Cloud, which also has a low metallicity (see Sect.~\ref{ss:extragalhotcores}). The higher level of UV irradiation in the Large Magellanic Cloud, related to its prominent high-mass star formation activity, may play a key role in this chemical difference with the outer Galaxy.

\subsection{Extragalactic hot cores}
\label{ss:extragalhotcores}

The star formation rate of the Large and Small Magellanic Clouds (LMC and SMC) is $\sim$15 and $\sim$50 times lower, respectively, than in the Milky Way \citep[0.07--0.16~M$_\odot$~yr$^{-1}$ and 0.02--0.05~M$_\odot$~yr$^{-1}$, respectively, vs. $\sim$1.7~M$_\odot$~yr$^{-1}$;][]{For18,Licquia15}. The metallicity of the LMC and SMC are also lower by factors $\sim$2 and $\sim$4, respectively, compared to the Solar neighborhood \citep[][]{Massey03}. The presence of one of the simplest COMs, methanol, in the LMC has been known since the detection of Class II methanol masers \citep[][]{Sinclair92}, which are exclusively associated with regions forming high-mass stars \citep[][]{Menten91,Minier03}. However, subsequent surveys in the LMC revealed an occurrence rate of this type of masers 45 times lower than in the Galaxy \citep[][]{Green08}. The difference in star formation rate would account for a factor $\sim$15 only. This may suggest that methanol is underabundant in the LMC compared to our Galaxy, which may be related to their difference in metallicity.

Thermal emission of methanol and methyl acetylene, CH$_3$CCH, was detected in the LMC with single-dish telescopes toward the prominent star-forming region N159W \citep[][]{Heikila99} and, in the case of methanol, toward N113 as well \citep[][]{Wang09}. There were in both cases too few transitions to derive rotational temperatures. Methanol was also tentatively detected in the ice mantles of dust grains in the infrared toward a few high-mass YSOs of the LMC, with abundances relative to water lower than in our Galaxy, which may be due to higher temperatures that reduce the efficiency of CO hydrogenation \citep[][]{Shimonishi16a,Acharyya15}. 

The advent of ALMA in the 2010s enabled the detection of hot cores in the Magellanic clouds. \citet{Shimonishi16b} reported compact molecular emission around the high-mass YSO ST11 in the LMC with a SO$_2$ rotational temperature higher than 100~K but they concluded from their nondetection of methanol that the latter is at least 2--3 orders of magnitude less abundant than in Galactic hot cores. In contrast, two hot cores with clear emission of three COMs (methanol, dimethyl ether, and methyl formate) were identified in N113 in the LMC with methanol rotational temperatures of $\sim$130~K, methanol abundances more than one order of magnitude higher than in ST11, and abundances of dimethyl ether and methyl formate with respect to methanol of 10--20\% and $\sim$5\%, respectively \citep[][]{Sewilo18}. The authors argued that, after accounting for the difference in metallicity, the COM abundances in the LMC hot cores are similar to those found at the lower end of the abundance distribution of Galactic hot cores. However, the number of sources is still too limited to draw firm conclusions. Compact emission of methanol and methyl cyanide was also detected toward a third hot core in the LMC, ST16, with methanol tracing two temperature components (140~K and 60~K) and methyl cyanide only the lukewarm one \citep[][]{Shimonishi20}. Two additional hot cores, N105-2~A and N105-2~B, were detected via their CH$_3$OH, CH$_3$OCH$_3$, and CH$_3$CN emission with ALMA in the star forming region N105 \citep[][]{Sewilo22}. The methanol abundances in ST16, N105-2~A, and N105-2~B lie between that of the ST11 and N113 hot cores, which suggests that methanol is overall underabundant in LMC hot cores compared to Galactic ones, even after scaling for metallicity. Simulations suggest indeed that the abundance of COMs in low-metallicity environments do not simply scale with metallicity, the temperature during the prestellar phase also plays a role \citep[][; see Sect.~\ref{ss:hotcores}]{Acharyya18}.
Spatial inhomogeneities in the LMC metallicity due to tidal interactions with the lower-metallicity SMC may also explain part of the dispersion of the COM abundances in the LMC hot cores detected so far.

Methanol was detected at even lower metallicity in the vicinity of a high-mass YSO in the SMC, with a low rotational temperature of $\sim$12~K and spatially not associated with any infrared source \citep[][]{Shimonishi18}. It seems to trace cold and dense material. Given the concomitant detection of SO$_2$ and SiO, shock chemistry may be the origin of this methanol emission.

\section{Galactic environments}
\label{s:galenv}

\subsection{Photodissociation regions}
\label{ss:pdrs}

Photodissociation Regions (PDRs) are the mostly neutral outer layers of molecular clouds exposed to an external far-ultraviolet (FUV) radiation field \citep[e.g.,][, and references therein]{Hollenbach91}. The radiation field is in the 6--13.6 eV range, and the incident FUV flux can be measured in units of equivalent average local interstellar flux \citep[$G_0$ = 1.6 $\times$ 10$^{3}$ erg s$^{-1}$ cm$^{-2}$;][]{Habing68}.

\begin{figure*}[!t]
    \centering
    \includegraphics[width=10cm,angle=-90]{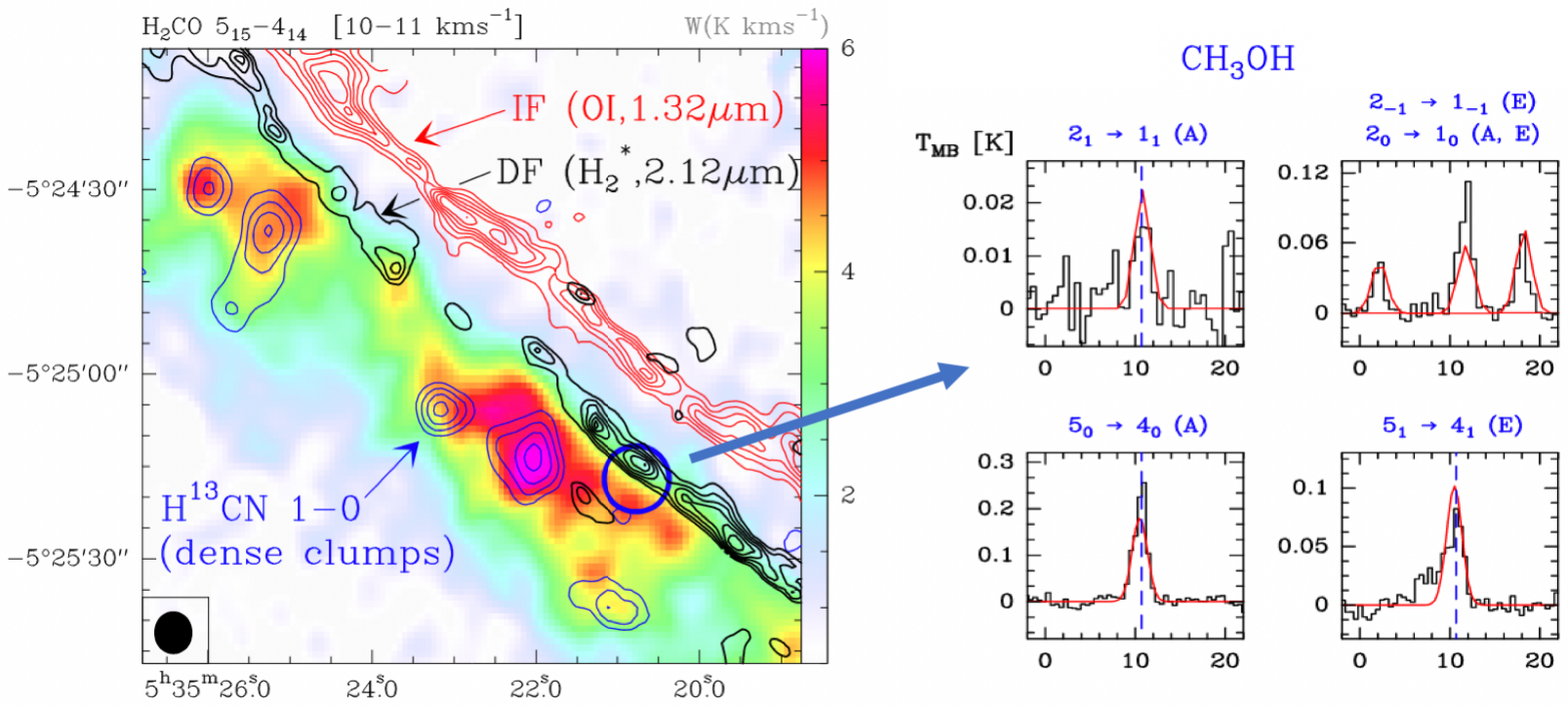}
    \vspace{-2.5cm}
    \caption{Adapted from \citet{Cuadrado17}.
    \textit{Left panel}: Stratification of the PDR in the
    Orion bar. The colour scale and black contours are  for the H$_2$CO and H$_2$ emissions, respectively, as mapped with the IRAM-30m. 
    The blue ring indicates where CH$_3$OH lines have been detected. \textit{Right panels}: Examples of methanol spectra.}
    \label{fig:cuadrado-methanol}
\end{figure*}

A typical region with a FUV flux higher than 10$^{4}$ $G_0$ is the edge of the Orion bar \citep[e.g.,] [, and references therein]{Tielens93,Go16}. Figure \ref{fig:cuadrado-methanol} shows the atomic and molecular stratification as mapped using the IRAM 30-m telescope by \citet{Cuadrado17}. The authors detected CH$_3$CN in the gas phase with a temperature of $\sim$100~K, while narrow ($\sim$2~km~s$^{-1}$) CH$_3$OH emission is clearly detected where the gas is more shielded and the temperature decreases to values lower than 50 K \citep{Cuadrado17}. The abundances of both species relative to H$_2$ fall in the range 10$^{-10}$--10$^{-9}$. On the other hand, a classical example of PDR illuminated by a relatively weak-FUV radiation field ($\sim$ 60 $G_0$) is the Horsehead nebula, also in Orion \citep[e.g.,] [, and references therein]{Go09}. \citet{Guzman13} detected methanol in the region with a temperature of about 20-30 K using the IRAM 30-m antenna, deriving abundances with respect to H$_2$ of about 10$^{-10}$. These abundances have been proposed to be produced by UV photodesorption from ices. However, while VUV photodesorption has been found to be an efficient mechanism for CH$_3$CN \citep{Basalgete21}, this is not the case for methanol, which has been found to be photo-dissociated almost entirely upon VUV illumination \citep{Cruz-Diaz16,Bertin16}.
As a follow-up, \citet{Guzman14} reported also CH$_3$CHO detections toward the Horsehead PDR with, in this case, relative abundances of $\sim$ 10$^{-11}$.

\begin{figure*}[!t]
    \centering
    \hspace*{-1.5ex}\includegraphics[width=0.46\paperwidth,angle=-90]{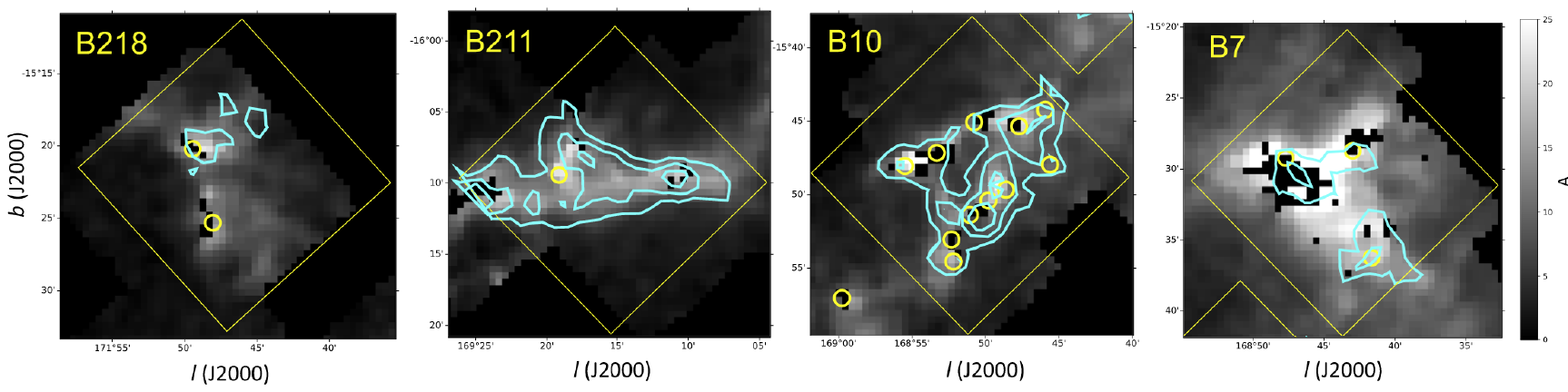}
    \vspace{-3.5cm}
    \caption{CH$_3$OH emission at 3~mm (cyan contours) obtained in different regions of the Taurus molecular cloud with the 12-m ARO single-dish telescope, overlaid on extinction maps \citep[][]{Scibelli20}.}
    \label{fig:methanol}
\end{figure*}

Moving toward low-mass star formation, \citet{Bouvier20} observed methanol emission at 1~mm and 3~mm with the IRAM 30-m antenna toward the closest analogue to the Sun’s birth environment, the Orion Molecular Cloud 2/3 (OMC-2/3) filament. These authors observed methanol emission over a few thousands au scale, not originating from the protostellar hot corinos but, instead, from the molecular cloud, suggesting that this emission comes from its PDR illuminated by the OB stars of the region. Given that methanol may be photo-dissociated when UV photodesorbed \citep{Cruz-Diaz16,Bertin16}, it could alternatively be transferred from the ices to the gas phase by chemical reactive desorption in regions with moderate H$_2$ volume densities and visual extinctions \citep[][; see Fig.~\ref{fig:methanol}]{jimenez16,vasyunin17,Scibelli20}. Finally, sputtering by cosmic rays could also be an efficient mechanism for the production of COMs (including CH$_3$OH) in cold sources \citep[][; see also Section$\,$\ref{ss:prestellar}]{dartois19,wakelam21}. To conclude, all these results highlight the importance of deep searches for COMs in extended PDR using single-dish telescopes. 

\subsection{Diffuse and translucent clouds}
\label{ss:translucent}

Translucent molecular clouds have a visual extinction, $A_{\rm v}$, between 1~mag and 5~mag \cite[][]{Snow06}. They are less shielded against the interstellar radiation field than dense clouds and yet, several COMs have been detected in such low-density environments ($n_{\mathrm{H}_2}$ $\sim$ 250--2500~cm$^{-3}$), in absorption against more distant, bright continuum sources. CH$_3$CN, CH$_3$OH, CH$_3$CHO, and NH$_2$CHO were identified with the GBT and ALMA in the Galactic center region and the former two COMs in the Scutum arm of the Galactic disk as well \citep[][]{Corby16,Thiel17}. While these COMs were initially reported as detected in diffuse clouds by \citet{Thiel17}, \citet{Liszt18} argued that these clouds cannot be diffuse and \citet{Thiel19} subsequently established their translucent nature. The COM relative abundances were found by \citet{Thiel17} to be similar to the abundances of COMs in a $z=0.89$ spiral galaxy that were detected in absorption against the quasar PKS 1830--211 (see Sect.~\ref{ss:highz}). This suggests that the processes that lead to molecular complexity have remained the same since the time when the Universe was half of its age. As recently proposed by \citet{garcia-sanchez22}, the emergence of COMs in the ISM may be a natural consequence of the complex system that is interstellar chemistry.

COMs were also searched for in diffuse molecular clouds \citep[$A_{\rm v} < 1$~mag,][]{Snow06}. Upper limits obtained with the Plateau de Bure interferometer were reported for CH$_3$CN, HC$_5$N, and CH$_3$OH \citep[][]{Liszt01,Liszt08}. They imply that the former two COMs are underabundant with respect to HCO$^+$ by at least a factor of a few compared to TMC-1 while the upper limits are not much constraining for methanol. More recently, \citet{Liszt18} showed that CH$_3$CN is surprizingly ubiquitous at $A_{\rm v} = 1$~mag, with an abundance ratio relative to HCN of $\sim$0.015, roughly a factor of two smaller than in TMC-1. The upper limits reported for methanol imply a ratio CH$_3$OH/CH$_3$CN $<$ 3--7 in the observed diffuse clouds while this ratio is 2--3.5 in TMC-1 \citep[][]{Ohishi92,Gratier16}. Therefore whether or not the chemistry of diffuse clouds leads to different ratios of O-bearing to N-bearing COMs compared to TMC-1 is still an open question. 

\subsection{Galactic Center molecular clouds}
\label{ss:galcent}
The Galactic Center region comprises about the inner 500 pc of the Milky Way and hosts slightly less than 5\% of the total molecular gas of the Galaxy \citep[about 3$\times$10$^7$$\,$M$_\odot$;][]{dahmen98}. This gas is distributed in a population of several dozen Giant Molecular Clouds (GMCs) forming the so called Central Molecular Zone (CMZ) of about 200$\,$pc \citep[][]{morris96}. Unlike in the Galactic disk, the molecular gas in the CMZ is hot \citep[gas kinetic temperatures ranging between 60 and $>$100~K;][]{morris83,huettemeister93,Ginsburg16}, dense \citep[average densities of a few 10$^4$$\,$cm$^{-3}$;][]{guesten83,morris96}, and turbulent \citep[linewidths of the molecular emission ranging from 15 to 50$\,$km$\,$s$^{-1}$;][]{bally87}, while the temperature of the dust is rather low \citep[with $T_{\rm dust}<20$~K;][]{rodriguez04}. The dense gas distribution is asymmetrical, with three times more mass at positive longitudes than at negative ones \citep[][]{launhardt02}. The origin of this asymmetry is unknown, although one possibility is that this asymmetry is produced by an acoustic instability induced in the Galactic bar potential \citep[][]{krumholz15,Mills17}. 

The Galactic Center region is an extremely harsh environment for star formation and interstellar chemistry since the molecular gas is under the effects of low-velocity shocks, strong UV radiation, X-rays, and cosmic rays \citep[e.g.,][]{martin-pintado97,koyama09}. These energetic phenomena affect the chemical composition of the gas, which shows different molecular tracers depending on the dense cloud location in the CMZ. Molecules such as SiO and HNCO are widespread throughout the CMZ, indicating that shock chemistry is prevalent throughout the Galactic Center region \citep[][]{martin-pintado97,jones12}. Certain locations are however depleted in SiO and HNCO but rich in radicals such as HCO and molecular ions such as HOC$^+$ or CO$^+$, which indicates that UV/X-ray photo-chemistry is at work \citep[][]{martin08,armijos20}. 

Despite these extreme conditions, the GMCs in the Galactic Center region show an active COM chemistry as revealed by the detection of molecules such as CH$_3$OH, C$_2$H$_5$OH, CH$_3$CHO, CH$_3$CN, CH$_2$NH, CH$_3$OCHO, CH$_3$OCH$_3$ or CH$_3$COOH 
%toward multiple GMCs in the CMZ 
with single-dish telescopes \citep[][]{martin-pintado01,requena06,requena08,zeng18}. The abundance ratios of these COMs with respect to CH$_3$OH are rather uniform for Galactic Center molecular clouds and Galactic hot cores, which suggests a similar average chemical composition across the Galaxy \citep[][]{requena06}. These authors established that frequent ($\sim$10$^5\,$years) shocks with velocities $>$6$\,$km$\,$s$^{-1}$ are required to explain the high abundances of COMs in Galactic Center GMCs. The emission of some of these COMs is indeed extended toward some of these clouds as, e.g., the Sgr~B2 molecular cloud. Examples of COMs showing extended emission include CH$_3$OH, CH$_3$CN, CH$_3$CCH, CH$_3$CHO, or C$_2$H$_5$OH \citep[][]{martin-pintado01,chengalur03,jones08,jones12}. There seems to be consensus about the idea of COMs detected in the Galactic Center region to be formed largely on dust grains and injected into the gas phase by large-scale shocks \citep[][]{martin-pintado01,requena06,Li20}.

Among the dense GMCs found in the Galactic Center region, the Sgr~B2 molecular cloud complex stands out as the most massive (it contains 10\% of the total dense gas in the CMZ) and active star-forming cloud. This cloud hosts the Sgr~B2(N) high-mass star-forming protocluster which contains several hot cores and has been an excellent target for the discovery of new COMs in the ISM (see Sects.$\,$\ref{ss:hotcores} and \ref{ss:comp-sgrb2}). In contrast to other high-mass star-forming regions in the Galactic disk, the high level of chemical complexity of the gas in the Sgr~B2 molecular cloud is not restricted to the Sgr~B2(N) hot cores but it is also found at larger scales. For instance, this is demonstrated by the extended emission of methylamine CH$_3$NH$_2$, glycolaldehyde CH$_2$(OH)CHO, and ethylene glycol (CH$_2$OH)$_2$ mapped toward this cloud \citep[][]{Li17,Li20}. An interesting location in the Sgr~B2 cloud complex is the peak of the HNCO and CH$_3$OH emission observed toward the northeastern region of Sgr~B2(N), the G+0.693--0.027 cloud. This cloud is quiescent since it does not present any signs of high-mass star formation in the form of UCHII regions, water masers, or IR/(sub-)millimeter sources \citep[][]{zeng20}. However, it is a rich source in molecular emission including COMs, where new molecular species of all families (C-, N-, S-, O-, and P-bearing) have recently been reported. This has been possible thanks to a high-sensitivity spectroscopic survey carried out with the IRAM 30\,m and Yebes 40\,m telescopes in the 7\,mm, 3\,mm, 2\,mm, and 1\,mm atmospheric windows. The survey has yielded the detection of molecules of prebiotic interest such as hydroxylamine \citep[NH$_2$OH;][]{rivilla20}, monothioformic acid \citep[HC(O)SH;][]{rodriguez21}, ethanolamine \citep[NH$_2$CH$_2$CH$_2$OH;][]{rivilla21}, vinyl amine and ethyl amine \citep[C$_2$H$_3$NH$_2$ and C$_2$H$_5$NH$_2$;][]{zeng21}, or n-propanol \citep[n-C$_3$H$_7$OH;][]{jimenez22b}. All these species have been proposed as important prebiotic precursors in theories of the origin of life \citep[][]{ruiz-mirazo14,becker19,jimenez20}. A comparison of the COM chemical composition of G+0.693--0.027 and a hot core in Sgr~B2(N) will be presented and discussed in Sect.~\ref{s:comparison}.

\subsection{Peculiar environments}
\label{ss:peculiar}

CK Vul is the remnant of an eruption that manifested itself in the form of three outbursts observed by European astronomers in 1670--1672 \citep[][]{Shara85}. The nature of this explosion, maybe a stellar merger, is still unclear \cite[][]{Banerjee20,Kaminski21}. The remnant consists of a large, expanding, bipolar nebula of recombining gas with a dynamical time consistent with the time elapsed since the eruption, and a more compact, also bipolar and expanding region that is prominent in rotational emission of molecules. Single-dish (IRAM 30~m, APEX) and interferometric (ALMA, SMA, VLA) observations revealed the presence of COMs in the inner parts of the remnant: CH$_3$OH, CH$_3$NH$_2$, NH$_2$CHO, and CH$_3$CN \citep[][]{Kaminski17,Kaminski20,Eyres18}. The spatial distribution of their emission differs: like NH$_3$ and CH$_2$NH, the former two COMs show an extended, bipolar morphology with two bright spots at the tips, while the emission of CH$_3$CN and NH$_2$CHO is dominated by a compact central structure that may be a disk, also traced by the submm dust emission and by C$^{17}$O, SO, SO$_2$, AlF, PN, HC$_3$N, HNCO, HCO$^+$, and N$_2$H$^+$ \citep[see Fig. 2 of][ and Fig. 4 of \citeauthor{Eyres18} \citeyear{Eyres18}]{Kaminski20}. The emission of SiO, SiS, HCN, HNC, H$_2$CO, and CS shows attributes of both morphologies. 

The presence of SiO suggests that the molecular component of the remnant originates in sputtering of dust grains through shocks \citep[][]{Schilke97,Caselli97,JimenezSerra08}. This may in turn imply that the observed COMs formed in the icy grain mantles and then desorbed under the impact of such shocks. The low excitation and kinetic temperatures that characterize the molecular emission \citep[10--20~K,][]{Kaminski17,Kaminski20} would then suggest that efficient cooling occurred after desorption. The H$_2$ densities estimated by \citet{Kaminski20} for the region traced by the molecular emission are on the order of 10$^4$--10$^6$~cm$^{-3}$, which implies freeze-out timescales on the order of $5 \times 10^5$--$5 \times 10^3$~yr at a temperature of 20~K \citep[][]{Charnley01}, much longer than the age of the remnant. Besides, \citet{Kaminski20} estimated that the amount of dust shielding is sufficient to protect the molecules from destruction by the interstellar radiation field. However, such a shock desorption scenario would still need to explain when the COMs formed in the grain mantles. Formation before the eruption in 1670 seems excluded by the low $^{12}$C/$^{13}$C isotopic ratio of methanol that points to material heavily processed by nuclear burning \citep[][]{Kaminski20}. An alternative scenario would be the formation of COMs in the gas phase through endothermic reactions at high temperatures triggered by shocks \citep[][]{Kaminski17}. Chemical modelling is needed to shed light on the formation of COMs in such a dynamic environment.

\section{Extragalactic environments}
\label{s:extragal}

\subsection{COM inventory and distribution in nearby galaxies}
\label{ss:nearbygal}

The first extragalactic detections of CH$_3$OH (in NGC253 and IC342), CH$_3$CN (in NGC253), and CH$_3$CCH (in NGC253 and M82) were reported with the IRAM 30 m telescope more than three decades ago \citep[][]{Henkel87,Mauersberger91}. The same three COMs (and only those) were detected in subsequent single-dish spectral line surveys of the nuclear region of both starburst galaxies NGC253 and M82, which made the derivation of their rotational temperatures possible, showing that CH$_3$OH and CH$_3$CN trace somewhat lower temperatures in the former source compared to the latter \citep[$\sim$10~K versus $\sim$30~K, ][]{Martin06,Aladro11}. A spectral line survey of eight nearby active galaxies (starburst and/or harboring an active galactic nucleus -- AGN) led to the additional tentative detections of CH$_3$CHO and HC$_5$N in some of the sources \citep[][]{Aladro15}. CH$_3$CCH was detected in the starburst sources but not in the AGNs, and the authors concluded that it may trace PDRs. The abundance of CH$_3$CN was found to be lowest in M82, maybe because this galaxy is heavily pervaded by UV radiation. Rotational emission from within vibrationally excited states of CH$_3$CN and HC$_3$N was detected in the ultra-luminous infrared galaxy Arp~220 with the SMA \citep[][]{Martin11}. By analogy to the Milky Way, this emission was interpreted as tracing hot cores in the nucleus of Arp~220.

Spatially resolved observations of the spiral galaxies IC342 and Maffei~2 have shown that the emission of methanol (and HNCO) correlates well with the expected locations of bar-induced orbital shocks \citep[][]{Meier05,Meier12}.
Methanol (but neither CH$_3$CCH nor CH$_3$CN) was also detected in a single-dish spectral line survey of the spiral galaxy M51 \citep[][]{Watanabe14}. The detection was made toward two positions in a spiral arm. The derived rotational temperatures are typical of galactic cold dark clouds  ($\sim$10~K), suggesting that the emission is widespread and not dominated by hot cores. The authors argued that the methanol desorption may have been triggered by large-scale shocks, like in IC342 and Maffei~2.

The methanol emission of giant molecular clouds in the AGN NGC1068 was studied with ALMA, revealing that the abundance ratio of methanol to $^{13}$CO is enhanced in the spiral arms compared to the bar end, with kinematical indications that the enhancement may be related to weak shocks \citep[][]{Tosaki17}. The central $\sim$200~pc region of NGC253 was also resolved in eight star forming clumps with ALMA \citep[][]{Ando17}. Methanol was detected in six of them; the reason for the nondetection in the other two clumps remains unclear. Finally, the ALMA Comprehensive High-resolution Extragalactic Molecular Inventory (ALCHEMI) large program will provide the spectrum of NGC253 over an unprecedented nearly contiguous frequency range of 289~GHz at a resolution of 28~pc \citep[][]{Martin21}. One aim of this survey is to explore the chemistry of COMs in the central molecular zone of NGC253. The analysis of the full data set is still in progress but the ACA part of the survey has already allowed the team to detect new extragalactic COMs (ethanol C$_2$H$_5$OH, propynal  HCCCHO, the three $^{13}$C isotopologues of CH$_3$CCH) and to confirm the detection of methylamine CH$_3$NH$_2$ \citep[][]{Martin21}.
Ethanol was found to have an abundance relative to methanol similar to that of Galactic hot cores and a bit higher than in GMCs of the Galactic Center region, suggesting that hot cores may contribute significantly to the unresolved emission traced with the ACA. The full high-angular-resolution data set will certainly bring new insights into this.

\subsection{COMs at high redshifts ($z$)}
\label{ss:highz}

The extragalactic sources mentioned in Sect.~\ref{ss:nearbygal} are nearby galaxies located at distances $\leq$80$\,$Mpc (see also Sect.~\ref{ss:extragalhotcores} for COM detections in the Magellanic clouds). However, COMs are not a unique feature of our nearby Universe but they are also found at high redshifts. This is the case of the $z=0.89$ face-on spiral galaxy located in front of the radio-loud quasar PKS~1830-211 with redshift $z=2.5$ \citep[][]{lidman99}. PKS~1830-211 is gravitationally lensed by the $z = 0.89$ spiral galaxy producing a 1$"$ Einstein ring and two images of the quasar toward the northeast (NE) and southwest (SW) of the ring \citep[][]{wiklind98,winn02}. One of the molecular-rich spiral arms of the $z$ = 0.89 galaxy intercepts the line of sight of the SW image of the quasar generating strong absorption in the molecular spectra at millimeter wavelengths \citep[][]{wiklind98,menten99,muller06}. 

Several spectral line surveys have been carried out at millimeter wavelengths to determine the chemical composition of the interstellar medium within the $z=0.89$ spiral galaxy. Since the galaxy's molecular spectra are redshifted, the frequencies of the molecular lines appear at frequencies decreased by a factor $(1+z)$. A first survey carried out with ATCA at 7\,mm (corresponding to 4\,mm wavelengths in the rest frame) discovered 28 molecular species plus eight isotopologues \citep[][]{muller11}. These included COM precursors such as CH$_2$CO, CH$_2$CN or CH$_2$NH, and COMs such as CH$_3$OH, CH$_3$CN, CH$_3$CCH, CH$_3$CHO, and CH$_3$NH$_2$. The measured relative abundances lie in-between those of typical Galactic diffuse clouds and translucent clouds (when considering the simple molecules), and they are very similar to those found in the nuclear region of the starburst galaxy NGC253 \citep[][]{muller11}.
PKS~1830-211 was also observed in the frame of an ALMA Early Science project, detecting a total of 42 different molecular species and 14 rare isotopologues. In this case, the prebiotic COM NH$_2$CHO was reported \citep[][]{muller14}. 

More recently, \citet{tercero20} carried out a high-sensitivity spectral line survey toward PKS~1830-211 again at 7\,mm using the new HEMT Q-band receiver available at the Yebes 40\,m telescope. In this new dataset, molecular species such as C$_3$H$^+$, C$_3$N, HCOOH, H$_2$CN and H$_2$NC, together with the COMs vinyl cyanide (C$_2$H$_3$CN) and methyl mercaptan (CH$_3$SH), were reported for the first time toward an extragalactic source  \citep[][]{tercero20,cabezas21}. Like for the COMs present in the moderately dense ($n$(H$_2$)$\sim$10$^3$--10$^4$$\,$cm$^{-3}$) and warm envelope ($T_{\rm kin}$$\sim$100$\,$K) of Sgr B2, the COMs detected toward the PKS~1830-211 absorber are likely formed by non-thermal processes such as shocks and/or  enhanced UV/X-ray/cosmic-ray fluxes in the surrounding medium \citep[see][]{tercero20}.

We finally note that other molecular absorbers have been identified in the millimeter wavelength range \citep[][]{combes08}. However, the absorber galaxy located in front of PKS 1830-211 not only is the strongest and richest in COMs but also the one located at the highest redshift \citep[][]{muller11}. It is also the source where the first extragalactic detection of a molecule not yet detected in our Galaxy has been made \citep[C$_2$H$_3$$^+$,][]{Muller24}.

\section{Comparison of chemical composition across environments}
\label{s:comparison}

\subsection{Comparison across different stages of Sun-like star formation}
\label{ss:comp-galenv}

\begin{figure}[!t]
    \centering
    \includegraphics[width=1.0\textwidth]{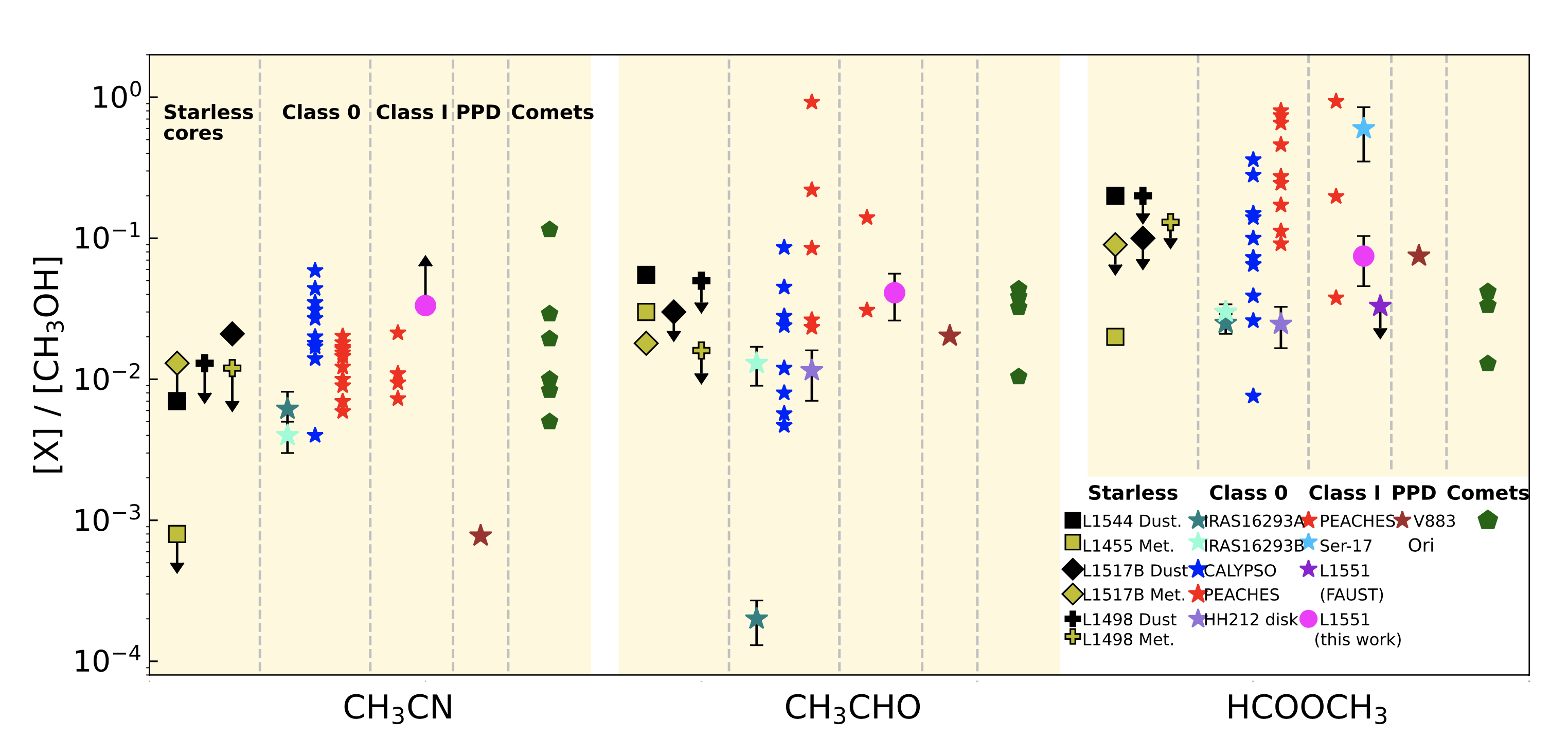}
    \caption{Adapted from \citet{Mercimek22}. Column density ratios of CH$_3$CN, CH$_3$CHO, and CH$_3$OCHO with respect to methanol  measured in Class 0, Class I sources, protoplanetary disks (PPD), and comets. The  references for all the measurements can be found in \citet{Mercimek22}. Among them, the outputs from the ALMA PILS \citep{Jorgensen16}, ALMA FAUST \citep{Bianchi20}, ALMA PEACHES \citep{Yang21}, and NOEMA CALYPSO \citep{Belloche20} projects are labeled in the legend in the bottom right. Additional data are taken from \citet{jimenez16,jimenez21} for starless cores, \citet{Punanova18} for L1544, \citet{jimenez21} for L1498, and
    \citet{megias23}  for L1517B.}
    \label{fig:comparison}
\end{figure}

Beside the Large Programs PILS \citep{Jorgensen16}, ASAI \citep{Lefloch18}, CALYPSO \citep{Belloche20}, FAUST \citep{Codella21}, and PEACHES \citep{Yang21} (see Sect.~\ref{ss:hotcorinos}), several smaller projects on Class~0 and Class~I hot corinos have been performed using first single-dish and then interferometric observations. In this case, less COMs are reported, mainly CH$_3$OH, CH$_3$CHO, CH$_3$OCHO, and CH$_3$OCH$_3$ \citep[e.g.,][]{Taquet15,Imai16,Marcelino18,Bergner19,Nazari21,MartinDomenech21,Chahine22}.

It is tempting to compare these measurements with the earliest (starless cores) and later (protoplanetary disks and comets) stages of the Sun-like star formation. Figure \ref{fig:comparison} reports a comparison of the COM abundances with respect to methanol in different hot corinos around Class 0 and I protostars \citep[adapted from ][]{Mercimek22}, and toward starless cores, protoplanetary disks, and comets. Although the scatter of the points is large, the plot suggests a sort of continuity when moving from starless cores to protoplanetary disks and comets. Note that, although the statistics are poor and some outliers are present, the COM abundances relative to methanol in starless cores are in agreement with those in protostars. In other words, the data so far available support a similar organic chemical composition, within an order of magnitude, during the stages of the low-mass star formation process. This in turn suggests that at least part of the material of the Solar System has been inherited from the earliest stages. More robust statistics are needed, in light of the lesson provided by the shocked gas in L1157-B1, where high-spatial resolution maps indicate that different COMs can trace different environments (see Sect. \ref{ss:outflows}). Thus, COM abundance ratios based on images with a $\sim$ 10 au spatial resolution would be instructive. 

To summarise, the formation of complex organics starts at the earliest stages of the formation of a Sun-like star and the corresponding planetary system. Organic chemistry is already at work during the prestellar phase, then blooms around protostars, where COMs show very high abundances in the gas phase. Nevertheless, the level of chemical complexity, as traced by COMs seems to be constant. More measurements in protoplanetary disks and comets are clearly needed to assess if this level of complexity remains constant in protoplanetary disks, where planets, asteroids, and comets form.

\subsection{Comparison between Sgr~B2(N) and G+0.693}
\label{ss:comp-sgrb2}

\subsubsection{Comparison between the observed COM abundances}
\label{ss:comp-sgrb2-s1}

\begin{figure}[!t]
    \centering
    \includegraphics[angle=0,width=1.0\textwidth]{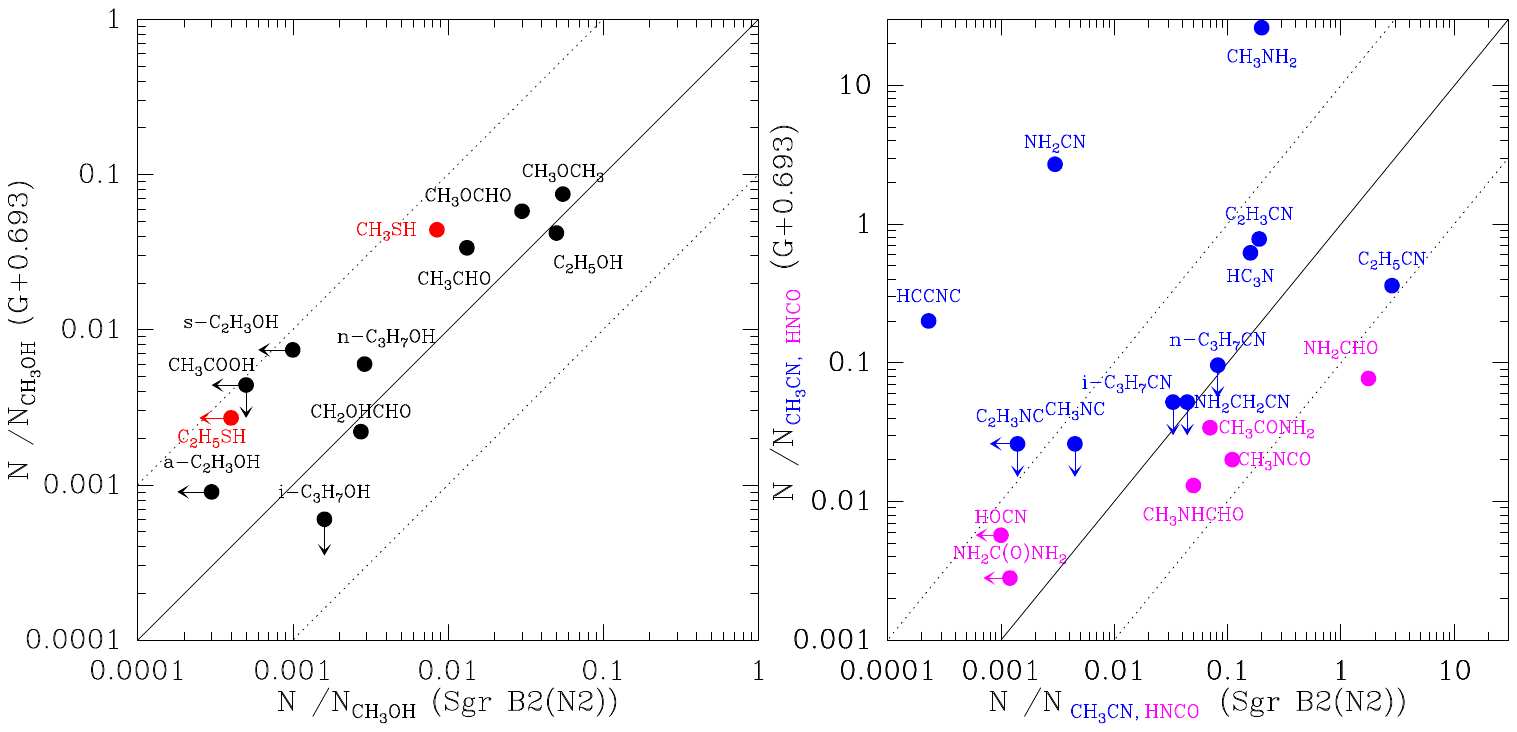}
    \caption{Left panel:  Comparison of the column density ratios of COMs measured toward the hot core Sgr B2(N2) and the Galactic Center GMC G+0.693, with respect to CH$_3$OH. Points in black refer to O-bearing COMs while points in red represent S-bearing COMs. 
    Right panel: Comparison of the column density ratios of N-bearing COMs measured toward the same sources. The N-bearing COMs are split into two chemical families, the molecules with the -CN/-NC radicals (whose ratios are calculated with respect to the column density of CH$_3$CN; see blue points), and the species with the -NCO functional group (with ratios calculated with respect to HNCO; points in magenta). Arrows indicate upper limits. Solid lines denote the 1:1 relation in both panels, while the dotted lines refer to column density ratios a factor 10 higher or lower with respect to the 1:1 relation. The column density ratios for Sgr B2(N2) have been extracted from: \citet{belloche14,belloche16,Belloche17,Belloche19,belloche22,mueller16,richard18,Bonfand19,martin-drumel19,willis20,sanz-novo20} and A. Belloche (private communication). The column density ratios for G+0.693 have been extracted from: \citet{requena06,zeng18,zeng23,jimenez20,jimenez22b,rodriguez21,rodriguez21b,sanz-novo22}.}
    \label{compSgrB2G0693}
\end{figure}

Among all the sources discussed in this Chapter, two of them stand as the richest in COM composition: the Sgr~B2(N2) hot core and the G+0.693 molecular cloud (see Sects.~\ref{ss:hotcores} and \ref{ss:galcent}). Both sources are located in the same environment, the Galactic Center region, and both belong to the same molecular cloud complex, Sgr~B2. Thus, it is reasonable to think that the chemistry of both sources started from the same initial conditions. However, their chemistry may have evolved differently given their different physical conditions and history. Indeed, while the hot core presents high densities \citep[$\geq$10$^7$$\,$cm$^{-3}$;][]{Bonfand19} and similar gas and dust temperatures (gas and dust are thermally coupled, with $T_{\rm gas}$$\sim$$T_{\rm dust}$$\sim$150--200$\,$K), for G+0.693 the gas density is of a few 10$^4$$\,$cm$^{-3}$ and the gas and dust are thermally decoupled \citep[$T_{\rm gas}$=70-150$\,$K vs. $T_{\rm dust}$$\leq$20$\,$K;][]{rodriguez00,zeng18,zeng20}. It has been recently proposed that G+0.693 may represent an early stage of star cluster formation in the Sgr~B2 molecular cloud, since narrow emission of deuterated species such as N$_2$D$^+$ and DCO$^+$ has been detected toward this source \citep{Colzi22}. These deuterated species would be probing a cold core just formed at the interacting shocked layer of a cloud-cloud collision \citep{zeng20}. 

To check whether the different physical conditions imprint a different chemistry in Sgr~B2(N2) and G+0.693, we compare the column density ratios of COMs detected toward both sources with respect to CH$_3$OH in the left panel of Fig.$\,$\ref{compSgrB2G0693} and with respect to HNCO and CH$_3$CN in the right panel. The comparison is performed for different COM families, the O- and S-bearing COMs (black and red points in the left panel, respectively), and the COMs with the -CN/-NC and -NCO functional groups (blue and magenta points in the right panel, respectively). O-bearing COMs tend to be more abundant relative to methanol in G+0.693 than toward Sgr~B2(N2) (within a factor of 10), especially the unsaturated species\footnote{Unsaturated molecules are here understood as hydrogen-poor compounds \citep{Herbst09}.} (see left panel). This trend for unsaturated species to be particularly enhanced in G+0.693 was already pointed out by \citet{requena06} for O-bearing COMs, and confirmed by \citet[][]{zeng18} for cyanides. As shown in the right panel, N-bearing COMs with the --CN and --NC functional groups are in general more abundant relative to methyl cyanide in G+0.693 than in Sgr~B2(N2) (see blue points). Interestingly, the abundance of HCCNC, NH$_2$CN and CH$_3$NH$_2$ are strongly enhanced toward G+0.693 with column density ratios with respect to CH$_3$CN more than two orders of magnitude higher than in Sgr B2(N2). For COMs with the --NCO functional group, however, the measured abundances relative to HNCO tend to be higher (within a factor of $\sim$10) in Sgr~B2(N2) with the exception of urea and HOCN (see magenta points in the right panel). Urea presents a column density ratio with respect to HNCO at least a factor of $\sim$2 higher in G+0.693; and HOCN is an unsaturated isomer of HNCO. Since the rest of --NCO species reported in Fig.~\ref{compSgrB2G0693} are more saturated (e.g., formamide NH$_2$CHO, acetamide CH$_3$CONH$_2$, or methyl isocyanate CH$_3$NCO), this trend is consistent with the idea of unsaturated species being more efficiently formed (or less efficiently destroyed) in the G+0.693 molecular cloud than in Sgr B2(N2). 

The S-bearing COMs CH$_3$SH and C$_2$H$_5$SH (see red points in the left panel) present higher column density ratios with respect to methanol in G+0.693 than in Sgr B2(N2). This thus suggests a higher S/O abundance ratio toward this cloud, suggesting that large amounts of sulfur are available toward G+0.693, although we do not know the main reservoir of sulfur. This may be the reason why new sulfur-bearing COMs such as monothioformic acid (HC(O)SH) and C$_2$H$_5$SH have been discovered and confirmed, respectively, toward G+0.693 \citep{rodriguez21} but not detected toward Sgr~B2(N)'s hot cores so far. 

\subsubsection{Comparison with predictions of chemical models}
\label{ss:comp-sgrb2-s2}

As mentioned in Sect.~\ref{ss:comp-sgrb2-s1}, the G+0.693 molecular cloud and the Sgr B2(N2) hot core are located in the same molecular cloud complex. It is thus reasonable to think that both objects shared a similar initial chemical composition. But is it possible to track back those initial conditions for the chemical composition of the Sgr B2 molecular cloud complex?

The chemistry of the Sgr B2(N2) hot molecular core has been modelled in detail from the initial molecular cloud conditions to the hot core warm-up phase \citep[see, e.g.,][]{garrod13,Garrod22}. In the models of \citet{Garrod22}, certain molecules such as C$_2$H$_5$OH, NH$_2$CN, or CH$_3$NH$_2$ are formed almost solely during the initial molecular cloud stage at dust temperatures between 10 K and 15 K. Therefore, the comparison of the abundances of C$_2$H$_5$OH, NH$_2$CN, or CH$_3$NH$_2$ between the two sources, can help us constrain the dominant chemical processes and relevant timescales for the formation of these molecules. As shown in  Fig.~\ref{compSgrB2G0693}, C$_2$H$_5$OH presents a similar column density ratio with respect to methanol toward both sources, which supports the idea that this molecule formed early-on in the parental molecular cloud. However, NH$_2$CN and CH$_3$NH$_2$ are clearly more abundant (relative to methyl cyanide) in the G+0.693 cloud than toward Sgr B2(N2) by several orders of magnitude, which indicates that these species, if formed early-on in the parental molecular cloud, experienced further chemical processing and thus are more sensitive to changes in the physical conditions of the source. Another possibility is that the molecule that we used to normalize their column densities, CH$_3$CN, may have experienced a different chemical processing in the two sources. Its formation history depends a lot on the evolution timescale in the models of \citet{Garrod22}.

The physical evolution of both sources is indeed very different. In the case of Sgr B2(N2), the initial dense core underwent gravitational collapse reaching probably low temperatures ($T_{\rm kin}$=$T_{\rm dust}$$\leq$20$\,$K) and, once the protostar was formed, the surrounding envelope was progressively warmed up \citep[][]{garrod13,Garrod22}. The warm-up of the envelope caused the release of the ice mantles of dust grains into the gas phase, yielding the rich chemistry observed toward the Sgr B2(N2) hot core. In these models, saturated molecules form on grains more efficiently than unsaturated ones, which would explain the higher molecular column density ratios measured between saturated and unsaturated species in Sgr B2(N2) (Fig.~\ref{compSgrB2G0693}). An illustrative example is provided by the N-bearing COMs C$_2$H$_3$CN and C$_2$H$_5$CN, with the saturated species C$_2$H$_5$CN being more abundant in Sgr B2(N2) than the unsaturated one, C$_2$H$_3$CN, consistent with the predictions of \citet{Garrod22}'s models, while it is the opposite in G+0.693.

We note that in these models a standard cosmic-ray ionisation rate of 1.3$\times$10$^{-17}$$\,$s$^{-1}$ is considered \citep[][]{garrod13,Garrod22}. However, cosmic rays are known to be enhanced by several orders of magnitude in the Galactic Center region \citep[by factors $\geq$100--1000; see e.g.][]{goto14}, which could potentially affect the chemistry of the hot core. \citet{Bonfand19} investigated the impact of cosmic rays on the chemistry of the Sgr B2(N) hot cores, finding that an enhanced cosmic-ray ionisation rate (by a factor of $\sim$50) reproduces best the COM abundances measured toward the Sgr B2 hot cores. This suggests that the high H$_2$ gas volume density and H$_2$ column density measured in these objects \citep[$n$(H$_2$$>$10$^7$$\,$cm$^{-3}$ and $N$(H$_2$)$>$10$^{24}$$\,$cm$^{-2}$;][]{Bonfand19} have significantly attenuated the cosmic-ray ionisation rate in the innermost regions of the hot cores \citep[see][]{padovani09}. 

In contrast, the H$_2$ gas volume density toward the G+0.693 molecular cloud is of a few 10$^4$$\,$cm$^{-3}$ \citep{guesten04,rodriguez00,zeng20}, and thus cosmic rays are not expected to be attenuated. In addition, G+0.693 is undergoing a cloud-cloud collision, and therefore gas and dust in this cloud are affected by the fast changes in density and temperature induced by low-velocity shocks. The sputtering of dust grains by the shocks erode the grain mantles and/or partially destroy the grain cores, injecting large amounts of molecular material into the gas phase \citep[][]{Caselli97,JimenezSerra08}. SiO, an excellent tracer of shocked gas, as well as other Si-bearing molecules \citep[see][; the latter reports the first detection of interstellar SiC$_2$]{martin-pintado97,Massalkhi23}, are indeed observed toward G+0.693, supporting the presence of low-velocity shocks in this cloud. 

In a scenario of low-velocity shocks with a high cosmic-ray ionisation rate, molecular ions are expected to be abundant, which would trigger an active ion-neutral gas-phase chemistry assisted by high kinetic temperatures of the gas \citep[][]{Ginsburg16,zeng18}. As proposed by \citet{zeng18}, this would be the chemical mechanism by which unsaturated molecules such as C$_2$H$_3$CN are more abundant than their saturated counterparts such as C$_2$H$_5$CN in G+0.693. Alternatively, shocks and cosmic rays are expected to enhance the abundances of atomic H and C, enabling gas-phase reactions with these atoms and fast atom addition/abstraction reactions on dust grains \citep{requena08}. These two scenarios need to be tested in the future with dedicated chemical models for the physical conditions of the G+0.693 molecular cloud. 

Finally, it is important to stress that the conditions of low-velocity shocks and enhanced cosmic-ray ionisation rates are not unique to the G+0.693 molecular cloud but they are also found toward other Galactic Center GMCs \citep[e.g.,][]{nanase15}. This naturally explains the widespread emission of COMs measured across the Central Molecular Zone in the Galactic Center region \citep{martin-pintado01,requena06}.

\section{Conclusions and outlook}
\label{s:conclusions}

In this Chapter we have shown that COMs are ubiquitous in the ISM, since they are found in a wide variety of astronomical environments under different physical conditions. No matter where we look at, whether the source is cold (as e.g. TMC-1) or hot (as Sgr~B2(N)), or whether it is located in the Galactic disk (as IRAS~16293--2422) or in the Galactic Center region (as G+0.693), COMs are present. As mentioned in Sects.$\,$\ref{ss:prestellar}, \ref{ss:hotcores}, and \ref{ss:galcent}, the large number of new COM detections has been possible thanks to the advent of instrumentation with higher sensitivity, higher angular resolution, and broader spectral coverage. Indeed, the high angular resolution of ALMA, along with its unbeatable sensitivity, has allowed us to reduce the line blending and line confusion in the crowded spectra of hot sources such as hot cores and hot corinos, enabling the discovery of new COMs including some of prebiotic interest \citep[e.g., glycolonitrile and urea;][]{Zeng19,Belloche19}. Line confusion is one of the limiting factors for the discovery of new COMs since, even with ALMA, we have probably reached the confusion limit in the observed spectra of the hot corino IRAS~16293--2422B and the hot cores in Sgr~B2(N) (Sects.$\,$\ref{ss:hotcorinos} and \ref{ss:hotcores}). Systematic surveys toward large samples of hot corinos and hot cores would be desirable to find new sources with rich COM chemistry and with lower levels of line blending and line confusion. Alternatively, observations at lower frequencies ($\leq$80$\,$GHz) can aid in reducing the level of line confusion toward IRAS~16293--2422B and Sgr~B2(N) because i) the spectra are cleaner of transitions from lighter molecules, and ii) the line density in velocity space for a given molecule becomes smaller. The problem of carrying low-frequency observations toward hot sources is that the molecular line intensities are expected to be weaker than at millimeter/sub-millimeter wavelengths since the molecular spectra peak at the latter wavelengths for temperatures of $\sim$100-300$\,$K. Therefore, higher sensitivity instrumentation operating at frequencies $\leq$80$\,$GHz are needed to probe the weak emission of COMs in hot sources at centimeter wavelengths.  

There are several instruments and facilities that will become operational at frequencies below 80$\,$GHz in the near future. The Band 1 receiver of ALMA has become available in 2023 and Band 2 will become available in 2026, covering the frequency ranges between 35 and 50~GHz and 70 and 116~GHz, respectively, with unprecedented sensitivity and high angular resolution. To cover even lower frequencies, in the northern hemisphere the VLA will undergo a major upgrade into the next generation VLA (ngVLA), which will consist of a main array with 244 18$\,$m-diameter antennas spread over $\sim$8,860$\,$km, and a second array with 19 6$\,$m-diameter dishes located at the center of the main array to collect short-spacing data. The frequency coverage will go from 1.2 GHz to 116 GHz. In the southern hemisphere, the Square Kilometer Array (SKA) will have two observatories, the SKA-low array in Australia covering frequencies from 50 MHz to 350 MHz, and the SKA-mid array in South Africa covering frequencies from 350 MHz to 14(30) GHz. SKA1-mid will cover frequencies from 350 MHz to 14 GHz during its Phase 1 while it is planned that in a second phase, SKA2-mid will cover frequencies up to 30 GHz. Due to its frequency coverage, SKA-mid is better suited for observations of COMs. Both ngVLA and SKA-mid will target COMs with a particular focus on those of biogenic nature.

Another way of circumventing the problem of line confusion is by targeting sources with low excitation temperatures. Since the excitation temperatures are low, only the lowest energy levels of a given molecule are populated, reducing significantly the confusion of the observed spectra. This is the case of cold dark clouds such as TMC-1 or the massive GMCs in the Galactic Center such as G+0.693. While the low excitation temperatures of COMs in TMC-1 are due to the low kinetic temperatures of the gas ($T_{\rm kin}$$\leq$10$\,$K), the low excitation temperatures of the COM emission in Galactic Center GMCs is produced by the subthermal excitation of the lines at H$_2$ densities (of $\sim$10$^4$$\,$cm$^{-3}$) that are lower than the critical densities of the COM rotational transitions ($\geq$10$^5$$\,$cm$^{-3}$). This approach has been very successful in the past few years, where spectroscopic surveys carried out at low frequencies with the GBT, Yebes 40~m, and IRAM 30~m telescopes toward TMC-1 and G+0.693 have yielded a plethora of new COM detections (see Sects.$\,$\ref{ss:prestellar} and \ref{ss:galcent}). It is however unclear whether line confusion has been reached already in existing spectral line surveys. Future steps could involve observations of Galactic Center GMCs carried out in absorption against bright radio continuum sources with facilities such as the SKA \citep[see][]{jimenez22a}.

It is important to stress the essential role of molecular spectroscopy in the discovery of new COMs in the ISM. In many cases, the millimeter spectra of the targeted molecular species is not available and thus, it requires specially-designed laboratory experiments. For large species such as COMs, this is not straightforward because they are typically thermolabile and hygroscopic and the samples cannot be prepared easily. Fortunately, new sample preparation and analysis techniques have recently been developed, which enables the spectroscopic characterization of large complex organics. For example, the ultrafast laser vaporization technique avoids the decomposition of large molecules such as sugars  \citep[see the cases of ribose and erythrulose in][]{cocinero12,insausti21}. 

Finally, a better understanding of the chemistry of COMs can help us to find out which COMs are more likely to form under interstellar conditions with high enough abundance so that they can be detected in the ISM. Joint efforts in observational astronomy, molecular spectroscopy, gas-phase and ice COM synthesis laboratory experiments, quantum chemical calculations, and astrochemical modelling will be needed in the near future to exploit the full capabilities of instruments such as the SKA and ngVLA and enhance our understanding of the chemical complexity of the ISM.

\vspace*{2ex}
\noindent\textit{Acknowledgments:} A.B. thanks C.-H. Rosie Chen for discussions about the Magellanic clouds. C. C. acknowledges E. Bianchi,
M. De Simone, and L. Podio for instructive discussions on chemistry at work in protostellar and protoplanetary disks. I.J.-S. thanks Jes\'us Mart\'{\i}n-Pintado for discussions on the chemistry of COMs in Galactic Center molecular clouds. 
%\appendix

%\section{Sample Appendix Section}
%\label{s:appendix}

\bibliographystyle{elsarticle} 
\bibliography{hbk-astrochem-coms}

\begin{thebibliography}{356}
\expandafter\ifx\csname natexlab\endcsname\relax\def\natexlab#1{#1}\fi
\providecommand{\url}[1]{\texttt{#1}}
\providecommand{\href}[2]{#2}
\providecommand{\path}[1]{#1}
\providecommand{\DOIprefix}{doi:}
\providecommand{\ArXivprefix}{arXiv:}
\providecommand{\URLprefix}{URL: }
\providecommand{\Pubmedprefix}{pmid:}
\providecommand{\doi}[1]{\href{http://dx.doi.org/#1}{\path{#1}}}
\providecommand{\Pubmed}[1]{\href{pmid:#1}{\path{#1}}}
\providecommand{\bibinfo}[2]{#2}
\ifx\xfnm\relax \def\xfnm[#1]{\unskip,\space#1}\fi
%Type = Article
\bibitem[{{Acharyya} and {Herbst}(2015)}]{Acharyya15}
\bibinfo{author}{{Acharyya}, K.}, \bibinfo{author}{{Herbst}, E.},
  \bibinfo{year}{2015}.
\newblock \bibinfo{title}{{Molecular Development in the Large Magellanic
  Cloud}}.
\newblock \bibinfo{journal}{\apj} \bibinfo{volume}{812}, \bibinfo{pages}{142}.
\newblock \DOIprefix\doi{10.1088/0004-637X/812/2/142}.
%Type = Article
\bibitem[{{Acharyya} and {Herbst}(2018)}]{Acharyya18}
\bibinfo{author}{{Acharyya}, K.}, \bibinfo{author}{{Herbst}, E.},
  \bibinfo{year}{2018}.
\newblock \bibinfo{title}{{Hot Cores in Magellanic Clouds}}.
\newblock \bibinfo{journal}{\apj} \bibinfo{volume}{859}, \bibinfo{pages}{51}.
\newblock \DOIprefix\doi{10.3847/1538-4357/aabaf2}.
%Type = Article
\bibitem[{{Ag{\'u}ndez} et~al.(2023){Ag{\'u}ndez}, {Loison}, {Hickson},
  {Wakelam}, {Fuentetaja}, {Cabezas}, {Marcelino}, {Tercero}, {de Vicente} and
  {Cernicharo}}]{agundez23}
\bibinfo{author}{{Ag{\'u}ndez}, M.}, \bibinfo{author}{{Loison}, J.C.},
  \bibinfo{author}{{Hickson}, K.M.}, \bibinfo{author}{{Wakelam}, V.},
  \bibinfo{author}{{Fuentetaja}, R.}, \bibinfo{author}{{Cabezas}, C.},
  \bibinfo{author}{{Marcelino}, N.}, \bibinfo{author}{{Tercero}, B.},
  \bibinfo{author}{{de Vicente}, P.}, \bibinfo{author}{{Cernicharo}, J.},
  \bibinfo{year}{2023}.
\newblock \bibinfo{title}{{Detection of ethanol, acetone, and propanal in TMC-1
  New O-bearing complex organics in cold sources}}.
\newblock \bibinfo{journal}{\aap} \bibinfo{volume}{673}, \bibinfo{pages}{A34}.
\newblock \DOIprefix\doi{10.1051/0004-6361/202346076},
  \href{http://arxiv.org/abs/2303.16121}{{\tt arXiv:2303.16121}}.
%Type = Article
\bibitem[{{Ag{\'u}ndez} et~al.(2019){Ag{\'u}ndez}, {Marcelino}, {Cernicharo},
  {Roueff} and {Tafalla}}]{agundez19}
\bibinfo{author}{{Ag{\'u}ndez}, M.}, \bibinfo{author}{{Marcelino}, N.},
  \bibinfo{author}{{Cernicharo}, J.}, \bibinfo{author}{{Roueff}, E.},
  \bibinfo{author}{{Tafalla}, M.}, \bibinfo{year}{2019}.
\newblock \bibinfo{title}{{A sensitive {\ensuremath{\lambda}} 3 mm line survey
  of L483. A broad view of the chemical composition of a core around a Class 0
  object}}.
\newblock \bibinfo{journal}{\aap} \bibinfo{volume}{625}, \bibinfo{pages}{A147}.
\newblock \DOIprefix\doi{10.1051/0004-6361/201935164},
  \href{http://arxiv.org/abs/1904.06565}{{\tt arXiv:1904.06565}}.
%Type = Article
\bibitem[{{Ag{\'u}ndez} et~al.(2021){Ag{\'u}ndez}, {Marcelino}, {Tercero},
  {Cabezas}, {de Vicente} and {Cernicharo}}]{agundez21}
\bibinfo{author}{{Ag{\'u}ndez}, M.}, \bibinfo{author}{{Marcelino}, N.},
  \bibinfo{author}{{Tercero}, B.}, \bibinfo{author}{{Cabezas}, C.},
  \bibinfo{author}{{de Vicente}, P.}, \bibinfo{author}{{Cernicharo}, J.},
  \bibinfo{year}{2021}.
\newblock \bibinfo{title}{{O-bearing complex organic molecules at the
  cyanopolyyne peak of TMC-1: Detection of C$_{2}$H$_{3}$CHO, C$_{2}$H$_{3}$OH,
  HCOOCH$_{3}$, and CH$_{3}$OCH$_{3}$}}.
\newblock \bibinfo{journal}{\aap} \bibinfo{volume}{649}, \bibinfo{pages}{L4}.
\newblock \DOIprefix\doi{10.1051/0004-6361/202140978},
  \href{http://arxiv.org/abs/2104.11506}{{\tt arXiv:2104.11506}}.
%Type = Article
\bibitem[{{Aikawa} et~al.(2020){Aikawa}, {Furuya}, {Yamamoto} and
  {Sakai}}]{aikawa20}
\bibinfo{author}{{Aikawa}, Y.}, \bibinfo{author}{{Furuya}, K.},
  \bibinfo{author}{{Yamamoto}, S.}, \bibinfo{author}{{Sakai}, N.},
  \bibinfo{year}{2020}.
\newblock \bibinfo{title}{{Chemical Variation among Protostellar Cores:
  Dependence on Prestellar Core Conditions}}.
\newblock \bibinfo{journal}{\apj} \bibinfo{volume}{897}, \bibinfo{pages}{110}.
\newblock \DOIprefix\doi{10.3847/1538-4357/ab994a},
  \href{http://arxiv.org/abs/2006.11696}{{\tt arXiv:2006.11696}}.
%Type = Article
\bibitem[{{Aikawa} et~al.(2008){Aikawa}, {Wakelam}, {Garrod} and
  {Herbst}}]{aikawa08}
\bibinfo{author}{{Aikawa}, Y.}, \bibinfo{author}{{Wakelam}, V.},
  \bibinfo{author}{{Garrod}, R.T.}, \bibinfo{author}{{Herbst}, E.},
  \bibinfo{year}{2008}.
\newblock \bibinfo{title}{{Molecular Evolution and Star Formation: From
  Prestellar Cores to Protostellar Cores}}.
\newblock \bibinfo{journal}{\apj} \bibinfo{volume}{674},
  \bibinfo{pages}{984--996}.
\newblock \DOIprefix\doi{10.1086/524096},
  \href{http://arxiv.org/abs/0710.0712}{{\tt arXiv:0710.0712}}.
%Type = Article
\bibitem[{{Akiyama} et~al.(2019){Akiyama}, {Vorobyov}, {Liu}, {Dong}, {de
  Leon}, {Liu} and {Tamura}}]{Akiyama19}
\bibinfo{author}{{Akiyama}, E.}, \bibinfo{author}{{Vorobyov}, E.I.},
  \bibinfo{author}{{Liu}, H.B.}, \bibinfo{author}{{Dong}, R.},
  \bibinfo{author}{{de Leon}, J.}, \bibinfo{author}{{Liu}, S.Y.},
  \bibinfo{author}{{Tamura}, M.}, \bibinfo{year}{2019}.
\newblock \bibinfo{title}{{A Tail Structure Associated with a Protoplanetary
  Disk around SU Aurigae}}.
\newblock \bibinfo{journal}{\aj} \bibinfo{volume}{157}, \bibinfo{pages}{165}.
\newblock \DOIprefix\doi{10.3847/1538-3881/ab0ae4},
  \href{http://arxiv.org/abs/1902.10306}{{\tt arXiv:1902.10306}}.
%Type = Article
\bibitem[{{Aladro} et~al.(2011){Aladro}, {Mart{\'\i}n}, {Mart{\'\i}n-Pintado},
  {Mauersberger}, {Henkel}, {Oca{\~n}a Flaquer} and
  {Amo-Baladr{\'o}n}}]{Aladro11}
\bibinfo{author}{{Aladro}, R.}, \bibinfo{author}{{Mart{\'\i}n}, S.},
  \bibinfo{author}{{Mart{\'\i}n-Pintado}, J.}, \bibinfo{author}{{Mauersberger},
  R.}, \bibinfo{author}{{Henkel}, C.}, \bibinfo{author}{{Oca{\~n}a Flaquer},
  B.}, \bibinfo{author}{{Amo-Baladr{\'o}n}, M.A.}, \bibinfo{year}{2011}.
\newblock \bibinfo{title}{{A {\ensuremath{\lambda}} = 1.3 mm and 2 mm molecular
  line survey towards M 82}}.
\newblock \bibinfo{journal}{\aap} \bibinfo{volume}{535}, \bibinfo{pages}{A84}.
\newblock \DOIprefix\doi{10.1051/0004-6361/201117397},
  \href{http://arxiv.org/abs/1109.6236}{{\tt arXiv:1109.6236}}.
%Type = Article
\bibitem[{{Aladro} et~al.(2015){Aladro}, {Mart{\'\i}n}, {Riquelme}, {Henkel},
  {Mauersberger}, {Mart{\'\i}n-Pintado}, {Wei{\ss}}, {Lefevre}, {Kramer},
  {Requena-Torres} and {Armijos-Abenda{\~n}o}}]{Aladro15}
\bibinfo{author}{{Aladro}, R.}, \bibinfo{author}{{Mart{\'\i}n}, S.},
  \bibinfo{author}{{Riquelme}, D.}, \bibinfo{author}{{Henkel}, C.},
  \bibinfo{author}{{Mauersberger}, R.}, \bibinfo{author}{{Mart{\'\i}n-Pintado},
  J.}, \bibinfo{author}{{Wei{\ss}}, A.}, \bibinfo{author}{{Lefevre}, C.},
  \bibinfo{author}{{Kramer}, C.}, \bibinfo{author}{{Requena-Torres}, M.A.},
  \bibinfo{author}{{Armijos-Abenda{\~n}o}, R.J.}, \bibinfo{year}{2015}.
\newblock \bibinfo{title}{{Lambda = 3 mm line survey of nearby active
  galaxies}}.
\newblock \bibinfo{journal}{\aap} \bibinfo{volume}{579}, \bibinfo{pages}{A101}.
\newblock \DOIprefix\doi{10.1051/0004-6361/201424918},
  \href{http://arxiv.org/abs/1504.03743}{{\tt arXiv:1504.03743}}.
%Type = Article
\bibitem[{{Altwegg} et~al.(2016){Altwegg}, {Balsiger}, {Bar-Nun}, {Berthelier},
  {Bieler}, {Bochsler}, {Briois}, {Calmonte}, {Combi}, {Cottin}, {De Keyser},
  {Dhooghe}, {Fiethe}, {Fuselier}, {Gasc}, {Gombosi}, {Hansen}, {Haessig}, {Ja
  ckel}, {Kopp}, {Korth}, {Le Roy}, {Mall}, {Marty}, {Mousis}, {Owen}, {Reme},
  {Rubin}, {Semon}, {Tzou}, {Waite} and {Wurz}}]{altwegg2016}
\bibinfo{author}{{Altwegg}, K.}, \bibinfo{author}{{Balsiger}, H.},
  \bibinfo{author}{{Bar-Nun}, A.}, \bibinfo{author}{{Berthelier}, J.J.},
  \bibinfo{author}{{Bieler}, A.}, \bibinfo{author}{{Bochsler}, P.},
  \bibinfo{author}{{Briois}, C.}, \bibinfo{author}{{Calmonte}, U.},
  \bibinfo{author}{{Combi}, M.R.}, \bibinfo{author}{{Cottin}, H.},
  \bibinfo{author}{{De Keyser}, J.}, \bibinfo{author}{{Dhooghe}, F.},
  \bibinfo{author}{{Fiethe}, B.}, \bibinfo{author}{{Fuselier}, S.A.},
  \bibinfo{author}{{Gasc}, S.}, \bibinfo{author}{{Gombosi}, T.I.},
  \bibinfo{author}{{Hansen}, K.C.}, \bibinfo{author}{{Haessig}, M.},
  \bibinfo{author}{{Ja ckel}, A.}, \bibinfo{author}{{Kopp}, E.},
  \bibinfo{author}{{Korth}, A.}, \bibinfo{author}{{Le Roy}, L.},
  \bibinfo{author}{{Mall}, U.}, \bibinfo{author}{{Marty}, B.},
  \bibinfo{author}{{Mousis}, O.}, \bibinfo{author}{{Owen}, T.},
  \bibinfo{author}{{Reme}, H.}, \bibinfo{author}{{Rubin}, M.},
  \bibinfo{author}{{Semon}, T.}, \bibinfo{author}{{Tzou}, C.Y.},
  \bibinfo{author}{{Waite}, J.H.}, \bibinfo{author}{{Wurz}, P.},
  \bibinfo{year}{2016}.
\newblock \bibinfo{title}{{Prebiotic chemicals--amino acid and phosphorus--in
  the coma of comet 67P/Churyumov-Gerasimenko}}.
\newblock \bibinfo{journal}{Science Advances} \bibinfo{volume}{2},
  \bibinfo{pages}{e1600285--e1600285}.
\newblock \DOIprefix\doi{10.1126/sciadv.1600285}.
%Type = Article
\bibitem[{{Altwegg} et~al.(2019){Altwegg}, {Balsiger} and
  {Fuselier}}]{Altwegg19}
\bibinfo{author}{{Altwegg}, K.}, \bibinfo{author}{{Balsiger}, H.},
  \bibinfo{author}{{Fuselier}, S.A.}, \bibinfo{year}{2019}.
\newblock \bibinfo{title}{{Cometary Chemistry and the Origin of Icy Solar
  System Bodies: The View After Rosetta}}.
\newblock \bibinfo{journal}{\araa} \bibinfo{volume}{57},
  \bibinfo{pages}{113--155}.
\newblock \DOIprefix\doi{10.1146/annurev-astro-091918-104409},
  \href{http://arxiv.org/abs/1908.04046}{{\tt arXiv:1908.04046}}.
%Type = Article
\bibitem[{{Ando} et~al.(2017){Ando}, {Nakanishi}, {Kohno}, {Izumi},
  {Mart{\'\i}n}, {Harada}, {Takano}, {Kuno}, {Nakai}, {Sugai}, {Sorai},
  {Tosaki}, {Matsubayashi}, {Nakajima}, {Nishimura} and {Tamura}}]{Ando17}
\bibinfo{author}{{Ando}, R.}, \bibinfo{author}{{Nakanishi}, K.},
  \bibinfo{author}{{Kohno}, K.}, \bibinfo{author}{{Izumi}, T.},
  \bibinfo{author}{{Mart{\'\i}n}, S.}, \bibinfo{author}{{Harada}, N.},
  \bibinfo{author}{{Takano}, S.}, \bibinfo{author}{{Kuno}, N.},
  \bibinfo{author}{{Nakai}, N.}, \bibinfo{author}{{Sugai}, H.},
  \bibinfo{author}{{Sorai}, K.}, \bibinfo{author}{{Tosaki}, T.},
  \bibinfo{author}{{Matsubayashi}, K.}, \bibinfo{author}{{Nakajima}, T.},
  \bibinfo{author}{{Nishimura}, Y.}, \bibinfo{author}{{Tamura}, Y.},
  \bibinfo{year}{2017}.
\newblock \bibinfo{title}{{Diverse Nuclear Star-forming Activities in the Heart
  of NGC 253 Resolved with 10-pc-scale ALMA Images}}.
\newblock \bibinfo{journal}{\apj} \bibinfo{volume}{849}, \bibinfo{pages}{81}.
\newblock \DOIprefix\doi{10.3847/1538-4357/aa8fd4},
  \href{http://arxiv.org/abs/1710.01432}{{\tt arXiv:1710.01432}}.
%Type = Article
\bibitem[{{Andr{\'e}} et~al.(2019){Andr{\'e}}, {Arzoumanian}, {K{\"o}nyves},
  {Shimajiri} and {Palmeirim}}]{andre19}
\bibinfo{author}{{Andr{\'e}}, P.}, \bibinfo{author}{{Arzoumanian}, D.},
  \bibinfo{author}{{K{\"o}nyves}, V.}, \bibinfo{author}{{Shimajiri}, Y.},
  \bibinfo{author}{{Palmeirim}, P.}, \bibinfo{year}{2019}.
\newblock \bibinfo{title}{{The role of molecular filaments in the origin of the
  prestellar core mass function and stellar initial mass function}}.
\newblock \bibinfo{journal}{\aap} \bibinfo{volume}{629}, \bibinfo{pages}{L4}.
\newblock \DOIprefix\doi{10.1051/0004-6361/201935915},
  \href{http://arxiv.org/abs/1907.13448}{{\tt arXiv:1907.13448}}.
%Type = Article
\bibitem[{{Andr{\'e}} et~al.(2010){Andr{\'e}}, {Men'shchikov}, {Bontemps},
  {K{\"o}nyves}, {Motte}, {Schneider}, {Didelon}, {Minier}, {Saraceno},
  {Ward-Thompson}, {di Francesco}, {White}, {Molinari}, {Testi}, {Abergel},
  {Griffin}, {Henning}, {Royer}, {Mer{\'\i}n}, {Vavrek}, {Attard},
  {Arzoumanian}, {Wilson}, {Ade}, {Aussel}, {Baluteau}, {Benedettini},
  {Bernard}, {Blommaert}, {Cambr{\'e}sy}, {Cox}, {di Giorgio}, {Hargrave},
  {Hennemann}, {Huang}, {Kirk}, {Krause}, {Launhardt}, {Leeks}, {Le Pennec},
  {Li}, {Martin}, {Maury}, {Olofsson}, {Omont}, {Peretto}, {Pezzuto}, {Prusti},
  {Roussel}, {Russeil}, {Sauvage}, {Sibthorpe}, {Sicilia-Aguilar}, {Spinoglio},
  {Waelkens}, {Woodcraft} and {Zavagno}}]{andre10}
\bibinfo{author}{{Andr{\'e}}, P.}, \bibinfo{author}{{Men'shchikov}, A.},
  \bibinfo{author}{{Bontemps}, S.}, \bibinfo{author}{{K{\"o}nyves}, V.},
  \bibinfo{author}{{Motte}, F.}, \bibinfo{author}{{Schneider}, N.},
  \bibinfo{author}{{Didelon}, P.}, \bibinfo{author}{{Minier}, V.},
  \bibinfo{author}{{Saraceno}, P.}, \bibinfo{author}{{Ward-Thompson}, D.},
  \bibinfo{author}{{di Francesco}, J.}, \bibinfo{author}{{White}, G.},
  \bibinfo{author}{{Molinari}, S.}, \bibinfo{author}{{Testi}, L.},
  \bibinfo{author}{{Abergel}, A.}, \bibinfo{author}{{Griffin}, M.},
  \bibinfo{author}{{Henning}, T.}, \bibinfo{author}{{Royer}, P.},
  \bibinfo{author}{{Mer{\'\i}n}, B.}, \bibinfo{author}{{Vavrek}, R.},
  \bibinfo{author}{{Attard}, M.}, \bibinfo{author}{{Arzoumanian}, D.},
  \bibinfo{author}{{Wilson}, C.D.}, \bibinfo{author}{{Ade}, P.},
  \bibinfo{author}{{Aussel}, H.}, \bibinfo{author}{{Baluteau}, J.P.},
  \bibinfo{author}{{Benedettini}, M.}, \bibinfo{author}{{Bernard}, J.P.},
  \bibinfo{author}{{Blommaert}, J.A.D.L.}, \bibinfo{author}{{Cambr{\'e}sy},
  L.}, \bibinfo{author}{{Cox}, P.}, \bibinfo{author}{{di Giorgio}, A.},
  \bibinfo{author}{{Hargrave}, P.}, \bibinfo{author}{{Hennemann}, M.},
  \bibinfo{author}{{Huang}, M.}, \bibinfo{author}{{Kirk}, J.},
  \bibinfo{author}{{Krause}, O.}, \bibinfo{author}{{Launhardt}, R.},
  \bibinfo{author}{{Leeks}, S.}, \bibinfo{author}{{Le Pennec}, J.},
  \bibinfo{author}{{Li}, J.Z.}, \bibinfo{author}{{Martin}, P.G.},
  \bibinfo{author}{{Maury}, A.}, \bibinfo{author}{{Olofsson}, G.},
  \bibinfo{author}{{Omont}, A.}, \bibinfo{author}{{Peretto}, N.},
  \bibinfo{author}{{Pezzuto}, S.}, \bibinfo{author}{{Prusti}, T.},
  \bibinfo{author}{{Roussel}, H.}, \bibinfo{author}{{Russeil}, D.},
  \bibinfo{author}{{Sauvage}, M.}, \bibinfo{author}{{Sibthorpe}, B.},
  \bibinfo{author}{{Sicilia-Aguilar}, A.}, \bibinfo{author}{{Spinoglio}, L.},
  \bibinfo{author}{{Waelkens}, C.}, \bibinfo{author}{{Woodcraft}, A.},
  \bibinfo{author}{{Zavagno}, A.}, \bibinfo{year}{2010}.
\newblock \bibinfo{title}{{From filamentary clouds to prestellar cores to the
  stellar IMF: Initial highlights from the Herschel Gould Belt Survey}}.
\newblock \bibinfo{journal}{\aap} \bibinfo{volume}{518}, \bibinfo{pages}{L102}.
\newblock \DOIprefix\doi{10.1051/0004-6361/201014666},
  \href{http://arxiv.org/abs/1005.2618}{{\tt arXiv:1005.2618}}.
%Type = Article
\bibitem[{{Arce} et~al.(2008){Arce}, {Santiago-Garc{\'{\i}}a}, {J{\o}rgensen},
  {Tafalla} and {Bachiller}}]{Arce08}
\bibinfo{author}{{Arce}, H.G.}, \bibinfo{author}{{Santiago-Garc{\'{\i}}a}, J.},
  \bibinfo{author}{{J{\o}rgensen}, J.K.}, \bibinfo{author}{{Tafalla}, M.},
  \bibinfo{author}{{Bachiller}, R.}, \bibinfo{year}{2008}.
\newblock \bibinfo{title}{{Complex Molecules in the L1157 Molecular Outflow}}.
\newblock \bibinfo{journal}{\apjl} \bibinfo{volume}{681}, \bibinfo{pages}{L21}.
\newblock \DOIprefix\doi{10.1086/590110},
  \href{http://arxiv.org/abs/0805.2550}{{\tt arXiv:0805.2550}}.
%Type = Inproceedings
\bibitem[{{Arce} et~al.(2007){Arce}, {Shepherd}, {Gueth}, {Lee}, {Bachiller},
  {Rosen} and {Beuther}}]{Arce07}
\bibinfo{author}{{Arce}, H.G.}, \bibinfo{author}{{Shepherd}, D.},
  \bibinfo{author}{{Gueth}, F.}, \bibinfo{author}{{Lee}, C.F.},
  \bibinfo{author}{{Bachiller}, R.}, \bibinfo{author}{{Rosen}, A.},
  \bibinfo{author}{{Beuther}, H.}, \bibinfo{year}{2007}.
\newblock \bibinfo{title}{{Molecular Outflows in Low- and High-Mass
  Star-forming Regions}}, in: \bibinfo{editor}{{Reipurth}, B.},
  \bibinfo{editor}{{Jewitt}, D.}, \bibinfo{editor}{{Keil}, K.} (Eds.),
  \bibinfo{booktitle}{Protostars and Planets V}, p. \bibinfo{pages}{245}.
\newblock \href{http://arxiv.org/abs/astro-ph/0603071}{{\tt
  arXiv:astro-ph/0603071}}.
%Type = Article
\bibitem[{{Armijos-Abenda{\~n}o} et~al.(2020){Armijos-Abenda{\~n}o},
  {Mart{\'\i}n-Pintado}, {L{\'o}pez}, {Llerena}, {Harada}, {Requena-Torres},
  {Mart{\'\i}n}, {Rivilla}, {Riquelme} and {Aldas}}]{armijos20}
\bibinfo{author}{{Armijos-Abenda{\~n}o}, J.},
  \bibinfo{author}{{Mart{\'\i}n-Pintado}, J.}, \bibinfo{author}{{L{\'o}pez},
  E.}, \bibinfo{author}{{Llerena}, M.}, \bibinfo{author}{{Harada}, N.},
  \bibinfo{author}{{Requena-Torres}, M.A.}, \bibinfo{author}{{Mart{\'\i}n},
  S.}, \bibinfo{author}{{Rivilla}, V.M.}, \bibinfo{author}{{Riquelme}, D.},
  \bibinfo{author}{{Aldas}, F.}, \bibinfo{year}{2020}.
\newblock \bibinfo{title}{{On the Effects of UV Photons/X-Rays on the Chemistry
  of the Sgr B2 Cloud}}.
\newblock \bibinfo{journal}{\apj} \bibinfo{volume}{895}, \bibinfo{pages}{57}.
\newblock \DOIprefix\doi{10.3847/1538-4357/ab8d34},
  \href{http://arxiv.org/abs/2004.12900}{{\tt arXiv:2004.12900}}.
%Type = Article
\bibitem[{{Bachiller} et~al.(1998){Bachiller}, {Codella}, {Colomer}, {Liechti}
  and {Walmsley}}]{Bachiller98}
\bibinfo{author}{{Bachiller}, R.}, \bibinfo{author}{{Codella}, C.},
  \bibinfo{author}{{Colomer}, F.}, \bibinfo{author}{{Liechti}, S.},
  \bibinfo{author}{{Walmsley}, C.M.}, \bibinfo{year}{1998}.
\newblock \bibinfo{title}{{Methanol in protostellar outflows. Single-dish and
  interferometric maps of NGC 1333/IRAS 2}}.
\newblock \bibinfo{journal}{\aap} \bibinfo{volume}{335},
  \bibinfo{pages}{266--276}.
%Type = Article
\bibitem[{{Bachiller} et~al.(2001){Bachiller}, {P{\'e}rez Guti{\'e}rrez},
  {Kumar} and {Tafalla}}]{Bachiller01}
\bibinfo{author}{{Bachiller}, R.}, \bibinfo{author}{{P{\'e}rez Guti{\'e}rrez},
  M.}, \bibinfo{author}{{Kumar}, M.S.N.}, \bibinfo{author}{{Tafalla}, M.},
  \bibinfo{year}{2001}.
\newblock \bibinfo{title}{{Chemically active outflow L 1157}}.
\newblock \bibinfo{journal}{\aap} \bibinfo{volume}{372},
  \bibinfo{pages}{899--912}.
\newblock \DOIprefix\doi{10.1051/0004-6361:20010519}.
%Type = Article
\bibitem[{{Bacmann} et~al.(2012){Bacmann}, {Taquet}, {Faure}, {Kahane} and
  {Ceccarelli}}]{bacmann12}
\bibinfo{author}{{Bacmann}, A.}, \bibinfo{author}{{Taquet}, V.},
  \bibinfo{author}{{Faure}, A.}, \bibinfo{author}{{Kahane}, C.},
  \bibinfo{author}{{Ceccarelli}, C.}, \bibinfo{year}{2012}.
\newblock \bibinfo{title}{{Detection of complex organic molecules in a
  prestellar core: a new challenge for astrochemical models}}.
\newblock \bibinfo{journal}{\aap} \bibinfo{volume}{541}, \bibinfo{pages}{L12}.
\newblock \DOIprefix\doi{10.1051/0004-6361/201219207}.
%Type = Article
\bibitem[{{Baek} et~al.(2022){Baek}, {Lee}, {Hirota}, {Kim} and {Kim}}]{Baek22}
\bibinfo{author}{{Baek}, G.}, \bibinfo{author}{{Lee}, J.E.},
  \bibinfo{author}{{Hirota}, T.}, \bibinfo{author}{{Kim}, K.T.},
  \bibinfo{author}{{Kim}, M.K.}, \bibinfo{year}{2022}.
\newblock \bibinfo{title}{{Complex Organic Molecules Detected in 12 High-mass
  Star-forming Regions with Atacama Large Millimeter/submillimeter Array}}.
\newblock \bibinfo{journal}{\apj} \bibinfo{volume}{939}, \bibinfo{pages}{84}.
\newblock \DOIprefix\doi{10.3847/1538-4357/ac81d3},
  \href{http://arxiv.org/abs/2207.08223}{{\tt arXiv:2207.08223}}.
%Type = Article
\bibitem[{{Ball} et~al.(1970){Ball}, {Gottlieb}, {Lilley} and
  {Radford}}]{Ball70}
\bibinfo{author}{{Ball}, J.A.}, \bibinfo{author}{{Gottlieb}, C.A.},
  \bibinfo{author}{{Lilley}, A.E.}, \bibinfo{author}{{Radford}, H.E.},
  \bibinfo{year}{1970}.
\newblock \bibinfo{title}{{Detection of Methyl Alcohol in Sagittarius}}.
\newblock \bibinfo{journal}{\apjl} \bibinfo{volume}{162},
  \bibinfo{pages}{L203}.
\newblock \DOIprefix\doi{10.1086/180654}.
%Type = Inproceedings
\bibitem[{{Bally} et~al.(2007){Bally}, {Reipurth} and {Davis}}]{Bally07}
\bibinfo{author}{{Bally}, J.}, \bibinfo{author}{{Reipurth}, B.},
  \bibinfo{author}{{Davis}, C.J.}, \bibinfo{year}{2007}.
\newblock \bibinfo{title}{{Observations of Jets and Outflows from Young
  Stars}}, in: \bibinfo{editor}{{Reipurth}, B.}, \bibinfo{editor}{{Jewitt},
  D.}, \bibinfo{editor}{{Keil}, K.} (Eds.), \bibinfo{booktitle}{Protostars and
  Planets V}, p. \bibinfo{pages}{215}.
%Type = Article
\bibitem[{{Bally} et~al.(1987){Bally}, {Stark}, {Wilson} and
  {Henkel}}]{bally87}
\bibinfo{author}{{Bally}, J.}, \bibinfo{author}{{Stark}, A.A.},
  \bibinfo{author}{{Wilson}, R.W.}, \bibinfo{author}{{Henkel}, C.},
  \bibinfo{year}{1987}.
\newblock \bibinfo{title}{{Galactic Center Molecular Clouds. I. Spatial and
  Spatial Velocity Maps}}.
\newblock \bibinfo{journal}{\apjs} \bibinfo{volume}{65}, \bibinfo{pages}{13}.
\newblock \DOIprefix\doi{10.1086/191217}.
%Type = Article
\bibitem[{{Balucani} et~al.(2015){Balucani}, {Ceccarelli} and
  {Taquet}}]{balucani15}
\bibinfo{author}{{Balucani}, N.}, \bibinfo{author}{{Ceccarelli}, C.},
  \bibinfo{author}{{Taquet}, V.}, \bibinfo{year}{2015}.
\newblock \bibinfo{title}{{Formation of complex organic molecules in cold
  objects: the role of gas-phase reactions.}}
\newblock \bibinfo{journal}{\mnras} \bibinfo{volume}{449},
  \bibinfo{pages}{L16--L20}.
\newblock \DOIprefix\doi{10.1093/mnrasl/slv009},
  \href{http://arxiv.org/abs/1501.03668}{{\tt arXiv:1501.03668}}.
%Type = Article
\bibitem[{{Banerjee} et~al.(2020){Banerjee}, {Geballe}, {Evans}, {Shahbandeh},
  {Woodward}, {Gehrz}, {Eyres}, {Starrfield} and {Zijlstra}}]{Banerjee20}
\bibinfo{author}{{Banerjee}, D.P.K.}, \bibinfo{author}{{Geballe}, T.R.},
  \bibinfo{author}{{Evans}, A.}, \bibinfo{author}{{Shahbandeh}, M.},
  \bibinfo{author}{{Woodward}, C.E.}, \bibinfo{author}{{Gehrz}, R.D.},
  \bibinfo{author}{{Eyres}, S.P.S.}, \bibinfo{author}{{Starrfield}, S.},
  \bibinfo{author}{{Zijlstra}, A.}, \bibinfo{year}{2020}.
\newblock \bibinfo{title}{{Near-infrared Spectroscopy of CK Vulpeculae:
  Revealing a Remarkably Powerful Blast from the Past}}.
\newblock \bibinfo{journal}{\apjl} \bibinfo{volume}{904}, \bibinfo{pages}{L23}.
\newblock \DOIprefix\doi{10.3847/2041-8213/abc885},
  \href{http://arxiv.org/abs/2011.02939}{{\tt arXiv:2011.02939}}.
%Type = Article
\bibitem[{{Basalg{\`e}te} et~al.(2021){Basalg{\`e}te}, {Oca{\~n}a},
  {F{\'e}raud}, {Romanzin}, {Philippe}, {Michaut}, {Fillion} and
  {Bertin}}]{Basalgete21}
\bibinfo{author}{{Basalg{\`e}te}, R.}, \bibinfo{author}{{Oca{\~n}a}, A.J.},
  \bibinfo{author}{{F{\'e}raud}, G.}, \bibinfo{author}{{Romanzin}, C.},
  \bibinfo{author}{{Philippe}, L.}, \bibinfo{author}{{Michaut}, X.},
  \bibinfo{author}{{Fillion}, J.H.}, \bibinfo{author}{{Bertin}, M.},
  \bibinfo{year}{2021}.
\newblock \bibinfo{title}{{Photodesorption of Acetonitrile CH$_{3}$CN in
  UV-irradiated Regions of the Interstellar Medium: Experimental Evidence}}.
\newblock \bibinfo{journal}{\apj} \bibinfo{volume}{922}, \bibinfo{pages}{213}.
\newblock \DOIprefix\doi{10.3847/1538-4357/ac2d93},
  \href{http://arxiv.org/abs/2110.04021}{{\tt arXiv:2110.04021}}.
%Type = Article
\bibitem[{Becker et~al.(2019)Becker, Feldmann, Wiedemann, Okamura, Schneider,
  Iwan, Crisp, Rossa, Amatov and Carell}]{becker19}
\bibinfo{author}{Becker, S.}, \bibinfo{author}{Feldmann, J.},
  \bibinfo{author}{Wiedemann, S.}, \bibinfo{author}{Okamura, H.},
  \bibinfo{author}{Schneider, C.}, \bibinfo{author}{Iwan, K.},
  \bibinfo{author}{Crisp, A.}, \bibinfo{author}{Rossa, M.},
  \bibinfo{author}{Amatov, T.}, \bibinfo{author}{Carell, T.},
  \bibinfo{year}{2019}.
\newblock \bibinfo{title}{Unified prebiotically plausible synthesis of
  pyrimidine and purine rna ribonucleotides}.
\newblock \bibinfo{journal}{Science} \bibinfo{volume}{366},
  \bibinfo{pages}{76--82}.
\newblock \DOIprefix\doi{10.1126/science.aax2747}.
%Type = Article
\bibitem[{{Belloche} et~al.(2014){Belloche}, {Garrod}, {M{\"u}ller} and
  {Menten}}]{belloche14}
\bibinfo{author}{{Belloche}, A.}, \bibinfo{author}{{Garrod}, R.T.},
  \bibinfo{author}{{M{\"u}ller}, H.S.P.}, \bibinfo{author}{{Menten}, K.M.},
  \bibinfo{year}{2014}.
\newblock \bibinfo{title}{{Detection of a branched alkyl molecule in the
  interstellar medium: iso-propyl cyanide}}.
\newblock \bibinfo{journal}{Science} \bibinfo{volume}{345},
  \bibinfo{pages}{1584--1587}.
\newblock \DOIprefix\doi{10.1126/science.1256678},
  \href{http://arxiv.org/abs/1410.2607}{{\tt arXiv:1410.2607}}.
%Type = Article
\bibitem[{{Belloche} et~al.(2019){Belloche}, {Garrod}, {M{\"u}ller}, {Menten},
  {Medvedev}, {Thomas} and {Kisiel}}]{Belloche19}
\bibinfo{author}{{Belloche}, A.}, \bibinfo{author}{{Garrod}, R.T.},
  \bibinfo{author}{{M{\"u}ller}, H.S.P.}, \bibinfo{author}{{Menten}, K.M.},
  \bibinfo{author}{{Medvedev}, I.}, \bibinfo{author}{{Thomas}, J.},
  \bibinfo{author}{{Kisiel}, Z.}, \bibinfo{year}{2019}.
\newblock \bibinfo{title}{{Re-exploring Molecular Complexity with ALMA
  (ReMoCA): interstellar detection of urea}}.
\newblock \bibinfo{journal}{\aap} \bibinfo{volume}{628}, \bibinfo{pages}{A10}.
\newblock \DOIprefix\doi{10.1051/0004-6361/201935428},
  \href{http://arxiv.org/abs/1906.04614}{{\tt arXiv:1906.04614}}.
%Type = Article
\bibitem[{{Belloche} et~al.(2022){Belloche}, {Garrod}, {Zingsheim},
  {M{\"u}ller} and {Menten}}]{belloche22}
\bibinfo{author}{{Belloche}, A.}, \bibinfo{author}{{Garrod}, R.T.},
  \bibinfo{author}{{Zingsheim}, O.}, \bibinfo{author}{{M{\"u}ller}, H.S.P.},
  \bibinfo{author}{{Menten}, K.M.}, \bibinfo{year}{2022}.
\newblock \bibinfo{title}{{Interstellar detection and chemical modeling of
  iso-propanol and its normal isomer}}.
\newblock \bibinfo{journal}{\aap} \bibinfo{volume}{662}, \bibinfo{pages}{A110}.
\newblock \DOIprefix\doi{10.1051/0004-6361/202243575},
  \href{http://arxiv.org/abs/2204.09912}{{\tt arXiv:2204.09912}}.
%Type = Article
\bibitem[{{Belloche} et~al.(2020){Belloche}, {Maury}, {Maret}, {Anderl},
  {Bacmann}, {Andr{\'e}}, {Bontemps}, {Cabrit}, {Codella}, {Gaudel}, {Gueth},
  {Lef{\`e}vre}, {Lefloch}, {Podio} and {Testi}}]{Belloche20}
\bibinfo{author}{{Belloche}, A.}, \bibinfo{author}{{Maury}, A.J.},
  \bibinfo{author}{{Maret}, S.}, \bibinfo{author}{{Anderl}, S.},
  \bibinfo{author}{{Bacmann}, A.}, \bibinfo{author}{{Andr{\'e}}, P.},
  \bibinfo{author}{{Bontemps}, S.}, \bibinfo{author}{{Cabrit}, S.},
  \bibinfo{author}{{Codella}, C.}, \bibinfo{author}{{Gaudel}, M.},
  \bibinfo{author}{{Gueth}, F.}, \bibinfo{author}{{Lef{\`e}vre}, C.},
  \bibinfo{author}{{Lefloch}, B.}, \bibinfo{author}{{Podio}, L.},
  \bibinfo{author}{{Testi}, L.}, \bibinfo{year}{2020}.
\newblock \bibinfo{title}{{Questioning the spatial origin of complex organic
  molecules in young protostars with the CALYPSO survey}}.
\newblock \bibinfo{journal}{\aap} \bibinfo{volume}{635}, \bibinfo{pages}{A198}.
\newblock \DOIprefix\doi{10.1051/0004-6361/201937352},
  \href{http://arxiv.org/abs/2002.00592}{{\tt arXiv:2002.00592}}.
%Type = Article
\bibitem[{{Belloche} et~al.(2008){Belloche}, {Menten}, {Comito}, {M{\"u}ller},
  {Schilke}, {Ott}, {Thorwirth} and {Hieret}}]{belloche08}
\bibinfo{author}{{Belloche}, A.}, \bibinfo{author}{{Menten}, K.M.},
  \bibinfo{author}{{Comito}, C.}, \bibinfo{author}{{M{\"u}ller}, H.S.P.},
  \bibinfo{author}{{Schilke}, P.}, \bibinfo{author}{{Ott}, J.},
  \bibinfo{author}{{Thorwirth}, S.}, \bibinfo{author}{{Hieret}, C.},
  \bibinfo{year}{2008}.
\newblock \bibinfo{title}{{Detection of amino acetonitrile in Sgr B2(N)}}.
\newblock \bibinfo{journal}{\aap} \bibinfo{volume}{482},
  \bibinfo{pages}{179--196}.
\newblock \DOIprefix\doi{10.1051/0004-6361:20079203},
  \href{http://arxiv.org/abs/0801.3219}{{\tt arXiv:0801.3219}}.
%Type = Article
\bibitem[{{Belloche} et~al.(2017){Belloche}, {Meshcheryakov}, {Garrod},
  {Ilyushin}, {Alekseev}, {Motiyenko}, {Margul{\`e}s}, {M{\"u}ller} and
  {Menten}}]{Belloche17}
\bibinfo{author}{{Belloche}, A.}, \bibinfo{author}{{Meshcheryakov}, A.A.},
  \bibinfo{author}{{Garrod}, R.T.}, \bibinfo{author}{{Ilyushin}, V.V.},
  \bibinfo{author}{{Alekseev}, E.A.}, \bibinfo{author}{{Motiyenko}, R.A.},
  \bibinfo{author}{{Margul{\`e}s}, L.}, \bibinfo{author}{{M{\"u}ller}, H.S.P.},
  \bibinfo{author}{{Menten}, K.M.}, \bibinfo{year}{2017}.
\newblock \bibinfo{title}{{Rotational spectroscopy, tentative interstellar
  detection, and chemical modeling of N-methylformamide}}.
\newblock \bibinfo{journal}{\aap} \bibinfo{volume}{601}, \bibinfo{pages}{A49}.
\newblock \DOIprefix\doi{10.1051/0004-6361/201629724},
  \href{http://arxiv.org/abs/1701.04640}{{\tt arXiv:1701.04640}}.
%Type = Article
\bibitem[{{Belloche} et~al.(2016){Belloche}, {M{\"u}ller}, {Garrod} and
  {Menten}}]{belloche16}
\bibinfo{author}{{Belloche}, A.}, \bibinfo{author}{{M{\"u}ller}, H.S.P.},
  \bibinfo{author}{{Garrod}, R.T.}, \bibinfo{author}{{Menten}, K.M.},
  \bibinfo{year}{2016}.
\newblock \bibinfo{title}{{Exploring molecular complexity with ALMA (EMoCA):
  Deuterated complex organic molecules in Sagittarius B2(N2)}}.
\newblock \bibinfo{journal}{\aap} \bibinfo{volume}{587}, \bibinfo{pages}{A91}.
\newblock \DOIprefix\doi{10.1051/0004-6361/201527268},
  \href{http://arxiv.org/abs/1511.05721}{{\tt arXiv:1511.05721}}.
%Type = Article
\bibitem[{{Belloche} et~al.(2013){Belloche}, {M{\"u}ller}, {Menten}, {Schilke}
  and {Comito}}]{belloche13}
\bibinfo{author}{{Belloche}, A.}, \bibinfo{author}{{M{\"u}ller}, H.S.P.},
  \bibinfo{author}{{Menten}, K.M.}, \bibinfo{author}{{Schilke}, P.},
  \bibinfo{author}{{Comito}, C.}, \bibinfo{year}{2013}.
\newblock \bibinfo{title}{{Complex organic molecules in the interstellar
  medium: IRAM 30 m line survey of Sagittarius B2(N) and (M)}}.
\newblock \bibinfo{journal}{\aap} \bibinfo{volume}{559}, \bibinfo{pages}{A47}.
\newblock \DOIprefix\doi{10.1051/0004-6361/201321096},
  \href{http://arxiv.org/abs/1308.5062}{{\tt arXiv:1308.5062}}.
%Type = Article
\bibitem[{{Belloche} et~al.(2011){Belloche}, {Schuller}, {Parise}, {Andr{\'e}},
  {Hatchell}, {J{\o}rgensen}, {Bontemps}, {Wei{\ss}}, {Menten} and
  {Muders}}]{belloche11}
\bibinfo{author}{{Belloche}, A.}, \bibinfo{author}{{Schuller}, F.},
  \bibinfo{author}{{Parise}, B.}, \bibinfo{author}{{Andr{\'e}}, P.},
  \bibinfo{author}{{Hatchell}, J.}, \bibinfo{author}{{J{\o}rgensen}, J.K.},
  \bibinfo{author}{{Bontemps}, S.}, \bibinfo{author}{{Wei{\ss}}, A.},
  \bibinfo{author}{{Menten}, K.M.}, \bibinfo{author}{{Muders}, D.},
  \bibinfo{year}{2011}.
\newblock \bibinfo{title}{{The end of star formation in Chamaeleon I?. A LABOCA
  census of starless and protostellar cores}}.
\newblock \bibinfo{journal}{\aap} \bibinfo{volume}{527}, \bibinfo{pages}{A145}.
\newblock \DOIprefix\doi{10.1051/0004-6361/201015733},
  \href{http://arxiv.org/abs/1101.0718}{{\tt arXiv:1101.0718}}.
%Type = Article
\bibitem[{{Beltr{\'a}n} et~al.(2018){Beltr{\'a}n}, {Cesaroni}, {Rivilla},
  {S{\'a}nchez-Monge}, {Moscadelli}, {Ahmadi}, {Allen}, {Beuther}, {Etoka},
  {Galli}, {Galv{\'a}n-Madrid}, {Goddi}, {Johnston}, {Klaassen},
  {K{\"o}lligan}, {Kuiper}, {Kumar}, {Maud}, {Mottram}, {Peters}, {Schilke},
  {Testi}, {van der Tak} and {Walmsley}}]{Beltran18}
\bibinfo{author}{{Beltr{\'a}n}, M.T.}, \bibinfo{author}{{Cesaroni}, R.},
  \bibinfo{author}{{Rivilla}, V.M.}, \bibinfo{author}{{S{\'a}nchez-Monge},
  {\'A}.}, \bibinfo{author}{{Moscadelli}, L.}, \bibinfo{author}{{Ahmadi}, A.},
  \bibinfo{author}{{Allen}, V.}, \bibinfo{author}{{Beuther}, H.},
  \bibinfo{author}{{Etoka}, S.}, \bibinfo{author}{{Galli}, D.},
  \bibinfo{author}{{Galv{\'a}n-Madrid}, R.}, \bibinfo{author}{{Goddi}, C.},
  \bibinfo{author}{{Johnston}, K.G.}, \bibinfo{author}{{Klaassen}, P.D.},
  \bibinfo{author}{{K{\"o}lligan}, A.}, \bibinfo{author}{{Kuiper}, R.},
  \bibinfo{author}{{Kumar}, M.S.N.}, \bibinfo{author}{{Maud}, L.T.},
  \bibinfo{author}{{Mottram}, J.C.}, \bibinfo{author}{{Peters}, T.},
  \bibinfo{author}{{Schilke}, P.}, \bibinfo{author}{{Testi}, L.},
  \bibinfo{author}{{van der Tak}, F.}, \bibinfo{author}{{Walmsley}, C.M.},
  \bibinfo{year}{2018}.
\newblock \bibinfo{title}{{Accelerating infall and rotational spin-up in the
  hot molecular core G31.41+0.31}}.
\newblock \bibinfo{journal}{\aap} \bibinfo{volume}{615}, \bibinfo{pages}{A141}.
\newblock \DOIprefix\doi{10.1051/0004-6361/201832811},
  \href{http://arxiv.org/abs/1803.05300}{{\tt arXiv:1803.05300}}.
%Type = Article
\bibitem[{{Beltr{\'a}n} et~al.(2009){Beltr{\'a}n}, {Codella}, {Viti}, {Neri}
  and {Cesaroni}}]{beltran09}
\bibinfo{author}{{Beltr{\'a}n}, M.T.}, \bibinfo{author}{{Codella}, C.},
  \bibinfo{author}{{Viti}, S.}, \bibinfo{author}{{Neri}, R.},
  \bibinfo{author}{{Cesaroni}, R.}, \bibinfo{year}{2009}.
\newblock \bibinfo{title}{{First Detection of Glycolaldehyde Outside the
  Galactic Center}}.
\newblock \bibinfo{journal}{\apjl} \bibinfo{volume}{690},
  \bibinfo{pages}{L93--L96}.
\newblock \DOIprefix\doi{10.1088/0004-637X/690/2/L93},
  \href{http://arxiv.org/abs/0811.3821}{{\tt arXiv:0811.3821}}.
%Type = Article
\bibitem[{{Beltr{\'a}n} et~al.(2022){Beltr{\'a}n}, {Rivilla}, {Cesaroni},
  {Galli}, {Moscadelli}, {Ahmadi}, {Beuther}, {Etoka}, {Goddi}, {Klaassen},
  {Kuiper}, {Kumar}, {Lorenzani}, {Peters}, {S{\'a}nchez-Monge}, {Schilke},
  {van der Tak} and {Vig}}]{beltran22}
\bibinfo{author}{{Beltr{\'a}n}, M.T.}, \bibinfo{author}{{Rivilla}, V.M.},
  \bibinfo{author}{{Cesaroni}, R.}, \bibinfo{author}{{Galli}, D.},
  \bibinfo{author}{{Moscadelli}, L.}, \bibinfo{author}{{Ahmadi}, A.},
  \bibinfo{author}{{Beuther}, H.}, \bibinfo{author}{{Etoka}, S.},
  \bibinfo{author}{{Goddi}, C.}, \bibinfo{author}{{Klaassen}, P.D.},
  \bibinfo{author}{{Kuiper}, R.}, \bibinfo{author}{{Kumar}, M.S.N.},
  \bibinfo{author}{{Lorenzani}, A.}, \bibinfo{author}{{Peters}, T.},
  \bibinfo{author}{{S{\'a}nchez-Monge}, {\'A}.}, \bibinfo{author}{{Schilke},
  P.}, \bibinfo{author}{{van der Tak}, F.}, \bibinfo{author}{{Vig}, S.},
  \bibinfo{year}{2022}.
\newblock \bibinfo{title}{{The sharp ALMA view of infall and outflow in the
  massive protocluster G31.41+0.31}}.
\newblock \bibinfo{journal}{\aap} \bibinfo{volume}{659}, \bibinfo{pages}{A81}.
\newblock \DOIprefix\doi{10.1051/0004-6361/202142703},
  \href{http://arxiv.org/abs/2201.10438}{{\tt arXiv:2201.10438}}.
%Type = Article
\bibitem[{{Benedettini} et~al.(2012){Benedettini}, {Busquet}, {Lefloch},
  {Codella}, {Cabrit}, {Ceccarelli}, {Giannini}, {Nisini}, {Vasta},
  {Cernicharo}, {Lorenzani} and {di Giorgio}}]{Benedettini12}
\bibinfo{author}{{Benedettini}, M.}, \bibinfo{author}{{Busquet}, G.},
  \bibinfo{author}{{Lefloch}, B.}, \bibinfo{author}{{Codella}, C.},
  \bibinfo{author}{{Cabrit}, S.}, \bibinfo{author}{{Ceccarelli}, C.},
  \bibinfo{author}{{Giannini}, T.}, \bibinfo{author}{{Nisini}, B.},
  \bibinfo{author}{{Vasta}, M.}, \bibinfo{author}{{Cernicharo}, J.},
  \bibinfo{author}{{Lorenzani}, A.}, \bibinfo{author}{{di Giorgio}, A.M.},
  \bibinfo{year}{2012}.
\newblock \bibinfo{title}{{The CHESS survey of the L1157-B1 shock: the
  dissociative jet shock as revealed by Herschel-PACS}}.
\newblock \bibinfo{journal}{\aap} \bibinfo{volume}{539}, \bibinfo{pages}{L3}.
\newblock \DOIprefix\doi{10.1051/0004-6361/201118732},
  \href{http://arxiv.org/abs/1202.1451}{{\tt arXiv:1202.1451}}.
%Type = Article
\bibitem[{{Bergin} and {Langer}(1997)}]{bergin97}
\bibinfo{author}{{Bergin}, E.A.}, \bibinfo{author}{{Langer}, W.D.},
  \bibinfo{year}{1997}.
\newblock \bibinfo{title}{{Chemical Evolution in Preprotostellar and
  Protostellar Cores}}.
\newblock \bibinfo{journal}{\apj} \bibinfo{volume}{486},
  \bibinfo{pages}{316--328}.
\newblock \DOIprefix\doi{10.1086/304510}.
%Type = Article
\bibitem[{{Bergner} et~al.(2018){Bergner}, {Guzm{\'a}n}, {{\"O}berg}, {Loomis}
  and {Pegues}}]{Bergner18}
\bibinfo{author}{{Bergner}, J.B.}, \bibinfo{author}{{Guzm{\'a}n}, V.G.},
  \bibinfo{author}{{{\"O}berg}, K.I.}, \bibinfo{author}{{Loomis}, R.A.},
  \bibinfo{author}{{Pegues}, J.}, \bibinfo{year}{2018}.
\newblock \bibinfo{title}{{A Survey of CH$_{3}$CN and HC$_{3}$N in
  Protoplanetary Disks}}.
\newblock \bibinfo{journal}{\apj} \bibinfo{volume}{857}, \bibinfo{pages}{69}.
\newblock \DOIprefix\doi{10.3847/1538-4357/aab664},
  \href{http://arxiv.org/abs/1803.04986}{{\tt arXiv:1803.04986}}.
%Type = Article
\bibitem[{{Bergner} et~al.(2019){Bergner}, {Mart{\'\i}n-Dom{\'e}nech},
  {{\"O}berg}, {J{\o}rgensen}, {Artur de la Villarmois} and
  {Brinch}}]{Bergner19}
\bibinfo{author}{{Bergner}, J.B.}, \bibinfo{author}{{Mart{\'\i}n-Dom{\'e}nech},
  R.}, \bibinfo{author}{{{\"O}berg}, K.I.}, \bibinfo{author}{{J{\o}rgensen},
  J.K.}, \bibinfo{author}{{Artur de la Villarmois}, E.},
  \bibinfo{author}{{Brinch}, C.}, \bibinfo{year}{2019}.
\newblock \bibinfo{title}{{Organic Complexity in Protostellar Disk
  Candidates}}.
\newblock \bibinfo{journal}{ACS Earth and Space Chemistry} \bibinfo{volume}{3},
  \bibinfo{pages}{1564--1575}.
\newblock \DOIprefix\doi{10.1021/acsearthspacechem.9b00059},
  \href{http://arxiv.org/abs/1907.07791}{{\tt arXiv:1907.07791}}.
%Type = Article
\bibitem[{{Bertin} et~al.(2016){Bertin}, {Romanzin}, {Doronin}, {Philippe},
  {Jeseck}, {Ligterink}, {Linnartz}, {Michaut} and {Fillion}}]{Bertin16}
\bibinfo{author}{{Bertin}, M.}, \bibinfo{author}{{Romanzin}, C.},
  \bibinfo{author}{{Doronin}, M.}, \bibinfo{author}{{Philippe}, L.},
  \bibinfo{author}{{Jeseck}, P.}, \bibinfo{author}{{Ligterink}, N.},
  \bibinfo{author}{{Linnartz}, H.}, \bibinfo{author}{{Michaut}, X.},
  \bibinfo{author}{{Fillion}, J.H.}, \bibinfo{year}{2016}.
\newblock \bibinfo{title}{{UV Photodesorption of Methanol in Pure and CO-rich
  Ices: Desorption Rates of the Intact Molecule and of the Photofragments}}.
\newblock \bibinfo{journal}{\apjl} \bibinfo{volume}{817}, \bibinfo{pages}{L12}.
\newblock \DOIprefix\doi{10.3847/2041-8205/817/2/L12},
  \href{http://arxiv.org/abs/1601.07027}{{\tt arXiv:1601.07027}}.
%Type = Article
\bibitem[{{Beuther} et~al.(2018){Beuther}, {Mottram}, {Ahmadi}, {Bosco},
  {Linz}, {Henning}, {Klaassen}, {Winters}, {Maud}, {Kuiper}, {Semenov},
  {Gieser}, {Peters}, {Urquhart}, {Pudritz}, {Ragan}, {Feng}, {Keto},
  {Leurini}, {Cesaroni}, {Beltran}, {Palau}, {S{\'a}nchez-Monge},
  {Galvan-Madrid}, {Zhang}, {Schilke}, {Wyrowski}, {Johnston}, {Longmore},
  {Lumsden}, {Hoare}, {Menten} and {Csengeri}}]{Beuther18}
\bibinfo{author}{{Beuther}, H.}, \bibinfo{author}{{Mottram}, J.C.},
  \bibinfo{author}{{Ahmadi}, A.}, \bibinfo{author}{{Bosco}, F.},
  \bibinfo{author}{{Linz}, H.}, \bibinfo{author}{{Henning}, T.},
  \bibinfo{author}{{Klaassen}, P.}, \bibinfo{author}{{Winters}, J.M.},
  \bibinfo{author}{{Maud}, L.T.}, \bibinfo{author}{{Kuiper}, R.},
  \bibinfo{author}{{Semenov}, D.}, \bibinfo{author}{{Gieser}, C.},
  \bibinfo{author}{{Peters}, T.}, \bibinfo{author}{{Urquhart}, J.S.},
  \bibinfo{author}{{Pudritz}, R.}, \bibinfo{author}{{Ragan}, S.E.},
  \bibinfo{author}{{Feng}, S.}, \bibinfo{author}{{Keto}, E.},
  \bibinfo{author}{{Leurini}, S.}, \bibinfo{author}{{Cesaroni}, R.},
  \bibinfo{author}{{Beltran}, M.}, \bibinfo{author}{{Palau}, A.},
  \bibinfo{author}{{S{\'a}nchez-Monge}, {\'A}.},
  \bibinfo{author}{{Galvan-Madrid}, R.}, \bibinfo{author}{{Zhang}, Q.},
  \bibinfo{author}{{Schilke}, P.}, \bibinfo{author}{{Wyrowski}, F.},
  \bibinfo{author}{{Johnston}, K.G.}, \bibinfo{author}{{Longmore}, S.N.},
  \bibinfo{author}{{Lumsden}, S.}, \bibinfo{author}{{Hoare}, M.},
  \bibinfo{author}{{Menten}, K.M.}, \bibinfo{author}{{Csengeri}, T.},
  \bibinfo{year}{2018}.
\newblock \bibinfo{title}{{Fragmentation and disk formation during high-mass
  star formation. IRAM NOEMA (Northern Extended Millimeter Array) large program
  CORE}}.
\newblock \bibinfo{journal}{\aap} \bibinfo{volume}{617}, \bibinfo{pages}{A100}.
\newblock \DOIprefix\doi{10.1051/0004-6361/201833021},
  \href{http://arxiv.org/abs/1805.01191}{{\tt arXiv:1805.01191}}.
%Type = Article
\bibitem[{{Bianchi} et~al.(2019b){Bianchi}, {Ceccarelli}, {Codella},
  {Enrique-Romero}, {Favre} and {Lefloch}}]{Bianchi19b}
\bibinfo{author}{{Bianchi}, E.}, \bibinfo{author}{{Ceccarelli}, C.},
  \bibinfo{author}{{Codella}, C.}, \bibinfo{author}{{Enrique-Romero}, J.},
  \bibinfo{author}{{Favre}, C.}, \bibinfo{author}{{Lefloch}, B.},
  \bibinfo{year}{2019b}.
\newblock \bibinfo{title}{{Astrochemistry as a Tool To Follow Protostellar
  Evolution: The Class I Stage}}.
\newblock \bibinfo{journal}{ACS Earth and Space Chemistry} \bibinfo{volume}{3},
  \bibinfo{pages}{2659--2674}.
\newblock \DOIprefix\doi{10.1021/acsearthspacechem.9b00158},
  \href{http://arxiv.org/abs/1911.08991}{{\tt arXiv:1911.08991}}.
%Type = Article
\bibitem[{{Bianchi} et~al.(2020){Bianchi}, {Chandler}, {Ceccarelli}, {Codella},
  {Sakai}, {L{\'o}pez-Sepulcre}, {Maud}, {Moellenbrock}, {Svoboda}, {Watanabe},
  {Sakai}, {M{\'e}nard}, {Aikawa}, {Alves}, {Balucani}, {Bouvier}, {Caselli},
  {Caux}, {Charnley}, {Choudhury}, {De Simone}, {Dulieu}, {Dur{\'a}n}, {Evans},
  {Favre}, {Fedele}, {Feng}, {Fontani}, {Francis}, {Hama}, {Hanawa}, {Herbst},
  {Hirota}, {Imai}, {Isella}, {Jim{\'e}nez-Serra}, {Johnstone}, {Kahane},
  {Lefloch}, {Loinard}, {Maureira}, {Mercimek}, {Miotello}, {Mori}, {Nakatani},
  {Nomura}, {Oba}, {Ohashi}, {Okoda}, {Ospina-Zamudio}, {Oya}, {Pineda},
  {Podio}, {Rimola}, {Cox}, {Shirley}, {Taquet}, {Testi}, {Vastel}, {Viti},
  {Watanabe}, {Witzel}, {Xue}, {Zhang}, {Zhao} and {Yamamoto}}]{Bianchi20}
\bibinfo{author}{{Bianchi}, E.}, \bibinfo{author}{{Chandler}, C.J.},
  \bibinfo{author}{{Ceccarelli}, C.}, \bibinfo{author}{{Codella}, C.},
  \bibinfo{author}{{Sakai}, N.}, \bibinfo{author}{{L{\'o}pez-Sepulcre}, A.},
  \bibinfo{author}{{Maud}, L.T.}, \bibinfo{author}{{Moellenbrock}, G.},
  \bibinfo{author}{{Svoboda}, B.}, \bibinfo{author}{{Watanabe}, Y.},
  \bibinfo{author}{{Sakai}, T.}, \bibinfo{author}{{M{\'e}nard}, F.},
  \bibinfo{author}{{Aikawa}, Y.}, \bibinfo{author}{{Alves}, F.},
  \bibinfo{author}{{Balucani}, N.}, \bibinfo{author}{{Bouvier}, M.},
  \bibinfo{author}{{Caselli}, P.}, \bibinfo{author}{{Caux}, E.},
  \bibinfo{author}{{Charnley}, S.}, \bibinfo{author}{{Choudhury}, S.},
  \bibinfo{author}{{De Simone}, M.}, \bibinfo{author}{{Dulieu}, F.},
  \bibinfo{author}{{Dur{\'a}n}, A.}, \bibinfo{author}{{Evans}, L.},
  \bibinfo{author}{{Favre}, C.}, \bibinfo{author}{{Fedele}, D.},
  \bibinfo{author}{{Feng}, S.}, \bibinfo{author}{{Fontani}, F.},
  \bibinfo{author}{{Francis}, L.}, \bibinfo{author}{{Hama}, T.},
  \bibinfo{author}{{Hanawa}, T.}, \bibinfo{author}{{Herbst}, E.},
  \bibinfo{author}{{Hirota}, T.}, \bibinfo{author}{{Imai}, M.},
  \bibinfo{author}{{Isella}, A.}, \bibinfo{author}{{Jim{\'e}nez-Serra}, I.},
  \bibinfo{author}{{Johnstone}, D.}, \bibinfo{author}{{Kahane}, C.},
  \bibinfo{author}{{Lefloch}, B.}, \bibinfo{author}{{Loinard}, L.},
  \bibinfo{author}{{Maureira}, M.J.}, \bibinfo{author}{{Mercimek}, S.},
  \bibinfo{author}{{Miotello}, A.}, \bibinfo{author}{{Mori}, S.},
  \bibinfo{author}{{Nakatani}, R.}, \bibinfo{author}{{Nomura}, H.},
  \bibinfo{author}{{Oba}, Y.}, \bibinfo{author}{{Ohashi}, S.},
  \bibinfo{author}{{Okoda}, Y.}, \bibinfo{author}{{Ospina-Zamudio}, J.},
  \bibinfo{author}{{Oya}, Y.}, \bibinfo{author}{{Pineda}, J.},
  \bibinfo{author}{{Podio}, L.}, \bibinfo{author}{{Rimola}, A.},
  \bibinfo{author}{{Cox}, D.S.}, \bibinfo{author}{{Shirley}, Y.},
  \bibinfo{author}{{Taquet}, V.}, \bibinfo{author}{{Testi}, L.},
  \bibinfo{author}{{Vastel}, C.}, \bibinfo{author}{{Viti}, S.},
  \bibinfo{author}{{Watanabe}, N.}, \bibinfo{author}{{Witzel}, A.},
  \bibinfo{author}{{Xue}, C.}, \bibinfo{author}{{Zhang}, Y.},
  \bibinfo{author}{{Zhao}, B.}, \bibinfo{author}{{Yamamoto}, S.},
  \bibinfo{year}{2020}.
\newblock \bibinfo{title}{{FAUST I. The hot corino at the heart of the
  prototypical Class I protostar L1551 IRS5}}.
\newblock \bibinfo{journal}{\mnras} \bibinfo{volume}{498},
  \bibinfo{pages}{L87--L92}.
\newblock \DOIprefix\doi{10.1093/mnrasl/slaa130},
  \href{http://arxiv.org/abs/2007.10275}{{\tt arXiv:2007.10275}}.
%Type = Article
\bibitem[{{Bianchi} et~al.(2019a){Bianchi}, {Codella}, {Ceccarelli}, {Vazart},
  {Bachiller}, {Balucani}, {Bouvier}, {De Simone}, {Enrique-Romero}, {Kahane},
  {Lefloch}, {L{\'o}pez-Sepulcre}, {Ospina-Zamudio}, {Podio} and
  {Taquet}}]{Bianchi19a}
\bibinfo{author}{{Bianchi}, E.}, \bibinfo{author}{{Codella}, C.},
  \bibinfo{author}{{Ceccarelli}, C.}, \bibinfo{author}{{Vazart}, F.},
  \bibinfo{author}{{Bachiller}, R.}, \bibinfo{author}{{Balucani}, N.},
  \bibinfo{author}{{Bouvier}, M.}, \bibinfo{author}{{De Simone}, M.},
  \bibinfo{author}{{Enrique-Romero}, J.}, \bibinfo{author}{{Kahane}, C.},
  \bibinfo{author}{{Lefloch}, B.}, \bibinfo{author}{{L{\'o}pez-Sepulcre}, A.},
  \bibinfo{author}{{Ospina-Zamudio}, J.}, \bibinfo{author}{{Podio}, L.},
  \bibinfo{author}{{Taquet}, V.}, \bibinfo{year}{2019a}.
\newblock \bibinfo{title}{{The census of interstellar complex organic molecules
  in the Class I hot corino of SVS13-A}}.
\newblock \bibinfo{journal}{\mnras} \bibinfo{volume}{483},
  \bibinfo{pages}{1850--1861}.
\newblock \DOIprefix\doi{10.1093/mnras/sty2915},
  \href{http://arxiv.org/abs/1810.11411}{{\tt arXiv:1810.11411}}.
%Type = Article
\bibitem[{{Bianchi} et~al.(2022){Bianchi}, {L{\'o}pez-Sepulcre}, {Ceccarelli},
  {Codella}, {Podio}, {Bouvier} and {Enrique-Romero}}]{Bianchi22}
\bibinfo{author}{{Bianchi}, E.}, \bibinfo{author}{{L{\'o}pez-Sepulcre}, A.},
  \bibinfo{author}{{Ceccarelli}, C.}, \bibinfo{author}{{Codella}, C.},
  \bibinfo{author}{{Podio}, L.}, \bibinfo{author}{{Bouvier}, M.},
  \bibinfo{author}{{Enrique-Romero}, J.}, \bibinfo{year}{2022}.
\newblock \bibinfo{title}{{The Two Hot Corinos of the SVS13-A Protostellar
  Binary System: Counterposed Siblings}}.
\newblock \bibinfo{journal}{\apjl} \bibinfo{volume}{928}, \bibinfo{pages}{L3}.
\newblock \DOIprefix\doi{10.3847/2041-8213/ac5a56},
  \href{http://arxiv.org/abs/2203.03412}{{\tt arXiv:2203.03412}}.
%Type = Article
\bibitem[{{Bizzocchi} et~al.(2014){Bizzocchi}, {Caselli}, {Spezzano} and
  {Leonardo}}]{bizzocchi14}
\bibinfo{author}{{Bizzocchi}, L.}, \bibinfo{author}{{Caselli}, P.},
  \bibinfo{author}{{Spezzano}, S.}, \bibinfo{author}{{Leonardo}, E.},
  \bibinfo{year}{2014}.
\newblock \bibinfo{title}{{Deuterated methanol in the pre-stellar core L1544}}.
\newblock \bibinfo{journal}{\aap} \bibinfo{volume}{569}, \bibinfo{pages}{A27}.
\newblock \DOIprefix\doi{10.1051/0004-6361/201423858},
  \href{http://arxiv.org/abs/1408.2491}{{\tt arXiv:1408.2491}}.
%Type = Article
\bibitem[{{Blake} et~al.(1995){Blake}, {Sandell}, {van Dishoeck}, {Groesbeck},
  {Mundy} and {Aspin}}]{Blake95}
\bibinfo{author}{{Blake}, G.A.}, \bibinfo{author}{{Sandell}, G.},
  \bibinfo{author}{{van Dishoeck}, E.F.}, \bibinfo{author}{{Groesbeck}, T.D.},
  \bibinfo{author}{{Mundy}, L.G.}, \bibinfo{author}{{Aspin}, C.},
  \bibinfo{year}{1995}.
\newblock \bibinfo{title}{{A Molecular Line Study of NGC 1333/IRAS 4}}.
\newblock \bibinfo{journal}{\apj} \bibinfo{volume}{441}, \bibinfo{pages}{689}.
\newblock \DOIprefix\doi{10.1086/175392}.
%Type = Article
\bibitem[{{Blake} et~al.(1986){Blake}, {Sutton}, {Masson} and
  {Phillips}}]{blake86}
\bibinfo{author}{{Blake}, G.A.}, \bibinfo{author}{{Sutton}, E.C.},
  \bibinfo{author}{{Masson}, C.R.}, \bibinfo{author}{{Phillips}, T.G.},
  \bibinfo{year}{1986}.
\newblock \bibinfo{title}{{The Rotational Emission-Line Spectrum of Orion A
  between 247 and 263 GHz}}.
\newblock \bibinfo{journal}{\apjs} \bibinfo{volume}{60}, \bibinfo{pages}{357}.
\newblock \DOIprefix\doi{10.1086/191090}.
%Type = Article
\bibitem[{{Blake} et~al.(1987){Blake}, {Sutton}, {Masson} and
  {Phillips}}]{blake87}
\bibinfo{author}{{Blake}, G.A.}, \bibinfo{author}{{Sutton}, E.C.},
  \bibinfo{author}{{Masson}, C.R.}, \bibinfo{author}{{Phillips}, T.G.},
  \bibinfo{year}{1987}.
\newblock \bibinfo{title}{{Molecular Abundances in OMC-1: The Chemical
  Composition of Interstellar Molecular Clouds and the Influence of Massive
  Star Formation}}.
\newblock \bibinfo{journal}{\apj} \bibinfo{volume}{315}, \bibinfo{pages}{621}.
\newblock \DOIprefix\doi{10.1086/165165}.
%Type = Article
\bibitem[{{B{\o}gelund} et~al.(2019){B{\o}gelund}, {Barr}, {Taquet},
  {Ligterink}, {Persson}, {Hogerheijde} and {van Dishoeck}}]{Bogelund19}
\bibinfo{author}{{B{\o}gelund}, E.G.}, \bibinfo{author}{{Barr}, A.G.},
  \bibinfo{author}{{Taquet}, V.}, \bibinfo{author}{{Ligterink}, N.F.W.},
  \bibinfo{author}{{Persson}, M.V.}, \bibinfo{author}{{Hogerheijde}, M.R.},
  \bibinfo{author}{{van Dishoeck}, E.F.}, \bibinfo{year}{2019}.
\newblock \bibinfo{title}{{Molecular complexity on disc scales uncovered by
  ALMA. Chemical composition of the high-mass protostar AFGL 4176}}.
\newblock \bibinfo{journal}{\aap} \bibinfo{volume}{628}, \bibinfo{pages}{A2}.
\newblock \DOIprefix\doi{10.1051/0004-6361/201834527},
  \href{http://arxiv.org/abs/1906.06156}{{\tt arXiv:1906.06156}}.
%Type = Article
\bibitem[{{Bonfand} et~al.(2019){Bonfand}, {Belloche}, {Garrod}, {Menten},
  {Willis}, {St{\'e}phan} and {M{\"u}ller}}]{Bonfand19}
\bibinfo{author}{{Bonfand}, M.}, \bibinfo{author}{{Belloche}, A.},
  \bibinfo{author}{{Garrod}, R.T.}, \bibinfo{author}{{Menten}, K.M.},
  \bibinfo{author}{{Willis}, E.}, \bibinfo{author}{{St{\'e}phan}, G.},
  \bibinfo{author}{{M{\"u}ller}, H.S.P.}, \bibinfo{year}{2019}.
\newblock \bibinfo{title}{{The complex chemistry of hot cores in Sgr B2(N):
  influence of cosmic-ray ionization and thermal history}}.
\newblock \bibinfo{journal}{\aap} \bibinfo{volume}{628}, \bibinfo{pages}{A27}.
\newblock \DOIprefix\doi{10.1051/0004-6361/201935523},
  \href{http://arxiv.org/abs/1906.04695}{{\tt arXiv:1906.04695}}.
%Type = Article
\bibitem[{{Bonfand} et~al.(2017){Bonfand}, {Belloche}, {Menten}, {Garrod} and
  {M{\"u}ller}}]{Bonfand17}
\bibinfo{author}{{Bonfand}, M.}, \bibinfo{author}{{Belloche}, A.},
  \bibinfo{author}{{Menten}, K.M.}, \bibinfo{author}{{Garrod}, R.T.},
  \bibinfo{author}{{M{\"u}ller}, H.S.P.}, \bibinfo{year}{2017}.
\newblock \bibinfo{title}{{Exploring molecular complexity with ALMA (EMoCA):
  Detection of three new hot cores in Sagittarius B2(N)}}.
\newblock \bibinfo{journal}{\aap} \bibinfo{volume}{604}, \bibinfo{pages}{A60}.
\newblock \DOIprefix\doi{10.1051/0004-6361/201730648},
  \href{http://arxiv.org/abs/1703.09544}{{\tt arXiv:1703.09544}}.
%Type = Article
\bibitem[{{Booth} et~al.(2021){Booth}, {Walsh}, {Terwisscha van Scheltinga},
  {van Dishoeck}, {Ilee}, {Hogerheijde}, {Kama} and {Nomura}}]{Booth21}
\bibinfo{author}{{Booth}, A.S.}, \bibinfo{author}{{Walsh}, C.},
  \bibinfo{author}{{Terwisscha van Scheltinga}, J.}, \bibinfo{author}{{van
  Dishoeck}, E.F.}, \bibinfo{author}{{Ilee}, J.D.},
  \bibinfo{author}{{Hogerheijde}, M.R.}, \bibinfo{author}{{Kama}, M.},
  \bibinfo{author}{{Nomura}, H.}, \bibinfo{year}{2021}.
\newblock \bibinfo{title}{{An inherited complex organic molecule reservoir in a
  warm planet-hosting disk}}.
\newblock \bibinfo{journal}{Nature Astronomy} \bibinfo{volume}{5},
  \bibinfo{pages}{684--690}.
\newblock \DOIprefix\doi{10.1038/s41550-021-01352-w},
  \href{http://arxiv.org/abs/2104.08348}{{\tt arXiv:2104.08348}}.
%Type = Article
\bibitem[{{Bottinelli} et~al.(2004a){Bottinelli}, {Ceccarelli}, {Lefloch},
  {Williams}, {Castets}, {Caux}, {Cazaux}, {Maret}, {Parise} and
  {Tielens}}]{bottinelli04a}
\bibinfo{author}{{Bottinelli}, S.}, \bibinfo{author}{{Ceccarelli}, C.},
  \bibinfo{author}{{Lefloch}, B.}, \bibinfo{author}{{Williams}, J.P.},
  \bibinfo{author}{{Castets}, A.}, \bibinfo{author}{{Caux}, E.},
  \bibinfo{author}{{Cazaux}, S.}, \bibinfo{author}{{Maret}, S.},
  \bibinfo{author}{{Parise}, B.}, \bibinfo{author}{{Tielens}, A.G.G.M.},
  \bibinfo{year}{2004}a.
\newblock \bibinfo{title}{{Complex Molecules in the Hot Core of the Low-Mass
  Protostar NGC 1333 IRAS 4A}}.
\newblock \bibinfo{journal}{\apj} \bibinfo{volume}{615},
  \bibinfo{pages}{354--358}.
\newblock \DOIprefix\doi{10.1086/423952},
  \href{http://arxiv.org/abs/astro-ph/0407154}{{\tt arXiv:astro-ph/0407154}}.
%Type = Article
\bibitem[{{Bottinelli} et~al.(2004b){Bottinelli}, {Ceccarelli}, {Neri},
  {Williams}, {Caux}, {Cazaux}, {Lefloch}, {Maret} and
  {Tielens}}]{bottinelli04b}
\bibinfo{author}{{Bottinelli}, S.}, \bibinfo{author}{{Ceccarelli}, C.},
  \bibinfo{author}{{Neri}, R.}, \bibinfo{author}{{Williams}, J.P.},
  \bibinfo{author}{{Caux}, E.}, \bibinfo{author}{{Cazaux}, S.},
  \bibinfo{author}{{Lefloch}, B.}, \bibinfo{author}{{Maret}, S.},
  \bibinfo{author}{{Tielens}, A.G.G.M.}, \bibinfo{year}{2004}b.
\newblock \bibinfo{title}{{Near-Arcsecond Resolution Observations of the Hot
  Corino of the Solar-Type Protostar IRAS 16293-2422}}.
\newblock \bibinfo{journal}{\apjl} \bibinfo{volume}{617},
  \bibinfo{pages}{L69--L72}.
\newblock \DOIprefix\doi{10.1086/426964},
  \href{http://arxiv.org/abs/astro-ph/0410601}{{\tt arXiv:astro-ph/0410601}}.
%Type = Article
\bibitem[{{Bouvier} et~al.(2020){Bouvier}, {L{\'o}pez-Sepulcre}, {Ceccarelli},
  {Kahane}, {Imai}, {Sakai}, {Yamamoto} and {Dagdigian}}]{Bouvier20}
\bibinfo{author}{{Bouvier}, M.}, \bibinfo{author}{{L{\'o}pez-Sepulcre}, A.},
  \bibinfo{author}{{Ceccarelli}, C.}, \bibinfo{author}{{Kahane}, C.},
  \bibinfo{author}{{Imai}, M.}, \bibinfo{author}{{Sakai}, N.},
  \bibinfo{author}{{Yamamoto}, S.}, \bibinfo{author}{{Dagdigian}, P.J.},
  \bibinfo{year}{2020}.
\newblock \bibinfo{title}{{Hunting for hot corinos and WCCC sources in the
  OMC-2/3 filament}}.
\newblock \bibinfo{journal}{\aap} \bibinfo{volume}{636}, \bibinfo{pages}{A19}.
\newblock \DOIprefix\doi{10.1051/0004-6361/201937164},
  \href{http://arxiv.org/abs/2003.06198}{{\tt arXiv:2003.06198}}.
%Type = Article
\bibitem[{{Brogan} et~al.(2016){Brogan}, {Hunter}, {Cyganowski}, {Chandler},
  {Friesen} and {Indebetouw}}]{Brogan16}
\bibinfo{author}{{Brogan}, C.L.}, \bibinfo{author}{{Hunter}, T.R.},
  \bibinfo{author}{{Cyganowski}, C.J.}, \bibinfo{author}{{Chandler}, C.J.},
  \bibinfo{author}{{Friesen}, R.}, \bibinfo{author}{{Indebetouw}, R.},
  \bibinfo{year}{2016}.
\newblock \bibinfo{title}{{The Massive Protostellar Cluster NGC 6334I at 220 au
  Resolution: Discovery of Further Multiplicity, Diversity, and a Hot
  Multi-core}}.
\newblock \bibinfo{journal}{\apj} \bibinfo{volume}{832}, \bibinfo{pages}{187}.
\newblock \DOIprefix\doi{10.3847/0004-637X/832/2/187},
  \href{http://arxiv.org/abs/1609.07470}{{\tt arXiv:1609.07470}}.
%Type = Article
\bibitem[{{Brunken} et~al.(2022){Brunken}, {Booth}, {Leemker}, {Nazari}, {van
  der Marel} and {van Dishoeck}}]{Brunken22}
\bibinfo{author}{{Brunken}, N.G.C.}, \bibinfo{author}{{Booth}, A.S.},
  \bibinfo{author}{{Leemker}, M.}, \bibinfo{author}{{Nazari}, P.},
  \bibinfo{author}{{van der Marel}, N.}, \bibinfo{author}{{van Dishoeck},
  E.F.}, \bibinfo{year}{2022}.
\newblock \bibinfo{title}{{A major asymmetric ice trap in a planet-forming
  disk. III. First detection of dimethyl ether}}.
\newblock \bibinfo{journal}{\aap} \bibinfo{volume}{659}, \bibinfo{pages}{A29}.
\newblock \DOIprefix\doi{10.1051/0004-6361/202142981}.
%Type = Article
\bibitem[{{Burkhardt} et~al.(2021){Burkhardt}, {Long Kelvin Lee}, {Bryan
  Changala}, {Shingledecker}, {Cooke}, {Loomis}, {Wei}, {Charnley}, {Herbst},
  {McCarthy} and {McGuire}}]{Burkhardt21}
\bibinfo{author}{{Burkhardt}, A.M.}, \bibinfo{author}{{Long Kelvin Lee}, K.},
  \bibinfo{author}{{Bryan Changala}, P.}, \bibinfo{author}{{Shingledecker},
  C.N.}, \bibinfo{author}{{Cooke}, I.R.}, \bibinfo{author}{{Loomis}, R.A.},
  \bibinfo{author}{{Wei}, H.}, \bibinfo{author}{{Charnley}, S.B.},
  \bibinfo{author}{{Herbst}, E.}, \bibinfo{author}{{McCarthy}, M.C.},
  \bibinfo{author}{{McGuire}, B.A.}, \bibinfo{year}{2021}.
\newblock \bibinfo{title}{{Discovery of the Pure Polycyclic Aromatic
  Hydrocarbon Indene (c-C9H8) with GOTHAM Observations of TMC-1}}.
\newblock \bibinfo{journal}{\apjl} \bibinfo{volume}{913}, \bibinfo{pages}{L18}.
\newblock \DOIprefix\doi{10.3847/2041-8213/abfd3a},
  \href{http://arxiv.org/abs/2104.15117}{{\tt arXiv:2104.15117}}.
%Type = Article
\bibitem[{{Busch} et~al.(2022){Busch}, {Belloche}, {Garrod}, {M{\"u}ller} and
  {Menten}}]{Busch22}
\bibinfo{author}{{Busch}, L.A.}, \bibinfo{author}{{Belloche}, A.},
  \bibinfo{author}{{Garrod}, R.T.}, \bibinfo{author}{{M{\"u}ller}, H.S.P.},
  \bibinfo{author}{{Menten}, K.M.}, \bibinfo{year}{2022}.
\newblock \bibinfo{title}{{Resolving desorption of complex organic molecules in
  a hot core. Transition from non-thermal to thermal desorption or two-step
  thermal desorption?}}
\newblock \bibinfo{journal}{\aap} \bibinfo{volume}{665}, \bibinfo{pages}{A96}.
\newblock \DOIprefix\doi{10.1051/0004-6361/202243383},
  \href{http://arxiv.org/abs/2206.11174}{{\tt arXiv:2206.11174}}.
%Type = Article
\bibitem[{{Busquet} et~al.(2014){Busquet}, {Lefloch}, {Benedettini},
  {Ceccarelli}, {Codella}, {Cabrit}, {Nisini}, {Viti}, {G{\'o}mez-Ruiz},
  {Gusdorf}, {di Giorgio} and {Wiesenfeld}}]{Busquet14}
\bibinfo{author}{{Busquet}, G.}, \bibinfo{author}{{Lefloch}, B.},
  \bibinfo{author}{{Benedettini}, M.}, \bibinfo{author}{{Ceccarelli}, C.},
  \bibinfo{author}{{Codella}, C.}, \bibinfo{author}{{Cabrit}, S.},
  \bibinfo{author}{{Nisini}, B.}, \bibinfo{author}{{Viti}, S.},
  \bibinfo{author}{{G{\'o}mez-Ruiz}, A.I.}, \bibinfo{author}{{Gusdorf}, A.},
  \bibinfo{author}{{di Giorgio}, A.M.}, \bibinfo{author}{{Wiesenfeld}, L.},
  \bibinfo{year}{2014}.
\newblock \bibinfo{title}{{The CHESS survey of the L1157-B1 bow-shock: high and
  low excitation water vapor}}.
\newblock \bibinfo{journal}{\aap} \bibinfo{volume}{561}, \bibinfo{pages}{A120}.
\newblock \DOIprefix\doi{10.1051/0004-6361/201322347},
  \href{http://arxiv.org/abs/1311.2840}{{\tt arXiv:1311.2840}}.
%Type = Article
\bibitem[{{Cabezas} et~al.(2021){Cabezas}, {Ag{\'u}ndez}, {Marcelino},
  {Tercero}, {Cuadrado} and {Cernicharo}}]{cabezas21}
\bibinfo{author}{{Cabezas}, C.}, \bibinfo{author}{{Ag{\'u}ndez}, M.},
  \bibinfo{author}{{Marcelino}, N.}, \bibinfo{author}{{Tercero}, B.},
  \bibinfo{author}{{Cuadrado}, S.}, \bibinfo{author}{{Cernicharo}, J.},
  \bibinfo{year}{2021}.
\newblock \bibinfo{title}{{Interstellar detection of the simplest aminocarbyne
  H$_{2}$NC: an ignored but abundant molecule}}.
\newblock \bibinfo{journal}{\aap} \bibinfo{volume}{654}, \bibinfo{pages}{A45}.
\newblock \DOIprefix\doi{10.1051/0004-6361/202141491},
  \href{http://arxiv.org/abs/2107.08389}{{\tt arXiv:2107.08389}}.
%Type = Article
\bibitem[{{Calcutt} et~al.(2018){Calcutt}, {J{\o}rgensen}, {M{\"u}ller},
  {Kristensen}, {Coutens}, {Bourke}, {Garrod}, {Persson}, {van der Wiel}, {van
  Dishoeck} and {Wampfler}}]{Calcutt18}
\bibinfo{author}{{Calcutt}, H.}, \bibinfo{author}{{J{\o}rgensen}, J.K.},
  \bibinfo{author}{{M{\"u}ller}, H.S.P.}, \bibinfo{author}{{Kristensen}, L.E.},
  \bibinfo{author}{{Coutens}, A.}, \bibinfo{author}{{Bourke}, T.L.},
  \bibinfo{author}{{Garrod}, R.T.}, \bibinfo{author}{{Persson}, M.V.},
  \bibinfo{author}{{van der Wiel}, M.H.D.}, \bibinfo{author}{{van Dishoeck},
  E.F.}, \bibinfo{author}{{Wampfler}, S.F.}, \bibinfo{year}{2018}.
\newblock \bibinfo{title}{{The ALMA-PILS survey: complex nitriles towards IRAS
  16293-2422}}.
\newblock \bibinfo{journal}{\aap} \bibinfo{volume}{616}, \bibinfo{pages}{A90}.
\newblock \DOIprefix\doi{10.1051/0004-6361/201732289},
  \href{http://arxiv.org/abs/1804.09210}{{\tt arXiv:1804.09210}}.
%Type = Article
\bibitem[{{Calcutt} et~al.(2014){Calcutt}, {Viti}, {Codella}, {Beltr{\'a}n},
  {Fontani} and {Woods}}]{calcutt14}
\bibinfo{author}{{Calcutt}, H.}, \bibinfo{author}{{Viti}, S.},
  \bibinfo{author}{{Codella}, C.}, \bibinfo{author}{{Beltr{\'a}n}, M.T.},
  \bibinfo{author}{{Fontani}, F.}, \bibinfo{author}{{Woods}, P.M.},
  \bibinfo{year}{2014}.
\newblock \bibinfo{title}{{A high-resolution study of complex organic molecules
  in hot cores}}.
\newblock \bibinfo{journal}{\mnras} \bibinfo{volume}{443},
  \bibinfo{pages}{3157--3173}.
\newblock \DOIprefix\doi{10.1093/mnras/stu1363},
  \href{http://arxiv.org/abs/1407.1661}{{\tt arXiv:1407.1661}}.
%Type = Article
\bibitem[{{Caselli} et~al.(2002a){Caselli}, {Benson}, {Myers} and
  {Tafalla}}]{caselli02c}
\bibinfo{author}{{Caselli}, P.}, \bibinfo{author}{{Benson}, P.J.},
  \bibinfo{author}{{Myers}, P.C.}, \bibinfo{author}{{Tafalla}, M.},
  \bibinfo{year}{2002}a.
\newblock \bibinfo{title}{{Dense Cores in Dark Clouds. XIV. N$_{2}$H$^{+}$
  (1-0) Maps of Dense Cloud Cores}}.
\newblock \bibinfo{journal}{\apj} \bibinfo{volume}{572},
  \bibinfo{pages}{238--263}.
\newblock \DOIprefix\doi{10.1086/340195},
  \href{http://arxiv.org/abs/astro-ph/0202173}{{\tt arXiv:astro-ph/0202173}}.
%Type = Article
\bibitem[{{Caselli} et~al.(1997){Caselli}, {Hartquist} and
  {Havnes}}]{Caselli97}
\bibinfo{author}{{Caselli}, P.}, \bibinfo{author}{{Hartquist}, T.W.},
  \bibinfo{author}{{Havnes}, O.}, \bibinfo{year}{1997}.
\newblock \bibinfo{title}{{Grain-grain collisions and sputtering in oblique
  C-type shocks.}}
\newblock \bibinfo{journal}{\aap} \bibinfo{volume}{322},
  \bibinfo{pages}{296--301}.
%Type = Article
\bibitem[{{Caselli} et~al.(1999){Caselli}, {Walmsley}, {Tafalla}, {Dore} and
  {Myers}}]{caselli99}
\bibinfo{author}{{Caselli}, P.}, \bibinfo{author}{{Walmsley}, C.M.},
  \bibinfo{author}{{Tafalla}, M.}, \bibinfo{author}{{Dore}, L.},
  \bibinfo{author}{{Myers}, P.C.}, \bibinfo{year}{1999}.
\newblock \bibinfo{title}{{CO Depletion in the Starless Cloud Core L1544}}.
\newblock \bibinfo{journal}{\apjl} \bibinfo{volume}{523},
  \bibinfo{pages}{L165--L169}.
\newblock \DOIprefix\doi{10.1086/312280}.
%Type = Article
\bibitem[{{Caselli} et~al.(2002b){Caselli}, {Walmsley}, {Zucconi}, {Tafalla},
  {Dore} and {Myers}}]{caselli02a}
\bibinfo{author}{{Caselli}, P.}, \bibinfo{author}{{Walmsley}, C.M.},
  \bibinfo{author}{{Zucconi}, A.}, \bibinfo{author}{{Tafalla}, M.},
  \bibinfo{author}{{Dore}, L.}, \bibinfo{author}{{Myers}, P.C.},
  \bibinfo{year}{2002}b.
\newblock \bibinfo{title}{{Molecular Ions in L1544. I. Kinematics}}.
\newblock \bibinfo{journal}{\apj} \bibinfo{volume}{565},
  \bibinfo{pages}{331--343}.
\newblock \DOIprefix\doi{10.1086/324301},
  \href{http://arxiv.org/abs/astro-ph/0109021}{{\tt arXiv:astro-ph/0109021}}.
%Type = Article
\bibitem[{{Caselli} et~al.(2002c){Caselli}, {Walmsley}, {Zucconi}, {Tafalla},
  {Dore} and {Myers}}]{caselli02b}
\bibinfo{author}{{Caselli}, P.}, \bibinfo{author}{{Walmsley}, C.M.},
  \bibinfo{author}{{Zucconi}, A.}, \bibinfo{author}{{Tafalla}, M.},
  \bibinfo{author}{{Dore}, L.}, \bibinfo{author}{{Myers}, P.C.},
  \bibinfo{year}{2002}c.
\newblock \bibinfo{title}{{Molecular Ions in L1544. II. The Ionization
  Degree}}.
\newblock \bibinfo{journal}{\apj} \bibinfo{volume}{565},
  \bibinfo{pages}{344--358}.
\newblock \DOIprefix\doi{10.1086/324302},
  \href{http://arxiv.org/abs/astro-ph/0109023}{{\tt arXiv:astro-ph/0109023}}.
%Type = Article
\bibitem[{{Cazaux} et~al.(2003){Cazaux}, {Tielens}, {Ceccarelli}, {Castets},
  {Wakelam}, {Caux}, {Parise} and {Teyssier}}]{cazaux03}
\bibinfo{author}{{Cazaux}, S.}, \bibinfo{author}{{Tielens}, A.G.G.M.},
  \bibinfo{author}{{Ceccarelli}, C.}, \bibinfo{author}{{Castets}, A.},
  \bibinfo{author}{{Wakelam}, V.}, \bibinfo{author}{{Caux}, E.},
  \bibinfo{author}{{Parise}, B.}, \bibinfo{author}{{Teyssier}, D.},
  \bibinfo{year}{2003}.
\newblock \bibinfo{title}{{The Hot Core around the Low-mass Protostar IRAS
  16293-2422: Scoundrels Rule!}}
\newblock \bibinfo{journal}{\apjl} \bibinfo{volume}{593},
  \bibinfo{pages}{L51--L55}.
\newblock \DOIprefix\doi{10.1086/378038}.
%Type = Article
\bibitem[{{Ceccarelli} et~al.(2017){Ceccarelli}, {Caselli}, {Fontani}, {Neri},
  {L{\'o}pez-Sepulcre}, {Codella}, {Feng}, {Jim{\'e}nez-Serra}, {Lefloch},
  {Pineda}, {Vastel}, {Alves}, {Bachiller}, {Balucani}, {Bianchi}, {Bizzocchi},
  {Bottinelli}, {Caux}, {Chac{\'o}n-Tanarro}, {Choudhury}, {Coutens}, {Dulieu},
  {Favre}, {Hily-Blant}, {Holdship}, {Kahane}, {Jaber Al-Edhari}, {Laas},
  {Ospina}, {Oya}, {Podio}, {Pon}, {Punanova}, {Quenard}, {Rimola}, {Sakai},
  {Sims}, {Spezzano}, {Taquet}, {Testi}, {Theul{\'e}}, {Ugliengo}, {Vasyunin},
  {Viti}, {Wiesenfeld} and {Yamamoto}}]{Ceccarelli17}
\bibinfo{author}{{Ceccarelli}, C.}, \bibinfo{author}{{Caselli}, P.},
  \bibinfo{author}{{Fontani}, F.}, \bibinfo{author}{{Neri}, R.},
  \bibinfo{author}{{L{\'o}pez-Sepulcre}, A.}, \bibinfo{author}{{Codella}, C.},
  \bibinfo{author}{{Feng}, S.}, \bibinfo{author}{{Jim{\'e}nez-Serra}, I.},
  \bibinfo{author}{{Lefloch}, B.}, \bibinfo{author}{{Pineda}, J.E.},
  \bibinfo{author}{{Vastel}, C.}, \bibinfo{author}{{Alves}, F.},
  \bibinfo{author}{{Bachiller}, R.}, \bibinfo{author}{{Balucani}, N.},
  \bibinfo{author}{{Bianchi}, E.}, \bibinfo{author}{{Bizzocchi}, L.},
  \bibinfo{author}{{Bottinelli}, S.}, \bibinfo{author}{{Caux}, E.},
  \bibinfo{author}{{Chac{\'o}n-Tanarro}, A.}, \bibinfo{author}{{Choudhury},
  R.}, \bibinfo{author}{{Coutens}, A.}, \bibinfo{author}{{Dulieu}, F.},
  \bibinfo{author}{{Favre}, C.}, \bibinfo{author}{{Hily-Blant}, P.},
  \bibinfo{author}{{Holdship}, J.}, \bibinfo{author}{{Kahane}, C.},
  \bibinfo{author}{{Jaber Al-Edhari}, A.}, \bibinfo{author}{{Laas}, J.},
  \bibinfo{author}{{Ospina}, J.}, \bibinfo{author}{{Oya}, Y.},
  \bibinfo{author}{{Podio}, L.}, \bibinfo{author}{{Pon}, A.},
  \bibinfo{author}{{Punanova}, A.}, \bibinfo{author}{{Quenard}, D.},
  \bibinfo{author}{{Rimola}, A.}, \bibinfo{author}{{Sakai}, N.},
  \bibinfo{author}{{Sims}, I.R.}, \bibinfo{author}{{Spezzano}, S.},
  \bibinfo{author}{{Taquet}, V.}, \bibinfo{author}{{Testi}, L.},
  \bibinfo{author}{{Theul{\'e}}, P.}, \bibinfo{author}{{Ugliengo}, P.},
  \bibinfo{author}{{Vasyunin}, A.I.}, \bibinfo{author}{{Viti}, S.},
  \bibinfo{author}{{Wiesenfeld}, L.}, \bibinfo{author}{{Yamamoto}, S.},
  \bibinfo{year}{2017}.
\newblock \bibinfo{title}{{Seeds Of Life In Space (SOLIS): The Organic
  Composition Diversity at 300-1000 au Scale in Solar-type Star-forming
  Regions}}.
\newblock \bibinfo{journal}{\apj} \bibinfo{volume}{850}, \bibinfo{pages}{176}.
\newblock \DOIprefix\doi{10.3847/1538-4357/aa961d},
  \href{http://arxiv.org/abs/1710.10437}{{\tt arXiv:1710.10437}}.
%Type = Inproceedings
\bibitem[{{Ceccarelli} et~al.(2023){Ceccarelli}, {Codella}, {Balucani},
  {Bockelee-Morvan}, {Herbst}, {Vastel}, {Caselli}, {Favre}, {Lefloch}, {Oberg}
  and {Yamamoto}}]{Ceccarelli2023}
\bibinfo{author}{{Ceccarelli}, C.}, \bibinfo{author}{{Codella}, C.},
  \bibinfo{author}{{Balucani}, N.}, \bibinfo{author}{{Bockelee-Morvan}, D.},
  \bibinfo{author}{{Herbst}, E.}, \bibinfo{author}{{Vastel}, C.},
  \bibinfo{author}{{Caselli}, P.}, \bibinfo{author}{{Favre}, C.},
  \bibinfo{author}{{Lefloch}, B.}, \bibinfo{author}{{Oberg}, K.},
  \bibinfo{author}{{Yamamoto}, S.}, \bibinfo{year}{2023}.
\newblock \bibinfo{title}{{Organic Chemistry in the First Phases of Solar-Type
  Protostars}}, in: \bibinfo{editor}{{Inutsuka}, S.},
  \bibinfo{editor}{{Aikawa}, Y.}, \bibinfo{editor}{{Muto}, T.},
  \bibinfo{editor}{{Tomida}, K.}, \bibinfo{editor}{{Tamura}, M.} (Eds.),
  \bibinfo{booktitle}{Protostars and Planets VII}, p. \bibinfo{pages}{379}.
%Type = Article
\bibitem[{{Cernicharo} et~al.(2021){Cernicharo}, {Ag{\'u}ndez}, {Cabezas},
  {Tercero}, {Marcelino}, {Pardo} and {de Vicente}}]{cernicharo21}
\bibinfo{author}{{Cernicharo}, J.}, \bibinfo{author}{{Ag{\'u}ndez}, M.},
  \bibinfo{author}{{Cabezas}, C.}, \bibinfo{author}{{Tercero}, B.},
  \bibinfo{author}{{Marcelino}, N.}, \bibinfo{author}{{Pardo}, J.R.},
  \bibinfo{author}{{de Vicente}, P.}, \bibinfo{year}{2021}.
\newblock \bibinfo{title}{{Pure hydrocarbon cycles in TMC-1: Discovery of
  ethynyl cyclopropenylidene, cyclopentadiene, and indene}}.
\newblock \bibinfo{journal}{\aap} \bibinfo{volume}{649}, \bibinfo{pages}{L15}.
\newblock \DOIprefix\doi{10.1051/0004-6361/202141156},
  \href{http://arxiv.org/abs/2104.13991}{{\tt arXiv:2104.13991}}.
%Type = Article
\bibitem[{{Cernicharo} et~al.(2023){Cernicharo}, {Tercero}, {Marcelino},
  {Ag{\'u}ndez} and {de Vicente}}]{cernicharo23}
\bibinfo{author}{{Cernicharo}, J.}, \bibinfo{author}{{Tercero}, B.},
  \bibinfo{author}{{Marcelino}, N.}, \bibinfo{author}{{Ag{\'u}ndez}, M.},
  \bibinfo{author}{{de Vicente}, P.}, \bibinfo{year}{2023}.
\newblock \bibinfo{title}{{The spatial distribution of an aromatic molecule,
  C$_{6}$H$_{5}$CN, in the cold dark cloud TMC-1}}.
\newblock \bibinfo{journal}{\aap} \bibinfo{volume}{674}, \bibinfo{pages}{L4}.
\newblock \DOIprefix\doi{10.1051/0004-6361/202346722},
  \href{http://arxiv.org/abs/2305.15315}{{\tt arXiv:2305.15315}}.
%Type = Article
\bibitem[{{Cesaroni} et~al.(2014){Cesaroni}, {Galli}, {Neri} and
  {Walmsley}}]{cesaroni14}
\bibinfo{author}{{Cesaroni}, R.}, \bibinfo{author}{{Galli}, D.},
  \bibinfo{author}{{Neri}, R.}, \bibinfo{author}{{Walmsley}, C.M.},
  \bibinfo{year}{2014}.
\newblock \bibinfo{title}{{Imaging the disk around IRAS 20126+4104 at
  subarcsecond resolution}}.
\newblock \bibinfo{journal}{\aap} \bibinfo{volume}{566}, \bibinfo{pages}{A73}.
\newblock \DOIprefix\doi{10.1051/0004-6361/201323065}.
%Type = Article
\bibitem[{{Chahine} et~al.(2022){Chahine}, {L{\'o}pez-Sepulcre}, {Neri},
  {Ceccarelli}, {Mercimek}, {Codella}, {Bouvier}, {Bianchi}, {Favre}, {Podio},
  {Alves}, {Sakai} and {Yamamoto}}]{Chahine22}
\bibinfo{author}{{Chahine}, L.}, \bibinfo{author}{{L{\'o}pez-Sepulcre}, A.},
  \bibinfo{author}{{Neri}, R.}, \bibinfo{author}{{Ceccarelli}, C.},
  \bibinfo{author}{{Mercimek}, S.}, \bibinfo{author}{{Codella}, C.},
  \bibinfo{author}{{Bouvier}, M.}, \bibinfo{author}{{Bianchi}, E.},
  \bibinfo{author}{{Favre}, C.}, \bibinfo{author}{{Podio}, L.},
  \bibinfo{author}{{Alves}, F.O.}, \bibinfo{author}{{Sakai}, N.},
  \bibinfo{author}{{Yamamoto}, S.}, \bibinfo{year}{2022}.
\newblock \bibinfo{title}{{Organic chemistry in the protosolar analogue
  HOPS-108: Environment matters}}.
\newblock \bibinfo{journal}{\aap} \bibinfo{volume}{657}, \bibinfo{pages}{A78}.
\newblock \DOIprefix\doi{10.1051/0004-6361/202141811},
  \href{http://arxiv.org/abs/2112.08077}{{\tt arXiv:2112.08077}}.
%Type = Article
\bibitem[{{Charnley} et~al.(2001){Charnley}, {Rodgers} and
  {Ehrenfreund}}]{Charnley01}
\bibinfo{author}{{Charnley}, S.B.}, \bibinfo{author}{{Rodgers}, S.D.},
  \bibinfo{author}{{Ehrenfreund}, P.}, \bibinfo{year}{2001}.
\newblock \bibinfo{title}{{Gas-grain chemical models of star-forming molecular
  clouds as constrained by ISO and SWAS observations}}.
\newblock \bibinfo{journal}{\aap} \bibinfo{volume}{378},
  \bibinfo{pages}{1024--1036}.
\newblock \DOIprefix\doi{10.1051/0004-6361:20011193}.
%Type = Article
\bibitem[{{Chengalur} and {Kanekar}(2003)}]{chengalur03}
\bibinfo{author}{{Chengalur}, J.N.}, \bibinfo{author}{{Kanekar}, N.},
  \bibinfo{year}{2003}.
\newblock \bibinfo{title}{{Widespread acetaldehyde near the Galactic Centre}}.
\newblock \bibinfo{journal}{\aap} \bibinfo{volume}{403},
  \bibinfo{pages}{L43--L46}.
\newblock \DOIprefix\doi{10.1051/0004-6361:20030577},
  \href{http://arxiv.org/abs/astro-ph/0304406}{{\tt arXiv:astro-ph/0304406}}.
%Type = Article
\bibitem[{{Chuang} et~al.(2016){Chuang}, {Fedoseev}, {Ioppolo}, {van Dishoeck}
  and {Linnartz}}]{chuang16}
\bibinfo{author}{{Chuang}, K.J.}, \bibinfo{author}{{Fedoseev}, G.},
  \bibinfo{author}{{Ioppolo}, S.}, \bibinfo{author}{{van Dishoeck}, E.F.},
  \bibinfo{author}{{Linnartz}, H.}, \bibinfo{year}{2016}.
\newblock \bibinfo{title}{{H-atom addition and abstraction reactions in mixed
  CO, H$_{2}$CO and CH$_{3}$OH ices - an extended view on complex organic
  molecule formation}}.
\newblock \bibinfo{journal}{\mnras} \bibinfo{volume}{455},
  \bibinfo{pages}{1702--1712}.
\newblock \DOIprefix\doi{10.1093/mnras/stv2288},
  \href{http://arxiv.org/abs/1606.01049}{{\tt arXiv:1606.01049}}.
%Type = Article
\bibitem[{{Cocinero} et~al.(2012){Cocinero}, {Lesarri}, {{\'E}cija},
  {Basterretxea}, {Grabow}, {Fern{\'a}ndez} and {Casta{\~n}o}}]{cocinero12}
\bibinfo{author}{{Cocinero}, E.J.}, \bibinfo{author}{{Lesarri}, A.},
  \bibinfo{author}{{{\'E}cija}, P.}, \bibinfo{author}{{Basterretxea}, F.J.},
  \bibinfo{author}{{Grabow}, J.U.}, \bibinfo{author}{{Fern{\'a}ndez}, J.A.},
  \bibinfo{author}{{Casta{\~n}o}, F.}, \bibinfo{year}{2012}.
\newblock \bibinfo{title}{{Ribose Found in the Gas Phase}}.
\newblock \bibinfo{journal}{Angewandte Chemie} \bibinfo{volume}{124},
  \bibinfo{pages}{3173--3178}.
\newblock \DOIprefix\doi{10.1002/ange.201107973}.
%Type = Article
\bibitem[{{Codella} et~al.(2005){Codella}, {Bachiller}, {Benedettini},
  {Caselli}, {Viti} and {Wakelam}}]{Codella05}
\bibinfo{author}{{Codella}, C.}, \bibinfo{author}{{Bachiller}, R.},
  \bibinfo{author}{{Benedettini}, M.}, \bibinfo{author}{{Caselli}, P.},
  \bibinfo{author}{{Viti}, S.}, \bibinfo{author}{{Wakelam}, V.},
  \bibinfo{year}{2005}.
\newblock \bibinfo{title}{{Chemical differentiation along the CepA-East
  outflows}}.
\newblock \bibinfo{journal}{\mnras} \bibinfo{volume}{361},
  \bibinfo{pages}{244--258}.
\newblock \DOIprefix\doi{10.1111/j.1365-2966.2005.09165.x},
  \href{http://arxiv.org/abs/astro-ph/0505168}{{\tt arXiv:astro-ph/0505168}}.
%Type = Article
\bibitem[{{Codella} et~al.(2018){Codella}, {Bianchi}, {Tabone}, {Lee},
  {Cabrit}, {Ceccarelli}, {Podio}, {Bacciotti}, {Bachiller}, {Chapillon},
  {Gueth}, {Gusdorf}, {Lefloch}, {Leurini}, {Pineau des For{\^e}ts}, {Rygl} and
  {Tafalla}}]{Codella18}
\bibinfo{author}{{Codella}, C.}, \bibinfo{author}{{Bianchi}, E.},
  \bibinfo{author}{{Tabone}, B.}, \bibinfo{author}{{Lee}, C.F.},
  \bibinfo{author}{{Cabrit}, S.}, \bibinfo{author}{{Ceccarelli}, C.},
  \bibinfo{author}{{Podio}, L.}, \bibinfo{author}{{Bacciotti}, F.},
  \bibinfo{author}{{Bachiller}, R.}, \bibinfo{author}{{Chapillon}, E.},
  \bibinfo{author}{{Gueth}, F.}, \bibinfo{author}{{Gusdorf}, A.},
  \bibinfo{author}{{Lefloch}, B.}, \bibinfo{author}{{Leurini}, S.},
  \bibinfo{author}{{Pineau des For{\^e}ts}, G.}, \bibinfo{author}{{Rygl},
  K.L.J.}, \bibinfo{author}{{Tafalla}, M.}, \bibinfo{year}{2018}.
\newblock \bibinfo{title}{{Water and interstellar complex organics associated
  with the HH 212 protostellar disc. On disc atmospheres, disc winds, and
  accretion shocks}}.
\newblock \bibinfo{journal}{\aap} \bibinfo{volume}{617}, \bibinfo{pages}{A10}.
\newblock \DOIprefix\doi{10.1051/0004-6361/201832765},
  \href{http://arxiv.org/abs/1806.07967}{{\tt arXiv:1806.07967}}.
%Type = Article
\bibitem[{{Codella} et~al.(2017){Codella}, {Ceccarelli}, {Caselli}, {Balucani},
  {Barone}, {Fontani}, {Lefloch}, {Podio}, {Viti}, {Feng}, {Bachiller},
  {Bianchi}, {Dulieu}, {Jim{\'e}nez-Serra}, {Holdship}, {Neri}, {Pineda},
  {Pon}, {Sims}, {Spezzano}, {Vasyunin}, {Alves}, {Bizzocchi}, {Bottinelli},
  {Caux}, {Chac{\'o}n-Tanarro}, {Choudhury}, {Coutens}, {Favre}, {Hily-Blant},
  {Kahane}, {Jaber Al-Edhari}, {Laas}, {L{\'o}pez-Sepulcre}, {Ospina}, {Oya},
  {Punanova}, {Puzzarini}, {Quenard}, {Rimola}, {Sakai}, {Skouteris}, {Taquet},
  {Testi}, {Theul{\'e}}, {Ugliengo}, {Vastel}, {Vazart}, {Wiesenfeld} and
  {Yamamoto}}]{Codella17}
\bibinfo{author}{{Codella}, C.}, \bibinfo{author}{{Ceccarelli}, C.},
  \bibinfo{author}{{Caselli}, P.}, \bibinfo{author}{{Balucani}, N.},
  \bibinfo{author}{{Barone}, V.}, \bibinfo{author}{{Fontani}, F.},
  \bibinfo{author}{{Lefloch}, B.}, \bibinfo{author}{{Podio}, L.},
  \bibinfo{author}{{Viti}, S.}, \bibinfo{author}{{Feng}, S.},
  \bibinfo{author}{{Bachiller}, R.}, \bibinfo{author}{{Bianchi}, E.},
  \bibinfo{author}{{Dulieu}, F.}, \bibinfo{author}{{Jim{\'e}nez-Serra}, I.},
  \bibinfo{author}{{Holdship}, J.}, \bibinfo{author}{{Neri}, R.},
  \bibinfo{author}{{Pineda}, J.E.}, \bibinfo{author}{{Pon}, A.},
  \bibinfo{author}{{Sims}, I.}, \bibinfo{author}{{Spezzano}, S.},
  \bibinfo{author}{{Vasyunin}, A.I.}, \bibinfo{author}{{Alves}, F.},
  \bibinfo{author}{{Bizzocchi}, L.}, \bibinfo{author}{{Bottinelli}, S.},
  \bibinfo{author}{{Caux}, E.}, \bibinfo{author}{{Chac{\'o}n-Tanarro}, A.},
  \bibinfo{author}{{Choudhury}, R.}, \bibinfo{author}{{Coutens}, A.},
  \bibinfo{author}{{Favre}, C.}, \bibinfo{author}{{Hily-Blant}, P.},
  \bibinfo{author}{{Kahane}, C.}, \bibinfo{author}{{Jaber Al-Edhari}, A.},
  \bibinfo{author}{{Laas}, J.}, \bibinfo{author}{{L{\'o}pez-Sepulcre}, A.},
  \bibinfo{author}{{Ospina}, J.}, \bibinfo{author}{{Oya}, Y.},
  \bibinfo{author}{{Punanova}, A.}, \bibinfo{author}{{Puzzarini}, C.},
  \bibinfo{author}{{Quenard}, D.}, \bibinfo{author}{{Rimola}, A.},
  \bibinfo{author}{{Sakai}, N.}, \bibinfo{author}{{Skouteris}, D.},
  \bibinfo{author}{{Taquet}, V.}, \bibinfo{author}{{Testi}, L.},
  \bibinfo{author}{{Theul{\'e}}, P.}, \bibinfo{author}{{Ugliengo}, P.},
  \bibinfo{author}{{Vastel}, C.}, \bibinfo{author}{{Vazart}, F.},
  \bibinfo{author}{{Wiesenfeld}, L.}, \bibinfo{author}{{Yamamoto}, S.},
  \bibinfo{year}{2017}.
\newblock \bibinfo{title}{{Seeds of Life in Space (SOLIS). II. Formamide in
  protostellar shocks: Evidence for gas-phase formation}}.
\newblock \bibinfo{journal}{\aap} \bibinfo{volume}{605}, \bibinfo{pages}{L3}.
\newblock \DOIprefix\doi{10.1051/0004-6361/201731249},
  \href{http://arxiv.org/abs/1708.04663}{{\tt arXiv:1708.04663}}.
%Type = Article
\bibitem[{{Codella} et~al.(2021){Codella}, {Ceccarelli}, {Chandler}, {Sakai},
  {Yamamoto} and {FAUST Team}}]{Codella21}
\bibinfo{author}{{Codella}, C.}, \bibinfo{author}{{Ceccarelli}, C.},
  \bibinfo{author}{{Chandler}, C.}, \bibinfo{author}{{Sakai}, N.},
  \bibinfo{author}{{Yamamoto}, S.}, \bibinfo{author}{{FAUST Team}},
  \bibinfo{year}{2021}.
\newblock \bibinfo{title}{{Enlightening the chemistry of infalling envelopes
  and accretion disks around Sun-like protostars: the ALMA FAUST project}}.
\newblock \bibinfo{journal}{Frontiers in Astronomy and Space Sciences}
  \bibinfo{volume}{8}, \bibinfo{pages}{227}.
\newblock \DOIprefix\doi{10.3389/fspas.2021.782006},
  \href{http://arxiv.org/abs/2111.14121}{{\tt arXiv:2111.14121}}.
%Type = Article
\bibitem[{{Codella} et~al.(2019){Codella}, {Ceccarelli}, {Lee}, {Bianchi},
  {Balucani}, {Podio}, {Cabrit}, {Gueth}, {Gusdorf}, {Lefloch} and
  {Tabone}}]{Codella19}
\bibinfo{author}{{Codella}, C.}, \bibinfo{author}{{Ceccarelli}, C.},
  \bibinfo{author}{{Lee}, C.F.}, \bibinfo{author}{{Bianchi}, E.},
  \bibinfo{author}{{Balucani}, N.}, \bibinfo{author}{{Podio}, L.},
  \bibinfo{author}{{Cabrit}, S.}, \bibinfo{author}{{Gueth}, F.},
  \bibinfo{author}{{Gusdorf}, A.}, \bibinfo{author}{{Lefloch}, B.},
  \bibinfo{author}{{Tabone}, B.}, \bibinfo{year}{2019}.
\newblock \bibinfo{title}{{The HH 212 Interstellar Laboratory: Astrochemistry
  as a Tool To Reveal Protostellar Disks on Solar System Scales around a Rising
  Sun}}.
\newblock \bibinfo{journal}{ACS Earth and Space Chemistry} \bibinfo{volume}{3},
  \bibinfo{pages}{2110--2121}.
\newblock \DOIprefix\doi{10.1021/acsearthspacechem.9b00136},
  \href{http://arxiv.org/abs/1910.04442}{{\tt arXiv:1910.04442}}.
%Type = Article
\bibitem[{{Codella} et~al.(2015){Codella}, {Fontani}, {Ceccarelli}, {Podio},
  {Viti}, {Bachiller}, {Benedettini} and {Lefloch}}]{Codella15}
\bibinfo{author}{{Codella}, C.}, \bibinfo{author}{{Fontani}, F.},
  \bibinfo{author}{{Ceccarelli}, C.}, \bibinfo{author}{{Podio}, L.},
  \bibinfo{author}{{Viti}, S.}, \bibinfo{author}{{Bachiller}, R.},
  \bibinfo{author}{{Benedettini}, M.}, \bibinfo{author}{{Lefloch}, B.},
  \bibinfo{year}{2015}.
\newblock \bibinfo{title}{{Astrochemistry at work in the L1157-B1 shock:
  acetaldehyde formation}}.
\newblock \bibinfo{journal}{\mnras} \bibinfo{volume}{449},
  \bibinfo{pages}{L11--L15}.
\newblock \DOIprefix\doi{10.1093/mnrasl/slu204},
  \href{http://arxiv.org/abs/1412.8318}{{\tt arXiv:1412.8318}}.
%Type = Article
\bibitem[{{Coletta} et~al.(2020){Coletta}, {Fontani}, {Rivilla}, {Mininni},
  {Colzi}, {S{\'a}nchez-Monge} and {Beltr{\'a}n}}]{Coletta20}
\bibinfo{author}{{Coletta}, A.}, \bibinfo{author}{{Fontani}, F.},
  \bibinfo{author}{{Rivilla}, V.M.}, \bibinfo{author}{{Mininni}, C.},
  \bibinfo{author}{{Colzi}, L.}, \bibinfo{author}{{S{\'a}nchez-Monge}, {\'A}.},
  \bibinfo{author}{{Beltr{\'a}n}, M.T.}, \bibinfo{year}{2020}.
\newblock \bibinfo{title}{{Evolutionary study of complex organic molecules in
  high-mass star-forming regions}}.
\newblock \bibinfo{journal}{\aap} \bibinfo{volume}{641}, \bibinfo{pages}{A54}.
\newblock \DOIprefix\doi{10.1051/0004-6361/202038212},
  \href{http://arxiv.org/abs/2006.15413}{{\tt arXiv:2006.15413}}.
%Type = Article
\bibitem[{{Colzi} et~al.(2022){Colzi}, {Mart{\'\i}n-Pintado}, {Rivilla},
  {Jim{\'e}nez-Serra}, {Zeng}, {Rodr{\'\i}guez-Almeida}, {Rico-Villas},
  {Mart{\'\i}n} and {Requena-Torres}}]{Colzi22}
\bibinfo{author}{{Colzi}, L.}, \bibinfo{author}{{Mart{\'\i}n-Pintado}, J.},
  \bibinfo{author}{{Rivilla}, V.M.}, \bibinfo{author}{{Jim{\'e}nez-Serra}, I.},
  \bibinfo{author}{{Zeng}, S.}, \bibinfo{author}{{Rodr{\'\i}guez-Almeida},
  L.F.}, \bibinfo{author}{{Rico-Villas}, F.}, \bibinfo{author}{{Mart{\'\i}n},
  S.}, \bibinfo{author}{{Requena-Torres}, M.A.}, \bibinfo{year}{2022}.
\newblock \bibinfo{title}{{Deuterium Fractionation as a Multiphase Component
  Tracer in the Galactic Center}}.
\newblock \bibinfo{journal}{\apjl} \bibinfo{volume}{926}, \bibinfo{pages}{L22}.
\newblock \DOIprefix\doi{10.3847/2041-8213/ac52ac},
  \href{http://arxiv.org/abs/2202.04111}{{\tt arXiv:2202.04111}}.
%Type = Article
\bibitem[{{Colzi} et~al.(2021){Colzi}, {Rivilla}, {Beltr{\'a}n},
  {Jim{\'e}nez-Serra}, {Mininni}, {Melosso}, {Cesaroni}, {Fontani},
  {Lorenzani}, {S{\'a}nchez-Monge}, {Viti}, {Schilke}, {Testi}, {Alonso} and
  {Kolesnikov{\'a}}}]{Colzi21}
\bibinfo{author}{{Colzi}, L.}, \bibinfo{author}{{Rivilla}, V.M.},
  \bibinfo{author}{{Beltr{\'a}n}, M.T.}, \bibinfo{author}{{Jim{\'e}nez-Serra},
  I.}, \bibinfo{author}{{Mininni}, C.}, \bibinfo{author}{{Melosso}, M.},
  \bibinfo{author}{{Cesaroni}, R.}, \bibinfo{author}{{Fontani}, F.},
  \bibinfo{author}{{Lorenzani}, A.}, \bibinfo{author}{{S{\'a}nchez-Monge}, A.},
  \bibinfo{author}{{Viti}, S.}, \bibinfo{author}{{Schilke}, P.},
  \bibinfo{author}{{Testi}, L.}, \bibinfo{author}{{Alonso}, E.R.},
  \bibinfo{author}{{Kolesnikov{\'a}}, L.}, \bibinfo{year}{2021}.
\newblock \bibinfo{title}{{The GUAPOS project. II. A comprehensive study of
  peptide-like bond molecules}}.
\newblock \bibinfo{journal}{\aap} \bibinfo{volume}{653}, \bibinfo{pages}{A129}.
\newblock \DOIprefix\doi{10.1051/0004-6361/202141573},
  \href{http://arxiv.org/abs/2107.11258}{{\tt arXiv:2107.11258}}.
%Type = Article
\bibitem[{{Combes}(2008)}]{combes08}
\bibinfo{author}{{Combes}, F.}, \bibinfo{year}{2008}.
\newblock \bibinfo{title}{{Molecular absorptions in high-z objects}}.
\newblock \bibinfo{journal}{\apss} \bibinfo{volume}{313},
  \bibinfo{pages}{321--326}.
\newblock \DOIprefix\doi{10.1007/s10509-007-9632-3},
  \href{http://arxiv.org/abs/astro-ph/0701894}{{\tt arXiv:astro-ph/0701894}}.
%Type = Article
\bibitem[{{Comito} et~al.(2005){Comito}, {Schilke}, {Phillips}, {Lis}, {Motte}
  and {Mehringer}}]{comito05}
\bibinfo{author}{{Comito}, C.}, \bibinfo{author}{{Schilke}, P.},
  \bibinfo{author}{{Phillips}, T.G.}, \bibinfo{author}{{Lis}, D.C.},
  \bibinfo{author}{{Motte}, F.}, \bibinfo{author}{{Mehringer}, D.},
  \bibinfo{year}{2005}.
\newblock \bibinfo{title}{{A Molecular Line Survey of Orion KL in the 350
  Micron Band}}.
\newblock \bibinfo{journal}{\apjs} \bibinfo{volume}{156},
  \bibinfo{pages}{127--167}.
\newblock \DOIprefix\doi{10.1086/425996}.
%Type = Phdthesis
\bibitem[{{Corby}(2016)}]{Corby16}
\bibinfo{author}{{Corby}, J.}, \bibinfo{year}{2016}.
\newblock \bibinfo{title}{{Astrochemistry in The Age of Broadband Radio
  Astronomy}}.
\newblock Ph.D. thesis. University of Virginia.
%Type = Article
\bibitem[{{Coutens} et~al.(2022){Coutens}, {Loison}, {Boulanger}, {Caux},
  {M{\"u}ller}, {Wakelam}, {Manigand} and {J{\o}rgensen}}]{coutens22}
\bibinfo{author}{{Coutens}, A.}, \bibinfo{author}{{Loison}, J.C.},
  \bibinfo{author}{{Boulanger}, A.}, \bibinfo{author}{{Caux}, E.},
  \bibinfo{author}{{M{\"u}ller}, H.S.P.}, \bibinfo{author}{{Wakelam}, V.},
  \bibinfo{author}{{Manigand}, S.}, \bibinfo{author}{{J{\o}rgensen}, J.K.},
  \bibinfo{year}{2022}.
\newblock \bibinfo{title}{{The ALMA-PILS survey: First tentative detection of
  3-hydroxypropenal (HOCHCHCHO) in the interstellar medium and chemical
  modeling of the C$_{3}$H$_{4}$O$_{2}$ isomers}}.
\newblock \bibinfo{journal}{\aap} \bibinfo{volume}{660}, \bibinfo{pages}{L6}.
\newblock \DOIprefix\doi{10.1051/0004-6361/202243038},
  \href{http://arxiv.org/abs/2203.14119}{{\tt arXiv:2203.14119}}.
%Type = Article
\bibitem[{{Crapsi} et~al.(2005){Crapsi}, {Caselli}, {Walmsley}, {Myers},
  {Tafalla}, {Lee} and {Bourke}}]{crapsi05}
\bibinfo{author}{{Crapsi}, A.}, \bibinfo{author}{{Caselli}, P.},
  \bibinfo{author}{{Walmsley}, C.M.}, \bibinfo{author}{{Myers}, P.C.},
  \bibinfo{author}{{Tafalla}, M.}, \bibinfo{author}{{Lee}, C.W.},
  \bibinfo{author}{{Bourke}, T.L.}, \bibinfo{year}{2005}.
\newblock \bibinfo{title}{{Probing the Evolutionary Status of Starless Cores
  through N$_{2}$H$^{+}$ and N$_{2}$D$^{+}$ Observations}}.
\newblock \bibinfo{journal}{\apj} \bibinfo{volume}{619},
  \bibinfo{pages}{379--406}.
\newblock \DOIprefix\doi{10.1086/426472},
  \href{http://arxiv.org/abs/astro-ph/0409529}{{\tt arXiv:astro-ph/0409529}}.
%Type = Article
\bibitem[{{Cronin} and {Pizzarello}(1983)}]{cronin1983}
\bibinfo{author}{{Cronin}, J.R.}, \bibinfo{author}{{Pizzarello}, S.},
  \bibinfo{year}{1983}.
\newblock \bibinfo{title}{{Amino acids in meteorites}}.
\newblock \bibinfo{journal}{Advances in Space Research} \bibinfo{volume}{3},
  \bibinfo{pages}{5--18}.
\newblock \DOIprefix\doi{10.1016/0273-1177(83)90036-4}.
%Type = Article
\bibitem[{{Cruz-Diaz} et~al.(2016){Cruz-Diaz}, {Mart{\'\i}n-Dom{\'e}nech},
  {Mu{\~n}oz Caro} and {Chen}}]{Cruz-Diaz16}
\bibinfo{author}{{Cruz-Diaz}, G.A.},
  \bibinfo{author}{{Mart{\'\i}n-Dom{\'e}nech}, R.}, \bibinfo{author}{{Mu{\~n}oz
  Caro}, G.M.}, \bibinfo{author}{{Chen}, Y.J.}, \bibinfo{year}{2016}.
\newblock \bibinfo{title}{{Negligible photodesorption of methanol ice and
  active photon-induced desorption of its irradiation products}}.
\newblock \bibinfo{journal}{\aap} \bibinfo{volume}{592}, \bibinfo{pages}{A68}.
\newblock \DOIprefix\doi{10.1051/0004-6361/201526761},
  \href{http://arxiv.org/abs/1605.01767}{{\tt arXiv:1605.01767}}.
%Type = Article
\bibitem[{{Csengeri} et~al.(2018){Csengeri}, {Bontemps}, {Wyrowski},
  {Belloche}, {Menten}, {Leurini}, {Beuther}, {Bronfman}, {Commer{\c{c}}on},
  {Chapillon}, {Longmore}, {Palau}, {Tan} and {Urquhart}}]{Csengeri18}
\bibinfo{author}{{Csengeri}, T.}, \bibinfo{author}{{Bontemps}, S.},
  \bibinfo{author}{{Wyrowski}, F.}, \bibinfo{author}{{Belloche}, A.},
  \bibinfo{author}{{Menten}, K.M.}, \bibinfo{author}{{Leurini}, S.},
  \bibinfo{author}{{Beuther}, H.}, \bibinfo{author}{{Bronfman}, L.},
  \bibinfo{author}{{Commer{\c{c}}on}, B.}, \bibinfo{author}{{Chapillon}, E.},
  \bibinfo{author}{{Longmore}, S.}, \bibinfo{author}{{Palau}, A.},
  \bibinfo{author}{{Tan}, J.C.}, \bibinfo{author}{{Urquhart}, J.S.},
  \bibinfo{year}{2018}.
\newblock \bibinfo{title}{{Search for high-mass protostars with ALMA revealed
  up to kilo-parsec scales (SPARKS). I. Indication for a centrifugal barrier in
  the environment of a single high-mass envelope}}.
\newblock \bibinfo{journal}{\aap} \bibinfo{volume}{617}, \bibinfo{pages}{A89}.
\newblock \DOIprefix\doi{10.1051/0004-6361/201832753},
  \href{http://arxiv.org/abs/1804.06482}{{\tt arXiv:1804.06482}}.
%Type = Article
\bibitem[{{Cuadrado} et~al.(2017){Cuadrado}, {Goicoechea}, {Cernicharo},
  {Fuente}, {Pety} and {Tercero}}]{Cuadrado17}
\bibinfo{author}{{Cuadrado}, S.}, \bibinfo{author}{{Goicoechea}, J.R.},
  \bibinfo{author}{{Cernicharo}, J.}, \bibinfo{author}{{Fuente}, A.},
  \bibinfo{author}{{Pety}, J.}, \bibinfo{author}{{Tercero}, B.},
  \bibinfo{year}{2017}.
\newblock \bibinfo{title}{{Complex organic molecules in strongly UV-irradiated
  gas}}.
\newblock \bibinfo{journal}{\aap} \bibinfo{volume}{603}, \bibinfo{pages}{A124}.
\newblock \DOIprefix\doi{10.1051/0004-6361/201730459},
  \href{http://arxiv.org/abs/1705.06612}{{\tt arXiv:1705.06612}}.
%Type = Article
\bibitem[{{Cummins} et~al.(1986){Cummins}, {Linke} and {Thaddeus}}]{Cummins86}
\bibinfo{author}{{Cummins}, S.E.}, \bibinfo{author}{{Linke}, R.A.},
  \bibinfo{author}{{Thaddeus}, P.}, \bibinfo{year}{1986}.
\newblock \bibinfo{title}{{A Survey of the Millimeter-Wave Spectrum of
  Sagittarius B2}}.
\newblock \bibinfo{journal}{\apjs} \bibinfo{volume}{60}, \bibinfo{pages}{819}.
\newblock \DOIprefix\doi{10.1086/191102}.
%Type = Article
\bibitem[{{Dahmen} et~al.(1998){Dahmen}, {Huttemeister}, {Wilson} and
  {Mauersberger}}]{dahmen98}
\bibinfo{author}{{Dahmen}, G.}, \bibinfo{author}{{Huttemeister}, S.},
  \bibinfo{author}{{Wilson}, T.L.}, \bibinfo{author}{{Mauersberger}, R.},
  \bibinfo{year}{1998}.
\newblock \bibinfo{title}{{Molecular gas in the Galactic center region. II. Gas
  mass and N\_, = H\_2/I\_\^(12)CO conversion based on a C\^(18)O(J = 1 -> 0)
  survey}}.
\newblock \bibinfo{journal}{\aap} \bibinfo{volume}{331},
  \bibinfo{pages}{959--976}.
\newblock \href{http://arxiv.org/abs/astro-ph/9711117}{{\tt
  arXiv:astro-ph/9711117}}.
%Type = Article
\bibitem[{{Dartois} et~al.(2019){Dartois}, {Chabot}, {Id Barkach}, {Rothard},
  {Aug{\'e}}, {Agnihotri}, {Domaracka} and {Boduch}}]{dartois19}
\bibinfo{author}{{Dartois}, E.}, \bibinfo{author}{{Chabot}, M.},
  \bibinfo{author}{{Id Barkach}, T.}, \bibinfo{author}{{Rothard}, H.},
  \bibinfo{author}{{Aug{\'e}}, B.}, \bibinfo{author}{{Agnihotri}, A.N.},
  \bibinfo{author}{{Domaracka}, A.}, \bibinfo{author}{{Boduch}, P.},
  \bibinfo{year}{2019}.
\newblock \bibinfo{title}{{Non-thermal desorption of complex organic molecules.
  Efficient CH$_{3}$OH and CH$_{3}$COOCH$_{3}$ sputtering by cosmic rays
  (Corrigendum)}}.
\newblock \bibinfo{journal}{\aap} \bibinfo{volume}{628}, \bibinfo{pages}{C2}.
\newblock \DOIprefix\doi{10.1051/0004-6361/201834787e}.
%Type = Article
\bibitem[{{De Simone} et~al.(2020a){De Simone}, {Ceccarelli}, {Codella},
  {Svoboda}, {Chandler}, {Bouvier}, {Yamamoto}, {Sakai}, {Caselli}, {Favre},
  {Loinard}, {Lefloch}, {Liu}, {L{\'o}pez-Sepulcre}, {Pineda}, {Taquet} and
  {Testi}}]{Desimone20b}
\bibinfo{author}{{De Simone}, M.}, \bibinfo{author}{{Ceccarelli}, C.},
  \bibinfo{author}{{Codella}, C.}, \bibinfo{author}{{Svoboda}, B.E.},
  \bibinfo{author}{{Chandler}, C.}, \bibinfo{author}{{Bouvier}, M.},
  \bibinfo{author}{{Yamamoto}, S.}, \bibinfo{author}{{Sakai}, N.},
  \bibinfo{author}{{Caselli}, P.}, \bibinfo{author}{{Favre}, C.},
  \bibinfo{author}{{Loinard}, L.}, \bibinfo{author}{{Lefloch}, B.},
  \bibinfo{author}{{Liu}, H.B.}, \bibinfo{author}{{L{\'o}pez-Sepulcre}, A.},
  \bibinfo{author}{{Pineda}, J.E.}, \bibinfo{author}{{Taquet}, V.},
  \bibinfo{author}{{Testi}, L.}, \bibinfo{year}{2020}a.
\newblock \bibinfo{title}{{Hot Corinos Chemical Diversity: Myth or Reality?}}
\newblock \bibinfo{journal}{\apjl} \bibinfo{volume}{896}, \bibinfo{pages}{L3}.
\newblock \DOIprefix\doi{10.3847/2041-8213/ab8d41},
  \href{http://arxiv.org/abs/2006.04484}{{\tt arXiv:2006.04484}}.
%Type = Article
\bibitem[{{De Simone} et~al.(2020b){De Simone}, {Codella}, {Ceccarelli},
  {L{\'o}pez-Sepulcre}, {Witzel}, {Neri}, {Balucani}, {Caselli}, {Favre},
  {Fontani}, {Lefloch}, {Ospina-Zamudio}, {Pineda} and {Taquet}}]{Desimone20a}
\bibinfo{author}{{De Simone}, M.}, \bibinfo{author}{{Codella}, C.},
  \bibinfo{author}{{Ceccarelli}, C.}, \bibinfo{author}{{L{\'o}pez-Sepulcre},
  A.}, \bibinfo{author}{{Witzel}, A.}, \bibinfo{author}{{Neri}, R.},
  \bibinfo{author}{{Balucani}, N.}, \bibinfo{author}{{Caselli}, P.},
  \bibinfo{author}{{Favre}, C.}, \bibinfo{author}{{Fontani}, F.},
  \bibinfo{author}{{Lefloch}, B.}, \bibinfo{author}{{Ospina-Zamudio}, J.},
  \bibinfo{author}{{Pineda}, J.E.}, \bibinfo{author}{{Taquet}, V.},
  \bibinfo{year}{2020}b.
\newblock \bibinfo{title}{{Seeds of Life in Space (SOLIS). X. Interstellar
  complex organic molecules in the NGC 1333 IRAS 4A outflows}}.
\newblock \bibinfo{journal}{\aap} \bibinfo{volume}{640}, \bibinfo{pages}{A75}.
\newblock \DOIprefix\doi{10.1051/0004-6361/201937004},
  \href{http://arxiv.org/abs/2006.09925}{{\tt arXiv:2006.09925}}.
%Type = Article
\bibitem[{{De Simone} et~al.(2017){De Simone}, {Codella}, {Testi}, {Belloche},
  {Maury}, {Anderl}, {Andr{\'e}}, {Maret} and {Podio}}]{DeSimone17}
\bibinfo{author}{{De Simone}, M.}, \bibinfo{author}{{Codella}, C.},
  \bibinfo{author}{{Testi}, L.}, \bibinfo{author}{{Belloche}, A.},
  \bibinfo{author}{{Maury}, A.J.}, \bibinfo{author}{{Anderl}, S.},
  \bibinfo{author}{{Andr{\'e}}, P.}, \bibinfo{author}{{Maret}, S.},
  \bibinfo{author}{{Podio}, L.}, \bibinfo{year}{2017}.
\newblock \bibinfo{title}{{Glycolaldehyde in Perseus young solar analogs}}.
\newblock \bibinfo{journal}{\aap} \bibinfo{volume}{599}, \bibinfo{pages}{A121}.
\newblock \DOIprefix\doi{10.1051/0004-6361/201630049},
  \href{http://arxiv.org/abs/1701.00724}{{\tt arXiv:1701.00724}}.
%Type = Article
\bibitem[{{De Vries} and {Myers}(2005)}]{devries05}
\bibinfo{author}{{De Vries}, C.H.}, \bibinfo{author}{{Myers}, P.C.},
  \bibinfo{year}{2005}.
\newblock \bibinfo{title}{{Molecular Line Profile Fitting with Analytic
  Radiative Transfer Models}}.
\newblock \bibinfo{journal}{\apj} \bibinfo{volume}{620},
  \bibinfo{pages}{800--815}.
\newblock \DOIprefix\doi{10.1086/427141},
  \href{http://arxiv.org/abs/astro-ph/0410748}{{\tt arXiv:astro-ph/0410748}}.
%Type = Article
\bibitem[{{Douglas} et~al.(2022){Douglas}, {Lucas}, {Walsh}, {West}, {Blitz}
  and {Heard}}]{Douglas22}
\bibinfo{author}{{Douglas}, K.M.}, \bibinfo{author}{{Lucas}, D.I.},
  \bibinfo{author}{{Walsh}, C.}, \bibinfo{author}{{West}, N.A.},
  \bibinfo{author}{{Blitz}, M.A.}, \bibinfo{author}{{Heard}, D.E.},
  \bibinfo{year}{2022}.
\newblock \bibinfo{title}{{The Gas-phase Reaction of NH$_{2}$ with Formaldehyde
  (CH$_{2}$O) is not a Source of Formamide (NH$_{2}$CHO) in Interstellar
  Environments}}.
\newblock \bibinfo{journal}{\apjl} \bibinfo{volume}{937}, \bibinfo{pages}{L16}.
\newblock \DOIprefix\doi{10.3847/2041-8213/ac8cef},
  \href{http://arxiv.org/abs/2208.12658}{{\tt arXiv:2208.12658}}.
%Type = Article
\bibitem[{{Drozdovskaya} et~al.(2018){Drozdovskaya}, {van Dishoeck},
  {J{\o}rgensen}, {Calmonte}, {van der Wiel}, {Coutens}, {Calcutt},
  {M{\"u}ller}, {Bjerkeli}, {Persson}, {Wampfler} and {Altwegg}}]{Droz18}
\bibinfo{author}{{Drozdovskaya}, M.N.}, \bibinfo{author}{{van Dishoeck}, E.F.},
  \bibinfo{author}{{J{\o}rgensen}, J.K.}, \bibinfo{author}{{Calmonte}, U.},
  \bibinfo{author}{{van der Wiel}, M.H.D.}, \bibinfo{author}{{Coutens}, A.},
  \bibinfo{author}{{Calcutt}, H.}, \bibinfo{author}{{M{\"u}ller}, H.S.P.},
  \bibinfo{author}{{Bjerkeli}, P.}, \bibinfo{author}{{Persson}, M.V.},
  \bibinfo{author}{{Wampfler}, S.F.}, \bibinfo{author}{{Altwegg}, K.},
  \bibinfo{year}{2018}.
\newblock \bibinfo{title}{{The ALMA-PILS survey: the sulphur connection between
  protostars and comets: IRAS 16293-2422 B and 67P/Churyumov-Gerasimenko}}.
\newblock \bibinfo{journal}{\mnras} \bibinfo{volume}{476},
  \bibinfo{pages}{4949--4964}.
\newblock \DOIprefix\doi{10.1093/mnras/sty462},
  \href{http://arxiv.org/abs/1802.02977}{{\tt arXiv:1802.02977}}.
%Type = Article
\bibitem[{{Drozdovskaya} et~al.(2019){Drozdovskaya}, {van Dishoeck}, {Rubin},
  {J{\o}rgensen} and {Altwegg}}]{Droz19}
\bibinfo{author}{{Drozdovskaya}, M.N.}, \bibinfo{author}{{van Dishoeck}, E.F.},
  \bibinfo{author}{{Rubin}, M.}, \bibinfo{author}{{J{\o}rgensen}, J.K.},
  \bibinfo{author}{{Altwegg}, K.}, \bibinfo{year}{2019}.
\newblock \bibinfo{title}{{Ingredients for solar-like systems: protostar IRAS
  16293-2422 B versus comet 67P/Churyumov-Gerasimenko}}.
\newblock \bibinfo{journal}{\mnras} \bibinfo{volume}{490},
  \bibinfo{pages}{50--79}.
\newblock \DOIprefix\doi{10.1093/mnras/stz2430},
  \href{http://arxiv.org/abs/1908.11290}{{\tt arXiv:1908.11290}}.
%Type = Article
\bibitem[{{Dulieu} et~al.(2019){Dulieu}, {Nguyen}, {Congiu}, {Baouche} and
  {Taquet}}]{dulieu19}
\bibinfo{author}{{Dulieu}, F.}, \bibinfo{author}{{Nguyen}, T.},
  \bibinfo{author}{{Congiu}, E.}, \bibinfo{author}{{Baouche}, S.},
  \bibinfo{author}{{Taquet}, V.}, \bibinfo{year}{2019}.
\newblock \bibinfo{title}{{Efficient formation route of the prebiotic molecule
  formamide on interstellar dust grains}}.
\newblock \bibinfo{journal}{\mnras} \bibinfo{volume}{484},
  \bibinfo{pages}{L119--L123}.
\newblock \DOIprefix\doi{10.1093/mnrasl/slz013}.
%Type = Article
\bibitem[{{El-Abd} et~al.(2019){El-Abd}, {Brogan}, {Hunter}, {Willis}, {Garrod}
  and {McGuire}}]{ElAbd19}
\bibinfo{author}{{El-Abd}, S.J.}, \bibinfo{author}{{Brogan}, C.L.},
  \bibinfo{author}{{Hunter}, T.R.}, \bibinfo{author}{{Willis}, E.R.},
  \bibinfo{author}{{Garrod}, R.T.}, \bibinfo{author}{{McGuire}, B.A.},
  \bibinfo{year}{2019}.
\newblock \bibinfo{title}{{Interstellar Glycolaldehyde, Methyl Formate, and
  Acetic Acid. I. A Bimodal Abundance Pattern in Star-forming Regions}}.
\newblock \bibinfo{journal}{\apj} \bibinfo{volume}{883}, \bibinfo{pages}{129}.
\newblock \DOIprefix\doi{10.3847/1538-4357/ab3646},
  \href{http://arxiv.org/abs/1907.13551}{{\tt arXiv:1907.13551}}.
%Type = Article
\bibitem[{{Eyres} et~al.(2018){Eyres}, {Evans}, {Zijlstra}, {Avison}, {Gehrz},
  {Hajduk}, {Starrfield}, {Mohamed}, {Woodward} and {Wagner}}]{Eyres18}
\bibinfo{author}{{Eyres}, S.P.S.}, \bibinfo{author}{{Evans}, A.},
  \bibinfo{author}{{Zijlstra}, A.}, \bibinfo{author}{{Avison}, A.},
  \bibinfo{author}{{Gehrz}, R.D.}, \bibinfo{author}{{Hajduk}, M.},
  \bibinfo{author}{{Starrfield}, S.}, \bibinfo{author}{{Mohamed}, S.},
  \bibinfo{author}{{Woodward}, C.E.}, \bibinfo{author}{{Wagner}, R.M.},
  \bibinfo{year}{2018}.
\newblock \bibinfo{title}{{ALMA reveals the aftermath of a white dwarf-brown
  dwarf merger in CK Vulpeculae}}.
\newblock \bibinfo{journal}{\mnras} \bibinfo{volume}{481},
  \bibinfo{pages}{4931--4939}.
\newblock \DOIprefix\doi{10.1093/mnras/sty2554},
  \href{http://arxiv.org/abs/1809.05849}{{\tt arXiv:1809.05849}}.
%Type = Article
\bibitem[{{Favre} et~al.(2011){Favre}, {Despois}, {Brouillet}, {Baudry},
  {Combes}, {Gu{\'e}lin}, {Wootten} and {Wlodarczak}}]{favre11}
\bibinfo{author}{{Favre}, C.}, \bibinfo{author}{{Despois}, D.},
  \bibinfo{author}{{Brouillet}, N.}, \bibinfo{author}{{Baudry}, A.},
  \bibinfo{author}{{Combes}, F.}, \bibinfo{author}{{Gu{\'e}lin}, M.},
  \bibinfo{author}{{Wootten}, A.}, \bibinfo{author}{{Wlodarczak}, G.},
  \bibinfo{year}{2011}.
\newblock \bibinfo{title}{{HCOOCH$_{3}$ as a probe of temperature and structure
  in Orion-KL}}.
\newblock \bibinfo{journal}{\aap} \bibinfo{volume}{532}, \bibinfo{pages}{A32}.
\newblock \DOIprefix\doi{10.1051/0004-6361/201015345}.
%Type = Article
\bibitem[{{Fontani} et~al.(2014){Fontani}, {Codella}, {Ceccarelli}, {Lefloch},
  {Viti} and {Benedettini}}]{Fontani14}
\bibinfo{author}{{Fontani}, F.}, \bibinfo{author}{{Codella}, C.},
  \bibinfo{author}{{Ceccarelli}, C.}, \bibinfo{author}{{Lefloch}, B.},
  \bibinfo{author}{{Viti}, S.}, \bibinfo{author}{{Benedettini}, M.},
  \bibinfo{year}{2014}.
\newblock \bibinfo{title}{{The L1157-B1 Astrochemical Laboratory: Measuring the
  True Formaldehyde Deuteration on Grain Mantles}}.
\newblock \bibinfo{journal}{\apjl} \bibinfo{volume}{788}, \bibinfo{pages}{L43}.
\newblock \DOIprefix\doi{10.1088/2041-8205/788/2/L43}.
%Type = Article
\bibitem[{{For} et~al.(2018){For}, {Staveley-Smith}, {Hurley-Walker},
  {Franzen}, {Kapi{\'n}ska}, {Filipovi{\'c}}, {Collier}, {Wu}, {Grieve},
  {Callingham}, {Bell}, {Bernardi}, {Bowman}, {Briggs}, {Cappallo},
  {Deshpande}, {Dwarakanath}, {Gaensler}, {Greenhill}, {Hancock}, {Hazelton},
  {Hindson}, {Johnston-Hollitt}, {Kaplan}, {Lenc}, {Lonsdale}, {McKinley},
  {McWhirter}, {Mitchell}, {Morales}, {Morgan}, {Morgan}, {Oberoi}, {Offringa},
  {Ord}, {Prabu}, {Procopio}, {Shankar}, {Srivani}, {Subrahmanyan}, {Tingay},
  {Wayth}, {Webster}, {Williams}, {Williams} and {Zheng}}]{For18}
\bibinfo{author}{{For}, B.Q.}, \bibinfo{author}{{Staveley-Smith}, L.},
  \bibinfo{author}{{Hurley-Walker}, N.}, \bibinfo{author}{{Franzen}, T.},
  \bibinfo{author}{{Kapi{\'n}ska}, A.D.}, \bibinfo{author}{{Filipovi{\'c}},
  M.D.}, \bibinfo{author}{{Collier}, J.D.}, \bibinfo{author}{{Wu}, C.},
  \bibinfo{author}{{Grieve}, K.}, \bibinfo{author}{{Callingham}, J.R.},
  \bibinfo{author}{{Bell}, M.E.}, \bibinfo{author}{{Bernardi}, G.},
  \bibinfo{author}{{Bowman}, J.D.}, \bibinfo{author}{{Briggs}, F.},
  \bibinfo{author}{{Cappallo}, R.J.}, \bibinfo{author}{{Deshpande}, A.A.},
  \bibinfo{author}{{Dwarakanath}, K.S.}, \bibinfo{author}{{Gaensler}, B.M.},
  \bibinfo{author}{{Greenhill}, L.J.}, \bibinfo{author}{{Hancock}, P.},
  \bibinfo{author}{{Hazelton}, B.J.}, \bibinfo{author}{{Hindson}, L.},
  \bibinfo{author}{{Johnston-Hollitt}, M.}, \bibinfo{author}{{Kaplan}, D.L.},
  \bibinfo{author}{{Lenc}, E.}, \bibinfo{author}{{Lonsdale}, C.J.},
  \bibinfo{author}{{McKinley}, B.}, \bibinfo{author}{{McWhirter}, S.R.},
  \bibinfo{author}{{Mitchell}, D.A.}, \bibinfo{author}{{Morales}, M.F.},
  \bibinfo{author}{{Morgan}, E.}, \bibinfo{author}{{Morgan}, J.},
  \bibinfo{author}{{Oberoi}, D.}, \bibinfo{author}{{Offringa}, A.},
  \bibinfo{author}{{Ord}, S.M.}, \bibinfo{author}{{Prabu}, T.},
  \bibinfo{author}{{Procopio}, P.}, \bibinfo{author}{{Shankar}, N.U.},
  \bibinfo{author}{{Srivani}, K.S.}, \bibinfo{author}{{Subrahmanyan}, R.},
  \bibinfo{author}{{Tingay}, S.J.}, \bibinfo{author}{{Wayth}, R.B.},
  \bibinfo{author}{{Webster}, R.L.}, \bibinfo{author}{{Williams}, A.},
  \bibinfo{author}{{Williams}, C.L.}, \bibinfo{author}{{Zheng}, Q.},
  \bibinfo{year}{2018}.
\newblock \bibinfo{title}{{A multifrequency radio continuum study of the
  Magellanic Clouds - I. Overall structure and star formation rates}}.
\newblock \bibinfo{journal}{\mnras} \bibinfo{volume}{480},
  \bibinfo{pages}{2743--2756}.
\newblock \DOIprefix\doi{10.1093/mnras/sty1960}.
%Type = Inproceedings
\bibitem[{{Frank} et~al.(2014){Frank}, {Ray}, {Cabrit}, {Hartigan}, {Arce},
  {Bacciotti}, {Bally}, {Benisty}, {Eisl{\"o}ffel}, {G{\"u}del}, {Lebedev},
  {Nisini} and {Raga}}]{Frank14}
\bibinfo{author}{{Frank}, A.}, \bibinfo{author}{{Ray}, T.P.},
  \bibinfo{author}{{Cabrit}, S.}, \bibinfo{author}{{Hartigan}, P.},
  \bibinfo{author}{{Arce}, H.G.}, \bibinfo{author}{{Bacciotti}, F.},
  \bibinfo{author}{{Bally}, J.}, \bibinfo{author}{{Benisty}, M.},
  \bibinfo{author}{{Eisl{\"o}ffel}, J.}, \bibinfo{author}{{G{\"u}del}, M.},
  \bibinfo{author}{{Lebedev}, S.}, \bibinfo{author}{{Nisini}, B.},
  \bibinfo{author}{{Raga}, A.}, \bibinfo{year}{2014}.
\newblock \bibinfo{title}{{Jets and Outflows from Star to Cloud: Observations
  Confront Theory}}, in: \bibinfo{editor}{{Beuther}, H.},
  \bibinfo{editor}{{Klessen}, R.S.}, \bibinfo{editor}{{Dullemond}, C.P.},
  \bibinfo{editor}{{Henning}, T.} (Eds.), \bibinfo{booktitle}{Protostars and
  Planets VI}, p. \bibinfo{pages}{451}.
\newblock \DOIprefix\doi{10.2458/azu\_uapress\_9780816531240-ch020},
  \href{http://arxiv.org/abs/1402.3553}{{\tt arXiv:1402.3553}}.
%Type = Article
\bibitem[{{Fuente} et~al.(2014){Fuente}, {Cernicharo}, {Caselli}, {McCoey},
  {Johnstone}, {Fich}, {van Kempen}, {Palau}, {Y{\i}ld{\i}z}, {Tercero} and
  {L{\'o}pez}}]{Fuente14}
\bibinfo{author}{{Fuente}, A.}, \bibinfo{author}{{Cernicharo}, J.},
  \bibinfo{author}{{Caselli}, P.}, \bibinfo{author}{{McCoey}, C.},
  \bibinfo{author}{{Johnstone}, D.}, \bibinfo{author}{{Fich}, M.},
  \bibinfo{author}{{van Kempen}, T.}, \bibinfo{author}{{Palau}, A.},
  \bibinfo{author}{{Y{\i}ld{\i}z}, U.A.}, \bibinfo{author}{{Tercero}, B.},
  \bibinfo{author}{{L{\'o}pez}, A.}, \bibinfo{year}{2014}.
\newblock \bibinfo{title}{{The hot core towards the intermediate-mass protostar
  NGC 7129 FIRS 2. Chemical similarities with Orion KL}}.
\newblock \bibinfo{journal}{\aap} \bibinfo{volume}{568}, \bibinfo{pages}{A65}.
\newblock \DOIprefix\doi{10.1051/0004-6361/201323074},
  \href{http://arxiv.org/abs/1405.4639}{{\tt arXiv:1405.4639}}.
%Type = Article
\bibitem[{{Furukawa} et~al.(2019){Furukawa}, {Chikaraishi}, {Ohkouchi},
  {Ogawa}, {Glavin}, {Dworkin}, {Abe} and {Nakamura}}]{furukawa2019}
\bibinfo{author}{{Furukawa}, Y.}, \bibinfo{author}{{Chikaraishi}, Y.},
  \bibinfo{author}{{Ohkouchi}, N.}, \bibinfo{author}{{Ogawa}, N.O.},
  \bibinfo{author}{{Glavin}, D.P.}, \bibinfo{author}{{Dworkin}, J.P.},
  \bibinfo{author}{{Abe}, C.}, \bibinfo{author}{{Nakamura}, T.},
  \bibinfo{year}{2019}.
\newblock \bibinfo{title}{{Extraterrestrial ribose and other sugars in
  primitive meteorites}}.
\newblock \bibinfo{journal}{Proceedings of the National Academy of Science}
  \bibinfo{volume}{116}, \bibinfo{pages}{24440--24445}.
\newblock \DOIprefix\doi{10.1073/pnas.1907169116}.
%Type = Article
\bibitem[{{Garay} and {Lizano}(1999)}]{Garay99}
\bibinfo{author}{{Garay}, G.}, \bibinfo{author}{{Lizano}, S.},
  \bibinfo{year}{1999}.
\newblock \bibinfo{title}{{Massive Stars: Their Environment and Formation}}.
\newblock \bibinfo{journal}{\pasp} \bibinfo{volume}{111},
  \bibinfo{pages}{1049--1087}.
\newblock \DOIprefix\doi{10.1086/316416},
  \href{http://arxiv.org/abs/astro-ph/9907293}{{\tt arXiv:astro-ph/9907293}}.
%Type = Article
\bibitem[{{Garc{\'\i}a de la Concepci{\'o}n} et~al.(2023){Garc{\'\i}a de la
  Concepci{\'o}n}, {Jim{\'e}nez-Serra}, {Rivilla}, {Colzi} and
  {Mart{\'\i}n-Pintado}}]{garcia-concepcion2023}
\bibinfo{author}{{Garc{\'\i}a de la Concepci{\'o}n}, J.},
  \bibinfo{author}{{Jim{\'e}nez-Serra}, I.}, \bibinfo{author}{{Rivilla}, V.M.},
  \bibinfo{author}{{Colzi}, L.}, \bibinfo{author}{{Mart{\'\i}n-Pintado}, J.},
  \bibinfo{year}{2023}.
\newblock \bibinfo{title}{{Paving the way to the synthesis of PAHs in dark
  molecular clouds: The formation of cyclopentadienyl radical
  (c-C$_{5}$H$_{5}$)}}.
\newblock \bibinfo{journal}{\aap} \bibinfo{volume}{673}, \bibinfo{pages}{A118}.
\newblock \DOIprefix\doi{10.1051/0004-6361/202345854}.
%Type = Article
\bibitem[{{Garc{\'\i}a-S{\'a}nchez} et~al.(2022){Garc{\'\i}a-S{\'a}nchez},
  {Jim{\'e}nez-Serra}, {Puente-S{\'a}nchez} and {Aguirre}}]{garcia-sanchez22}
\bibinfo{author}{{Garc{\'\i}a-S{\'a}nchez}, M.},
  \bibinfo{author}{{Jim{\'e}nez-Serra}, I.},
  \bibinfo{author}{{Puente-S{\'a}nchez}, F.}, \bibinfo{author}{{Aguirre}, J.},
  \bibinfo{year}{2022}.
\newblock \bibinfo{title}{{The emergence of interstellar molecular complexity
  explained by interacting networks}}.
\newblock \bibinfo{journal}{Proceedings of the National Academy of Science}
  \bibinfo{volume}{119}, \bibinfo{pages}{e2119734119}.
\newblock \DOIprefix\doi{10.1073/pnas.2119734119},
  \href{http://arxiv.org/abs/2207.14176}{{\tt arXiv:2207.14176}}.
%Type = Article
\bibitem[{{Garrod}(2013)}]{garrod13}
\bibinfo{author}{{Garrod}, R.T.}, \bibinfo{year}{2013}.
\newblock \bibinfo{title}{{A Three-phase Chemical Model of Hot Cores: The
  Formation of Glycine}}.
\newblock \bibinfo{journal}{\apj} \bibinfo{volume}{765}, \bibinfo{pages}{60}.
\newblock \DOIprefix\doi{10.1088/0004-637X/765/1/60},
  \href{http://arxiv.org/abs/1302.0688}{{\tt arXiv:1302.0688}}.
%Type = Article
\bibitem[{{Garrod} and {Herbst}(2006)}]{Garrod06}
\bibinfo{author}{{Garrod}, R.T.}, \bibinfo{author}{{Herbst}, E.},
  \bibinfo{year}{2006}.
\newblock \bibinfo{title}{{Formation of methyl formate and other organic
  species in the warm-up phase of hot molecular cores}}.
\newblock \bibinfo{journal}{\aap} \bibinfo{volume}{457},
  \bibinfo{pages}{927--936}.
\newblock \DOIprefix\doi{10.1051/0004-6361:20065560},
  \href{http://arxiv.org/abs/astro-ph/0607560}{{\tt arXiv:astro-ph/0607560}}.
%Type = Article
\bibitem[{{Garrod} et~al.(2022){Garrod}, {Jin}, {Matis}, {Jones}, {Willis} and
  {Herbst}}]{Garrod22}
\bibinfo{author}{{Garrod}, R.T.}, \bibinfo{author}{{Jin}, M.},
  \bibinfo{author}{{Matis}, K.A.}, \bibinfo{author}{{Jones}, D.},
  \bibinfo{author}{{Willis}, E.R.}, \bibinfo{author}{{Herbst}, E.},
  \bibinfo{year}{2022}.
\newblock \bibinfo{title}{{Formation of Complex Organic Molecules in Hot
  Molecular Cores through Nondiffusive Grain-surface and Ice-mantle
  Chemistry}}.
\newblock \bibinfo{journal}{\apjs} \bibinfo{volume}{259}, \bibinfo{pages}{1}.
\newblock \DOIprefix\doi{10.3847/1538-4365/ac3131},
  \href{http://arxiv.org/abs/2110.09743}{{\tt arXiv:2110.09743}}.
%Type = Article
\bibitem[{{Garufi} et~al.(2022){Garufi}, {Podio}, {Codella}, {Segura-Cox},
  {Vander Donckt}, {Mercimek}, {Bacciotti}, {Fedele}, {Kasper}, {Pineda},
  {Humphreys} and {Testi}}]{Garufi22}
\bibinfo{author}{{Garufi}, A.}, \bibinfo{author}{{Podio}, L.},
  \bibinfo{author}{{Codella}, C.}, \bibinfo{author}{{Segura-Cox}, D.},
  \bibinfo{author}{{Vander Donckt}, M.}, \bibinfo{author}{{Mercimek}, S.},
  \bibinfo{author}{{Bacciotti}, F.}, \bibinfo{author}{{Fedele}, D.},
  \bibinfo{author}{{Kasper}, M.}, \bibinfo{author}{{Pineda}, J.E.},
  \bibinfo{author}{{Humphreys}, E.}, \bibinfo{author}{{Testi}, L.},
  \bibinfo{year}{2022}.
\newblock \bibinfo{title}{{ALMA chemical survey of disk-outflow sources in
  Taurus (ALMA-DOT). VI. Accretion shocks in the disk of DG Tau and HL Tau}}.
\newblock \bibinfo{journal}{\aap} \bibinfo{volume}{658}, \bibinfo{pages}{A104}.
\newblock \DOIprefix\doi{10.1051/0004-6361/202141264},
  \href{http://arxiv.org/abs/2110.13820}{{\tt arXiv:2110.13820}}.
%Type = Article
\bibitem[{{Gieser} et~al.(2021){Gieser}, {Beuther}, {Semenov}, {Ahmadi},
  {Suri}, {M{\"o}ller}, {Beltr{\'a}n}, {Klaassen}, {Zhang}, {Urquhart},
  {Henning}, {Feng}, {Galv{\'a}n-Madrid}, {de Souza Magalh{\~a}es},
  {Moscadelli}, {Longmore}, {Leurini}, {Kuiper}, {Peters}, {Menten},
  {Csengeri}, {Fuller}, {Wyrowski}, {Lumsden}, {S{\'a}nchez-Monge}, {Maud},
  {Linz}, {Palau}, {Schilke}, {Pety}, {Pudritz}, {Winters} and
  {Pi{\'e}tu}}]{Gieser21}
\bibinfo{author}{{Gieser}, C.}, \bibinfo{author}{{Beuther}, H.},
  \bibinfo{author}{{Semenov}, D.}, \bibinfo{author}{{Ahmadi}, A.},
  \bibinfo{author}{{Suri}, S.}, \bibinfo{author}{{M{\"o}ller}, T.},
  \bibinfo{author}{{Beltr{\'a}n}, M.T.}, \bibinfo{author}{{Klaassen}, P.},
  \bibinfo{author}{{Zhang}, Q.}, \bibinfo{author}{{Urquhart}, J.S.},
  \bibinfo{author}{{Henning}, T.}, \bibinfo{author}{{Feng}, S.},
  \bibinfo{author}{{Galv{\'a}n-Madrid}, R.}, \bibinfo{author}{{de Souza
  Magalh{\~a}es}, V.}, \bibinfo{author}{{Moscadelli}, L.},
  \bibinfo{author}{{Longmore}, S.}, \bibinfo{author}{{Leurini}, S.},
  \bibinfo{author}{{Kuiper}, R.}, \bibinfo{author}{{Peters}, T.},
  \bibinfo{author}{{Menten}, K.M.}, \bibinfo{author}{{Csengeri}, T.},
  \bibinfo{author}{{Fuller}, G.}, \bibinfo{author}{{Wyrowski}, F.},
  \bibinfo{author}{{Lumsden}, S.}, \bibinfo{author}{{S{\'a}nchez-Monge},
  {\'A}.}, \bibinfo{author}{{Maud}, L.}, \bibinfo{author}{{Linz}, H.},
  \bibinfo{author}{{Palau}, A.}, \bibinfo{author}{{Schilke}, P.},
  \bibinfo{author}{{Pety}, J.}, \bibinfo{author}{{Pudritz}, R.},
  \bibinfo{author}{{Winters}, J.M.}, \bibinfo{author}{{Pi{\'e}tu}, V.},
  \bibinfo{year}{2021}.
\newblock \bibinfo{title}{{Physical and chemical structure of high-mass
  star-forming regions. Unraveling chemical complexity with CORE: the NOEMA
  large program}}.
\newblock \bibinfo{journal}{\aap} \bibinfo{volume}{648}, \bibinfo{pages}{A66}.
\newblock \DOIprefix\doi{10.1051/0004-6361/202039670},
  \href{http://arxiv.org/abs/2102.11676}{{\tt arXiv:2102.11676}}.
%Type = Article
\bibitem[{{Ginsburg} et~al.(2016){Ginsburg}, {Henkel}, {Ao}, {Riquelme},
  {Kauffmann}, {Pillai}, {Mills}, {Requena-Torres}, {Immer}, {Testi}, {Ott},
  {Bally}, {Battersby}, {Darling}, {Aalto}, {Stanke}, {Kendrew}, {Kruijssen},
  {Longmore}, {Dale}, {Guesten} and {Menten}}]{Ginsburg16}
\bibinfo{author}{{Ginsburg}, A.}, \bibinfo{author}{{Henkel}, C.},
  \bibinfo{author}{{Ao}, Y.}, \bibinfo{author}{{Riquelme}, D.},
  \bibinfo{author}{{Kauffmann}, J.}, \bibinfo{author}{{Pillai}, T.},
  \bibinfo{author}{{Mills}, E.A.C.}, \bibinfo{author}{{Requena-Torres}, M.A.},
  \bibinfo{author}{{Immer}, K.}, \bibinfo{author}{{Testi}, L.},
  \bibinfo{author}{{Ott}, J.}, \bibinfo{author}{{Bally}, J.},
  \bibinfo{author}{{Battersby}, C.}, \bibinfo{author}{{Darling}, J.},
  \bibinfo{author}{{Aalto}, S.}, \bibinfo{author}{{Stanke}, T.},
  \bibinfo{author}{{Kendrew}, S.}, \bibinfo{author}{{Kruijssen}, J.M.D.},
  \bibinfo{author}{{Longmore}, S.}, \bibinfo{author}{{Dale}, J.},
  \bibinfo{author}{{Guesten}, R.}, \bibinfo{author}{{Menten}, K.M.},
  \bibinfo{year}{2016}.
\newblock \bibinfo{title}{{Dense gas in the Galactic central molecular zone is
  warm and heated by turbulence}}.
\newblock \bibinfo{journal}{\aap} \bibinfo{volume}{586}, \bibinfo{pages}{A50}.
\newblock \DOIprefix\doi{10.1051/0004-6361/201526100},
  \href{http://arxiv.org/abs/1509.01583}{{\tt arXiv:1509.01583}}.
%Type = Article
\bibitem[{{Glavin} and {Dworkin}(2009)}]{glavin2009}
\bibinfo{author}{{Glavin}, D.P.}, \bibinfo{author}{{Dworkin}, J.P.},
  \bibinfo{year}{2009}.
\newblock \bibinfo{title}{{Enrichment in L-Isovaline by Aqueous Alteration on
  CI and CM Meteorite Parent Bodies}}.
\newblock \bibinfo{journal}{Meteoritics and Planetary Science Supplement}
  \bibinfo{volume}{72}, \bibinfo{pages}{5009}.
%Type = Article
\bibitem[{{Glavin} et~al.(2025){Glavin}, {Dworkin}, {Alexander}, {Aponte},
  {Baczynski}, {Barnes}, {Bechtel}, {Berger}, {Burton}, {Caselli}, {Chung},
  {Clemett}, {Cody}, {Dominguez}, {Elsila}, {Farnsworth}, {Foustoukos},
  {Freeman}, {Furukawa}, {Gainsforth}, {Graham}, {Grassi}, {Giuliano},
  {Hamilton}, {Haenecour}, {Heck}, {Hofmann}, {House}, {Huang}, {Kaplan},
  {Keller}, {Kim}, {Koga}, {Liss}, {McLain}, {Marcus}, {Matney}, {McCoy},
  {McIntosh}, {Mojarro}, {Naraoka}, {Nguyen}, {Nuevo}, {Nuth}, {Oba}, {Parker},
  {Peretyazhko}, {Sandford}, {Santos}, {Schmitt-Kopplin}, {Seguin}, {Simkus},
  {Shahid}, {Takano}, {Thomas-Keprta}, {Tripathi}, {Weiss}, {Zheng}, {Lunning},
  {Righter}, {Connolly} and {Lauretta}}]{glavin2025}
\bibinfo{author}{{Glavin}, D.P.}, \bibinfo{author}{{Dworkin}, J.P.},
  \bibinfo{author}{{Alexander}, C.M.O.}, \bibinfo{author}{{Aponte}, J.C.},
  \bibinfo{author}{{Baczynski}, A.A.}, \bibinfo{author}{{Barnes}, J.J.},
  \bibinfo{author}{{Bechtel}, H.A.}, \bibinfo{author}{{Berger}, E.L.},
  \bibinfo{author}{{Burton}, A.S.}, \bibinfo{author}{{Caselli}, P.},
  \bibinfo{author}{{Chung}, A.H.}, \bibinfo{author}{{Clemett}, S.J.},
  \bibinfo{author}{{Cody}, G.D.}, \bibinfo{author}{{Dominguez}, G.},
  \bibinfo{author}{{Elsila}, J.E.}, \bibinfo{author}{{Farnsworth}, K.K.},
  \bibinfo{author}{{Foustoukos}, D.I.}, \bibinfo{author}{{Freeman}, K.H.},
  \bibinfo{author}{{Furukawa}, Y.}, \bibinfo{author}{{Gainsforth}, Z.},
  \bibinfo{author}{{Graham}, H.V.}, \bibinfo{author}{{Grassi}, T.},
  \bibinfo{author}{{Giuliano}, B.M.}, \bibinfo{author}{{Hamilton}, V.E.},
  \bibinfo{author}{{Haenecour}, P.}, \bibinfo{author}{{Heck}, P.R.},
  \bibinfo{author}{{Hofmann}, A.E.}, \bibinfo{author}{{House}, C.H.},
  \bibinfo{author}{{Huang}, Y.}, \bibinfo{author}{{Kaplan}, H.H.},
  \bibinfo{author}{{Keller}, L.P.}, \bibinfo{author}{{Kim}, B.},
  \bibinfo{author}{{Koga}, T.}, \bibinfo{author}{{Liss}, M.},
  \bibinfo{author}{{McLain}, H.L.}, \bibinfo{author}{{Marcus}, M.A.},
  \bibinfo{author}{{Matney}, M.}, \bibinfo{author}{{McCoy}, T.J.},
  \bibinfo{author}{{McIntosh}, O.M.}, \bibinfo{author}{{Mojarro}, A.},
  \bibinfo{author}{{Naraoka}, H.}, \bibinfo{author}{{Nguyen}, A.N.},
  \bibinfo{author}{{Nuevo}, M.}, \bibinfo{author}{{Nuth}, J.A.},
  \bibinfo{author}{{Oba}, Y.}, \bibinfo{author}{{Parker}, E.T.},
  \bibinfo{author}{{Peretyazhko}, T.S.}, \bibinfo{author}{{Sandford}, S.A.},
  \bibinfo{author}{{Santos}, E.}, \bibinfo{author}{{Schmitt-Kopplin}, P.},
  \bibinfo{author}{{Seguin}, F.}, \bibinfo{author}{{Simkus}, D.N.},
  \bibinfo{author}{{Shahid}, A.}, \bibinfo{author}{{Takano}, Y.},
  \bibinfo{author}{{Thomas-Keprta}, K.L.}, \bibinfo{author}{{Tripathi}, H.},
  \bibinfo{author}{{Weiss}, G.}, \bibinfo{author}{{Zheng}, Y.},
  \bibinfo{author}{{Lunning}, N.G.}, \bibinfo{author}{{Righter}, K.},
  \bibinfo{author}{{Connolly}, H.C.}, \bibinfo{author}{{Lauretta}, D.S.},
  \bibinfo{year}{2025}.
\newblock \bibinfo{title}{{Abundant ammonia and nitrogen-rich soluble organic
  matter in samples from asteroid (101955) Bennu}}.
\newblock \bibinfo{journal}{Nature Astronomy}
  \DOIprefix\doi{10.1038/s41550-024-02472-9}.
%Type = Article
\bibitem[{{Goicoechea} et~al.(2009){Goicoechea}, {Compi{\`e}gne} and
  {Habart}}]{Go09}
\bibinfo{author}{{Goicoechea}, J.R.}, \bibinfo{author}{{Compi{\`e}gne}, M.},
  \bibinfo{author}{{Habart}, E.}, \bibinfo{year}{2009}.
\newblock \bibinfo{title}{{Far-Infrared Detection of Neutral Atomic Oxygen
  Toward the Horsehead Nebula}}.
\newblock \bibinfo{journal}{\apjl} \bibinfo{volume}{699},
  \bibinfo{pages}{L165--L168}.
\newblock \DOIprefix\doi{10.1088/0004-637X/699/2/L165},
  \href{http://arxiv.org/abs/0906.0691}{{\tt arXiv:0906.0691}}.
%Type = Article
\bibitem[{{Goicoechea} et~al.(2016){Goicoechea}, {Pety}, {Cuadrado},
  {Cernicharo}, {Chapillon}, {Fuente}, {Gerin}, {Joblin}, {Marcelino} and
  {Pilleri}}]{Go16}
\bibinfo{author}{{Goicoechea}, J.R.}, \bibinfo{author}{{Pety}, J.},
  \bibinfo{author}{{Cuadrado}, S.}, \bibinfo{author}{{Cernicharo}, J.},
  \bibinfo{author}{{Chapillon}, E.}, \bibinfo{author}{{Fuente}, A.},
  \bibinfo{author}{{Gerin}, M.}, \bibinfo{author}{{Joblin}, C.},
  \bibinfo{author}{{Marcelino}, N.}, \bibinfo{author}{{Pilleri}, P.},
  \bibinfo{year}{2016}.
\newblock \bibinfo{title}{{Compression and ablation of the photo-irradiated
  molecular cloud the Orion Bar}}.
\newblock \bibinfo{journal}{\nat} \bibinfo{volume}{537},
  \bibinfo{pages}{207--209}.
\newblock \DOIprefix\doi{10.1038/nature18957},
  \href{http://arxiv.org/abs/1608.06173}{{\tt arXiv:1608.06173}}.
%Type = Article
\bibitem[{{Goto} et~al.(2014){Goto}, {Geballe}, {Indriolo}, {Yusef-Zadeh},
  {Usuda}, {Henning} and {Oka}}]{goto14}
\bibinfo{author}{{Goto}, M.}, \bibinfo{author}{{Geballe}, T.R.},
  \bibinfo{author}{{Indriolo}, N.}, \bibinfo{author}{{Yusef-Zadeh}, F.},
  \bibinfo{author}{{Usuda}, T.}, \bibinfo{author}{{Henning}, T.},
  \bibinfo{author}{{Oka}, T.}, \bibinfo{year}{2014}.
\newblock \bibinfo{title}{{Infrared H\_3\^+ and CO Studies of the Galactic
  Core: GCIRS 3 and GCIRS 1W}}.
\newblock \bibinfo{journal}{\apj} \bibinfo{volume}{786}, \bibinfo{pages}{96}.
\newblock \DOIprefix\doi{10.1088/0004-637X/786/2/96},
  \href{http://arxiv.org/abs/1404.2271}{{\tt arXiv:1404.2271}}.
%Type = Article
\bibitem[{{Goto} et~al.(2021){Goto}, {Vasyunin}, {Giuliano},
  {Jim{\'e}nez-Serra}, {Caselli}, {Rom{\'a}n-Z{\'u}{\~n}iga} and
  {Alves}}]{goto21}
\bibinfo{author}{{Goto}, M.}, \bibinfo{author}{{Vasyunin}, A.I.},
  \bibinfo{author}{{Giuliano}, B.M.}, \bibinfo{author}{{Jim{\'e}nez-Serra},
  I.}, \bibinfo{author}{{Caselli}, P.},
  \bibinfo{author}{{Rom{\'a}n-Z{\'u}{\~n}iga}, C.G.}, \bibinfo{author}{{Alves},
  J.}, \bibinfo{year}{2021}.
\newblock \bibinfo{title}{{Water and methanol ice in L 1544}}.
\newblock \bibinfo{journal}{\aap} \bibinfo{volume}{651}, \bibinfo{pages}{A53}.
\newblock \DOIprefix\doi{10.1051/0004-6361/201936385},
  \href{http://arxiv.org/abs/2012.10883}{{\tt arXiv:2012.10883}}.
%Type = Article
\bibitem[{{Gratier} et~al.(2016){Gratier}, {Majumdar}, {Ohishi}, {Roueff},
  {Loison}, {Hickson} and {Wakelam}}]{Gratier16}
\bibinfo{author}{{Gratier}, P.}, \bibinfo{author}{{Majumdar}, L.},
  \bibinfo{author}{{Ohishi}, M.}, \bibinfo{author}{{Roueff}, E.},
  \bibinfo{author}{{Loison}, J.C.}, \bibinfo{author}{{Hickson}, K.M.},
  \bibinfo{author}{{Wakelam}, V.}, \bibinfo{year}{2016}.
\newblock \bibinfo{title}{{A New Reference Chemical Composition for TMC-1}}.
\newblock \bibinfo{journal}{\apjs} \bibinfo{volume}{225}, \bibinfo{pages}{25}.
\newblock \DOIprefix\doi{10.3847/0067-0049/225/2/25},
  \href{http://arxiv.org/abs/1610.00524}{{\tt arXiv:1610.00524}}.
%Type = Article
\bibitem[{{Green} et~al.(2008){Green}, {Caswell}, {Fuller}, {Breen}, {Brooks},
  {Burton}, {Chrysostomou}, {Cox}, {Diamond}, {Ellingsen}, {Gray}, {Hoare},
  {Masheder}, {McClure-Griffiths}, {Pestalozzi}, {Phillips}, {Quinn},
  {Thompson}, {Voronkov}, {Walsh}, {Ward-Thompson}, {Wong-McSweeney}, {Yates}
  and {Cohen}}]{Green08}
\bibinfo{author}{{Green}, J.A.}, \bibinfo{author}{{Caswell}, J.L.},
  \bibinfo{author}{{Fuller}, G.A.}, \bibinfo{author}{{Breen}, S.L.},
  \bibinfo{author}{{Brooks}, K.}, \bibinfo{author}{{Burton}, M.G.},
  \bibinfo{author}{{Chrysostomou}, A.}, \bibinfo{author}{{Cox}, J.},
  \bibinfo{author}{{Diamond}, P.J.}, \bibinfo{author}{{Ellingsen}, S.P.},
  \bibinfo{author}{{Gray}, M.D.}, \bibinfo{author}{{Hoare}, M.G.},
  \bibinfo{author}{{Masheder}, M.R.W.}, \bibinfo{author}{{McClure-Griffiths},
  N.}, \bibinfo{author}{{Pestalozzi}, M.}, \bibinfo{author}{{Phillips}, C.},
  \bibinfo{author}{{Quinn}, L.}, \bibinfo{author}{{Thompson}, M.A.},
  \bibinfo{author}{{Voronkov}, M.}, \bibinfo{author}{{Walsh}, A.},
  \bibinfo{author}{{Ward-Thompson}, D.}, \bibinfo{author}{{Wong-McSweeney},
  D.}, \bibinfo{author}{{Yates}, J.A.}, \bibinfo{author}{{Cohen}, R.J.},
  \bibinfo{year}{2008}.
\newblock \bibinfo{title}{{Multibeam maser survey of methanol and excited OH in
  the Magellanic Clouds: new detections and maser abundance estimates}}.
\newblock \bibinfo{journal}{\mnras} \bibinfo{volume}{385},
  \bibinfo{pages}{948--956}.
\newblock \DOIprefix\doi{10.1111/j.1365-2966.2008.12888.x},
  \href{http://arxiv.org/abs/0801.0384}{{\tt arXiv:0801.0384}}.
%Type = Article
\bibitem[{{Gueth} et~al.(1996){Gueth}, {Guilloteau} and {Bachiller}}]{Gueth96}
\bibinfo{author}{{Gueth}, F.}, \bibinfo{author}{{Guilloteau}, S.},
  \bibinfo{author}{{Bachiller}, R.}, \bibinfo{year}{1996}.
\newblock \bibinfo{title}{{A precessing jet in the L1157 molecular outflow.}}
\newblock \bibinfo{journal}{\aap} \bibinfo{volume}{307},
  \bibinfo{pages}{891--897}.
%Type = Article
\bibitem[{{Gueth} et~al.(1998){Gueth}, {Guilloteau} and {Bachiller}}]{Gueth98}
\bibinfo{author}{{Gueth}, F.}, \bibinfo{author}{{Guilloteau}, S.},
  \bibinfo{author}{{Bachiller}, R.}, \bibinfo{year}{1998}.
\newblock \bibinfo{title}{{SiO shocks in the L1157 molecular outflow}}.
\newblock \bibinfo{journal}{\aap} \bibinfo{volume}{333},
  \bibinfo{pages}{287--297}.
%Type = Article
\bibitem[{{Guillet} et~al.(2011){Guillet}, {Pineau Des For{\^e}ts} and
  {Jones}}]{Guillet11}
\bibinfo{author}{{Guillet}, V.}, \bibinfo{author}{{Pineau Des For{\^e}ts}, G.},
  \bibinfo{author}{{Jones}, A.P.}, \bibinfo{year}{2011}.
\newblock \bibinfo{title}{{Shocks in dense clouds. III. Dust processing and
  feedback effects in C-type shocks}}.
\newblock \bibinfo{journal}{\aap} \bibinfo{volume}{527}, \bibinfo{pages}{A123}.
\newblock \DOIprefix\doi{10.1051/0004-6361/201015973}.
%Type = Article
\bibitem[{{Gusdorf} et~al.(2008a){Gusdorf}, {Cabrit}, {Flower} and {Pineau Des
  For{\^e}ts}}]{Gusdorf08a}
\bibinfo{author}{{Gusdorf}, A.}, \bibinfo{author}{{Cabrit}, S.},
  \bibinfo{author}{{Flower}, D.R.}, \bibinfo{author}{{Pineau Des For{\^e}ts},
  G.}, \bibinfo{year}{2008a}.
\newblock \bibinfo{title}{{SiO line emission from C-type shock waves:
  interstellar jets and outflows}}.
\newblock \bibinfo{journal}{\aap} \bibinfo{volume}{482},
  \bibinfo{pages}{809--829}.
\newblock \DOIprefix\doi{10.1051/0004-6361:20078900},
  \href{http://arxiv.org/abs/0803.2791}{{\tt arXiv:0803.2791}}.
%Type = Article
\bibitem[{{Gusdorf} et~al.(2008b){Gusdorf}, {Pineau Des For{\^e}ts}, {Cabrit}
  and {Flower}}]{Gusdorf08b}
\bibinfo{author}{{Gusdorf}, A.}, \bibinfo{author}{{Pineau Des For{\^e}ts}, G.},
  \bibinfo{author}{{Cabrit}, S.}, \bibinfo{author}{{Flower}, D.R.},
  \bibinfo{year}{2008b}.
\newblock \bibinfo{title}{{SiO line emission from interstellar jets and
  outflows: silicon-containing mantles and non-stationary shock waves}}.
\newblock \bibinfo{journal}{\aap} \bibinfo{volume}{490},
  \bibinfo{pages}{695--706}.
\newblock \DOIprefix\doi{10.1051/0004-6361:200810443}.
%Type = Article
\bibitem[{{G{\"u}sten} and {Henkel}(1983)}]{guesten83}
\bibinfo{author}{{G{\"u}sten}, R.}, \bibinfo{author}{{Henkel}, C.},
  \bibinfo{year}{1983}.
\newblock \bibinfo{title}{{H$_{2}$ densities and masses of the molecular clouds
  close to the galactic center.}}
\newblock \bibinfo{journal}{\aap} \bibinfo{volume}{125},
  \bibinfo{pages}{136--145}.
%Type = Inproceedings
\bibitem[{{G{\"u}sten} and {Philipp}(2004)}]{guesten04}
\bibinfo{author}{{G{\"u}sten}, R.}, \bibinfo{author}{{Philipp}, S.D.},
  \bibinfo{year}{2004}.
\newblock \bibinfo{title}{{Galactic Center Molecular Clouds}}, in:
  \bibinfo{editor}{{Pfalzner}, S.}, \bibinfo{editor}{{Kramer}, C.},
  \bibinfo{editor}{{Staubmeier}, C.}, \bibinfo{editor}{{Heithausen}, A.}
  (Eds.), \bibinfo{booktitle}{The Dense Interstellar Medium in Galaxies}, p.
  \bibinfo{pages}{253}.
\newblock \DOIprefix\doi{10.1007/978-3-642-18902-9\_46},
  \href{http://arxiv.org/abs/astro-ph/0402019}{{\tt arXiv:astro-ph/0402019}}.
%Type = Article
\bibitem[{{Guzm{\'a}n} et~al.(2013){Guzm{\'a}n}, {Goicoechea}, {Pety},
  {Gratier}, {Gerin}, {Roueff}, {Le Petit}, {Le Bourlot} and
  {Faure}}]{Guzman13}
\bibinfo{author}{{Guzm{\'a}n}, V.V.}, \bibinfo{author}{{Goicoechea}, J.R.},
  \bibinfo{author}{{Pety}, J.}, \bibinfo{author}{{Gratier}, P.},
  \bibinfo{author}{{Gerin}, M.}, \bibinfo{author}{{Roueff}, E.},
  \bibinfo{author}{{Le Petit}, F.}, \bibinfo{author}{{Le Bourlot}, J.},
  \bibinfo{author}{{Faure}, A.}, \bibinfo{year}{2013}.
\newblock \bibinfo{title}{{The IRAM-30 m line survey of the Horsehead PDR. IV.
  Comparative chemistry of H$_{2}$CO and CH$_{3}$OH}}.
\newblock \bibinfo{journal}{\aap} \bibinfo{volume}{560}, \bibinfo{pages}{A73}.
\newblock \DOIprefix\doi{10.1051/0004-6361/201322460},
  \href{http://arxiv.org/abs/1310.6231}{{\tt arXiv:1310.6231}}.
%Type = Article
\bibitem[{{Guzm{\'a}n} et~al.(2014){Guzm{\'a}n}, {Pety}, {Gratier},
  {Goicoechea}, {Gerin}, {Roueff}, {Le Petit} and {Le Bourlot}}]{Guzman14}
\bibinfo{author}{{Guzm{\'a}n}, V.V.}, \bibinfo{author}{{Pety}, J.},
  \bibinfo{author}{{Gratier}, P.}, \bibinfo{author}{{Goicoechea}, J.R.},
  \bibinfo{author}{{Gerin}, M.}, \bibinfo{author}{{Roueff}, E.},
  \bibinfo{author}{{Le Petit}, F.}, \bibinfo{author}{{Le Bourlot}, J.},
  \bibinfo{year}{2014}.
\newblock \bibinfo{title}{{Chemical complexity in the Horsehead
  photodissociation region}}.
\newblock \bibinfo{journal}{Faraday Discussions} \bibinfo{volume}{168},
  \bibinfo{pages}{103--127}.
\newblock \DOIprefix\doi{10.1039/C3FD00114H},
  \href{http://arxiv.org/abs/1404.7798}{{\tt arXiv:1404.7798}}.
%Type = Article
\bibitem[{{Habing}(1968)}]{Habing68}
\bibinfo{author}{{Habing}, H.J.}, \bibinfo{year}{1968}.
\newblock \bibinfo{title}{{The interstellar radiation density between 912 A and
  2400 A}}.
\newblock \bibinfo{journal}{\bain} \bibinfo{volume}{19}, \bibinfo{pages}{421}.
%Type = Article
\bibitem[{{Harada} et~al.(2015){Harada}, {Riquelme}, {Viti},
  {Jim{\'e}nez-Serra}, {Requena-Torres}, {Menten}, {Mart{\'\i}n}, {Aladro},
  {Martin-Pintado} and {Hochg{\"u}rtel}}]{nanase15}
\bibinfo{author}{{Harada}, N.}, \bibinfo{author}{{Riquelme}, D.},
  \bibinfo{author}{{Viti}, S.}, \bibinfo{author}{{Jim{\'e}nez-Serra}, I.},
  \bibinfo{author}{{Requena-Torres}, M.A.}, \bibinfo{author}{{Menten}, K.M.},
  \bibinfo{author}{{Mart{\'\i}n}, S.}, \bibinfo{author}{{Aladro}, R.},
  \bibinfo{author}{{Martin-Pintado}, J.}, \bibinfo{author}{{Hochg{\"u}rtel},
  S.}, \bibinfo{year}{2015}.
\newblock \bibinfo{title}{{Chemical features in the circumnuclear disk of the
  Galactic center}}.
\newblock \bibinfo{journal}{\aap} \bibinfo{volume}{584}, \bibinfo{pages}{A102}.
\newblock \DOIprefix\doi{10.1051/0004-6361/201526994},
  \href{http://arxiv.org/abs/1510.02904}{{\tt arXiv:1510.02904}}.
%Type = Article
\bibitem[{{Heikkil{\"a}} et~al.(1999){Heikkil{\"a}}, {Johansson} and
  {Olofsson}}]{Heikila99}
\bibinfo{author}{{Heikkil{\"a}}, A.}, \bibinfo{author}{{Johansson}, L.E.B.},
  \bibinfo{author}{{Olofsson}, H.}, \bibinfo{year}{1999}.
\newblock \bibinfo{title}{{Molecular abundance variations in the Magellanic
  Clouds}}.
\newblock \bibinfo{journal}{\aap} \bibinfo{volume}{344},
  \bibinfo{pages}{817--847}.
%Type = Article
\bibitem[{{Henkel} et~al.(1987){Henkel}, {Jacq}, {Mauersberger}, {Menten} and
  {Steppe}}]{Henkel87}
\bibinfo{author}{{Henkel}, C.}, \bibinfo{author}{{Jacq}, T.},
  \bibinfo{author}{{Mauersberger}, R.}, \bibinfo{author}{{Menten}, K.M.},
  \bibinfo{author}{{Steppe}, H.}, \bibinfo{year}{1987}.
\newblock \bibinfo{title}{{The detection of extragalactic methanol.}}
\newblock \bibinfo{journal}{\aap} \bibinfo{volume}{188},
  \bibinfo{pages}{L1--L4}.
%Type = Article
\bibitem[{{Herbst} and {van Dishoeck}(2009)}]{Herbst09}
\bibinfo{author}{{Herbst}, E.}, \bibinfo{author}{{van Dishoeck}, E.F.},
  \bibinfo{year}{2009}.
\newblock \bibinfo{title}{{Complex Organic Interstellar Molecules}}.
\newblock \bibinfo{journal}{\araa} \bibinfo{volume}{47},
  \bibinfo{pages}{427--480}.
\newblock \DOIprefix\doi{10.1146/annurev-astro-082708-101654}.
%Type = Article
\bibitem[{{Holdship} et~al.(2019){Holdship}, {Rawlings}, {Viti}, {Balucani},
  {Skouteris} and {Williams}}]{holdship19}
\bibinfo{author}{{Holdship}, J.}, \bibinfo{author}{{Rawlings}, J.},
  \bibinfo{author}{{Viti}, S.}, \bibinfo{author}{{Balucani}, N.},
  \bibinfo{author}{{Skouteris}, D.}, \bibinfo{author}{{Williams}, D.},
  \bibinfo{year}{2019}.
\newblock \bibinfo{title}{{Investigating the Efficiency of Explosion Chemistry
  as a Source of Complex Organic Molecules in TMC-1}}.
\newblock \bibinfo{journal}{\apj} \bibinfo{volume}{878}, \bibinfo{pages}{65}.
\newblock \DOIprefix\doi{10.3847/1538-4357/ab1f7b},
  \href{http://arxiv.org/abs/1905.01901}{{\tt arXiv:1905.01901}}.
%Type = Article
\bibitem[{{Hollenbach} et~al.(1991){Hollenbach}, {Takahashi} and
  {Tielens}}]{Hollenbach91}
\bibinfo{author}{{Hollenbach}, D.J.}, \bibinfo{author}{{Takahashi}, T.},
  \bibinfo{author}{{Tielens}, A.G.G.M.}, \bibinfo{year}{1991}.
\newblock \bibinfo{title}{{Low-Density Photodissociation Regions}}.
\newblock \bibinfo{journal}{\apj} \bibinfo{volume}{377}, \bibinfo{pages}{192}.
\newblock \DOIprefix\doi{10.1086/170347}.
%Type = Article
\bibitem[{{Hunter} et~al.(2017){Hunter}, {Brogan}, {MacLeod}, {Cyganowski},
  {Chandler}, {Chibueze}, {Friesen}, {Indebetouw}, {Thesner} and
  {Young}}]{Hunter17}
\bibinfo{author}{{Hunter}, T.R.}, \bibinfo{author}{{Brogan}, C.L.},
  \bibinfo{author}{{MacLeod}, G.}, \bibinfo{author}{{Cyganowski}, C.J.},
  \bibinfo{author}{{Chandler}, C.J.}, \bibinfo{author}{{Chibueze}, J.O.},
  \bibinfo{author}{{Friesen}, R.}, \bibinfo{author}{{Indebetouw}, R.},
  \bibinfo{author}{{Thesner}, C.}, \bibinfo{author}{{Young}, K.H.},
  \bibinfo{year}{2017}.
\newblock \bibinfo{title}{{An Extraordinary Outburst in the Massive
  Protostellar System NGC6334I-MM1: Quadrupling of the Millimeter Continuum}}.
\newblock \bibinfo{journal}{Submillimeter Array Newsletter}
  \bibinfo{volume}{24}, \bibinfo{pages}{2--4}.
%Type = Article
\bibitem[{{H{\"u}ttemeister} et~al.(1993){H{\"u}ttemeister}, {Wilson}, {Bania}
  and {Martin-Pintado}}]{huettemeister93}
\bibinfo{author}{{H{\"u}ttemeister}, S.}, \bibinfo{author}{{Wilson}, T.L.},
  \bibinfo{author}{{Bania}, T.M.}, \bibinfo{author}{{Martin-Pintado}, J.},
  \bibinfo{year}{1993}.
\newblock \bibinfo{title}{{Kinetic temperatures in Galactic Center molecular
  clouds.}}
\newblock \bibinfo{journal}{\aap} \bibinfo{volume}{280},
  \bibinfo{pages}{255--267}.
%Type = Article
\bibitem[{{Ilee} et~al.(2021){Ilee}, {Walsh}, {Booth}, {Aikawa}, {Andrews},
  {Bae}, {Bergin}, {Bergner}, {Bosman}, {Cataldi}, {Cleeves}, {Czekala},
  {Guzm{\'a}n}, {Huang}, {Law}, {Le Gal}, {Loomis}, {M{\'e}nard}, {Nomura},
  {{\"O}berg}, {Qi}, {Schwarz}, {Teague}, {Tsukagoshi}, {Wilner}, {Yamato} and
  {Zhang}}]{Ilee21}
\bibinfo{author}{{Ilee}, J.D.}, \bibinfo{author}{{Walsh}, C.},
  \bibinfo{author}{{Booth}, A.S.}, \bibinfo{author}{{Aikawa}, Y.},
  \bibinfo{author}{{Andrews}, S.M.}, \bibinfo{author}{{Bae}, J.},
  \bibinfo{author}{{Bergin}, E.A.}, \bibinfo{author}{{Bergner}, J.B.},
  \bibinfo{author}{{Bosman}, A.D.}, \bibinfo{author}{{Cataldi}, G.},
  \bibinfo{author}{{Cleeves}, L.I.}, \bibinfo{author}{{Czekala}, I.},
  \bibinfo{author}{{Guzm{\'a}n}, V.V.}, \bibinfo{author}{{Huang}, J.},
  \bibinfo{author}{{Law}, C.J.}, \bibinfo{author}{{Le Gal}, R.},
  \bibinfo{author}{{Loomis}, R.A.}, \bibinfo{author}{{M{\'e}nard}, F.},
  \bibinfo{author}{{Nomura}, H.}, \bibinfo{author}{{{\"O}berg}, K.I.},
  \bibinfo{author}{{Qi}, C.}, \bibinfo{author}{{Schwarz}, K.R.},
  \bibinfo{author}{{Teague}, R.}, \bibinfo{author}{{Tsukagoshi}, T.},
  \bibinfo{author}{{Wilner}, D.J.}, \bibinfo{author}{{Yamato}, Y.},
  \bibinfo{author}{{Zhang}, K.}, \bibinfo{year}{2021}.
\newblock \bibinfo{title}{{Molecules with ALMA at Planet-forming Scales (MAPS).
  IX. Distribution and Properties of the Large Organic Molecules HC$_{3}$N,
  CH$_{3}$CN, and c-C$_{3}$H$_{2}$}}.
\newblock \bibinfo{journal}{\apjs} \bibinfo{volume}{257}, \bibinfo{pages}{9}.
\newblock \DOIprefix\doi{10.3847/1538-4365/ac1441},
  \href{http://arxiv.org/abs/2109.06319}{{\tt arXiv:2109.06319}}.
%Type = Article
\bibitem[{{Imai} et~al.(2016){Imai}, {Sakai}, {Oya}, {L{\'o}pez-Sepulcre},
  {Watanabe}, {Ceccarelli}, {Lefloch}, {Caux}, {Vastel}, {Kahane}, {Sakai},
  {Hirota}, {Aikawa} and {Yamamoto}}]{Imai16}
\bibinfo{author}{{Imai}, M.}, \bibinfo{author}{{Sakai}, N.},
  \bibinfo{author}{{Oya}, Y.}, \bibinfo{author}{{L{\'o}pez-Sepulcre}, A.},
  \bibinfo{author}{{Watanabe}, Y.}, \bibinfo{author}{{Ceccarelli}, C.},
  \bibinfo{author}{{Lefloch}, B.}, \bibinfo{author}{{Caux}, E.},
  \bibinfo{author}{{Vastel}, C.}, \bibinfo{author}{{Kahane}, C.},
  \bibinfo{author}{{Sakai}, T.}, \bibinfo{author}{{Hirota}, T.},
  \bibinfo{author}{{Aikawa}, Y.}, \bibinfo{author}{{Yamamoto}, S.},
  \bibinfo{year}{2016}.
\newblock \bibinfo{title}{{Discovery of a Hot Corino in the Bok Globule B335}}.
\newblock \bibinfo{journal}{\apjl} \bibinfo{volume}{830}, \bibinfo{pages}{L37}.
\newblock \DOIprefix\doi{10.3847/2041-8205/830/2/L37},
  \href{http://arxiv.org/abs/1610.03942}{{\tt arXiv:1610.03942}}.
%Type = Article
\bibitem[{{Insausti} et~al.(2021){Insausti}, {Alonso}, {Tercero}, {Santos},
  {Calabrese}, {Vogt}, {Corzana}, {Demaison}, {Cernicharo} and
  {Cocinero}}]{insausti21}
\bibinfo{author}{{Insausti}, A.}, \bibinfo{author}{{Alonso}, E.R.},
  \bibinfo{author}{{Tercero}, B.}, \bibinfo{author}{{Santos}, J.I.},
  \bibinfo{author}{{Calabrese}, C.}, \bibinfo{author}{{Vogt}, N.},
  \bibinfo{author}{{Corzana}, F.}, \bibinfo{author}{{Demaison}, J.},
  \bibinfo{author}{{Cernicharo}, J.}, \bibinfo{author}{{Cocinero}, E.J.},
  \bibinfo{year}{2021}.
\newblock \bibinfo{title}{{Laboratory Observation of, Astrochemical Search for,
  and Structure of Elusive Erythrulose in the Interstellar Medium}}.
\newblock \bibinfo{journal}{The Journal of Physical Chemistry Letters}
  \bibinfo{volume}{12}, \bibinfo{pages}{1352--1359}.
\newblock \DOIprefix\doi{10.1021/acs.jpclett.0c03050}.
%Type = Article
\bibitem[{{Ivlev} et~al.(2015){Ivlev}, {R{\"o}cker}, {Vasyunin} and
  {Caselli}}]{ivlev15}
\bibinfo{author}{{Ivlev}, A.V.}, \bibinfo{author}{{R{\"o}cker}, T.B.},
  \bibinfo{author}{{Vasyunin}, A.}, \bibinfo{author}{{Caselli}, P.},
  \bibinfo{year}{2015}.
\newblock \bibinfo{title}{{Impulsive Spot Heating and Thermal Explosion of
  Interstellar Grains Revisited}}.
\newblock \bibinfo{journal}{\apj} \bibinfo{volume}{805}, \bibinfo{pages}{59}.
\newblock \DOIprefix\doi{10.1088/0004-637X/805/1/59},
  \href{http://arxiv.org/abs/1503.05012}{{\tt arXiv:1503.05012}}.
%Type = Article
\bibitem[{{Jim{\'e}nez-Serra} et~al.(2008){Jim{\'e}nez-Serra}, {Caselli},
  {Mart{\'\i}n-Pintado} and {Hartquist}}]{JimenezSerra08}
\bibinfo{author}{{Jim{\'e}nez-Serra}, I.}, \bibinfo{author}{{Caselli}, P.},
  \bibinfo{author}{{Mart{\'\i}n-Pintado}, J.}, \bibinfo{author}{{Hartquist},
  T.W.}, \bibinfo{year}{2008}.
\newblock \bibinfo{title}{{Parametrization of C-shocks. Evolution of the
  sputtering of grains}}.
\newblock \bibinfo{journal}{\aap} \bibinfo{volume}{482},
  \bibinfo{pages}{549--559}.
\newblock \DOIprefix\doi{10.1051/0004-6361:20078054},
  \href{http://arxiv.org/abs/0802.0594}{{\tt arXiv:0802.0594}}.
%Type = Article
\bibitem[{{Jim{\'e}nez-Serra} et~al.(2022a){Jim{\'e}nez-Serra},
  {Mart{\'\i}n-Pintado}, {Insausti}, {Alonso}, {Cocinero} and
  {Bourke}}]{jimenez22a}
\bibinfo{author}{{Jim{\'e}nez-Serra}, I.},
  \bibinfo{author}{{Mart{\'\i}n-Pintado}, J.}, \bibinfo{author}{{Insausti},
  A.}, \bibinfo{author}{{Alonso}, E.R.}, \bibinfo{author}{{Cocinero}, E.J.},
  \bibinfo{author}{{Bourke}, T.L.}, \bibinfo{year}{2022}a.
\newblock \bibinfo{title}{{The SKA as a Prebiotic Molecule Detector}}.
\newblock \bibinfo{journal}{Frontiers in Astronomy and Space Sciences}
  \bibinfo{volume}{9}, \bibinfo{pages}{843766}.
\newblock \DOIprefix\doi{10.3389/fspas.2022.843766},
  \href{http://arxiv.org/abs/2203.00534}{{\tt arXiv:2203.00534}}.
%Type = Article
\bibitem[{{Jim{\'e}nez-Serra} et~al.(2020){Jim{\'e}nez-Serra},
  {Mart{\'\i}n-Pintado}, {Rivilla}, {Rodr{\'\i}guez-Almeida}, {Alonso Alonso},
  {Zeng}, {Cocinero}, {Mart{\'\i}n}, {Requena-Torres}, {Mart{\'\i}n-Domenech}
  and {Testi}}]{jimenez20}
\bibinfo{author}{{Jim{\'e}nez-Serra}, I.},
  \bibinfo{author}{{Mart{\'\i}n-Pintado}, J.}, \bibinfo{author}{{Rivilla},
  V.M.}, \bibinfo{author}{{Rodr{\'\i}guez-Almeida}, L.},
  \bibinfo{author}{{Alonso Alonso}, E.R.}, \bibinfo{author}{{Zeng}, S.},
  \bibinfo{author}{{Cocinero}, E.J.}, \bibinfo{author}{{Mart{\'\i}n}, S.},
  \bibinfo{author}{{Requena-Torres}, M.},
  \bibinfo{author}{{Mart{\'\i}n-Domenech}, R.}, \bibinfo{author}{{Testi}, L.},
  \bibinfo{year}{2020}.
\newblock \bibinfo{title}{{Toward the RNA-World in the Interstellar
  Medium{\textemdash}Detection of Urea and Search of 2-Amino-oxazole and Simple
  Sugars}}.
\newblock \bibinfo{journal}{Astrobiology} \bibinfo{volume}{20},
  \bibinfo{pages}{1048--1066}.
\newblock \DOIprefix\doi{10.1089/ast.2019.2125},
  \href{http://arxiv.org/abs/2004.07834}{{\tt arXiv:2004.07834}}.
%Type = Article
\bibitem[{{Jim{\'e}nez-Serra} et~al.(2005){Jim{\'e}nez-Serra},
  {Mart{\'\i}n-Pintado}, {Rodr{\'\i}guez-Franco} and
  {Mart{\'\i}n}}]{jimenezserra05}
\bibinfo{author}{{Jim{\'e}nez-Serra}, I.},
  \bibinfo{author}{{Mart{\'\i}n-Pintado}, J.},
  \bibinfo{author}{{Rodr{\'\i}guez-Franco}, A.},
  \bibinfo{author}{{Mart{\'\i}n}, S.}, \bibinfo{year}{2005}.
\newblock \bibinfo{title}{{Grain Evolution across the Shocks in the L1448-mm
  Outflow}}.
\newblock \bibinfo{journal}{\apjl} \bibinfo{volume}{627},
  \bibinfo{pages}{L121--L124}.
\newblock \DOIprefix\doi{10.1086/432467},
  \href{http://arxiv.org/abs/astro-ph/0506408}{{\tt arXiv:astro-ph/0506408}}.
%Type = Article
\bibitem[{{Jim{\'e}nez-Serra} et~al.(2022b){Jim{\'e}nez-Serra},
  {Rodr{\'\i}guez-Almeida}, {Mart{\'\i}n-Pintado}, {Rivilla}, {Melosso},
  {Zeng}, {Colzi}, {Kawashima}, {Hirota}, {Puzzarini}, {Tercero}, {de Vicente},
  {Rico-Villas}, {Requena-Torres} and {Mart{\'\i}n}}]{jimenez22b}
\bibinfo{author}{{Jim{\'e}nez-Serra}, I.},
  \bibinfo{author}{{Rodr{\'\i}guez-Almeida}, L.F.},
  \bibinfo{author}{{Mart{\'\i}n-Pintado}, J.}, \bibinfo{author}{{Rivilla},
  V.M.}, \bibinfo{author}{{Melosso}, M.}, \bibinfo{author}{{Zeng}, S.},
  \bibinfo{author}{{Colzi}, L.}, \bibinfo{author}{{Kawashima}, Y.},
  \bibinfo{author}{{Hirota}, E.}, \bibinfo{author}{{Puzzarini}, C.},
  \bibinfo{author}{{Tercero}, B.}, \bibinfo{author}{{de Vicente}, P.},
  \bibinfo{author}{{Rico-Villas}, F.}, \bibinfo{author}{{Requena-Torres},
  M.A.}, \bibinfo{author}{{Mart{\'\i}n}, S.}, \bibinfo{year}{2022}b.
\newblock \bibinfo{title}{{Precursors of fatty alcohols in the ISM: Discovery
  of n-propanol}}.
\newblock \bibinfo{journal}{\aap} \bibinfo{volume}{663}, \bibinfo{pages}{A181}.
\newblock \DOIprefix\doi{10.1051/0004-6361/202142699},
  \href{http://arxiv.org/abs/2204.08267}{{\tt arXiv:2204.08267}}.
%Type = Article
\bibitem[{{Jim{\'e}nez-Serra} et~al.(2016){Jim{\'e}nez-Serra}, {Vasyunin},
  {Caselli}, {Marcelino}, {Billot}, {Viti}, {Testi}, {Vastel}, {Lefloch} and
  {Bachiller}}]{jimenez16}
\bibinfo{author}{{Jim{\'e}nez-Serra}, I.}, \bibinfo{author}{{Vasyunin}, A.I.},
  \bibinfo{author}{{Caselli}, P.}, \bibinfo{author}{{Marcelino}, N.},
  \bibinfo{author}{{Billot}, N.}, \bibinfo{author}{{Viti}, S.},
  \bibinfo{author}{{Testi}, L.}, \bibinfo{author}{{Vastel}, C.},
  \bibinfo{author}{{Lefloch}, B.}, \bibinfo{author}{{Bachiller}, R.},
  \bibinfo{year}{2016}.
\newblock \bibinfo{title}{{The Spatial Distribution of Complex Organic
  Molecules in the L1544 Pre-stellar Core}}.
\newblock \bibinfo{journal}{\apjl} \bibinfo{volume}{830}, \bibinfo{pages}{L6}.
\newblock \DOIprefix\doi{10.3847/2041-8205/830/1/L6},
  \href{http://arxiv.org/abs/1609.05045}{{\tt arXiv:1609.05045}}.
%Type = Article
\bibitem[{{Jim{\'e}nez-Serra} et~al.(2021){Jim{\'e}nez-Serra}, {Vasyunin},
  {Spezzano}, {Caselli}, {Cosentino} and {Viti}}]{jimenez21}
\bibinfo{author}{{Jim{\'e}nez-Serra}, I.}, \bibinfo{author}{{Vasyunin}, A.I.},
  \bibinfo{author}{{Spezzano}, S.}, \bibinfo{author}{{Caselli}, P.},
  \bibinfo{author}{{Cosentino}, G.}, \bibinfo{author}{{Viti}, S.},
  \bibinfo{year}{2021}.
\newblock \bibinfo{title}{{The Complex Organic Molecular Content in the L1498
  Starless Core}}.
\newblock \bibinfo{journal}{\apj} \bibinfo{volume}{917}, \bibinfo{pages}{44}.
\newblock \DOIprefix\doi{10.3847/1538-4357/ac024c},
  \href{http://arxiv.org/abs/2105.08363}{{\tt arXiv:2105.08363}}.
%Type = Article
\bibitem[{{Jim{\'e}nez-Serra} et~al.(2012){Jim{\'e}nez-Serra}, {Zhang}, {Viti},
  {Mart{\'\i}n-Pintado} and {de Wit}}]{jimenez12}
\bibinfo{author}{{Jim{\'e}nez-Serra}, I.}, \bibinfo{author}{{Zhang}, Q.},
  \bibinfo{author}{{Viti}, S.}, \bibinfo{author}{{Mart{\'\i}n-Pintado}, J.},
  \bibinfo{author}{{de Wit}, W.J.}, \bibinfo{year}{2012}.
\newblock \bibinfo{title}{{Chemical Segregation toward Massive Hot Cores: The
  AFGL2591 Star-forming Region}}.
\newblock \bibinfo{journal}{\apj} \bibinfo{volume}{753}, \bibinfo{pages}{34}.
\newblock \DOIprefix\doi{10.1088/0004-637X/753/1/34},
  \href{http://arxiv.org/abs/1204.6335}{{\tt arXiv:1204.6335}}.
%Type = Article
\bibitem[{{Jin} and {Garrod}(2020)}]{jin20}
\bibinfo{author}{{Jin}, M.}, \bibinfo{author}{{Garrod}, R.T.},
  \bibinfo{year}{2020}.
\newblock \bibinfo{title}{{Formation of Complex Organic Molecules in Cold
  Interstellar Environments through Nondiffusive Grain-surface and Ice-mantle
  Chemistry}}.
\newblock \bibinfo{journal}{\apjs} \bibinfo{volume}{249}, \bibinfo{pages}{26}.
\newblock \DOIprefix\doi{10.3847/1538-4365/ab9ec8},
  \href{http://arxiv.org/abs/2006.11127}{{\tt arXiv:2006.11127}}.
%Type = Article
\bibitem[{{Johnston} et~al.(2020){Johnston}, {Hoare}, {Beuther}, {Linz},
  {Boley}, {Kuiper}, {Kee} and {Robitaille}}]{Johnston20}
\bibinfo{author}{{Johnston}, K.G.}, \bibinfo{author}{{Hoare}, M.G.},
  \bibinfo{author}{{Beuther}, H.}, \bibinfo{author}{{Linz}, H.},
  \bibinfo{author}{{Boley}, P.}, \bibinfo{author}{{Kuiper}, R.},
  \bibinfo{author}{{Kee}, N.D.}, \bibinfo{author}{{Robitaille}, T.P.},
  \bibinfo{year}{2020}.
\newblock \bibinfo{title}{{A Detailed View of the Circumstellar Environment and
  Disk of the Forming O-star AFGL 4176}}.
\newblock \bibinfo{journal}{\apj} \bibinfo{volume}{896}, \bibinfo{pages}{35}.
\newblock \DOIprefix\doi{10.3847/1538-4357/ab8adc},
  \href{http://arxiv.org/abs/2004.13739}{{\tt arXiv:2004.13739}}.
%Type = Article
\bibitem[{{Jones} et~al.(2008){Jones}, {Burton}, {Cunningham}, {Menten},
  {Schilke}, {Belloche}, {Leurini}, {Ott} and {Walsh}}]{jones08}
\bibinfo{author}{{Jones}, P.A.}, \bibinfo{author}{{Burton}, M.G.},
  \bibinfo{author}{{Cunningham}, M.R.}, \bibinfo{author}{{Menten}, K.M.},
  \bibinfo{author}{{Schilke}, P.}, \bibinfo{author}{{Belloche}, A.},
  \bibinfo{author}{{Leurini}, S.}, \bibinfo{author}{{Ott}, J.},
  \bibinfo{author}{{Walsh}, A.J.}, \bibinfo{year}{2008}.
\newblock \bibinfo{title}{{Spectral imaging of the Sagittarius B2 region in
  multiple 3-mm molecular lines with the Mopra telescope}}.
\newblock \bibinfo{journal}{\mnras} \bibinfo{volume}{386},
  \bibinfo{pages}{117--137}.
\newblock \DOIprefix\doi{10.1111/j.1365-2966.2008.13009.x},
  \href{http://arxiv.org/abs/0712.0218}{{\tt arXiv:0712.0218}}.
%Type = Article
\bibitem[{{Jones} et~al.(2012){Jones}, {Burton}, {Cunningham},
  {Requena-Torres}, {Menten}, {Schilke}, {Belloche}, {Leurini},
  {Mart{\'\i}n-Pintado}, {Ott} and {Walsh}}]{jones12}
\bibinfo{author}{{Jones}, P.A.}, \bibinfo{author}{{Burton}, M.G.},
  \bibinfo{author}{{Cunningham}, M.R.}, \bibinfo{author}{{Requena-Torres},
  M.A.}, \bibinfo{author}{{Menten}, K.M.}, \bibinfo{author}{{Schilke}, P.},
  \bibinfo{author}{{Belloche}, A.}, \bibinfo{author}{{Leurini}, S.},
  \bibinfo{author}{{Mart{\'\i}n-Pintado}, J.}, \bibinfo{author}{{Ott}, J.},
  \bibinfo{author}{{Walsh}, A.J.}, \bibinfo{year}{2012}.
\newblock \bibinfo{title}{{Spectral imaging of the Central Molecular Zone in
  multiple 3-mm molecular lines}}.
\newblock \bibinfo{journal}{\mnras} \bibinfo{volume}{419},
  \bibinfo{pages}{2961--2986}.
\newblock \DOIprefix\doi{10.1111/j.1365-2966.2011.19941.x},
  \href{http://arxiv.org/abs/1110.1421}{{\tt arXiv:1110.1421}}.
%Type = Article
\bibitem[{{J{\o}rgensen} et~al.(2020){J{\o}rgensen}, {Belloche} and
  {Garrod}}]{Jorgensen20}
\bibinfo{author}{{J{\o}rgensen}, J.K.}, \bibinfo{author}{{Belloche}, A.},
  \bibinfo{author}{{Garrod}, R.T.}, \bibinfo{year}{2020}.
\newblock \bibinfo{title}{{Astrochemistry During the Formation of Stars}}.
\newblock \bibinfo{journal}{\araa} \bibinfo{volume}{58},
  \bibinfo{pages}{727--778}.
\newblock \DOIprefix\doi{10.1146/annurev-astro-032620-021927},
  \href{http://arxiv.org/abs/2006.07071}{{\tt arXiv:2006.07071}}.
%Type = Article
\bibitem[{{J{\o}rgensen} et~al.(2004){J{\o}rgensen}, {Hogerheijde}, {Blake},
  {van Dishoeck}, {Mundy} and {Sch{\"o}ier}}]{Jorgensen04}
\bibinfo{author}{{J{\o}rgensen}, J.K.}, \bibinfo{author}{{Hogerheijde}, M.R.},
  \bibinfo{author}{{Blake}, G.A.}, \bibinfo{author}{{van Dishoeck}, E.F.},
  \bibinfo{author}{{Mundy}, L.G.}, \bibinfo{author}{{Sch{\"o}ier}, F.L.},
  \bibinfo{year}{2004}.
\newblock \bibinfo{title}{{The impact of shocks on the chemistry of molecular
  clouds. High resolution images of chemical differentiation along the NGC
  1333-IRAS 2A outflow}}.
\newblock \bibinfo{journal}{\aap} \bibinfo{volume}{415},
  \bibinfo{pages}{1021--1037}.
\newblock \DOIprefix\doi{10.1051/0004-6361:20034216},
  \href{http://arxiv.org/abs/astro-ph/0311132}{{\tt arXiv:astro-ph/0311132}}.
%Type = Article
\bibitem[{{J{\o}rgensen} et~al.(2018){J{\o}rgensen}, {M{\"u}ller}, {Calcutt},
  {Coutens}, {Drozdovskaya}, {{\"O}berg}, {Persson}, {Taquet}, {van Dishoeck}
  and {Wampfler}}]{Jorgensen18}
\bibinfo{author}{{J{\o}rgensen}, J.K.}, \bibinfo{author}{{M{\"u}ller}, H.S.P.},
  \bibinfo{author}{{Calcutt}, H.}, \bibinfo{author}{{Coutens}, A.},
  \bibinfo{author}{{Drozdovskaya}, M.N.}, \bibinfo{author}{{{\"O}berg}, K.I.},
  \bibinfo{author}{{Persson}, M.V.}, \bibinfo{author}{{Taquet}, V.},
  \bibinfo{author}{{van Dishoeck}, E.F.}, \bibinfo{author}{{Wampfler}, S.F.},
  \bibinfo{year}{2018}.
\newblock \bibinfo{title}{{The ALMA-PILS survey: isotopic composition of
  oxygen-containing complex organic molecules toward IRAS 16293-2422B}}.
\newblock \bibinfo{journal}{\aap} \bibinfo{volume}{620}, \bibinfo{pages}{A170}.
\newblock \DOIprefix\doi{10.1051/0004-6361/201731667},
  \href{http://arxiv.org/abs/1808.08753}{{\tt arXiv:1808.08753}}.
%Type = Article
\bibitem[{{J{\o}rgensen} et~al.(2016){J{\o}rgensen}, {van der Wiel}, {Coutens},
  {Lykke}, {M{\"u}ller}, {van Dishoeck}, {Calcutt}, {Bjerkeli}, {Bourke},
  {Drozdovskaya}, {Favre}, {Fayolle}, {Garrod}, {Jacobsen}, {{\"O}berg},
  {Persson} and {Wampfler}}]{Jorgensen16}
\bibinfo{author}{{J{\o}rgensen}, J.K.}, \bibinfo{author}{{van der Wiel},
  M.H.D.}, \bibinfo{author}{{Coutens}, A.}, \bibinfo{author}{{Lykke}, J.M.},
  \bibinfo{author}{{M{\"u}ller}, H.S.P.}, \bibinfo{author}{{van Dishoeck},
  E.F.}, \bibinfo{author}{{Calcutt}, H.}, \bibinfo{author}{{Bjerkeli}, P.},
  \bibinfo{author}{{Bourke}, T.L.}, \bibinfo{author}{{Drozdovskaya}, M.N.},
  \bibinfo{author}{{Favre}, C.}, \bibinfo{author}{{Fayolle}, E.C.},
  \bibinfo{author}{{Garrod}, R.T.}, \bibinfo{author}{{Jacobsen}, S.K.},
  \bibinfo{author}{{{\"O}berg}, K.I.}, \bibinfo{author}{{Persson}, M.V.},
  \bibinfo{author}{{Wampfler}, S.F.}, \bibinfo{year}{2016}.
\newblock \bibinfo{title}{{The ALMA Protostellar Interferometric Line Survey
  (PILS). First results from an unbiased submillimeter wavelength line survey
  of the Class 0 protostellar binary IRAS 16293-2422 with ALMA}}.
\newblock \bibinfo{journal}{\aap} \bibinfo{volume}{595}, \bibinfo{pages}{A117}.
\newblock \DOIprefix\doi{10.1051/0004-6361/201628648},
  \href{http://arxiv.org/abs/1607.08733}{{\tt arXiv:1607.08733}}.
%Type = Article
\bibitem[{{Kami{\'n}ski} et~al.(2017){Kami{\'n}ski}, {Menten}, {Tylenda},
  {Karakas}, {Belloche} and {Patel}}]{Kaminski17}
\bibinfo{author}{{Kami{\'n}ski}, T.}, \bibinfo{author}{{Menten}, K.M.},
  \bibinfo{author}{{Tylenda}, R.}, \bibinfo{author}{{Karakas}, A.},
  \bibinfo{author}{{Belloche}, A.}, \bibinfo{author}{{Patel}, N.A.},
  \bibinfo{year}{2017}.
\newblock \bibinfo{title}{{Organic molecules, ions, and rare isotopologues in
  the remnant of the stellar-merger candidate, CK Vulpeculae (Nova 1670)}}.
\newblock \bibinfo{journal}{\aap} \bibinfo{volume}{607}, \bibinfo{pages}{A78}.
\newblock \DOIprefix\doi{10.1051/0004-6361/201731287},
  \href{http://arxiv.org/abs/1708.02261}{{\tt arXiv:1708.02261}}.
%Type = Article
\bibitem[{{Kami{\'n}ski} et~al.(2020){Kami{\'n}ski}, {Menten}, {Tylenda},
  {Wong}, {Belloche}, {Mehner}, {Schmidt} and {Patel}}]{Kaminski20}
\bibinfo{author}{{Kami{\'n}ski}, T.}, \bibinfo{author}{{Menten}, K.M.},
  \bibinfo{author}{{Tylenda}, R.}, \bibinfo{author}{{Wong}, K.T.},
  \bibinfo{author}{{Belloche}, A.}, \bibinfo{author}{{Mehner}, A.},
  \bibinfo{author}{{Schmidt}, M.R.}, \bibinfo{author}{{Patel}, N.A.},
  \bibinfo{year}{2020}.
\newblock \bibinfo{title}{{Molecular remnant of Nova 1670 (CK Vulpeculae). I.
  Properties and enigmatic origin of the gas}}.
\newblock \bibinfo{journal}{\aap} \bibinfo{volume}{644}, \bibinfo{pages}{A59}.
\newblock \DOIprefix\doi{10.1051/0004-6361/202038648},
  \href{http://arxiv.org/abs/2006.10471}{{\tt arXiv:2006.10471}}.
%Type = Article
\bibitem[{{Kami{\'n}ski} et~al.(2021){Kami{\'n}ski}, {Steffen}, {Bujarrabal},
  {Tylenda}, {Menten} and {Hajduk}}]{Kaminski21}
\bibinfo{author}{{Kami{\'n}ski}, T.}, \bibinfo{author}{{Steffen}, W.},
  \bibinfo{author}{{Bujarrabal}, V.}, \bibinfo{author}{{Tylenda}, R.},
  \bibinfo{author}{{Menten}, K.M.}, \bibinfo{author}{{Hajduk}, M.},
  \bibinfo{year}{2021}.
\newblock \bibinfo{title}{{Molecular remnant of Nova 1670 (CK Vulpeculae). II.
  A three-dimensional view of the gas distribution and velocity field}}.
\newblock \bibinfo{journal}{\aap} \bibinfo{volume}{646}, \bibinfo{pages}{A1}.
\newblock \DOIprefix\doi{10.1051/0004-6361/202039634},
  \href{http://arxiv.org/abs/2010.05832}{{\tt arXiv:2010.05832}}.
%Type = Article
\bibitem[{{Keto} et~al.(2015){Keto}, {Caselli} and {Rawlings}}]{keto15}
\bibinfo{author}{{Keto}, E.}, \bibinfo{author}{{Caselli}, P.},
  \bibinfo{author}{{Rawlings}, J.}, \bibinfo{year}{2015}.
\newblock \bibinfo{title}{{The dynamics of collapsing cores and star
  formation}}.
\newblock \bibinfo{journal}{\mnras} \bibinfo{volume}{446},
  \bibinfo{pages}{3731--3740}.
\newblock \DOIprefix\doi{10.1093/mnras/stu2247},
  \href{http://arxiv.org/abs/1410.5889}{{\tt arXiv:1410.5889}}.
%Type = Article
\bibitem[{{Koga} and {Naraoka}(2017)}]{koga2017}
\bibinfo{author}{{Koga}, T.}, \bibinfo{author}{{Naraoka}, H.},
  \bibinfo{year}{2017}.
\newblock \bibinfo{title}{{A new family of extraterrestrial amino acids in the
  Murchison meteorite}}.
\newblock \bibinfo{journal}{Scientific Reports} \bibinfo{volume}{7},
  \bibinfo{pages}{636}.
\newblock \DOIprefix\doi{10.1038/s41598-017-00693-9}.
%Type = Article
\bibitem[{{Koyama} et~al.(2009){Koyama}, {Takikawa}, {Hyodo}, {Inui},
  {Nobukawa}, {Matsumoto} and {Tsuru}}]{koyama09}
\bibinfo{author}{{Koyama}, K.}, \bibinfo{author}{{Takikawa}, Y.},
  \bibinfo{author}{{Hyodo}, Y.}, \bibinfo{author}{{Inui}, T.},
  \bibinfo{author}{{Nobukawa}, M.}, \bibinfo{author}{{Matsumoto}, H.},
  \bibinfo{author}{{Tsuru}, T.G.}, \bibinfo{year}{2009}.
\newblock \bibinfo{title}{{Spatial Distribution of the Galactic Center Diffuse
  X-Rays and the Spectra of the Brightest 6.4keV Clumps}}.
\newblock \bibinfo{journal}{\pasj} \bibinfo{volume}{61}, \bibinfo{pages}{S255}.
\newblock \DOIprefix\doi{10.1093/pasj/61.sp1.S255}.
%Type = Article
\bibitem[{{Krumholz} and {Kruijssen}(2015)}]{krumholz15}
\bibinfo{author}{{Krumholz}, M.R.}, \bibinfo{author}{{Kruijssen}, J.M.D.},
  \bibinfo{year}{2015}.
\newblock \bibinfo{title}{{A dynamical model for the formation of gas rings and
  episodic starbursts near galactic centres}}.
\newblock \bibinfo{journal}{\mnras} \bibinfo{volume}{453},
  \bibinfo{pages}{739--757}.
\newblock \DOIprefix\doi{10.1093/mnras/stv1670},
  \href{http://arxiv.org/abs/1505.07111}{{\tt arXiv:1505.07111}}.
%Type = Article
\bibitem[{{Launhardt} et~al.(2002){Launhardt}, {Zylka} and
  {Mezger}}]{launhardt02}
\bibinfo{author}{{Launhardt}, R.}, \bibinfo{author}{{Zylka}, R.},
  \bibinfo{author}{{Mezger}, P.G.}, \bibinfo{year}{2002}.
\newblock \bibinfo{title}{{The nuclear bulge of the Galaxy. III. Large-scale
  physical characteristics of stars and interstellar matter}}.
\newblock \bibinfo{journal}{\aap} \bibinfo{volume}{384},
  \bibinfo{pages}{112--139}.
\newblock \DOIprefix\doi{10.1051/0004-6361:20020017},
  \href{http://arxiv.org/abs/astro-ph/0201294}{{\tt arXiv:astro-ph/0201294}}.
%Type = Article
\bibitem[{{Law} et~al.(2021){Law}, {Zhang}, {{\"O}berg}, {Galv{\'a}n-Madrid},
  {Keto}, {Liu} and {Ho}}]{Law21}
\bibinfo{author}{{Law}, C.J.}, \bibinfo{author}{{Zhang}, Q.},
  \bibinfo{author}{{{\"O}berg}, K.I.}, \bibinfo{author}{{Galv{\'a}n-Madrid},
  R.}, \bibinfo{author}{{Keto}, E.}, \bibinfo{author}{{Liu}, H.B.},
  \bibinfo{author}{{Ho}, P.T.P.}, \bibinfo{year}{2021}.
\newblock \bibinfo{title}{{Subarcsecond Imaging of the Complex Organic
  Chemistry in Massive Star-forming Region G10.6-0.4}}.
\newblock \bibinfo{journal}{\apj} \bibinfo{volume}{909}, \bibinfo{pages}{214}.
\newblock \DOIprefix\doi{10.3847/1538-4357/abdeb8},
  \href{http://arxiv.org/abs/2101.07801}{{\tt arXiv:2101.07801}}.
%Type = Article
\bibitem[{{Lee} et~al.(2022){Lee}, {Codella}, {Ceccarelli} and
  {L{\'o}pez-Sepulcre}}]{Lee22}
\bibinfo{author}{{Lee}, C.F.}, \bibinfo{author}{{Codella}, C.},
  \bibinfo{author}{{Ceccarelli}, C.}, \bibinfo{author}{{L{\'o}pez-Sepulcre},
  A.}, \bibinfo{year}{2022}.
\newblock \bibinfo{title}{{Stratified Distribution of Organic Molecules at the
  Planet-formation Scale in the HH 212 Disk Atmosphere}}.
\newblock \bibinfo{journal}{\apj} \bibinfo{volume}{937}, \bibinfo{pages}{10}.
\newblock \DOIprefix\doi{10.3847/1538-4357/ac8c28},
  \href{http://arxiv.org/abs/2208.10693}{{\tt arXiv:2208.10693}}.
%Type = Article
\bibitem[{{Lee} et~al.(2019a){Lee}, {Codella}, {Li} and {Liu}}]{Lee19}
\bibinfo{author}{{Lee}, C.F.}, \bibinfo{author}{{Codella}, C.},
  \bibinfo{author}{{Li}, Z.Y.}, \bibinfo{author}{{Liu}, S.Y.},
  \bibinfo{year}{2019}a.
\newblock \bibinfo{title}{{First Abundance Measurement of Organic Molecules in
  the Atmosphere of HH 212 Protostellar Disk}}.
\newblock \bibinfo{journal}{\apj} \bibinfo{volume}{876}, \bibinfo{pages}{63}.
\newblock \DOIprefix\doi{10.3847/1538-4357/ab15db},
  \href{http://arxiv.org/abs/1904.10572}{{\tt arXiv:1904.10572}}.
%Type = Article
\bibitem[{{Lee} et~al.(2017a){Lee}, {Ho}, {Li}, {Hirano}, {Zhang} and
  {Shang}}]{Lee17b}
\bibinfo{author}{{Lee}, C.F.}, \bibinfo{author}{{Ho}, P.T.P.},
  \bibinfo{author}{{Li}, Z.Y.}, \bibinfo{author}{{Hirano}, N.},
  \bibinfo{author}{{Zhang}, Q.}, \bibinfo{author}{{Shang}, H.},
  \bibinfo{year}{2017}a.
\newblock \bibinfo{title}{{A rotating protostellar jet launched from the
  innermost disk of HH 212}}.
\newblock \bibinfo{journal}{Nature Astronomy} \bibinfo{volume}{1},
  \bibinfo{pages}{0152}.
\newblock \DOIprefix\doi{10.1038/s41550-017-0152},
  \href{http://arxiv.org/abs/1706.06343}{{\tt arXiv:1706.06343}}.
%Type = Article
\bibitem[{{Lee} et~al.(2017b){Lee}, {Li}, {Ho}, {Hirano}, {Zhang} and
  {Shang}}]{Lee17a}
\bibinfo{author}{{Lee}, C.F.}, \bibinfo{author}{{Li}, Z.Y.},
  \bibinfo{author}{{Ho}, P.T.P.}, \bibinfo{author}{{Hirano}, N.},
  \bibinfo{author}{{Zhang}, Q.}, \bibinfo{author}{{Shang}, H.},
  \bibinfo{year}{2017}b.
\newblock \bibinfo{title}{{First detection of equatorial dark dust lane in a
  protostellar disk at submillimeter wavelength}}.
\newblock \bibinfo{journal}{Science Advances} \bibinfo{volume}{3},
  \bibinfo{pages}{e1602935}.
\newblock \DOIprefix\doi{10.1126/sciadv.1602935},
  \href{http://arxiv.org/abs/1704.08962}{{\tt arXiv:1704.08962}}.
%Type = Article
\bibitem[{{Lee} et~al.(2017c){Lee}, {Li}, {Ho}, {Hirano}, {Zhang} and
  {Shang}}]{Lee17c}
\bibinfo{author}{{Lee}, C.F.}, \bibinfo{author}{{Li}, Z.Y.},
  \bibinfo{author}{{Ho}, P.T.P.}, \bibinfo{author}{{Hirano}, N.},
  \bibinfo{author}{{Zhang}, Q.}, \bibinfo{author}{{Shang}, H.},
  \bibinfo{year}{2017}c.
\newblock \bibinfo{title}{{Formation and Atmosphere of Complex Organic
  Molecules of the HH 212 Protostellar Disk}}.
\newblock \bibinfo{journal}{\apj} \bibinfo{volume}{843}, \bibinfo{pages}{27}.
\newblock \DOIprefix\doi{10.3847/1538-4357/aa7757},
  \href{http://arxiv.org/abs/1706.06041}{{\tt arXiv:1706.06041}}.
%Type = Article
\bibitem[{{Lee} et~al.(2019b){Lee}, {Lee}, {Baek}, {Aikawa}, {Cieza}, {Yoon},
  {Herczeg}, {Johnstone} and {Casassus}}]{LeeJ19}
\bibinfo{author}{{Lee}, J.E.}, \bibinfo{author}{{Lee}, S.},
  \bibinfo{author}{{Baek}, G.}, \bibinfo{author}{{Aikawa}, Y.},
  \bibinfo{author}{{Cieza}, L.}, \bibinfo{author}{{Yoon}, S.Y.},
  \bibinfo{author}{{Herczeg}, G.}, \bibinfo{author}{{Johnstone}, D.},
  \bibinfo{author}{{Casassus}, S.}, \bibinfo{year}{2019}b.
\newblock \bibinfo{title}{{The ice composition in the disk around V883 Ori
  revealed by its stellar outburst}}.
\newblock \bibinfo{journal}{Nature Astronomy} \bibinfo{volume}{3},
  \bibinfo{pages}{314--319}.
\newblock \DOIprefix\doi{10.1038/s41550-018-0680-0},
  \href{http://arxiv.org/abs/1809.00353}{{\tt arXiv:1809.00353}}.
%Type = Article
\bibitem[{{Lefloch} et~al.(2018){Lefloch}, {Bachiller}, {Ceccarelli},
  {Cernicharo}, {Codella}, {Fuente}, {Kahane}, {L{\'o}pez-Sepulcre}, {Tafalla},
  {Vastel}, {Caux}, {Gonz{\'a}lez-Garc{\'{\i}}a}, {Bianchi}, {G{\'o}mez-Ruiz},
  {Holdship}, {Mendoza}, {Ospina-Zamudio}, {Podio}, {Qu{\'e}nard}, {Roueff},
  {Sakai}, {Viti}, {Yamamoto}, {Yoshida}, {Favre}, {Monfredini},
  {Quiti{\'a}n-Lara}, {Marcelino}, {Boechat-Roberty} and {Cabrit}}]{Lefloch18}
\bibinfo{author}{{Lefloch}, B.}, \bibinfo{author}{{Bachiller}, R.},
  \bibinfo{author}{{Ceccarelli}, C.}, \bibinfo{author}{{Cernicharo}, J.},
  \bibinfo{author}{{Codella}, C.}, \bibinfo{author}{{Fuente}, A.},
  \bibinfo{author}{{Kahane}, C.}, \bibinfo{author}{{L{\'o}pez-Sepulcre}, A.},
  \bibinfo{author}{{Tafalla}, M.}, \bibinfo{author}{{Vastel}, C.},
  \bibinfo{author}{{Caux}, E.}, \bibinfo{author}{{Gonz{\'a}lez-Garc{\'{\i}}a},
  M.}, \bibinfo{author}{{Bianchi}, E.}, \bibinfo{author}{{G{\'o}mez-Ruiz}, A.},
  \bibinfo{author}{{Holdship}, J.}, \bibinfo{author}{{Mendoza}, E.},
  \bibinfo{author}{{Ospina-Zamudio}, J.}, \bibinfo{author}{{Podio}, L.},
  \bibinfo{author}{{Qu{\'e}nard}, D.}, \bibinfo{author}{{Roueff}, E.},
  \bibinfo{author}{{Sakai}, N.}, \bibinfo{author}{{Viti}, S.},
  \bibinfo{author}{{Yamamoto}, S.}, \bibinfo{author}{{Yoshida}, K.},
  \bibinfo{author}{{Favre}, C.}, \bibinfo{author}{{Monfredini}, T.},
  \bibinfo{author}{{Quiti{\'a}n-Lara}, H.M.}, \bibinfo{author}{{Marcelino},
  N.}, \bibinfo{author}{{Boechat-Roberty}, H.M.}, \bibinfo{author}{{Cabrit},
  S.}, \bibinfo{year}{2018}.
\newblock \bibinfo{title}{{Astrochemical evolution along star formation:
  overview of the IRAM Large Program ASAI}}.
\newblock \bibinfo{journal}{\mnras} \bibinfo{volume}{477},
  \bibinfo{pages}{4792--4809}.
\newblock \DOIprefix\doi{10.1093/mnras/sty937},
  \href{http://arxiv.org/abs/1803.10292}{{\tt arXiv:1803.10292}}.
%Type = Article
\bibitem[{{Lefloch} et~al.(2012){Lefloch}, {Cabrit}, {Busquet}, {Codella},
  {Ceccarelli}, {Cernicharo}, {Pardo}, {Benedettini}, {Lis} and
  {Nisini}}]{Lefloch12}
\bibinfo{author}{{Lefloch}, B.}, \bibinfo{author}{{Cabrit}, S.},
  \bibinfo{author}{{Busquet}, G.}, \bibinfo{author}{{Codella}, C.},
  \bibinfo{author}{{Ceccarelli}, C.}, \bibinfo{author}{{Cernicharo}, J.},
  \bibinfo{author}{{Pardo}, J.R.}, \bibinfo{author}{{Benedettini}, M.},
  \bibinfo{author}{{Lis}, D.C.}, \bibinfo{author}{{Nisini}, B.},
  \bibinfo{year}{2012}.
\newblock \bibinfo{title}{{The CHESS Survey of the L1157-B1 Shock Region: CO
  Spectral Signatures of Jet-driven Bow Shocks}}.
\newblock \bibinfo{journal}{\apjl} \bibinfo{volume}{757}, \bibinfo{pages}{L25}.
\newblock \DOIprefix\doi{10.1088/2041-8205/757/2/L25},
  \href{http://arxiv.org/abs/1208.4140}{{\tt arXiv:1208.4140}}.
%Type = Article
\bibitem[{{Lefloch} et~al.(2017){Lefloch}, {Ceccarelli}, {Codella}, {Favre},
  {Podio}, {Vastel}, {Viti} and {Bachiller}}]{Lefloch17}
\bibinfo{author}{{Lefloch}, B.}, \bibinfo{author}{{Ceccarelli}, C.},
  \bibinfo{author}{{Codella}, C.}, \bibinfo{author}{{Favre}, C.},
  \bibinfo{author}{{Podio}, L.}, \bibinfo{author}{{Vastel}, C.},
  \bibinfo{author}{{Viti}, S.}, \bibinfo{author}{{Bachiller}, R.},
  \bibinfo{year}{2017}.
\newblock \bibinfo{title}{{L1157-B1, a factory of complex organic molecules in
  a solar-type star-forming region}}.
\newblock \bibinfo{journal}{\mnras} \bibinfo{volume}{469},
  \bibinfo{pages}{L73--L77}.
\newblock \DOIprefix\doi{10.1093/mnrasl/slx050},
  \href{http://arxiv.org/abs/1704.04646}{{\tt arXiv:1704.04646}}.
%Type = Article
\bibitem[{{Leurini} et~al.(2013){Leurini}, {Codella}, {Gusdorf}, {Zapata},
  {G{\'o}mez-Ruiz}, {Testi} and {Pillai}}]{Leurini13}
\bibinfo{author}{{Leurini}, S.}, \bibinfo{author}{{Codella}, C.},
  \bibinfo{author}{{Gusdorf}, A.}, \bibinfo{author}{{Zapata}, L.},
  \bibinfo{author}{{G{\'o}mez-Ruiz}, A.}, \bibinfo{author}{{Testi}, L.},
  \bibinfo{author}{{Pillai}, T.}, \bibinfo{year}{2013}.
\newblock \bibinfo{title}{{Evidence of a SiO collimated outflow from a massive
  YSO in IRAS 17233-3606}}.
\newblock \bibinfo{journal}{\aap} \bibinfo{volume}{554}, \bibinfo{pages}{A35}.
\newblock \DOIprefix\doi{10.1051/0004-6361/201118154},
  \href{http://arxiv.org/abs/1304.4401}{{\tt arXiv:1304.4401}}.
%Type = Article
\bibitem[{{Li} et~al.(2017){Li}, {Shen}, {Wang}, {Chen}, {Li}, {Wu}, {Dong},
  {Zhao}, {Gou}, {Wang}, {Li}, {Wang} and {Zheng}}]{Li17}
\bibinfo{author}{{Li}, J.}, \bibinfo{author}{{Shen}, Z.},
  \bibinfo{author}{{Wang}, J.}, \bibinfo{author}{{Chen}, X.},
  \bibinfo{author}{{Li}, D.}, \bibinfo{author}{{Wu}, Y.},
  \bibinfo{author}{{Dong}, J.}, \bibinfo{author}{{Zhao}, R.},
  \bibinfo{author}{{Gou}, W.}, \bibinfo{author}{{Wang}, J.},
  \bibinfo{author}{{Li}, S.}, \bibinfo{author}{{Wang}, B.},
  \bibinfo{author}{{Zheng}, X.}, \bibinfo{year}{2017}.
\newblock \bibinfo{title}{{Widespread Presence of Glycolaldehyde and Ethylene
  Glycol around Sagittarius B2}}.
\newblock \bibinfo{journal}{\apj} \bibinfo{volume}{849}, \bibinfo{pages}{115}.
\newblock \DOIprefix\doi{10.3847/1538-4357/aa9069},
  \href{http://arxiv.org/abs/1709.10247}{{\tt arXiv:1709.10247}}.
%Type = Article
\bibitem[{{Li} et~al.(2020){Li}, {Wang}, {Qiao}, {Quan}, {Fang}, {Du}, {Li},
  {Shen}, {Li}, {Li}, {Shi}, {Zhang} and {Zhang}}]{Li20}
\bibinfo{author}{{Li}, J.}, \bibinfo{author}{{Wang}, J.},
  \bibinfo{author}{{Qiao}, H.}, \bibinfo{author}{{Quan}, D.},
  \bibinfo{author}{{Fang}, M.}, \bibinfo{author}{{Du}, F.},
  \bibinfo{author}{{Li}, F.}, \bibinfo{author}{{Shen}, Z.},
  \bibinfo{author}{{Li}, S.}, \bibinfo{author}{{Li}, D.},
  \bibinfo{author}{{Shi}, Y.}, \bibinfo{author}{{Zhang}, Z.},
  \bibinfo{author}{{Zhang}, J.}, \bibinfo{year}{2020}.
\newblock \bibinfo{title}{{Mapping observations of complex organic molecules
  around Sagittarius B2 with the ARO 12 m telescope}}.
\newblock \bibinfo{journal}{\mnras} \bibinfo{volume}{492},
  \bibinfo{pages}{556--565}.
\newblock \DOIprefix\doi{10.1093/mnras/stz3337},
  \href{http://arxiv.org/abs/1911.10087}{{\tt arXiv:1911.10087}}.
%Type = Article
\bibitem[{{Licquia} and {Newman}(2015)}]{Licquia15}
\bibinfo{author}{{Licquia}, T.C.}, \bibinfo{author}{{Newman}, J.A.},
  \bibinfo{year}{2015}.
\newblock \bibinfo{title}{{Improved Estimates of the Milky Way's Stellar Mass
  and Star Formation Rate from Hierarchical Bayesian Meta-Analysis}}.
\newblock \bibinfo{journal}{\apj} \bibinfo{volume}{806}, \bibinfo{pages}{96}.
\newblock \DOIprefix\doi{10.1088/0004-637X/806/1/96},
  \href{http://arxiv.org/abs/1407.1078}{{\tt arXiv:1407.1078}}.
%Type = Article
\bibitem[{{Lidman} et~al.(1999){Lidman}, {Courbin}, {Meylan}, {Broadhurst},
  {Frye} and {Welch}}]{lidman99}
\bibinfo{author}{{Lidman}, C.}, \bibinfo{author}{{Courbin}, F.},
  \bibinfo{author}{{Meylan}, G.}, \bibinfo{author}{{Broadhurst}, T.},
  \bibinfo{author}{{Frye}, B.}, \bibinfo{author}{{Welch}, W.J.W.},
  \bibinfo{year}{1999}.
\newblock \bibinfo{title}{{The Redshift of the Gravitationally Lensed Radio
  Source PKS 1830-211}}.
\newblock \bibinfo{journal}{\apjl} \bibinfo{volume}{514},
  \bibinfo{pages}{L57--L60}.
\newblock \DOIprefix\doi{10.1086/311949},
  \href{http://arxiv.org/abs/astro-ph/9902317}{{\tt arXiv:astro-ph/9902317}}.
%Type = Article
\bibitem[{{Ligterink} et~al.(2021){Ligterink}, {Ahmadi}, {Coutens},
  {Tychoniec}, {Calcutt}, {van Dishoeck}, {Linnartz}, {J{\o}rgensen}, {Garrod}
  and {Bouwman}}]{Ligterink21}
\bibinfo{author}{{Ligterink}, N.F.W.}, \bibinfo{author}{{Ahmadi}, A.},
  \bibinfo{author}{{Coutens}, A.}, \bibinfo{author}{{Tychoniec}, {\L}.},
  \bibinfo{author}{{Calcutt}, H.}, \bibinfo{author}{{van Dishoeck}, E.F.},
  \bibinfo{author}{{Linnartz}, H.}, \bibinfo{author}{{J{\o}rgensen}, J.K.},
  \bibinfo{author}{{Garrod}, R.T.}, \bibinfo{author}{{Bouwman}, J.},
  \bibinfo{year}{2021}.
\newblock \bibinfo{title}{{The prebiotic molecular inventory of Serpens SMM1.
  I. An investigation of the isomers CH$_{3}$NCO and HOCH$_{2}$CN}}.
\newblock \bibinfo{journal}{\aap} \bibinfo{volume}{647}, \bibinfo{pages}{A87}.
\newblock \DOIprefix\doi{10.1051/0004-6361/202039619},
  \href{http://arxiv.org/abs/2012.15672}{{\tt arXiv:2012.15672}}.
%Type = Article
\bibitem[{{Ligterink} et~al.(2017){Ligterink}, {Coutens}, {Kofman},
  {M{\"u}ller}, {Garrod}, {Calcutt}, {Wampfler}, {J{\o}rgensen}, {Linnartz} and
  {van Dishoeck}}]{Ligterink17}
\bibinfo{author}{{Ligterink}, N.F.W.}, \bibinfo{author}{{Coutens}, A.},
  \bibinfo{author}{{Kofman}, V.}, \bibinfo{author}{{M{\"u}ller}, H.S.P.},
  \bibinfo{author}{{Garrod}, R.T.}, \bibinfo{author}{{Calcutt}, H.},
  \bibinfo{author}{{Wampfler}, S.F.}, \bibinfo{author}{{J{\o}rgensen}, J.K.},
  \bibinfo{author}{{Linnartz}, H.}, \bibinfo{author}{{van Dishoeck}, E.F.},
  \bibinfo{year}{2017}.
\newblock \bibinfo{title}{{The ALMA-PILS survey: detection of CH$_{3}$NCO
  towards the low-mass protostar IRAS 16293-2422 and laboratory constraints on
  its formation}}.
\newblock \bibinfo{journal}{\mnras} \bibinfo{volume}{469},
  \bibinfo{pages}{2219--2229}.
\newblock \DOIprefix\doi{10.1093/mnras/stx890},
  \href{http://arxiv.org/abs/1703.03252}{{\tt arXiv:1703.03252}}.
%Type = Article
\bibitem[{{Ligterink} et~al.(2020){Ligterink}, {El-Abd}, {Brogan}, {Hunter},
  {Remijan}, {Garrod} and {McGuire}}]{Ligterink20}
\bibinfo{author}{{Ligterink}, N.F.W.}, \bibinfo{author}{{El-Abd}, S.J.},
  \bibinfo{author}{{Brogan}, C.L.}, \bibinfo{author}{{Hunter}, T.R.},
  \bibinfo{author}{{Remijan}, A.J.}, \bibinfo{author}{{Garrod}, R.T.},
  \bibinfo{author}{{McGuire}, B.M.}, \bibinfo{year}{2020}.
\newblock \bibinfo{title}{{The Family of Amide Molecules toward NGC 6334I}}.
\newblock \bibinfo{journal}{\apj} \bibinfo{volume}{901}, \bibinfo{pages}{37}.
\newblock \DOIprefix\doi{10.3847/1538-4357/abad38},
  \href{http://arxiv.org/abs/2008.09157}{{\tt arXiv:2008.09157}}.
%Type = Article
\bibitem[{{Liszt} et~al.(2018){Liszt}, {Gerin}, {Beasley} and {Pety}}]{Liszt18}
\bibinfo{author}{{Liszt}, H.}, \bibinfo{author}{{Gerin}, M.},
  \bibinfo{author}{{Beasley}, A.}, \bibinfo{author}{{Pety}, J.},
  \bibinfo{year}{2018}.
\newblock \bibinfo{title}{{Chemical Complexity in Local Diffuse and Translucent
  Clouds: Ubiquitous Linear C$_{3}$H and CH$_{3}$CN, a Detection of HC$_{3}$N
  and an Upper Limit on the Abundance of CH$_{2}$CN}}.
\newblock \bibinfo{journal}{\apj} \bibinfo{volume}{856}, \bibinfo{pages}{151}.
\newblock \DOIprefix\doi{10.3847/1538-4357/aab208},
  \href{http://arxiv.org/abs/1802.08325}{{\tt arXiv:1802.08325}}.
%Type = Article
\bibitem[{{Liszt} and {Lucas}(2001)}]{Liszt01}
\bibinfo{author}{{Liszt}, H.}, \bibinfo{author}{{Lucas}, R.},
  \bibinfo{year}{2001}.
\newblock \bibinfo{title}{{Comparative chemistry of diffuse clouds. II. CN,
  HCN, HNC, CH$_{3}$CN \& N$_{2}$H$^{+}$}}.
\newblock \bibinfo{journal}{\aap} \bibinfo{volume}{370},
  \bibinfo{pages}{576--585}.
\newblock \DOIprefix\doi{10.1051/0004-6361:20010260},
  \href{http://arxiv.org/abs/astro-ph/0103247}{{\tt arXiv:astro-ph/0103247}}.
%Type = Article
\bibitem[{{Liszt} et~al.(2008){Liszt}, {Pety} and {Lucas}}]{Liszt08}
\bibinfo{author}{{Liszt}, H.S.}, \bibinfo{author}{{Pety}, J.},
  \bibinfo{author}{{Lucas}, R.}, \bibinfo{year}{2008}.
\newblock \bibinfo{title}{{Limits on chemical complexity in diffuse clouds:
  search for CH$_{3}$OH and HC$_{5}$N absorption}}.
\newblock \bibinfo{journal}{\aap} \bibinfo{volume}{486},
  \bibinfo{pages}{493--496}.
\newblock \DOIprefix\doi{10.1051/0004-6361:200809851}.
%Type = Article
\bibitem[{{Liu} et~al.(2021){Liu}, {Liu}, {Evans}, {Wang}, {Garay}, {Qin},
  {Li}, {Stutz}, {Goldsmith}, {Liu}, {Tej}, {Zhang}, {Juvela}, {Li}, {Wang},
  {Bronfman}, {Ren}, {Wu}, {Kim}, {Lee}, {Tatematsu}, {Cunningham}, {Liu},
  {Wu}, {Hirota}, {Lee}, {Li}, {Kang}, {Mardones}, {Ristorcelli}, {Zhang},
  {Luo}, {Toth}, {Yi}, {Yun}, {Peng}, {Li}, {Zhu}, {Shen}, {Baug}, {Dewangan},
  {Chakali}, {Liu}, {Xu}, {Wang}, {Zhang}, {Li}, {Zhang}, {Zhou}, {Tang},
  {Xue}, {Issac}, {Soam} and {{\'A}lvarez-Guti{\'e}rrez}}]{LiuHL21}
\bibinfo{author}{{Liu}, H.L.}, \bibinfo{author}{{Liu}, T.},
  \bibinfo{author}{{Evans}, Neal~J., I.}, \bibinfo{author}{{Wang}, K.},
  \bibinfo{author}{{Garay}, G.}, \bibinfo{author}{{Qin}, S.L.},
  \bibinfo{author}{{Li}, S.}, \bibinfo{author}{{Stutz}, A.},
  \bibinfo{author}{{Goldsmith}, P.F.}, \bibinfo{author}{{Liu}, S.Y.},
  \bibinfo{author}{{Tej}, A.}, \bibinfo{author}{{Zhang}, Q.},
  \bibinfo{author}{{Juvela}, M.}, \bibinfo{author}{{Li}, D.},
  \bibinfo{author}{{Wang}, J.Z.}, \bibinfo{author}{{Bronfman}, L.},
  \bibinfo{author}{{Ren}, Z.}, \bibinfo{author}{{Wu}, Y.F.},
  \bibinfo{author}{{Kim}, K.T.}, \bibinfo{author}{{Lee}, C.W.},
  \bibinfo{author}{{Tatematsu}, K.}, \bibinfo{author}{{Cunningham}, M.R.},
  \bibinfo{author}{{Liu}, X.C.}, \bibinfo{author}{{Wu}, J.W.},
  \bibinfo{author}{{Hirota}, T.}, \bibinfo{author}{{Lee}, J.E.},
  \bibinfo{author}{{Li}, P.S.}, \bibinfo{author}{{Kang}, S.J.},
  \bibinfo{author}{{Mardones}, D.}, \bibinfo{author}{{Ristorcelli}, I.},
  \bibinfo{author}{{Zhang}, Y.}, \bibinfo{author}{{Luo}, Q.Y.},
  \bibinfo{author}{{Toth}, L.V.}, \bibinfo{author}{{Yi}, H.w.},
  \bibinfo{author}{{Yun}, H.S.}, \bibinfo{author}{{Peng}, Y.P.},
  \bibinfo{author}{{Li}, J.}, \bibinfo{author}{{Zhu}, F.Y.},
  \bibinfo{author}{{Shen}, Z.Q.}, \bibinfo{author}{{Baug}, T.},
  \bibinfo{author}{{Dewangan}, L.K.}, \bibinfo{author}{{Chakali}, E.},
  \bibinfo{author}{{Liu}, R.}, \bibinfo{author}{{Xu}, F.W.},
  \bibinfo{author}{{Wang}, Y.}, \bibinfo{author}{{Zhang}, C.},
  \bibinfo{author}{{Li}, J.}, \bibinfo{author}{{Zhang}, C.},
  \bibinfo{author}{{Zhou}, J.}, \bibinfo{author}{{Tang}, M.},
  \bibinfo{author}{{Xue}, Q.}, \bibinfo{author}{{Issac}, N.},
  \bibinfo{author}{{Soam}, A.}, \bibinfo{author}{{{\'A}lvarez-Guti{\'e}rrez},
  R.H.}, \bibinfo{year}{2021}.
\newblock \bibinfo{title}{{ATOMS: ALMA three-millimeter observations of massive
  star-forming regions - III. Catalogues of candidate hot molecular cores and
  hyper/ultra compact H II regions}}.
\newblock \bibinfo{journal}{\mnras} \bibinfo{volume}{505},
  \bibinfo{pages}{2801--2818}.
\newblock \DOIprefix\doi{10.1093/mnras/stab1352},
  \href{http://arxiv.org/abs/2105.03554}{{\tt arXiv:2105.03554}}.
%Type = Article
\bibitem[{{Liu} et~al.(2022){Liu}, {Liu}, {Evans}, {Wang}, {Garay}, {Qin},
  {Li}, {Stutz}, {Goldsmith}, {Liu}, {Tej}, {Zhang}, {Juvela}, {Li}, {Wang},
  {Bronfman}, {Ren}, {Wu}, {Kim}, {Lee}, {Tatematsu}, {Cunningham}, {Liu},
  {Wu}, {Hirota}, {Lee}, {Li}, {Kang}, {Mardones}, {Ristorcelli}, {Zhang},
  {Luo}, {Toth}, {Yi}, {Yun}, {Peng}, {Li}, {Zhu}, {Shen}, {Baug}, {Dewangan},
  {Chakali}, {Liu}, {Xu}, {Wang}, {Zhang}, {Li}, {Zhang}, {Zhou}, {Tang},
  {Xue}, {Issac}, {Soam} and {{\'A}lvarez-Guti{\'e}rrez}}]{LiuHL22}
\bibinfo{author}{{Liu}, H.L.}, \bibinfo{author}{{Liu}, T.},
  \bibinfo{author}{{Evans}, Neal~J., I.}, \bibinfo{author}{{Wang}, K.},
  \bibinfo{author}{{Garay}, G.}, \bibinfo{author}{{Qin}, S.L.},
  \bibinfo{author}{{Li}, S.}, \bibinfo{author}{{Stutz}, A.},
  \bibinfo{author}{{Goldsmith}, P.F.}, \bibinfo{author}{{Liu}, S.Y.},
  \bibinfo{author}{{Tej}, A.}, \bibinfo{author}{{Zhang}, Q.},
  \bibinfo{author}{{Juvela}, M.}, \bibinfo{author}{{Li}, D.},
  \bibinfo{author}{{Wang}, J.Z.}, \bibinfo{author}{{Bronfman}, L.},
  \bibinfo{author}{{Ren}, Z.}, \bibinfo{author}{{Wu}, Y.F.},
  \bibinfo{author}{{Kim}, K.T.}, \bibinfo{author}{{Lee}, C.W.},
  \bibinfo{author}{{Tatematsu}, K.}, \bibinfo{author}{{Cunningham}, M.R.},
  \bibinfo{author}{{Liu}, X.C.}, \bibinfo{author}{{Wu}, J.W.},
  \bibinfo{author}{{Hirota}, T.}, \bibinfo{author}{{Lee}, J.E.},
  \bibinfo{author}{{Li}, P.S.}, \bibinfo{author}{{Kang}, S.J.},
  \bibinfo{author}{{Mardones}, D.}, \bibinfo{author}{{Ristorcelli}, I.},
  \bibinfo{author}{{Zhang}, Y.}, \bibinfo{author}{{Luo}, Q.Y.},
  \bibinfo{author}{{Toth}, L.V.}, \bibinfo{author}{{Yi}, H.w.},
  \bibinfo{author}{{Yun}, H.S.}, \bibinfo{author}{{Peng}, Y.P.},
  \bibinfo{author}{{Li}, J.}, \bibinfo{author}{{Zhu}, F.Y.},
  \bibinfo{author}{{Shen}, Z.Q.}, \bibinfo{author}{{Baug}, T.},
  \bibinfo{author}{{Dewangan}, L.K.}, \bibinfo{author}{{Chakali}, E.},
  \bibinfo{author}{{Liu}, R.}, \bibinfo{author}{{Xu}, F.W.},
  \bibinfo{author}{{Wang}, Y.}, \bibinfo{author}{{Zhang}, C.},
  \bibinfo{author}{{Li}, J.}, \bibinfo{author}{{Zhang}, C.},
  \bibinfo{author}{{Zhou}, J.}, \bibinfo{author}{{Tang}, M.},
  \bibinfo{author}{{Xue}, Q.}, \bibinfo{author}{{Issac}, N.},
  \bibinfo{author}{{Soam}, A.}, \bibinfo{author}{{{\'A}lvarez-Guti{\'e}rrez},
  R.H.}, \bibinfo{year}{2022}.
\newblock \bibinfo{title}{{Erratum: ATOMS: ALMA three-millimeter observations
  of massive star-forming regions - III. Catalogues of candidate hot molecular
  cores and hyper/ultra compact H II regions}}.
\newblock \bibinfo{journal}{\mnras} \bibinfo{volume}{511},
  \bibinfo{pages}{501--505}.
\newblock \DOIprefix\doi{10.1093/mnras/stac039}.
%Type = Article
\bibitem[{{Liu} et~al.(2020){Liu}, {Evans}, {Kim}, {Goldsmith}, {Liu}, {Zhang},
  {Tatematsu}, {Wang}, {Juvela}, {Bronfman}, {Cunningham}, {Garay}, {Hirota},
  {Lee}, {Kang}, {Li}, {Li}, {Mardones}, {Qin}, {Ristorcelli}, {Tej}, {Toth},
  {Wu}, {Wu}, {Yi}, {Yun}, {Liu}, {Peng}, {Li}, {Li}, {Lee}, {Shen}, {Baug},
  {Wang}, {Zhang}, {Issac}, {Zhu}, {Luo}, {Soam}, {Liu}, {Xu}, {Wang}, {Zhang},
  {Ren} and {Zhang}}]{Liu20}
\bibinfo{author}{{Liu}, T.}, \bibinfo{author}{{Evans}, N.J.},
  \bibinfo{author}{{Kim}, K.T.}, \bibinfo{author}{{Goldsmith}, P.F.},
  \bibinfo{author}{{Liu}, S.Y.}, \bibinfo{author}{{Zhang}, Q.},
  \bibinfo{author}{{Tatematsu}, K.}, \bibinfo{author}{{Wang}, K.},
  \bibinfo{author}{{Juvela}, M.}, \bibinfo{author}{{Bronfman}, L.},
  \bibinfo{author}{{Cunningham}, M.R.}, \bibinfo{author}{{Garay}, G.},
  \bibinfo{author}{{Hirota}, T.}, \bibinfo{author}{{Lee}, J.E.},
  \bibinfo{author}{{Kang}, S.J.}, \bibinfo{author}{{Li}, D.},
  \bibinfo{author}{{Li}, P.S.}, \bibinfo{author}{{Mardones}, D.},
  \bibinfo{author}{{Qin}, S.L.}, \bibinfo{author}{{Ristorcelli}, I.},
  \bibinfo{author}{{Tej}, A.}, \bibinfo{author}{{Toth}, L.V.},
  \bibinfo{author}{{Wu}, J.W.}, \bibinfo{author}{{Wu}, Y.F.},
  \bibinfo{author}{{Yi}, H.w.}, \bibinfo{author}{{Yun}, H.S.},
  \bibinfo{author}{{Liu}, H.L.}, \bibinfo{author}{{Peng}, Y.P.},
  \bibinfo{author}{{Li}, J.}, \bibinfo{author}{{Li}, S.H.},
  \bibinfo{author}{{Lee}, C.W.}, \bibinfo{author}{{Shen}, Z.Q.},
  \bibinfo{author}{{Baug}, T.}, \bibinfo{author}{{Wang}, J.Z.},
  \bibinfo{author}{{Zhang}, Y.}, \bibinfo{author}{{Issac}, N.},
  \bibinfo{author}{{Zhu}, F.Y.}, \bibinfo{author}{{Luo}, Q.Y.},
  \bibinfo{author}{{Soam}, A.}, \bibinfo{author}{{Liu}, X.C.},
  \bibinfo{author}{{Xu}, F.W.}, \bibinfo{author}{{Wang}, Y.},
  \bibinfo{author}{{Zhang}, C.}, \bibinfo{author}{{Ren}, Z.},
  \bibinfo{author}{{Zhang}, C.}, \bibinfo{year}{2020}.
\newblock \bibinfo{title}{{ATOMS: ALMA Three-millimeter Observations of Massive
  Star-forming regions - I. Survey description and a first look at
  G9.62+0.19}}.
\newblock \bibinfo{journal}{\mnras} \bibinfo{volume}{496},
  \bibinfo{pages}{2790--2820}.
\newblock \DOIprefix\doi{10.1093/mnras/staa1577},
  \href{http://arxiv.org/abs/2006.01549}{{\tt arXiv:2006.01549}}.
%Type = Article
\bibitem[{{Loomis} et~al.(2018){Loomis}, {Cleeves}, {{\"O}berg}, {Aikawa},
  {Bergner}, {Furuya}, {Guzman} and {Walsh}}]{Loomis18}
\bibinfo{author}{{Loomis}, R.A.}, \bibinfo{author}{{Cleeves}, L.I.},
  \bibinfo{author}{{{\"O}berg}, K.I.}, \bibinfo{author}{{Aikawa}, Y.},
  \bibinfo{author}{{Bergner}, J.}, \bibinfo{author}{{Furuya}, K.},
  \bibinfo{author}{{Guzman}, V.V.}, \bibinfo{author}{{Walsh}, C.},
  \bibinfo{year}{2018}.
\newblock \bibinfo{title}{{The Distribution and Excitation of CH$_{3}$CN in a
  Solar Nebula Analog}}.
\newblock \bibinfo{journal}{\apj} \bibinfo{volume}{859}, \bibinfo{pages}{131}.
\newblock \DOIprefix\doi{10.3847/1538-4357/aac169},
  \href{http://arxiv.org/abs/1805.01458}{{\tt arXiv:1805.01458}}.
%Type = Article
\bibitem[{{L{\'o}pez-Sepulcre} et~al.(2011){L{\'o}pez-Sepulcre}, {Walmsley},
  {Cesaroni}, {Codella}, {Schuller}, {Bronfman}, {Carey}, {Menten}, {Molinari}
  and {Noriega-Crespo}}]{Lopez11}
\bibinfo{author}{{L{\'o}pez-Sepulcre}, A.}, \bibinfo{author}{{Walmsley}, C.M.},
  \bibinfo{author}{{Cesaroni}, R.}, \bibinfo{author}{{Codella}, C.},
  \bibinfo{author}{{Schuller}, F.}, \bibinfo{author}{{Bronfman}, L.},
  \bibinfo{author}{{Carey}, S.J.}, \bibinfo{author}{{Menten}, K.M.},
  \bibinfo{author}{{Molinari}, S.}, \bibinfo{author}{{Noriega-Crespo}, A.},
  \bibinfo{year}{2011}.
\newblock \bibinfo{title}{{SiO outflows in high-mass star forming regions: A
  potential chemical clock?}}
\newblock \bibinfo{journal}{\aap} \bibinfo{volume}{526}, \bibinfo{pages}{L2}.
\newblock \DOIprefix\doi{10.1051/0004-6361/201015827},
  \href{http://arxiv.org/abs/1011.5419}{{\tt arXiv:1011.5419}}.
%Type = Article
\bibitem[{{Manigand} et~al.(2021){Manigand}, {Coutens}, {Loison}, {Wakelam},
  {Calcutt}, {M{\"u}ller}, {J{\o}rgensen}, {Taquet}, {Wampfler}, {Bourke},
  {Kulterer}, {van Dishoeck}, {Drozdovskaya} and {Ligterink}}]{Manigand21}
\bibinfo{author}{{Manigand}, S.}, \bibinfo{author}{{Coutens}, A.},
  \bibinfo{author}{{Loison}, J.C.}, \bibinfo{author}{{Wakelam}, V.},
  \bibinfo{author}{{Calcutt}, H.}, \bibinfo{author}{{M{\"u}ller}, H.S.P.},
  \bibinfo{author}{{J{\o}rgensen}, J.K.}, \bibinfo{author}{{Taquet}, V.},
  \bibinfo{author}{{Wampfler}, S.F.}, \bibinfo{author}{{Bourke}, T.L.},
  \bibinfo{author}{{Kulterer}, B.M.}, \bibinfo{author}{{van Dishoeck}, E.F.},
  \bibinfo{author}{{Drozdovskaya}, M.N.}, \bibinfo{author}{{Ligterink},
  N.F.W.}, \bibinfo{year}{2021}.
\newblock \bibinfo{title}{{The ALMA-PILS survey: first detection of the
  unsaturated 3-carbon molecules Propenal (C$_{2}$H$_{3}$CHO) and Propylene
  (C$_{3}$H$_{6}$) towards IRAS 16293{\textendash}2422 B}}.
\newblock \bibinfo{journal}{\aap} \bibinfo{volume}{645}, \bibinfo{pages}{A53}.
\newblock \DOIprefix\doi{10.1051/0004-6361/202038113},
  \href{http://arxiv.org/abs/2007.04000}{{\tt arXiv:2007.04000}}.
%Type = Article
\bibitem[{{Manigand} et~al.(2020){Manigand}, {J{\o}rgensen}, {Calcutt},
  {M{\"u}ller}, {Ligterink}, {Coutens}, {Drozdovskaya}, {van Dishoeck} and
  {Wampfler}}]{Manigand20}
\bibinfo{author}{{Manigand}, S.}, \bibinfo{author}{{J{\o}rgensen}, J.K.},
  \bibinfo{author}{{Calcutt}, H.}, \bibinfo{author}{{M{\"u}ller}, H.S.P.},
  \bibinfo{author}{{Ligterink}, N.F.W.}, \bibinfo{author}{{Coutens}, A.},
  \bibinfo{author}{{Drozdovskaya}, M.N.}, \bibinfo{author}{{van Dishoeck},
  E.F.}, \bibinfo{author}{{Wampfler}, S.F.}, \bibinfo{year}{2020}.
\newblock \bibinfo{title}{{The ALMA-PILS survey: inventory of complex organic
  molecules towards IRAS 16293-2422 A}}.
\newblock \bibinfo{journal}{\aap} \bibinfo{volume}{635}, \bibinfo{pages}{A48}.
\newblock \DOIprefix\doi{10.1051/0004-6361/201936299},
  \href{http://arxiv.org/abs/2001.06400}{{\tt arXiv:2001.06400}}.
%Type = Article
\bibitem[{{Marcelino} et~al.(2007){Marcelino}, {Cernicharo}, {Ag{\'u}ndez},
  {Roueff}, {Gerin}, {Mart{\'\i}n-Pintado}, {Mauersberger} and
  {Thum}}]{marcelino07}
\bibinfo{author}{{Marcelino}, N.}, \bibinfo{author}{{Cernicharo}, J.},
  \bibinfo{author}{{Ag{\'u}ndez}, M.}, \bibinfo{author}{{Roueff}, E.},
  \bibinfo{author}{{Gerin}, M.}, \bibinfo{author}{{Mart{\'\i}n-Pintado}, J.},
  \bibinfo{author}{{Mauersberger}, R.}, \bibinfo{author}{{Thum}, C.},
  \bibinfo{year}{2007}.
\newblock \bibinfo{title}{{Discovery of Interstellar Propylene
  (CH$_{2}$CHCH$_{3}$): Missing Links in Interstellar Gas-Phase Chemistry}}.
\newblock \bibinfo{journal}{\apjl} \bibinfo{volume}{665},
  \bibinfo{pages}{L127--L130}.
\newblock \DOIprefix\doi{10.1086/521398},
  \href{http://arxiv.org/abs/0707.1308}{{\tt arXiv:0707.1308}}.
%Type = Article
\bibitem[{{Marcelino} et~al.(2018){Marcelino}, {Gerin}, {Cernicharo}, {Fuente},
  {Wootten}, {Chapillon}, {Pety}, {Lis}, {Roueff}, {Commer{\c{c}}on} and
  {Ciardi}}]{Marcelino18}
\bibinfo{author}{{Marcelino}, N.}, \bibinfo{author}{{Gerin}, M.},
  \bibinfo{author}{{Cernicharo}, J.}, \bibinfo{author}{{Fuente}, A.},
  \bibinfo{author}{{Wootten}, H.A.}, \bibinfo{author}{{Chapillon}, E.},
  \bibinfo{author}{{Pety}, J.}, \bibinfo{author}{{Lis}, D.C.},
  \bibinfo{author}{{Roueff}, E.}, \bibinfo{author}{{Commer{\c{c}}on}, B.},
  \bibinfo{author}{{Ciardi}, A.}, \bibinfo{year}{2018}.
\newblock \bibinfo{title}{{ALMA observations of the young protostellar system
  Barnard 1b: Signatures of an incipient hot corino in B1b-S}}.
\newblock \bibinfo{journal}{\aap} \bibinfo{volume}{620}, \bibinfo{pages}{A80}.
\newblock \DOIprefix\doi{10.1051/0004-6361/201731955},
  \href{http://arxiv.org/abs/1809.08014}{{\tt arXiv:1809.08014}}.
%Type = Article
\bibitem[{{Mart{\'\i}n} et~al.(2011){Mart{\'\i}n}, {Krips},
  {Mart{\'\i}n-Pintado}, {Aalto}, {Zhao}, {Peck}, {Petitpas}, {Monje}, {Greve}
  and {An}}]{Martin11}
\bibinfo{author}{{Mart{\'\i}n}, S.}, \bibinfo{author}{{Krips}, M.},
  \bibinfo{author}{{Mart{\'\i}n-Pintado}, J.}, \bibinfo{author}{{Aalto}, S.},
  \bibinfo{author}{{Zhao}, J.H.}, \bibinfo{author}{{Peck}, A.B.},
  \bibinfo{author}{{Petitpas}, G.R.}, \bibinfo{author}{{Monje}, R.},
  \bibinfo{author}{{Greve}, T.R.}, \bibinfo{author}{{An}, T.},
  \bibinfo{year}{2011}.
\newblock \bibinfo{title}{{The Submillimeter Array 1.3 mm line survey of Arp
  220}}.
\newblock \bibinfo{journal}{\aap} \bibinfo{volume}{527}, \bibinfo{pages}{A36}.
\newblock \DOIprefix\doi{10.1051/0004-6361/201015855},
  \href{http://arxiv.org/abs/1012.3753}{{\tt arXiv:1012.3753}}.
%Type = Article
\bibitem[{{Mart{\'\i}n} et~al.(2021){Mart{\'\i}n}, {Mangum}, {Harada},
  {Costagliola}, {Sakamoto}, {Muller}, {Aladro}, {Tanaka}, {Yoshimura},
  {Nakanishi}, {Herrero-Illana}, {M{\"u}hle}, {Aalto}, {Behrens}, {Colzi},
  {Emig}, {Fuller}, {Garc{\'\i}a-Burillo}, {Greve}, {Henkel}, {Holdship},
  {Humire}, {Hunt}, {Izumi}, {Kohno}, {K{\"o}nig}, {Meier}, {Nakajima},
  {Nishimura}, {Padovani}, {Rivilla}, {Takano}, {van der Werf}, {Viti} and
  {Yan}}]{Martin21}
\bibinfo{author}{{Mart{\'\i}n}, S.}, \bibinfo{author}{{Mangum}, J.G.},
  \bibinfo{author}{{Harada}, N.}, \bibinfo{author}{{Costagliola}, F.},
  \bibinfo{author}{{Sakamoto}, K.}, \bibinfo{author}{{Muller}, S.},
  \bibinfo{author}{{Aladro}, R.}, \bibinfo{author}{{Tanaka}, K.},
  \bibinfo{author}{{Yoshimura}, Y.}, \bibinfo{author}{{Nakanishi}, K.},
  \bibinfo{author}{{Herrero-Illana}, R.}, \bibinfo{author}{{M{\"u}hle}, S.},
  \bibinfo{author}{{Aalto}, S.}, \bibinfo{author}{{Behrens}, E.},
  \bibinfo{author}{{Colzi}, L.}, \bibinfo{author}{{Emig}, K.L.},
  \bibinfo{author}{{Fuller}, G.A.}, \bibinfo{author}{{Garc{\'\i}a-Burillo},
  S.}, \bibinfo{author}{{Greve}, T.R.}, \bibinfo{author}{{Henkel}, C.},
  \bibinfo{author}{{Holdship}, J.}, \bibinfo{author}{{Humire}, P.},
  \bibinfo{author}{{Hunt}, L.}, \bibinfo{author}{{Izumi}, T.},
  \bibinfo{author}{{Kohno}, K.}, \bibinfo{author}{{K{\"o}nig}, S.},
  \bibinfo{author}{{Meier}, D.S.}, \bibinfo{author}{{Nakajima}, T.},
  \bibinfo{author}{{Nishimura}, Y.}, \bibinfo{author}{{Padovani}, M.},
  \bibinfo{author}{{Rivilla}, V.M.}, \bibinfo{author}{{Takano}, S.},
  \bibinfo{author}{{van der Werf}, P.P.}, \bibinfo{author}{{Viti}, S.},
  \bibinfo{author}{{Yan}, Y.T.}, \bibinfo{year}{2021}.
\newblock \bibinfo{title}{{ALCHEMI, an ALMA Comprehensive High-resolution
  Extragalactic Molecular Inventory. Survey presentation and first results from
  the ACA array}}.
\newblock \bibinfo{journal}{\aap} \bibinfo{volume}{656}, \bibinfo{pages}{A46}.
\newblock \DOIprefix\doi{10.1051/0004-6361/202141567},
  \href{http://arxiv.org/abs/2109.08638}{{\tt arXiv:2109.08638}}.
%Type = Article
\bibitem[{{Mart{\'\i}n} et~al.(2006){Mart{\'\i}n}, {Mauersberger},
  {Mart{\'\i}n-Pintado}, {Henkel} and {Garc{\'\i}a-Burillo}}]{Martin06}
\bibinfo{author}{{Mart{\'\i}n}, S.}, \bibinfo{author}{{Mauersberger}, R.},
  \bibinfo{author}{{Mart{\'\i}n-Pintado}, J.}, \bibinfo{author}{{Henkel}, C.},
  \bibinfo{author}{{Garc{\'\i}a-Burillo}, S.}, \bibinfo{year}{2006}.
\newblock \bibinfo{title}{{A 2 Millimeter Spectral Line Survey of the Starburst
  Galaxy NGC 253}}.
\newblock \bibinfo{journal}{\apjs} \bibinfo{volume}{164},
  \bibinfo{pages}{450--476}.
\newblock \DOIprefix\doi{10.1086/503297},
  \href{http://arxiv.org/abs/astro-ph/0602360}{{\tt arXiv:astro-ph/0602360}}.
%Type = Article
\bibitem[{{Mart{\'\i}n} et~al.(2008){Mart{\'\i}n}, {Requena-Torres},
  {Mart{\'\i}n-Pintado} and {Mauersberger}}]{martin08}
\bibinfo{author}{{Mart{\'\i}n}, S.}, \bibinfo{author}{{Requena-Torres}, M.A.},
  \bibinfo{author}{{Mart{\'\i}n-Pintado}, J.}, \bibinfo{author}{{Mauersberger},
  R.}, \bibinfo{year}{2008}.
\newblock \bibinfo{title}{{Tracing Shocks and Photodissociation in the Galactic
  Center Region}}.
\newblock \bibinfo{journal}{\apj} \bibinfo{volume}{678},
  \bibinfo{pages}{245--254}.
\newblock \DOIprefix\doi{10.1086/533409},
  \href{http://arxiv.org/abs/0801.3614}{{\tt arXiv:0801.3614}}.
%Type = Article
\bibitem[{{Mart{\'\i}n-Dom{\'e}nech} et~al.(2021){Mart{\'\i}n-Dom{\'e}nech},
  {Bergner}, {{\"O}berg}, {Carpenter}, {Law}, {Huang}, {J{\o}rgensen},
  {Schwarz} and {Wilner}}]{MartinDomenech21}
\bibinfo{author}{{Mart{\'\i}n-Dom{\'e}nech}, R.}, \bibinfo{author}{{Bergner},
  J.B.}, \bibinfo{author}{{{\"O}berg}, K.I.}, \bibinfo{author}{{Carpenter},
  J.}, \bibinfo{author}{{Law}, C.J.}, \bibinfo{author}{{Huang}, J.},
  \bibinfo{author}{{J{\o}rgensen}, J.K.}, \bibinfo{author}{{Schwarz}, K.},
  \bibinfo{author}{{Wilner}, D.J.}, \bibinfo{year}{2021}.
\newblock \bibinfo{title}{{Hot Corino Chemistry in the Class I Binary Source
  Ser-emb 11}}.
\newblock \bibinfo{journal}{\apj} \bibinfo{volume}{923}, \bibinfo{pages}{155}.
\newblock \DOIprefix\doi{10.3847/1538-4357/ac26b9},
  \href{http://arxiv.org/abs/2109.11512}{{\tt arXiv:2109.11512}}.
%Type = Article
\bibitem[{{Mart{\'\i}n-Dom{\'e}nech} et~al.(2017){Mart{\'\i}n-Dom{\'e}nech},
  {Rivilla}, {Jim{\'e}nez-Serra}, {Qu{\'e}nard}, {Testi} and
  {Mart{\'\i}n-Pintado}}]{MartinDomenech17}
\bibinfo{author}{{Mart{\'\i}n-Dom{\'e}nech}, R.}, \bibinfo{author}{{Rivilla},
  V.M.}, \bibinfo{author}{{Jim{\'e}nez-Serra}, I.},
  \bibinfo{author}{{Qu{\'e}nard}, D.}, \bibinfo{author}{{Testi}, L.},
  \bibinfo{author}{{Mart{\'\i}n-Pintado}, J.}, \bibinfo{year}{2017}.
\newblock \bibinfo{title}{{Detection of methyl isocyanate (CH$_{3}$NCO) in a
  solar-type protostar}}.
\newblock \bibinfo{journal}{\mnras} \bibinfo{volume}{469},
  \bibinfo{pages}{2230--2234}.
\newblock \DOIprefix\doi{10.1093/mnras/stx915},
  \href{http://arxiv.org/abs/1701.04376}{{\tt arXiv:1701.04376}}.
%Type = Article
\bibitem[{{Martin-Drumel} et~al.(2019){Martin-Drumel}, {Lee}, {Belloche},
  {Zingsheim}, {Thorwirth}, {M{\"u}ller}, {Lewen}, {Garrod}, {Menten},
  {McCarthy} and {Schlemmer}}]{martin-drumel19}
\bibinfo{author}{{Martin-Drumel}, M.A.}, \bibinfo{author}{{Lee}, K.L.K.},
  \bibinfo{author}{{Belloche}, A.}, \bibinfo{author}{{Zingsheim}, O.},
  \bibinfo{author}{{Thorwirth}, S.}, \bibinfo{author}{{M{\"u}ller}, H.S.P.},
  \bibinfo{author}{{Lewen}, F.}, \bibinfo{author}{{Garrod}, R.T.},
  \bibinfo{author}{{Menten}, K.M.}, \bibinfo{author}{{McCarthy}, M.C.},
  \bibinfo{author}{{Schlemmer}, S.}, \bibinfo{year}{2019}.
\newblock \bibinfo{title}{{Submillimeter spectroscopy and astronomical searches
  of vinyl mercaptan, C$_{2}$H$_{3}$SH}}.
\newblock \bibinfo{journal}{\aap} \bibinfo{volume}{623}, \bibinfo{pages}{A167}.
\newblock \DOIprefix\doi{10.1051/0004-6361/201935032},
  \href{http://arxiv.org/abs/1902.05833}{{\tt arXiv:1902.05833}}.
%Type = Article
\bibitem[{{Mart{\'\i}n-Pintado} et~al.(1997){Mart{\'\i}n-Pintado}, {de
  Vicente}, {Fuente} and {Planesas}}]{martin-pintado97}
\bibinfo{author}{{Mart{\'\i}n-Pintado}, J.}, \bibinfo{author}{{de Vicente},
  P.}, \bibinfo{author}{{Fuente}, A.}, \bibinfo{author}{{Planesas}, P.},
  \bibinfo{year}{1997}.
\newblock \bibinfo{title}{{SiO Emission from the Galactic Center Molecular
  Clouds}}.
\newblock \bibinfo{journal}{\apjl} \bibinfo{volume}{482},
  \bibinfo{pages}{L45--L48}.
\newblock \DOIprefix\doi{10.1086/310691},
  \href{http://arxiv.org/abs/astro-ph/9704006}{{\tt arXiv:astro-ph/9704006}}.
%Type = Article
\bibitem[{{Mart{\'\i}n-Pintado} et~al.(2001){Mart{\'\i}n-Pintado}, {Rizzo}, {de
  Vicente}, {Rodr{\'\i}guez-Fern{\'a}ndez} and {Fuente}}]{martin-pintado01}
\bibinfo{author}{{Mart{\'\i}n-Pintado}, J.}, \bibinfo{author}{{Rizzo}, J.R.},
  \bibinfo{author}{{de Vicente}, P.},
  \bibinfo{author}{{Rodr{\'\i}guez-Fern{\'a}ndez}, N.J.},
  \bibinfo{author}{{Fuente}, A.}, \bibinfo{year}{2001}.
\newblock \bibinfo{title}{{Large-Scale Grain Mantle Disruption in the Galactic
  Center}}.
\newblock \bibinfo{journal}{\apjl} \bibinfo{volume}{548},
  \bibinfo{pages}{L65--L68}.
\newblock \DOIprefix\doi{10.1086/318937},
  \href{http://arxiv.org/abs/astro-ph/0011512}{{\tt arXiv:astro-ph/0011512}}.
%Type = Article
\bibitem[{{Massalkhi} et~al.(2023){Massalkhi}, {Jim{\'e}nez-Serra},
  {Mart{\'\i}n-Pintado}, {Rivilla}, {Colzi}, {Zeng}, {Mart{\'\i}n}, {Tercero},
  {de Vicente} and {Requena-Torres}}]{Massalkhi23}
\bibinfo{author}{{Massalkhi}, S.}, \bibinfo{author}{{Jim{\'e}nez-Serra}, I.},
  \bibinfo{author}{{Mart{\'\i}n-Pintado}, J.}, \bibinfo{author}{{Rivilla},
  V.M.}, \bibinfo{author}{{Colzi}, L.}, \bibinfo{author}{{Zeng}, S.},
  \bibinfo{author}{{Mart{\'\i}n}, S.}, \bibinfo{author}{{Tercero}, B.},
  \bibinfo{author}{{de Vicente}, P.}, \bibinfo{author}{{Requena-Torres}, M.A.},
  \bibinfo{year}{2023}.
\newblock \bibinfo{title}{{The first detection of SiC$_{2}$ in the interstellar
  medium}}.
\newblock \bibinfo{journal}{\aap} \bibinfo{volume}{678}, \bibinfo{pages}{A45}.
\newblock \DOIprefix\doi{10.1051/0004-6361/202346822},
  \href{http://arxiv.org/abs/2308.09459}{{\tt arXiv:2308.09459}}.
%Type = Article
\bibitem[{{Massey}(2003)}]{Massey03}
\bibinfo{author}{{Massey}, P.}, \bibinfo{year}{2003}.
\newblock \bibinfo{title}{{MASSIVE STARS IN THE LOCAL GROUP: Implications for
  Stellar Evolution and Star Formation}}.
\newblock \bibinfo{journal}{\araa} \bibinfo{volume}{41},
  \bibinfo{pages}{15--56}.
\newblock \DOIprefix\doi{10.1146/annurev.astro.41.071601.170033}.
%Type = Article
\bibitem[{{Mauersberger} et~al.(1991){Mauersberger}, {Henkel}, {Walmsley},
  {Sage} and {Wiklind}}]{Mauersberger91}
\bibinfo{author}{{Mauersberger}, R.}, \bibinfo{author}{{Henkel}, C.},
  \bibinfo{author}{{Walmsley}, C.M.}, \bibinfo{author}{{Sage}, L.J.},
  \bibinfo{author}{{Wiklind}, T.}, \bibinfo{year}{1991}.
\newblock \bibinfo{title}{{Dense gas in nearby galaxies. V. Multilevel studies
  of CH3CCH and CH3CN.}}
\newblock \bibinfo{journal}{\aap} \bibinfo{volume}{247}, \bibinfo{pages}{307}.
%Type = Article
\bibitem[{{Maureira} et~al.(2022){Maureira}, {Gong}, {Pineda}, {Liu},
  {Silsbee}, {Caselli}, {Zamponi}, {Segura-Cox} and
  {Schmiedeke}}]{Maureira2022}
\bibinfo{author}{{Maureira}, M.J.}, \bibinfo{author}{{Gong}, M.},
  \bibinfo{author}{{Pineda}, J.E.}, \bibinfo{author}{{Liu}, H.B.},
  \bibinfo{author}{{Silsbee}, K.}, \bibinfo{author}{{Caselli}, P.},
  \bibinfo{author}{{Zamponi}, J.}, \bibinfo{author}{{Segura-Cox}, D.M.},
  \bibinfo{author}{{Schmiedeke}, A.}, \bibinfo{year}{2022}.
\newblock \bibinfo{title}{{Dust Hot Spots at 10 au Scales around the Class 0
  Binary IRAS 16293-2422 A: A Departure from the Passive Irradiation Model}}.
\newblock \bibinfo{journal}{\apjl} \bibinfo{volume}{941}, \bibinfo{pages}{L23}.
\newblock \DOIprefix\doi{10.3847/2041-8213/aca53a},
  \href{http://arxiv.org/abs/2212.08436}{{\tt arXiv:2212.08436}}.
%Type = Article
\bibitem[{{McGuire}(2022)}]{McGuire22}
\bibinfo{author}{{McGuire}, B.A.}, \bibinfo{year}{2022}.
\newblock \bibinfo{title}{{2021 Census of Interstellar, Circumstellar,
  Extragalactic, Protoplanetary Disk, and Exoplanetary Molecules}}.
\newblock \bibinfo{journal}{\apjs} \bibinfo{volume}{259}, \bibinfo{pages}{30}.
\newblock \DOIprefix\doi{10.3847/1538-4365/ac2a48},
  \href{http://arxiv.org/abs/2109.13848}{{\tt arXiv:2109.13848}}.
%Type = Article
\bibitem[{{McGuire} et~al.(2021){McGuire}, {Loomis}, {Burkhardt}, {Lee},
  {Shingledecker}, {Charnley}, {Cooke}, {Cordiner}, {Herbst}, {Kalenskii},
  {Siebert}, {Willis}, {Xue}, {Remijan} and {McCarthy}}]{mcguire21}
\bibinfo{author}{{McGuire}, B.A.}, \bibinfo{author}{{Loomis}, R.A.},
  \bibinfo{author}{{Burkhardt}, A.M.}, \bibinfo{author}{{Lee}, K.L.K.},
  \bibinfo{author}{{Shingledecker}, C.N.}, \bibinfo{author}{{Charnley}, S.B.},
  \bibinfo{author}{{Cooke}, I.R.}, \bibinfo{author}{{Cordiner}, M.A.},
  \bibinfo{author}{{Herbst}, E.}, \bibinfo{author}{{Kalenskii}, S.},
  \bibinfo{author}{{Siebert}, M.A.}, \bibinfo{author}{{Willis}, E.R.},
  \bibinfo{author}{{Xue}, C.}, \bibinfo{author}{{Remijan}, A.J.},
  \bibinfo{author}{{McCarthy}, M.C.}, \bibinfo{year}{2021}.
\newblock \bibinfo{title}{{Detection of two interstellar polycyclic aromatic
  hydrocarbons via spectral matched filtering}}.
\newblock \bibinfo{journal}{Science} \bibinfo{volume}{371},
  \bibinfo{pages}{1265--1269}.
\newblock \DOIprefix\doi{10.1126/science.abb7535},
  \href{http://arxiv.org/abs/2103.09984}{{\tt arXiv:2103.09984}}.
%Type = Article
\bibitem[{{Meg{\'\i}as} et~al.(2023){Meg{\'\i}as}, {Jim{\'e}nez-Serra},
  {Mart{\'\i}n-Pintado}, {Vasyunin}, {Spezzano}, {Caselli}, {Cosentino} and
  {Viti}}]{megias23}
\bibinfo{author}{{Meg{\'\i}as}, A.}, \bibinfo{author}{{Jim{\'e}nez-Serra}, I.},
  \bibinfo{author}{{Mart{\'\i}n-Pintado}, J.}, \bibinfo{author}{{Vasyunin},
  A.I.}, \bibinfo{author}{{Spezzano}, S.}, \bibinfo{author}{{Caselli}, P.},
  \bibinfo{author}{{Cosentino}, G.}, \bibinfo{author}{{Viti}, S.},
  \bibinfo{year}{2023}.
\newblock \bibinfo{title}{{The complex organic molecular content in the L1517B
  starless core}}.
\newblock \bibinfo{journal}{\mnras} \bibinfo{volume}{519},
  \bibinfo{pages}{1601--1617}.
\newblock \DOIprefix\doi{10.1093/mnras/stac3449},
  \href{http://arxiv.org/abs/2211.16119}{{\tt arXiv:2211.16119}}.
%Type = Article
\bibitem[{{Meier} and {Turner}(2005)}]{Meier05}
\bibinfo{author}{{Meier}, D.S.}, \bibinfo{author}{{Turner}, J.L.},
  \bibinfo{year}{2005}.
\newblock \bibinfo{title}{{Spatially Resolved Chemistry in Nearby Galaxies. I.
  The Center of IC 342}}.
\newblock \bibinfo{journal}{\apj} \bibinfo{volume}{618},
  \bibinfo{pages}{259--280}.
\newblock \DOIprefix\doi{10.1086/426499},
  \href{http://arxiv.org/abs/astro-ph/0410039}{{\tt arXiv:astro-ph/0410039}}.
%Type = Article
\bibitem[{{Meier} and {Turner}(2012)}]{Meier12}
\bibinfo{author}{{Meier}, D.S.}, \bibinfo{author}{{Turner}, J.L.},
  \bibinfo{year}{2012}.
\newblock \bibinfo{title}{{Spatially Resolved Chemistry in nearby Galaxies. II.
  The Nuclear Bar in Maffei 2}}.
\newblock \bibinfo{journal}{\apj} \bibinfo{volume}{755}, \bibinfo{pages}{104}.
\newblock \DOIprefix\doi{10.1088/0004-637X/755/2/104},
  \href{http://arxiv.org/abs/1206.4098}{{\tt arXiv:1206.4098}}.
%Type = Article
\bibitem[{{Menten}(1991)}]{Menten91}
\bibinfo{author}{{Menten}, K.M.}, \bibinfo{year}{1991}.
\newblock \bibinfo{title}{{The Discovery of a New, Very Strong, and Widespread
  Interstellar Methanol Maser Line}}.
\newblock \bibinfo{journal}{\apjl} \bibinfo{volume}{380}, \bibinfo{pages}{L75}.
\newblock \DOIprefix\doi{10.1086/186177}.
%Type = Inproceedings
\bibitem[{{Menten} et~al.(1999){Menten}, {Carilli} and {Reid}}]{menten99}
\bibinfo{author}{{Menten}, K.M.}, \bibinfo{author}{{Carilli}, C.L.},
  \bibinfo{author}{{Reid}, M.J.}, \bibinfo{year}{1999}.
\newblock \bibinfo{title}{{Interferometric Observations of Redshifted Molecular
  Absorption toward Gravitational Lenses}}, in: \bibinfo{editor}{{Carilli},
  C.L.}, \bibinfo{editor}{{Radford}, S.J.E.}, \bibinfo{editor}{{Menten}, K.M.},
  \bibinfo{editor}{{Langston}, G.I.} (Eds.), \bibinfo{booktitle}{Highly
  Redshifted Radio Lines}, p. \bibinfo{pages}{218}.
\newblock \href{http://arxiv.org/abs/astro-ph/9812178}{{\tt
  arXiv:astro-ph/9812178}}.
%Type = Article
\bibitem[{{Mercimek} et~al.(2022){Mercimek}, {Codella}, {Podio}, {Bianchi},
  {Chahine}, {Bouvier}, {L{\'o}pez-Sepulcre}, {Neri} and
  {Ceccarelli}}]{Mercimek22}
\bibinfo{author}{{Mercimek}, S.}, \bibinfo{author}{{Codella}, C.},
  \bibinfo{author}{{Podio}, L.}, \bibinfo{author}{{Bianchi}, E.},
  \bibinfo{author}{{Chahine}, L.}, \bibinfo{author}{{Bouvier}, M.},
  \bibinfo{author}{{L{\'o}pez-Sepulcre}, A.}, \bibinfo{author}{{Neri}, R.},
  \bibinfo{author}{{Ceccarelli}, C.}, \bibinfo{year}{2022}.
\newblock \bibinfo{title}{{Chemical survey of Class I protostars with the
  IRAM-30 m}}.
\newblock \bibinfo{journal}{\aap} \bibinfo{volume}{659}, \bibinfo{pages}{A67}.
\newblock \DOIprefix\doi{10.1051/0004-6361/202141790},
  \href{http://arxiv.org/abs/2111.07573}{{\tt arXiv:2111.07573}}.
%Type = Article
\bibitem[{{Mills}(2017)}]{Mills17}
\bibinfo{author}{{Mills}, E.A.C.}, \bibinfo{year}{2017}.
\newblock \bibinfo{title}{{The Milky Way's Central Molecular Zone}}.
\newblock \bibinfo{journal}{arXiv e-prints} ,
  \bibinfo{pages}{arXiv:1705.05332}\href{http://arxiv.org/abs/1705.05332}{{\tt
  arXiv:1705.05332}}.
%Type = Article
\bibitem[{{Minier} et~al.(2003){Minier}, {Ellingsen}, {Norris} and
  {Booth}}]{Minier03}
\bibinfo{author}{{Minier}, V.}, \bibinfo{author}{{Ellingsen}, S.P.},
  \bibinfo{author}{{Norris}, R.P.}, \bibinfo{author}{{Booth}, R.S.},
  \bibinfo{year}{2003}.
\newblock \bibinfo{title}{{The protostellar mass limit for 6.7 GHz methanol
  masers. I. A low-mass YSO survey}}.
\newblock \bibinfo{journal}{\aap} \bibinfo{volume}{403},
  \bibinfo{pages}{1095--1100}.
\newblock \DOIprefix\doi{10.1051/0004-6361:20030465}.
%Type = Article
\bibitem[{{Mininni} et~al.(2020){Mininni}, {Beltr{\'a}n}, {Rivilla},
  {S{\'a}nchez-Monge}, {Fontani}, {M{\"o}ller}, {Cesaroni}, {Schilke}, {Viti},
  {Jim{\'e}nez-Serra}, {Colzi}, {Lorenzani} and {Testi}}]{Mininni20}
\bibinfo{author}{{Mininni}, C.}, \bibinfo{author}{{Beltr{\'a}n}, M.T.},
  \bibinfo{author}{{Rivilla}, V.M.}, \bibinfo{author}{{S{\'a}nchez-Monge}, A.},
  \bibinfo{author}{{Fontani}, F.}, \bibinfo{author}{{M{\"o}ller}, T.},
  \bibinfo{author}{{Cesaroni}, R.}, \bibinfo{author}{{Schilke}, P.},
  \bibinfo{author}{{Viti}, S.}, \bibinfo{author}{{Jim{\'e}nez-Serra}, I.},
  \bibinfo{author}{{Colzi}, L.}, \bibinfo{author}{{Lorenzani}, A.},
  \bibinfo{author}{{Testi}, L.}, \bibinfo{year}{2020}.
\newblock \bibinfo{title}{{The GUAPOS project: G31.41+0.31 Unbiased ALMA
  sPectral Observational Survey. I. Isomers of C$_{2}$H$_{4}$O$_{2}$}}.
\newblock \bibinfo{journal}{\aap} \bibinfo{volume}{644}, \bibinfo{pages}{A84}.
\newblock \DOIprefix\doi{10.1051/0004-6361/202038966},
  \href{http://arxiv.org/abs/2009.13297}{{\tt arXiv:2009.13297}}.
%Type = Article
\bibitem[{{Minissale} et~al.(2016){Minissale}, {Moudens}, {Baouche},
  {Chaabouni} and {Dulieu}}]{minissale16}
\bibinfo{author}{{Minissale}, M.}, \bibinfo{author}{{Moudens}, A.},
  \bibinfo{author}{{Baouche}, S.}, \bibinfo{author}{{Chaabouni}, H.},
  \bibinfo{author}{{Dulieu}, F.}, \bibinfo{year}{2016}.
\newblock \bibinfo{title}{{Hydrogenation of CO-bearing species on grains:
  unexpected chemical desorption of CO}}.
\newblock \bibinfo{journal}{\mnras} \bibinfo{volume}{458},
  \bibinfo{pages}{2953--2961}.
\newblock \DOIprefix\doi{10.1093/mnras/stw373}.
%Type = Article
\bibitem[{{Morris} et~al.(1983){Morris}, {Polish}, {Zuckerman} and
  {Kaifu}}]{morris83}
\bibinfo{author}{{Morris}, M.}, \bibinfo{author}{{Polish}, N.},
  \bibinfo{author}{{Zuckerman}, B.}, \bibinfo{author}{{Kaifu}, N.},
  \bibinfo{year}{1983}.
\newblock \bibinfo{title}{{The temperature of molecular gas in the galactic
  center region.}}
\newblock \bibinfo{journal}{\aj} \bibinfo{volume}{88},
  \bibinfo{pages}{1228--1235}.
\newblock \DOIprefix\doi{10.1086/113413}.
%Type = Article
\bibitem[{{Morris} and {Serabyn}(1996)}]{morris96}
\bibinfo{author}{{Morris}, M.}, \bibinfo{author}{{Serabyn}, E.},
  \bibinfo{year}{1996}.
\newblock \bibinfo{title}{{The Galactic Center Environment}}.
\newblock \bibinfo{journal}{\araa} \bibinfo{volume}{34},
  \bibinfo{pages}{645--702}.
\newblock \DOIprefix\doi{10.1146/annurev.astro.34.1.645}.
%Type = Article
\bibitem[{{Motte} et~al.(2018){Motte}, {Bontemps} and {Louvet}}]{Motte18}
\bibinfo{author}{{Motte}, F.}, \bibinfo{author}{{Bontemps}, S.},
  \bibinfo{author}{{Louvet}, F.}, \bibinfo{year}{2018}.
\newblock \bibinfo{title}{{High-Mass Star and Massive Cluster Formation in the
  Milky Way}}.
\newblock \bibinfo{journal}{\araa} \bibinfo{volume}{56},
  \bibinfo{pages}{41--82}.
\newblock \DOIprefix\doi{10.1146/annurev-astro-091916-055235},
  \href{http://arxiv.org/abs/1706.00118}{{\tt arXiv:1706.00118}}.
%Type = Article
\bibitem[{{Mottram} et~al.(2020){Mottram}, {Beuther}, {Ahmadi}, {Klaassen},
  {Beltr{\'a}n}, {Csengeri}, {Feng}, {Gieser}, {Henning}, {Johnston}, {Kuiper},
  {Leurini}, {Linz}, {Longmore}, {Lumsden}, {Maud}, {Moscadelli}, {Palau},
  {Peters}, {Pudritz}, {Ragan}, {S{\'a}nchez-Monge}, {Semenov}, {Urquhart},
  {Winters} and {Zinnecker}}]{Mottram20}
\bibinfo{author}{{Mottram}, J.C.}, \bibinfo{author}{{Beuther}, H.},
  \bibinfo{author}{{Ahmadi}, A.}, \bibinfo{author}{{Klaassen}, P.D.},
  \bibinfo{author}{{Beltr{\'a}n}, M.T.}, \bibinfo{author}{{Csengeri}, T.},
  \bibinfo{author}{{Feng}, S.}, \bibinfo{author}{{Gieser}, C.},
  \bibinfo{author}{{Henning}, T.}, \bibinfo{author}{{Johnston}, K.G.},
  \bibinfo{author}{{Kuiper}, R.}, \bibinfo{author}{{Leurini}, S.},
  \bibinfo{author}{{Linz}, H.}, \bibinfo{author}{{Longmore}, S.N.},
  \bibinfo{author}{{Lumsden}, S.}, \bibinfo{author}{{Maud}, L.T.},
  \bibinfo{author}{{Moscadelli}, L.}, \bibinfo{author}{{Palau}, A.},
  \bibinfo{author}{{Peters}, T.}, \bibinfo{author}{{Pudritz}, R.E.},
  \bibinfo{author}{{Ragan}, S.E.}, \bibinfo{author}{{S{\'a}nchez-Monge},
  {\'A}.}, \bibinfo{author}{{Semenov}, D.}, \bibinfo{author}{{Urquhart}, J.S.},
  \bibinfo{author}{{Winters}, J.M.}, \bibinfo{author}{{Zinnecker}, H.},
  \bibinfo{year}{2020}.
\newblock \bibinfo{title}{{From clump to disc scales in W3 IRS4. A case study
  of the IRAM NOEMA large programme CORE}}.
\newblock \bibinfo{journal}{\aap} \bibinfo{volume}{636}, \bibinfo{pages}{A118}.
\newblock \DOIprefix\doi{10.1051/0004-6361/201834152}.
%Type = Article
\bibitem[{{M{\"u}ller} et~al.(2016){M{\"u}ller}, {Walters}, {Wehres},
  {Belloche}, {Wilkins}, {Liu}, {Vicente}, {Garrod}, {Menten}, {Lewen} and
  {Schlemmer}}]{mueller16}
\bibinfo{author}{{M{\"u}ller}, H.S.P.}, \bibinfo{author}{{Walters}, A.},
  \bibinfo{author}{{Wehres}, N.}, \bibinfo{author}{{Belloche}, A.},
  \bibinfo{author}{{Wilkins}, O.H.}, \bibinfo{author}{{Liu}, D.},
  \bibinfo{author}{{Vicente}, R.}, \bibinfo{author}{{Garrod}, R.T.},
  \bibinfo{author}{{Menten}, K.M.}, \bibinfo{author}{{Lewen}, F.},
  \bibinfo{author}{{Schlemmer}, S.}, \bibinfo{year}{2016}.
\newblock \bibinfo{title}{{Laboratory spectroscopic study and astronomical
  detection of vibrationally excited n-propyl cyanide}}.
\newblock \bibinfo{journal}{\aap} \bibinfo{volume}{595}, \bibinfo{pages}{A87}.
\newblock \DOIprefix\doi{10.1051/0004-6361/201629309},
  \href{http://arxiv.org/abs/1608.08129}{{\tt arXiv:1608.08129}}.
%Type = Article
\bibitem[{{Muller} et~al.(2011){Muller}, {Beelen}, {Gu{\'e}lin}, {Aalto},
  {Black}, {Combes}, {Curran}, {Theule} and {Longmore}}]{muller11}
\bibinfo{author}{{Muller}, S.}, \bibinfo{author}{{Beelen}, A.},
  \bibinfo{author}{{Gu{\'e}lin}, M.}, \bibinfo{author}{{Aalto}, S.},
  \bibinfo{author}{{Black}, J.H.}, \bibinfo{author}{{Combes}, F.},
  \bibinfo{author}{{Curran}, S.J.}, \bibinfo{author}{{Theule}, P.},
  \bibinfo{author}{{Longmore}, S.N.}, \bibinfo{year}{2011}.
\newblock \bibinfo{title}{{Molecules at z = 0.89. A 4-mm-rest-frame
  absorption-line survey toward PKS 1830-211}}.
\newblock \bibinfo{journal}{\aap} \bibinfo{volume}{535}, \bibinfo{pages}{A103}.
\newblock \DOIprefix\doi{10.1051/0004-6361/201117096},
  \href{http://arxiv.org/abs/1104.3361}{{\tt arXiv:1104.3361}}.
%Type = Article
\bibitem[{{Muller} et~al.(2014){Muller}, {Combes}, {Gu{\'e}lin}, {G{\'e}rin},
  {Aalto}, {Beelen}, {Black}, {Curran}, {Darling}, {V-Trung},
  {Garc{\'\i}a-Burillo}, {Henkel}, {Horellou}, {Mart{\'\i}n},
  {Mart{\'\i}-Vidal}, {Menten}, {Murphy}, {Ott}, {Wiklind} and
  {Zwaan}}]{muller14}
\bibinfo{author}{{Muller}, S.}, \bibinfo{author}{{Combes}, F.},
  \bibinfo{author}{{Gu{\'e}lin}, M.}, \bibinfo{author}{{G{\'e}rin}, M.},
  \bibinfo{author}{{Aalto}, S.}, \bibinfo{author}{{Beelen}, A.},
  \bibinfo{author}{{Black}, J.H.}, \bibinfo{author}{{Curran}, S.J.},
  \bibinfo{author}{{Darling}, J.}, \bibinfo{author}{{V-Trung}, D.},
  \bibinfo{author}{{Garc{\'\i}a-Burillo}, S.}, \bibinfo{author}{{Henkel}, C.},
  \bibinfo{author}{{Horellou}, C.}, \bibinfo{author}{{Mart{\'\i}n}, S.},
  \bibinfo{author}{{Mart{\'\i}-Vidal}, I.}, \bibinfo{author}{{Menten}, K.M.},
  \bibinfo{author}{{Murphy}, M.T.}, \bibinfo{author}{{Ott}, J.},
  \bibinfo{author}{{Wiklind}, T.}, \bibinfo{author}{{Zwaan}, M.A.},
  \bibinfo{year}{2014}.
\newblock \bibinfo{title}{{An ALMA Early Science survey of molecular absorption
  lines toward PKS 1830-211. Analysis of the absorption profiles}}.
\newblock \bibinfo{journal}{\aap} \bibinfo{volume}{566}, \bibinfo{pages}{A112}.
\newblock \DOIprefix\doi{10.1051/0004-6361/201423646},
  \href{http://arxiv.org/abs/1404.7667}{{\tt arXiv:1404.7667}}.
%Type = Article
\bibitem[{{Muller} et~al.(2006){Muller}, {Gu{\'e}lin}, {Dumke}, {Lucas} and
  {Combes}}]{muller06}
\bibinfo{author}{{Muller}, S.}, \bibinfo{author}{{Gu{\'e}lin}, M.},
  \bibinfo{author}{{Dumke}, M.}, \bibinfo{author}{{Lucas}, R.},
  \bibinfo{author}{{Combes}, F.}, \bibinfo{year}{2006}.
\newblock \bibinfo{title}{{Probing isotopic ratios at z = 0.89: molecular line
  absorption in front of the quasar PKS 1830-211}}.
\newblock \bibinfo{journal}{\aap} \bibinfo{volume}{458},
  \bibinfo{pages}{417--426}.
\newblock \DOIprefix\doi{10.1051/0004-6361:20065187},
  \href{http://arxiv.org/abs/astro-ph/0608105}{{\tt arXiv:astro-ph/0608105}}.
%Type = Article
\bibitem[{{Muller} et~al.(2024){Muller}, {Le Gal}, {Roueff}, {Black}, {Faure},
  {Gu{\'e}lin}, {Omont}, {G{\'e}rin}, {Combes} and {Aalto}}]{Muller24}
\bibinfo{author}{{Muller}, S.}, \bibinfo{author}{{Le Gal}, R.},
  \bibinfo{author}{{Roueff}, E.}, \bibinfo{author}{{Black}, J.H.},
  \bibinfo{author}{{Faure}, A.}, \bibinfo{author}{{Gu{\'e}lin}, M.},
  \bibinfo{author}{{Omont}, A.}, \bibinfo{author}{{G{\'e}rin}, M.},
  \bibinfo{author}{{Combes}, F.}, \bibinfo{author}{{Aalto}, S.},
  \bibinfo{year}{2024}.
\newblock \bibinfo{title}{{Protonated acetylene in the z = 0.89 molecular
  absorber toward PKS 1830-211}}.
\newblock \bibinfo{journal}{\aap} \bibinfo{volume}{683}, \bibinfo{pages}{A62}.
\newblock \DOIprefix\doi{10.1051/0004-6361/202348994},
  \href{http://arxiv.org/abs/2401.09975}{{\tt arXiv:2401.09975}}.
%Type = Article
\bibitem[{{Nagy} et~al.(2019){Nagy}, {Spezzano}, {Caselli}, {Vasyunin},
  {Tafalla}, {Bizzocchi}, {Prudenzano} and {Redaelli}}]{nagy19}
\bibinfo{author}{{Nagy}, Z.}, \bibinfo{author}{{Spezzano}, S.},
  \bibinfo{author}{{Caselli}, P.}, \bibinfo{author}{{Vasyunin}, A.},
  \bibinfo{author}{{Tafalla}, M.}, \bibinfo{author}{{Bizzocchi}, L.},
  \bibinfo{author}{{Prudenzano}, D.}, \bibinfo{author}{{Redaelli}, E.},
  \bibinfo{year}{2019}.
\newblock \bibinfo{title}{{The chemical structure of the very young starless
  core L1521E}}.
\newblock \bibinfo{journal}{\aap} \bibinfo{volume}{630}, \bibinfo{pages}{A136}.
\newblock \DOIprefix\doi{10.1051/0004-6361/201935568},
  \href{http://arxiv.org/abs/1904.01136}{{\tt arXiv:1904.01136}}.
%Type = Article
\bibitem[{{Nazari} et~al.(2022){Nazari}, {Meijerhof}, {van Gelder}, {Ahmadi},
  {van Dishoeck}, {Tabone}, {Langeroodi}, {Ligterink}, {Jaspers},
  {Beltr{\'a}n}, {Fuller}, {S{\'a}nchez-Monge} and {Schilke}}]{Nazari22}
\bibinfo{author}{{Nazari}, P.}, \bibinfo{author}{{Meijerhof}, J.D.},
  \bibinfo{author}{{van Gelder}, M.L.}, \bibinfo{author}{{Ahmadi}, A.},
  \bibinfo{author}{{van Dishoeck}, E.F.}, \bibinfo{author}{{Tabone}, B.},
  \bibinfo{author}{{Langeroodi}, D.}, \bibinfo{author}{{Ligterink}, N.F.W.},
  \bibinfo{author}{{Jaspers}, J.}, \bibinfo{author}{{Beltr{\'a}n}, M.T.},
  \bibinfo{author}{{Fuller}, G.A.}, \bibinfo{author}{{S{\'a}nchez-Monge},
  {\'A}.}, \bibinfo{author}{{Schilke}, P.}, \bibinfo{year}{2022}.
\newblock \bibinfo{title}{{N-bearing complex organics toward high-mass
  protostars. Constant ratios pointing to formation in similar pre-stellar
  conditions across a large mass range}}.
\newblock \bibinfo{journal}{\aap} \bibinfo{volume}{668}, \bibinfo{pages}{A109}.
\newblock \DOIprefix\doi{10.1051/0004-6361/202243788},
  \href{http://arxiv.org/abs/2208.11128}{{\tt arXiv:2208.11128}}.
%Type = Article
\bibitem[{{Nazari} et~al.(2021){Nazari}, {van Gelder}, {van Dishoeck},
  {Tabone}, {van't Hoff}, {Ligterink}, {Beuther}, {Boogert}, {Caratti o
  Garatti}, {Klaassen}, {Linnartz}, {Taquet} and {Tychoniec}}]{Nazari21}
\bibinfo{author}{{Nazari}, P.}, \bibinfo{author}{{van Gelder}, M.L.},
  \bibinfo{author}{{van Dishoeck}, E.F.}, \bibinfo{author}{{Tabone}, B.},
  \bibinfo{author}{{van't Hoff}, M.L.R.}, \bibinfo{author}{{Ligterink},
  N.F.W.}, \bibinfo{author}{{Beuther}, H.}, \bibinfo{author}{{Boogert},
  A.C.A.}, \bibinfo{author}{{Caratti o Garatti}, A.},
  \bibinfo{author}{{Klaassen}, P.D.}, \bibinfo{author}{{Linnartz}, H.},
  \bibinfo{author}{{Taquet}, V.}, \bibinfo{author}{{Tychoniec}, {\L}.},
  \bibinfo{year}{2021}.
\newblock \bibinfo{title}{{Complex organic molecules in low-mass protostars on
  Solar System scales. II. Nitrogen-bearing species}}.
\newblock \bibinfo{journal}{\aap} \bibinfo{volume}{650}, \bibinfo{pages}{A150}.
\newblock \DOIprefix\doi{10.1051/0004-6361/202039996},
  \href{http://arxiv.org/abs/2104.03326}{{\tt arXiv:2104.03326}}.
%Type = Article
\bibitem[{{Neill} et~al.(2012){Neill}, {Muckle}, {Zaleski}, {Steber}, {Pate},
  {Lattanzi}, {Spezzano}, {McCarthy} and {Remijan}}]{neill12}
\bibinfo{author}{{Neill}, J.L.}, \bibinfo{author}{{Muckle}, M.T.},
  \bibinfo{author}{{Zaleski}, D.P.}, \bibinfo{author}{{Steber}, A.L.},
  \bibinfo{author}{{Pate}, B.H.}, \bibinfo{author}{{Lattanzi}, V.},
  \bibinfo{author}{{Spezzano}, S.}, \bibinfo{author}{{McCarthy}, M.C.},
  \bibinfo{author}{{Remijan}, A.J.}, \bibinfo{year}{2012}.
\newblock \bibinfo{title}{{Laboratory and Tentative Interstellar Detection of
  Trans-Methyl Formate Using the Publicly Available Green Bank Telescope Primos
  Survey}}.
\newblock \bibinfo{journal}{\apj} \bibinfo{volume}{755}, \bibinfo{pages}{153}.
\newblock \DOIprefix\doi{10.1088/0004-637X/755/2/153},
  \href{http://arxiv.org/abs/1206.6021}{{\tt arXiv:1206.6021}}.
%Type = Article
\bibitem[{{Oba} et~al.(2023){Oba}, {Koga}, {Takano}, {Ogawa}, {Ohkouchi},
  {Sasaki}, {Sato}, {Glavin}, {Dworkin}, {Naraoka}, {Tachibana}, {Yurimoto},
  {Nakamura}, {Noguchi}, {Okazaki}, {Yabuta}, {Sakamoto}, {Yada}, {Nishimura},
  {Nakato}, {Miyazaki}, {Yogata}, {Abe}, {Okada}, {Usui}, {Yoshikawa}, {Saiki},
  {Tanaka}, {Terui}, {Nakazawa}, {Watanabe}, {Tsuda} and
  {Hayabusa2-initial-analysis SOM team}}]{oba2023}
\bibinfo{author}{{Oba}, Y.}, \bibinfo{author}{{Koga}, T.},
  \bibinfo{author}{{Takano}, Y.}, \bibinfo{author}{{Ogawa}, N.O.},
  \bibinfo{author}{{Ohkouchi}, N.}, \bibinfo{author}{{Sasaki}, K.},
  \bibinfo{author}{{Sato}, H.}, \bibinfo{author}{{Glavin}, D.P.},
  \bibinfo{author}{{Dworkin}, J.P.}, \bibinfo{author}{{Naraoka}, H.},
  \bibinfo{author}{{Tachibana}, S.}, \bibinfo{author}{{Yurimoto}, H.},
  \bibinfo{author}{{Nakamura}, T.}, \bibinfo{author}{{Noguchi}, T.},
  \bibinfo{author}{{Okazaki}, R.}, \bibinfo{author}{{Yabuta}, H.},
  \bibinfo{author}{{Sakamoto}, K.}, \bibinfo{author}{{Yada}, T.},
  \bibinfo{author}{{Nishimura}, M.}, \bibinfo{author}{{Nakato}, A.},
  \bibinfo{author}{{Miyazaki}, A.}, \bibinfo{author}{{Yogata}, K.},
  \bibinfo{author}{{Abe}, M.}, \bibinfo{author}{{Okada}, T.},
  \bibinfo{author}{{Usui}, T.}, \bibinfo{author}{{Yoshikawa}, M.},
  \bibinfo{author}{{Saiki}, T.}, \bibinfo{author}{{Tanaka}, S.},
  \bibinfo{author}{{Terui}, F.}, \bibinfo{author}{{Nakazawa}, S.},
  \bibinfo{author}{{Watanabe}, S.i.}, \bibinfo{author}{{Tsuda}, Y.},
  \bibinfo{author}{{Hayabusa2-initial-analysis SOM team}},
  \bibinfo{year}{2023}.
\newblock \bibinfo{title}{{Uracil in the carbonaceous asteroid (162173)
  Ryugu}}.
\newblock \bibinfo{journal}{Nature Communications} \bibinfo{volume}{14},
  \bibinfo{pages}{1292}.
\newblock \DOIprefix\doi{10.1038/s41467-023-36904-3}.
%Type = Article
\bibitem[{{{\"O}berg} et~al.(2010){{\"O}berg}, {Bottinelli}, {J{\o}rgensen} and
  {van Dishoeck}}]{oberg10}
\bibinfo{author}{{{\"O}berg}, K.I.}, \bibinfo{author}{{Bottinelli}, S.},
  \bibinfo{author}{{J{\o}rgensen}, J.K.}, \bibinfo{author}{{van Dishoeck},
  E.F.}, \bibinfo{year}{2010}.
\newblock \bibinfo{title}{{A Cold Complex Chemistry Toward the Low-mass
  Protostar B1-b: Evidence for Complex Molecule Production in Ices}}.
\newblock \bibinfo{journal}{\apj} \bibinfo{volume}{716},
  \bibinfo{pages}{825--834}.
\newblock \DOIprefix\doi{10.1088/0004-637X/716/1/825},
  \href{http://arxiv.org/abs/1005.0637}{{\tt arXiv:1005.0637}}.
%Type = Article
\bibitem[{{{\"O}berg} et~al.(2015){{\"O}berg}, {Guzm{\'a}n}, {Furuya}, {Qi},
  {Aikawa}, {Andrews}, {Loomis} and {Wilner}}]{Oberg15a}
\bibinfo{author}{{{\"O}berg}, K.I.}, \bibinfo{author}{{Guzm{\'a}n}, V.V.},
  \bibinfo{author}{{Furuya}, K.}, \bibinfo{author}{{Qi}, C.},
  \bibinfo{author}{{Aikawa}, Y.}, \bibinfo{author}{{Andrews}, S.M.},
  \bibinfo{author}{{Loomis}, R.}, \bibinfo{author}{{Wilner}, D.J.},
  \bibinfo{year}{2015}.
\newblock \bibinfo{title}{{The comet-like composition of a protoplanetary disk
  as revealed by complex cyanides}}.
\newblock \bibinfo{journal}{\nat} \bibinfo{volume}{520},
  \bibinfo{pages}{198--201}.
\newblock \DOIprefix\doi{10.1038/nature14276},
  \href{http://arxiv.org/abs/1505.06347}{{\tt arXiv:1505.06347}}.
%Type = Article
\bibitem[{{{\"O}berg} et~al.(2021){{\"O}berg}, {Guzm{\'a}n}, {Walsh}, {Aikawa},
  {Bergin}, {Law}, {Loomis}, {Alarc{\'o}n}, {Andrews}, {Bae}, {Bergner},
  {Boehler}, {Booth}, {Bosman}, {Calahan}, {Cataldi}, {Cleeves}, {Czekala},
  {Furuya}, {Huang}, {Ilee}, {Kurtovic}, {Le Gal}, {Liu}, {Long}, {M{\'e}nard},
  {Nomura}, {P{\'e}rez}, {Qi}, {Schwarz}, {Sierra}, {Teague}, {Tsukagoshi},
  {Yamato}, {van't Hoff}, {Waggoner}, {Wilner} and {Zhang}}]{Oberg21}
\bibinfo{author}{{{\"O}berg}, K.I.}, \bibinfo{author}{{Guzm{\'a}n}, V.V.},
  \bibinfo{author}{{Walsh}, C.}, \bibinfo{author}{{Aikawa}, Y.},
  \bibinfo{author}{{Bergin}, E.A.}, \bibinfo{author}{{Law}, C.J.},
  \bibinfo{author}{{Loomis}, R.A.}, \bibinfo{author}{{Alarc{\'o}n}, F.},
  \bibinfo{author}{{Andrews}, S.M.}, \bibinfo{author}{{Bae}, J.},
  \bibinfo{author}{{Bergner}, J.B.}, \bibinfo{author}{{Boehler}, Y.},
  \bibinfo{author}{{Booth}, A.S.}, \bibinfo{author}{{Bosman}, A.D.},
  \bibinfo{author}{{Calahan}, J.K.}, \bibinfo{author}{{Cataldi}, G.},
  \bibinfo{author}{{Cleeves}, L.I.}, \bibinfo{author}{{Czekala}, I.},
  \bibinfo{author}{{Furuya}, K.}, \bibinfo{author}{{Huang}, J.},
  \bibinfo{author}{{Ilee}, J.D.}, \bibinfo{author}{{Kurtovic}, N.T.},
  \bibinfo{author}{{Le Gal}, R.}, \bibinfo{author}{{Liu}, Y.},
  \bibinfo{author}{{Long}, F.}, \bibinfo{author}{{M{\'e}nard}, F.},
  \bibinfo{author}{{Nomura}, H.}, \bibinfo{author}{{P{\'e}rez}, L.M.},
  \bibinfo{author}{{Qi}, C.}, \bibinfo{author}{{Schwarz}, K.R.},
  \bibinfo{author}{{Sierra}, A.}, \bibinfo{author}{{Teague}, R.},
  \bibinfo{author}{{Tsukagoshi}, T.}, \bibinfo{author}{{Yamato}, Y.},
  \bibinfo{author}{{van't Hoff}, M.L.R.}, \bibinfo{author}{{Waggoner}, A.R.},
  \bibinfo{author}{{Wilner}, D.J.}, \bibinfo{author}{{Zhang}, K.},
  \bibinfo{year}{2021}.
\newblock \bibinfo{title}{{Molecules with ALMA at Planet-forming Scales (MAPS).
  I. Program Overview and Highlights}}.
\newblock \bibinfo{journal}{\apjs} \bibinfo{volume}{257}, \bibinfo{pages}{1}.
\newblock \DOIprefix\doi{10.3847/1538-4365/ac1432},
  \href{http://arxiv.org/abs/2109.06268}{{\tt arXiv:2109.06268}}.
%Type = Inproceedings
\bibitem[{{Ohishi} et~al.(1992){Ohishi}, {Irvine} and {Kaifu}}]{Ohishi92}
\bibinfo{author}{{Ohishi}, M.}, \bibinfo{author}{{Irvine}, W.M.},
  \bibinfo{author}{{Kaifu}, N.}, \bibinfo{year}{1992}.
\newblock \bibinfo{title}{{Molecular Abundance Variations among and Within
  Cold, Dark Molecular Clouds(rp)}}, in: \bibinfo{editor}{{Singh}, P.D.} (Ed.),
  \bibinfo{booktitle}{Astrochemistry of Cosmic Phenomena}, p.
  \bibinfo{pages}{171}.
%Type = Article
\bibitem[{{Osorio} et~al.(1999){Osorio}, {Lizano} and {D'Alessio}}]{Osorio99}
\bibinfo{author}{{Osorio}, M.}, \bibinfo{author}{{Lizano}, S.},
  \bibinfo{author}{{D'Alessio}, P.}, \bibinfo{year}{1999}.
\newblock \bibinfo{title}{{Hot Molecular Cores and the Formation of Massive
  Stars}}.
\newblock \bibinfo{journal}{\apj} \bibinfo{volume}{525},
  \bibinfo{pages}{808--820}.
\newblock \DOIprefix\doi{10.1086/307929}.
%Type = Article
\bibitem[{{Oya} et~al.(2016){Oya}, {Sakai}, {L{\'o}pez-Sepulcre}, {Watanabe},
  {Ceccarelli}, {Lefloch}, {Favre} and {Yamamoto}}]{Oya16}
\bibinfo{author}{{Oya}, Y.}, \bibinfo{author}{{Sakai}, N.},
  \bibinfo{author}{{L{\'o}pez-Sepulcre}, A.}, \bibinfo{author}{{Watanabe}, Y.},
  \bibinfo{author}{{Ceccarelli}, C.}, \bibinfo{author}{{Lefloch}, B.},
  \bibinfo{author}{{Favre}, C.}, \bibinfo{author}{{Yamamoto}, S.},
  \bibinfo{year}{2016}.
\newblock \bibinfo{title}{{Infalling-Rotating Motion and Associated Chemical
  Change in the Envelope of IRAS 16293-2422 Source A Studied with ALMA}}.
\newblock \bibinfo{journal}{\apj} \bibinfo{volume}{824}, \bibinfo{pages}{88}.
\newblock \DOIprefix\doi{10.3847/0004-637X/824/2/88},
  \href{http://arxiv.org/abs/1605.00340}{{\tt arXiv:1605.00340}}.
%Type = Article
\bibitem[{{Oya} et~al.(2017){Oya}, {Sakai}, {Watanabe}, {Higuchi}, {Hirota},
  {L{\'o}pez-Sepulcre}, {Sakai}, {Aikawa}, {Ceccarelli}, {Lefloch}, {Caux},
  {Vastel}, {Kahane} and {Yamamoto}}]{Oya17}
\bibinfo{author}{{Oya}, Y.}, \bibinfo{author}{{Sakai}, N.},
  \bibinfo{author}{{Watanabe}, Y.}, \bibinfo{author}{{Higuchi}, A.E.},
  \bibinfo{author}{{Hirota}, T.}, \bibinfo{author}{{L{\'o}pez-Sepulcre}, A.},
  \bibinfo{author}{{Sakai}, T.}, \bibinfo{author}{{Aikawa}, Y.},
  \bibinfo{author}{{Ceccarelli}, C.}, \bibinfo{author}{{Lefloch}, B.},
  \bibinfo{author}{{Caux}, E.}, \bibinfo{author}{{Vastel}, C.},
  \bibinfo{author}{{Kahane}, C.}, \bibinfo{author}{{Yamamoto}, S.},
  \bibinfo{year}{2017}.
\newblock \bibinfo{title}{{L483: Warm Carbon-chain Chemistry Source Harboring
  Hot Corino Activity}}.
\newblock \bibinfo{journal}{\apj} \bibinfo{volume}{837}, \bibinfo{pages}{174}.
\newblock \DOIprefix\doi{10.3847/1538-4357/aa6300},
  \href{http://arxiv.org/abs/1703.03653}{{\tt arXiv:1703.03653}}.
%Type = Article
\bibitem[{{Padovani} et~al.(2009){Padovani}, {Galli} and
  {Glassgold}}]{padovani09}
\bibinfo{author}{{Padovani}, M.}, \bibinfo{author}{{Galli}, D.},
  \bibinfo{author}{{Glassgold}, A.E.}, \bibinfo{year}{2009}.
\newblock \bibinfo{title}{{Cosmic-ray ionization of molecular clouds}}.
\newblock \bibinfo{journal}{\aap} \bibinfo{volume}{501},
  \bibinfo{pages}{619--631}.
\newblock \DOIprefix\doi{10.1051/0004-6361/200911794},
  \href{http://arxiv.org/abs/0904.4149}{{\tt arXiv:0904.4149}}.
%Type = Article
\bibitem[{{Palau} et~al.(2017){Palau}, {Walsh}, {S{\'a}nchez-Monge}, {Girart},
  {Cesaroni}, {Jim{\'e}nez-Serra}, {Fuente}, {Zapata} and {Neri}}]{palau17}
\bibinfo{author}{{Palau}, A.}, \bibinfo{author}{{Walsh}, C.},
  \bibinfo{author}{{S{\'a}nchez-Monge}, {\'A}.}, \bibinfo{author}{{Girart},
  J.M.}, \bibinfo{author}{{Cesaroni}, R.},
  \bibinfo{author}{{Jim{\'e}nez-Serra}, I.}, \bibinfo{author}{{Fuente}, A.},
  \bibinfo{author}{{Zapata}, L.A.}, \bibinfo{author}{{Neri}, R.},
  \bibinfo{year}{2017}.
\newblock \bibinfo{title}{{Complex organic molecules tracing shocks along the
  outflow cavity in the high-mass protostar IRAS 20126+4104}}.
\newblock \bibinfo{journal}{\mnras} \bibinfo{volume}{467},
  \bibinfo{pages}{2723--2752}.
\newblock \DOIprefix\doi{10.1093/mnras/stx004},
  \href{http://arxiv.org/abs/1701.04802}{{\tt arXiv:1701.04802}}.
%Type = Article
\bibitem[{{Patel} et~al.(2005){Patel}, {Curiel}, {Sridharan}, {Zhang},
  {Hunter}, {Ho}, {Torrelles}, {Moran}, {G{\'o}mez} and {Anglada}}]{patel05}
\bibinfo{author}{{Patel}, N.A.}, \bibinfo{author}{{Curiel}, S.},
  \bibinfo{author}{{Sridharan}, T.K.}, \bibinfo{author}{{Zhang}, Q.},
  \bibinfo{author}{{Hunter}, T.R.}, \bibinfo{author}{{Ho}, P.T.P.},
  \bibinfo{author}{{Torrelles}, J.M.}, \bibinfo{author}{{Moran}, J.M.},
  \bibinfo{author}{{G{\'o}mez}, J.F.}, \bibinfo{author}{{Anglada}, G.},
  \bibinfo{year}{2005}.
\newblock \bibinfo{title}{{A disk of dust and molecular gas around a high-mass
  protostar}}.
\newblock \bibinfo{journal}{\nat} \bibinfo{volume}{437},
  \bibinfo{pages}{109--111}.
\newblock \DOIprefix\doi{10.1038/nature04011},
  \href{http://arxiv.org/abs/astro-ph/0509637}{{\tt arXiv:astro-ph/0509637}}.
%Type = Article
\bibitem[{{Pineda} et~al.(2020){Pineda}, {Segura-Cox}, {Caselli}, {Cunningham},
  {Zhao}, {Schmiedeke}, {Maureira} and {Neri}}]{Pineda20}
\bibinfo{author}{{Pineda}, J.E.}, \bibinfo{author}{{Segura-Cox}, D.},
  \bibinfo{author}{{Caselli}, P.}, \bibinfo{author}{{Cunningham}, N.},
  \bibinfo{author}{{Zhao}, B.}, \bibinfo{author}{{Schmiedeke}, A.},
  \bibinfo{author}{{Maureira}, M.J.}, \bibinfo{author}{{Neri}, R.},
  \bibinfo{year}{2020}.
\newblock \bibinfo{title}{{A protostellar system fed by a streamer of 10,500 au
  length}}.
\newblock \bibinfo{journal}{Nature Astronomy} \bibinfo{volume}{4},
  \bibinfo{pages}{1158--1163}.
\newblock \DOIprefix\doi{10.1038/s41550-020-1150-z},
  \href{http://arxiv.org/abs/2007.13430}{{\tt arXiv:2007.13430}}.
%Type = Article
\bibitem[{{Podio} et~al.(2016){Podio}, {Codella}, {Gueth}, {Cabrit}, {Maury},
  {Tabone}, {Lef{\`e}vre}, {Anderl}, {Andr{\'e}}, {Belloche}, {Bontemps},
  {Hennebelle}, {Lefloch}, {Maret} and {Testi}}]{Podio16}
\bibinfo{author}{{Podio}, L.}, \bibinfo{author}{{Codella}, C.},
  \bibinfo{author}{{Gueth}, F.}, \bibinfo{author}{{Cabrit}, S.},
  \bibinfo{author}{{Maury}, A.}, \bibinfo{author}{{Tabone}, B.},
  \bibinfo{author}{{Lef{\`e}vre}, C.}, \bibinfo{author}{{Anderl}, S.},
  \bibinfo{author}{{Andr{\'e}}, P.}, \bibinfo{author}{{Belloche}, A.},
  \bibinfo{author}{{Bontemps}, S.}, \bibinfo{author}{{Hennebelle}, P.},
  \bibinfo{author}{{Lefloch}, B.}, \bibinfo{author}{{Maret}, S.},
  \bibinfo{author}{{Testi}, L.}, \bibinfo{year}{2016}.
\newblock \bibinfo{title}{{First image of the L1157 molecular jet by the
  CALYPSO IRAM-PdBI survey}}.
\newblock \bibinfo{journal}{\aap} \bibinfo{volume}{593}, \bibinfo{pages}{L4}.
\newblock \DOIprefix\doi{10.1051/0004-6361/201628876},
  \href{http://arxiv.org/abs/1608.05026}{{\tt arXiv:1608.05026}}.
%Type = Article
\bibitem[{{Podio} et~al.(2020){Podio}, {Garufi}, {Codella}, {Fedele},
  {Bianchi}, {Bacciotti}, {Ceccarelli}, {Favre}, {Mercimek}, {Rygl} and
  {Testi}}]{Podio20}
\bibinfo{author}{{Podio}, L.}, \bibinfo{author}{{Garufi}, A.},
  \bibinfo{author}{{Codella}, C.}, \bibinfo{author}{{Fedele}, D.},
  \bibinfo{author}{{Bianchi}, E.}, \bibinfo{author}{{Bacciotti}, F.},
  \bibinfo{author}{{Ceccarelli}, C.}, \bibinfo{author}{{Favre}, C.},
  \bibinfo{author}{{Mercimek}, S.}, \bibinfo{author}{{Rygl}, K.},
  \bibinfo{author}{{Testi}, L.}, \bibinfo{year}{2020}.
\newblock \bibinfo{title}{{ALMA chemical survey of disk-outflow sources in
  Taurus (ALMA-DOT). II. Vertical stratification of CO, CS, CN, H$_{2}$CO, and
  CH$_{3}$OH in a Class I disk}}.
\newblock \bibinfo{journal}{\aap} \bibinfo{volume}{642}, \bibinfo{pages}{L7}.
\newblock \DOIprefix\doi{10.1051/0004-6361/202038952},
  \href{http://arxiv.org/abs/2008.12648}{{\tt arXiv:2008.12648}}.
%Type = Article
\bibitem[{{Punanova} et~al.(2018){Punanova}, {Caselli}, {Feng},
  {Chac{\'o}n-Tanarro}, {Ceccarelli}, {Neri}, {Fontani}, {Jim{\'e}nez-Serra},
  {Vastel}, {Bizzocchi}, {Pon}, {Vasyunin}, {Spezzano}, {Hily-Blant}, {Testi},
  {Viti}, {Yamamoto}, {Alves}, {Bachiller}, {Balucani}, {Bianchi},
  {Bottinelli}, {Caux}, {Choudhury}, {Codella}, {Dulieu}, {Favre}, {Holdship},
  {Jaber Al-Edhari}, {Kahane}, {Laas}, {LeFloch}, {L{\'o}pez-Sepulcre},
  {Ospina-Zamudio}, {Oya}, {Pineda}, {Podio}, {Quenard}, {Rimola}, {Sakai},
  {Sims}, {Taquet}, {Theul{\'e}} and {Ugliengo}}]{Punanova18}
\bibinfo{author}{{Punanova}, A.}, \bibinfo{author}{{Caselli}, P.},
  \bibinfo{author}{{Feng}, S.}, \bibinfo{author}{{Chac{\'o}n-Tanarro}, A.},
  \bibinfo{author}{{Ceccarelli}, C.}, \bibinfo{author}{{Neri}, R.},
  \bibinfo{author}{{Fontani}, F.}, \bibinfo{author}{{Jim{\'e}nez-Serra}, I.},
  \bibinfo{author}{{Vastel}, C.}, \bibinfo{author}{{Bizzocchi}, L.},
  \bibinfo{author}{{Pon}, A.}, \bibinfo{author}{{Vasyunin}, A.I.},
  \bibinfo{author}{{Spezzano}, S.}, \bibinfo{author}{{Hily-Blant}, P.},
  \bibinfo{author}{{Testi}, L.}, \bibinfo{author}{{Viti}, S.},
  \bibinfo{author}{{Yamamoto}, S.}, \bibinfo{author}{{Alves}, F.},
  \bibinfo{author}{{Bachiller}, R.}, \bibinfo{author}{{Balucani}, N.},
  \bibinfo{author}{{Bianchi}, E.}, \bibinfo{author}{{Bottinelli}, S.},
  \bibinfo{author}{{Caux}, E.}, \bibinfo{author}{{Choudhury}, R.},
  \bibinfo{author}{{Codella}, C.}, \bibinfo{author}{{Dulieu}, F.},
  \bibinfo{author}{{Favre}, C.}, \bibinfo{author}{{Holdship}, J.},
  \bibinfo{author}{{Jaber Al-Edhari}, A.}, \bibinfo{author}{{Kahane}, C.},
  \bibinfo{author}{{Laas}, J.}, \bibinfo{author}{{LeFloch}, B.},
  \bibinfo{author}{{L{\'o}pez-Sepulcre}, A.},
  \bibinfo{author}{{Ospina-Zamudio}, J.}, \bibinfo{author}{{Oya}, Y.},
  \bibinfo{author}{{Pineda}, J.E.}, \bibinfo{author}{{Podio}, L.},
  \bibinfo{author}{{Quenard}, D.}, \bibinfo{author}{{Rimola}, A.},
  \bibinfo{author}{{Sakai}, N.}, \bibinfo{author}{{Sims}, I.R.},
  \bibinfo{author}{{Taquet}, V.}, \bibinfo{author}{{Theul{\'e}}, P.},
  \bibinfo{author}{{Ugliengo}, P.}, \bibinfo{year}{2018}.
\newblock \bibinfo{title}{{Seeds of Life in Space (SOLIS). III. Zooming Into
  the Methanol Peak of the Prestellar Core L1544}}.
\newblock \bibinfo{journal}{\apj} \bibinfo{volume}{855}, \bibinfo{pages}{112}.
\newblock \DOIprefix\doi{10.3847/1538-4357/aaad09},
  \href{http://arxiv.org/abs/1802.00859}{{\tt arXiv:1802.00859}}.
%Type = Article
\bibitem[{{Punanova} et~al.(2022){Punanova}, {Vasyunin}, {Caselli}, {Howard},
  {Spezzano}, {Shirley}, {Scibelli} and {Harju}}]{punanova22}
\bibinfo{author}{{Punanova}, A.}, \bibinfo{author}{{Vasyunin}, A.},
  \bibinfo{author}{{Caselli}, P.}, \bibinfo{author}{{Howard}, A.},
  \bibinfo{author}{{Spezzano}, S.}, \bibinfo{author}{{Shirley}, Y.},
  \bibinfo{author}{{Scibelli}, S.}, \bibinfo{author}{{Harju}, J.},
  \bibinfo{year}{2022}.
\newblock \bibinfo{title}{{Methanol Mapping in Cold Cores: Testing Model
  Predictions}}.
\newblock \bibinfo{journal}{\apj} \bibinfo{volume}{927}, \bibinfo{pages}{213}.
\newblock \DOIprefix\doi{10.3847/1538-4357/ac4e7d},
  \href{http://arxiv.org/abs/2112.04538}{{\tt arXiv:2112.04538}}.
%Type = Article
\bibitem[{{Qin} et~al.(2022){Qin}, {Liu}, {Liu}, {Goldsmith}, {Li}, {Zhang},
  {Liu}, {Wu}, {Bronfman}, {Juvela}, {Lee}, {Garay}, {Zhang}, {He}, {Hsu},
  {Shen}, {Lee}, {Wang}, {Tang}, {Tang}, {Zhang}, {Yue}, {Xue}, {Li}, {Peng},
  {Dutta}, {Ge}, {Xu}, {Chen}, {Baug}, {Dewangan} and {Tej}}]{Qin22}
\bibinfo{author}{{Qin}, S.L.}, \bibinfo{author}{{Liu}, T.},
  \bibinfo{author}{{Liu}, X.}, \bibinfo{author}{{Goldsmith}, P.F.},
  \bibinfo{author}{{Li}, D.}, \bibinfo{author}{{Zhang}, Q.},
  \bibinfo{author}{{Liu}, H.L.}, \bibinfo{author}{{Wu}, Y.},
  \bibinfo{author}{{Bronfman}, L.}, \bibinfo{author}{{Juvela}, M.},
  \bibinfo{author}{{Lee}, C.W.}, \bibinfo{author}{{Garay}, G.},
  \bibinfo{author}{{Zhang}, Y.}, \bibinfo{author}{{He}, J.},
  \bibinfo{author}{{Hsu}, S.Y.}, \bibinfo{author}{{Shen}, Z.Q.},
  \bibinfo{author}{{Lee}, J.E.}, \bibinfo{author}{{Wang}, K.},
  \bibinfo{author}{{Tang}, N.}, \bibinfo{author}{{Tang}, M.},
  \bibinfo{author}{{Zhang}, C.}, \bibinfo{author}{{Yue}, Y.},
  \bibinfo{author}{{Xue}, Q.}, \bibinfo{author}{{Li}, S.},
  \bibinfo{author}{{Peng}, Y.}, \bibinfo{author}{{Dutta}, S.},
  \bibinfo{author}{{Ge}, J.}, \bibinfo{author}{{Xu}, F.},
  \bibinfo{author}{{Chen}, L.F.}, \bibinfo{author}{{Baug}, T.},
  \bibinfo{author}{{Dewangan}, L.}, \bibinfo{author}{{Tej}, A.},
  \bibinfo{year}{2022}.
\newblock \bibinfo{title}{{ATOMS: ALMA Three-millimeter Observations of Massive
  Star-forming regions - VIII. A search for hot cores by using
  C$_{2}$H$_{5}$CN, CH$_{3}$OCHO, and CH$_{3}$OH lines}}.
\newblock \bibinfo{journal}{\mnras} \bibinfo{volume}{511},
  \bibinfo{pages}{3463--3476}.
\newblock \DOIprefix\doi{10.1093/mnras/stac219},
  \href{http://arxiv.org/abs/2201.10044}{{\tt arXiv:2201.10044}}.
%Type = Article
\bibitem[{{Rawlings} et~al.(2013){Rawlings}, {Williams}, {Viti} and
  {Cecchi-Pestellini}}]{rawlings13}
\bibinfo{author}{{Rawlings}, J.M.C.}, \bibinfo{author}{{Williams}, D.A.},
  \bibinfo{author}{{Viti}, S.}, \bibinfo{author}{{Cecchi-Pestellini}, C.},
  \bibinfo{year}{2013}.
\newblock \bibinfo{title}{{A radical route to interstellar propylene
  formation.}}
\newblock \bibinfo{journal}{\mnras} \bibinfo{volume}{436},
  \bibinfo{pages}{L59--L63}.
\newblock \DOIprefix\doi{10.1093/mnrasl/slt113},
  \href{http://arxiv.org/abs/1308.2559}{{\tt arXiv:1308.2559}}.
%Type = Article
\bibitem[{{Redaelli} et~al.(2019){Redaelli}, {Bizzocchi}, {Caselli},
  {Sipil{\"a}}, {Lattanzi}, {Giuliano} and {Spezzano}}]{redaelli19}
\bibinfo{author}{{Redaelli}, E.}, \bibinfo{author}{{Bizzocchi}, L.},
  \bibinfo{author}{{Caselli}, P.}, \bibinfo{author}{{Sipil{\"a}}, O.},
  \bibinfo{author}{{Lattanzi}, V.}, \bibinfo{author}{{Giuliano}, B.M.},
  \bibinfo{author}{{Spezzano}, S.}, \bibinfo{year}{2019}.
\newblock \bibinfo{title}{{High-sensitivity maps of molecular ions in L1544. I.
  Deuteration of N$_{2}$H$^{+}$ and HCO$^{+}$ and primary evidence of
  N$_{2}$D$^{+}$ depletion}}.
\newblock \bibinfo{journal}{\aap} \bibinfo{volume}{629}, \bibinfo{pages}{A15}.
\newblock \DOIprefix\doi{10.1051/0004-6361/201935314},
  \href{http://arxiv.org/abs/1907.08217}{{\tt arXiv:1907.08217}}.
%Type = Article
\bibitem[{{Requena-Torres} et~al.(2008){Requena-Torres}, {Mart{\'\i}n-Pintado},
  {Mart{\'\i}n} and {Morris}}]{requena08}
\bibinfo{author}{{Requena-Torres}, M.A.},
  \bibinfo{author}{{Mart{\'\i}n-Pintado}, J.}, \bibinfo{author}{{Mart{\'\i}n},
  S.}, \bibinfo{author}{{Morris}, M.R.}, \bibinfo{year}{2008}.
\newblock \bibinfo{title}{{The Galactic Center: The Largest Oxygen-bearing
  Organic Molecule Repository}}.
\newblock \bibinfo{journal}{\apj} \bibinfo{volume}{672},
  \bibinfo{pages}{352--360}.
\newblock \DOIprefix\doi{10.1086/523627},
  \href{http://arxiv.org/abs/0709.0542}{{\tt arXiv:0709.0542}}.
%Type = Article
\bibitem[{{Requena-Torres} et~al.(2006){Requena-Torres}, {Mart{\'\i}n-Pintado},
  {Rodr{\'\i}guez-Franco}, {Mart{\'\i}n}, {Rodr{\'\i}guez-Fern{\'a}ndez} and
  {de Vicente}}]{requena06}
\bibinfo{author}{{Requena-Torres}, M.A.},
  \bibinfo{author}{{Mart{\'\i}n-Pintado}, J.},
  \bibinfo{author}{{Rodr{\'\i}guez-Franco}, A.},
  \bibinfo{author}{{Mart{\'\i}n}, S.},
  \bibinfo{author}{{Rodr{\'\i}guez-Fern{\'a}ndez}, N.J.}, \bibinfo{author}{{de
  Vicente}, P.}, \bibinfo{year}{2006}.
\newblock \bibinfo{title}{{Organic molecules in the Galactic center. Hot core
  chemistry without hot cores}}.
\newblock \bibinfo{journal}{\aap} \bibinfo{volume}{455},
  \bibinfo{pages}{971--985}.
\newblock \DOIprefix\doi{10.1051/0004-6361:20065190},
  \href{http://arxiv.org/abs/astro-ph/0605031}{{\tt arXiv:astro-ph/0605031}}.
%Type = Article
\bibitem[{{Richard} et~al.(2018){Richard}, {Belloche}, {Margul{\`e}s},
  {Motiyenko}, {Menten}, {Garrod} and {M{\"u}ller}}]{richard18}
\bibinfo{author}{{Richard}, C.}, \bibinfo{author}{{Belloche}, A.},
  \bibinfo{author}{{Margul{\`e}s}, L.}, \bibinfo{author}{{Motiyenko}, R.A.},
  \bibinfo{author}{{Menten}, K.M.}, \bibinfo{author}{{Garrod}, R.T.},
  \bibinfo{author}{{M{\"u}ller}, H.S.P.}, \bibinfo{year}{2018}.
\newblock \bibinfo{title}{{Rotational spectrum of 3-aminopropionitrile and
  searches for it in Sagittarius B2(N)}}.
\newblock \bibinfo{journal}{Journal of Molecular Spectroscopy}
  \bibinfo{volume}{345}, \bibinfo{pages}{51--59}.
\newblock \DOIprefix\doi{10.1016/j.jms.2017.12.003}.
%Type = Article
\bibitem[{{Rivilla} et~al.(2021){Rivilla}, {Jim{\'e}nez-Serra},
  {Mart{\'\i}n-Pintado}, {Briones}, {Rodr{\'\i}guez-Almeida}, {Rico-Villas},
  {Tercero}, {Zeng}, {Colzi}, {de Vicente}, {Mart{\'\i}n} and
  {Requena-Torres}}]{rivilla21}
\bibinfo{author}{{Rivilla}, V.M.}, \bibinfo{author}{{Jim{\'e}nez-Serra}, I.},
  \bibinfo{author}{{Mart{\'\i}n-Pintado}, J.}, \bibinfo{author}{{Briones}, C.},
  \bibinfo{author}{{Rodr{\'\i}guez-Almeida}, L.F.},
  \bibinfo{author}{{Rico-Villas}, F.}, \bibinfo{author}{{Tercero}, B.},
  \bibinfo{author}{{Zeng}, S.}, \bibinfo{author}{{Colzi}, L.},
  \bibinfo{author}{{de Vicente}, P.}, \bibinfo{author}{{Mart{\'\i}n}, S.},
  \bibinfo{author}{{Requena-Torres}, M.A.}, \bibinfo{year}{2021}.
\newblock \bibinfo{title}{{Discovery in space of ethanolamine, the simplest
  phospholipid head group}}.
\newblock \bibinfo{journal}{Proceedings of the National Academy of Science}
  \bibinfo{volume}{118}, \bibinfo{pages}{e2101314118}.
\newblock \DOIprefix\doi{10.1073/pnas.2101314118},
  \href{http://arxiv.org/abs/2105.11141}{{\tt arXiv:2105.11141}}.
%Type = Article
\bibitem[{{Rivilla} et~al.(2020){Rivilla}, {Mart{\'\i}n-Pintado},
  {Jim{\'e}nez-Serra}, {Mart{\'\i}n}, {Rodr{\'\i}guez-Almeida},
  {Requena-Torres}, {Rico-Villas}, {Zeng} and {Briones}}]{rivilla20}
\bibinfo{author}{{Rivilla}, V.M.}, \bibinfo{author}{{Mart{\'\i}n-Pintado}, J.},
  \bibinfo{author}{{Jim{\'e}nez-Serra}, I.}, \bibinfo{author}{{Mart{\'\i}n},
  S.}, \bibinfo{author}{{Rodr{\'\i}guez-Almeida}, L.F.},
  \bibinfo{author}{{Requena-Torres}, M.A.}, \bibinfo{author}{{Rico-Villas},
  F.}, \bibinfo{author}{{Zeng}, S.}, \bibinfo{author}{{Briones}, C.},
  \bibinfo{year}{2020}.
\newblock \bibinfo{title}{{Prebiotic Precursors of the Primordial RNA World in
  Space: Detection of NH$_{2}$OH}}.
\newblock \bibinfo{journal}{\apjl} \bibinfo{volume}{899}, \bibinfo{pages}{L28}.
\newblock \DOIprefix\doi{10.3847/2041-8213/abac55},
  \href{http://arxiv.org/abs/2008.00228}{{\tt arXiv:2008.00228}}.
%Type = Article
\bibitem[{{Rodr{\'\i}guez-Almeida} et~al.(2021a){Rodr{\'\i}guez-Almeida},
  {Jim{\'e}nez-Serra}, {Rivilla}, {Mart{\'\i}n-Pintado}, {Zeng}, {Tercero}, {de
  Vicente}, {Colzi}, {Rico-Villas}, {Mart{\'\i}n} and
  {Requena-Torres}}]{rodriguez21}
\bibinfo{author}{{Rodr{\'\i}guez-Almeida}, L.F.},
  \bibinfo{author}{{Jim{\'e}nez-Serra}, I.}, \bibinfo{author}{{Rivilla}, V.M.},
  \bibinfo{author}{{Mart{\'\i}n-Pintado}, J.}, \bibinfo{author}{{Zeng}, S.},
  \bibinfo{author}{{Tercero}, B.}, \bibinfo{author}{{de Vicente}, P.},
  \bibinfo{author}{{Colzi}, L.}, \bibinfo{author}{{Rico-Villas}, F.},
  \bibinfo{author}{{Mart{\'\i}n}, S.}, \bibinfo{author}{{Requena-Torres},
  M.A.}, \bibinfo{year}{2021}a.
\newblock \bibinfo{title}{{Thiols in the Interstellar Medium: First Detection
  of HC(O)SH and Confirmation of C$_{2}$H$_{5}$SH}}.
\newblock \bibinfo{journal}{\apjl} \bibinfo{volume}{912}, \bibinfo{pages}{L11}.
\newblock \DOIprefix\doi{10.3847/2041-8213/abf7cb},
  \href{http://arxiv.org/abs/2104.08036}{{\tt arXiv:2104.08036}}.
%Type = Article
\bibitem[{{Rodr{\'\i}guez-Almeida} et~al.(2021b){Rodr{\'\i}guez-Almeida},
  {Rivilla}, {Jim{\'e}nez-Serra}, {Melosso}, {Colzi}, {Zeng}, {Tercero}, {de
  Vicente}, {Mart{\'\i}n}, {Requena-Torres}, {Rico-Villas} and
  {Mart{\'\i}n-Pintado}}]{rodriguez21b}
\bibinfo{author}{{Rodr{\'\i}guez-Almeida}, L.F.}, \bibinfo{author}{{Rivilla},
  V.M.}, \bibinfo{author}{{Jim{\'e}nez-Serra}, I.}, \bibinfo{author}{{Melosso},
  M.}, \bibinfo{author}{{Colzi}, L.}, \bibinfo{author}{{Zeng}, S.},
  \bibinfo{author}{{Tercero}, B.}, \bibinfo{author}{{de Vicente}, P.},
  \bibinfo{author}{{Mart{\'\i}n}, S.}, \bibinfo{author}{{Requena-Torres},
  M.A.}, \bibinfo{author}{{Rico-Villas}, F.},
  \bibinfo{author}{{Mart{\'\i}n-Pintado}, J.}, \bibinfo{year}{2021}b.
\newblock \bibinfo{title}{{First detection of C$_{2}$H$_{5}$NCO in the ISM and
  search of other isocyanates towards the G+0.693-0.027 molecular cloud}}.
\newblock \bibinfo{journal}{\aap} \bibinfo{volume}{654}, \bibinfo{pages}{L1}.
\newblock \DOIprefix\doi{10.1051/0004-6361/202141989},
  \href{http://arxiv.org/abs/2109.07889}{{\tt arXiv:2109.07889}}.
%Type = Article
\bibitem[{{Rodr{\'\i}guez-Fern{\'a}ndez}
  et~al.(2000){Rodr{\'\i}guez-Fern{\'a}ndez}, {Mart{\'\i}n-Pintado}, {de
  Vicente}, {Fuente}, {H{\"u}ttemeister}, {Wilson} and {Kunze}}]{rodriguez00}
\bibinfo{author}{{Rodr{\'\i}guez-Fern{\'a}ndez}, N.J.},
  \bibinfo{author}{{Mart{\'\i}n-Pintado}, J.}, \bibinfo{author}{{de Vicente},
  P.}, \bibinfo{author}{{Fuente}, A.}, \bibinfo{author}{{H{\"u}ttemeister},
  S.}, \bibinfo{author}{{Wilson}, T.L.}, \bibinfo{author}{{Kunze}, D.},
  \bibinfo{year}{2000}.
\newblock \bibinfo{title}{{Non-equilibrium H\_2 ortho-to-para ratio in two
  molecular clouds of the Galactic Center}}.
\newblock \bibinfo{journal}{\aap} \bibinfo{volume}{356},
  \bibinfo{pages}{695--704}.
\newblock \DOIprefix\doi{10.48550/arXiv.astro-ph/0002478},
  \href{http://arxiv.org/abs/astro-ph/0002478}{{\tt arXiv:astro-ph/0002478}}.
%Type = Article
\bibitem[{{Rodr{\'\i}guez-Fern{\'a}ndez}
  et~al.(2004){Rodr{\'\i}guez-Fern{\'a}ndez}, {Mart{\'\i}n-Pintado}, {Fuente}
  and {Wilson}}]{rodriguez04}
\bibinfo{author}{{Rodr{\'\i}guez-Fern{\'a}ndez}, N.J.},
  \bibinfo{author}{{Mart{\'\i}n-Pintado}, J.}, \bibinfo{author}{{Fuente}, A.},
  \bibinfo{author}{{Wilson}, T.L.}, \bibinfo{year}{2004}.
\newblock \bibinfo{title}{{ISO observations of the Galactic center interstellar
  medium. Neutral gas and dust}}.
\newblock \bibinfo{journal}{\aap} \bibinfo{volume}{427},
  \bibinfo{pages}{217--229}.
\newblock \DOIprefix\doi{10.1051/0004-6361:20041370},
  \href{http://arxiv.org/abs/astro-ph/0407479}{{\tt arXiv:astro-ph/0407479}}.
%Type = Article
\bibitem[{Ruiz-Mirazo et~al.(2014)Ruiz-Mirazo, Briones and de~la
  Escosura}]{ruiz-mirazo14}
\bibinfo{author}{Ruiz-Mirazo, K.}, \bibinfo{author}{Briones, C.},
  \bibinfo{author}{de~la Escosura, A.}, \bibinfo{year}{2014}.
\newblock \bibinfo{title}{Prebiotic systems chemistry: New perspectives for the
  origins of life}.
\newblock \bibinfo{journal}{Chem. Rev.} \bibinfo{volume}{114},
  \bibinfo{pages}{285--366}.
\newblock \DOIprefix\doi{10.1021/cr2004844}.
%Type = Article
\bibitem[{{Sakai} et~al.(2017){Sakai}, {Oya}, {Higuchi}, {Aikawa}, {Hanawa},
  {Ceccarelli}, {Lefloch}, {L{\'o}pez-Sepulcre}, {Watanabe}, {Sakai}, {Hirota},
  {Caux}, {Vastel}, {Kahane} and {Yamamoto}}]{Sakai17}
\bibinfo{author}{{Sakai}, N.}, \bibinfo{author}{{Oya}, Y.},
  \bibinfo{author}{{Higuchi}, A.E.}, \bibinfo{author}{{Aikawa}, Y.},
  \bibinfo{author}{{Hanawa}, T.}, \bibinfo{author}{{Ceccarelli}, C.},
  \bibinfo{author}{{Lefloch}, B.}, \bibinfo{author}{{L{\'o}pez-Sepulcre}, A.},
  \bibinfo{author}{{Watanabe}, Y.}, \bibinfo{author}{{Sakai}, T.},
  \bibinfo{author}{{Hirota}, T.}, \bibinfo{author}{{Caux}, E.},
  \bibinfo{author}{{Vastel}, C.}, \bibinfo{author}{{Kahane}, C.},
  \bibinfo{author}{{Yamamoto}, S.}, \bibinfo{year}{2017}.
\newblock \bibinfo{title}{{Vertical structure of the transition zone from
  infalling rotating envelope to disc in the Class 0 protostar, IRAS
  04368+2557}}.
\newblock \bibinfo{journal}{\mnras} \bibinfo{volume}{467},
  \bibinfo{pages}{L76--L80}.
\newblock \DOIprefix\doi{10.1093/mnrasl/slx002}.
%Type = Article
\bibitem[{{Sakai} et~al.(2014a){Sakai}, {Oya}, {Sakai}, {Watanabe}, {Hirota},
  {Ceccarelli}, {Kahane}, {Lopez-Sepulcre}, {Lefloch}, {Vastel}, {Bottinelli},
  {Caux}, {Coutens}, {Aikawa}, {Takakuwa}, {Ohashi}, {Yen} and
  {Yamamoto}}]{Sakai14b}
\bibinfo{author}{{Sakai}, N.}, \bibinfo{author}{{Oya}, Y.},
  \bibinfo{author}{{Sakai}, T.}, \bibinfo{author}{{Watanabe}, Y.},
  \bibinfo{author}{{Hirota}, T.}, \bibinfo{author}{{Ceccarelli}, C.},
  \bibinfo{author}{{Kahane}, C.}, \bibinfo{author}{{Lopez-Sepulcre}, A.},
  \bibinfo{author}{{Lefloch}, B.}, \bibinfo{author}{{Vastel}, C.},
  \bibinfo{author}{{Bottinelli}, S.}, \bibinfo{author}{{Caux}, E.},
  \bibinfo{author}{{Coutens}, A.}, \bibinfo{author}{{Aikawa}, Y.},
  \bibinfo{author}{{Takakuwa}, S.}, \bibinfo{author}{{Ohashi}, N.},
  \bibinfo{author}{{Yen}, H.W.}, \bibinfo{author}{{Yamamoto}, S.},
  \bibinfo{year}{2014}a.
\newblock \bibinfo{title}{{A Chemical View of Protostellar-disk Formation in
  L1527}}.
\newblock \bibinfo{journal}{\apjl} \bibinfo{volume}{791}, \bibinfo{pages}{L38}.
\newblock \DOIprefix\doi{10.1088/2041-8205/791/2/L38}.
%Type = Article
\bibitem[{{Sakai} et~al.(2009){Sakai}, {Sakai}, {Hirota}, {Burton} and
  {Yamamoto}}]{sakai09}
\bibinfo{author}{{Sakai}, N.}, \bibinfo{author}{{Sakai}, T.},
  \bibinfo{author}{{Hirota}, T.}, \bibinfo{author}{{Burton}, M.},
  \bibinfo{author}{{Yamamoto}, S.}, \bibinfo{year}{2009}.
\newblock \bibinfo{title}{{Discovery of the Second Warm Carbon-Chain-Chemistry
  Source, IRAS15398 - 3359 in Lupus}}.
\newblock \bibinfo{journal}{\apj} \bibinfo{volume}{697},
  \bibinfo{pages}{769--786}.
\newblock \DOIprefix\doi{10.1088/0004-637X/697/1/769}.
%Type = Article
\bibitem[{{Sakai} et~al.(2014b){Sakai}, {Sakai}, {Hirota}, {Watanabe},
  {Ceccarelli}, {Kahane}, {Bottinelli}, {Caux}, {Demyk}, {Vastel}, {Coutens},
  {Taquet}, {Ohashi}, {Takakuwa}, {Yen}, {Aikawa} and {Yamamoto}}]{Sakai14a}
\bibinfo{author}{{Sakai}, N.}, \bibinfo{author}{{Sakai}, T.},
  \bibinfo{author}{{Hirota}, T.}, \bibinfo{author}{{Watanabe}, Y.},
  \bibinfo{author}{{Ceccarelli}, C.}, \bibinfo{author}{{Kahane}, C.},
  \bibinfo{author}{{Bottinelli}, S.}, \bibinfo{author}{{Caux}, E.},
  \bibinfo{author}{{Demyk}, K.}, \bibinfo{author}{{Vastel}, C.},
  \bibinfo{author}{{Coutens}, A.}, \bibinfo{author}{{Taquet}, V.},
  \bibinfo{author}{{Ohashi}, N.}, \bibinfo{author}{{Takakuwa}, S.},
  \bibinfo{author}{{Yen}, H.W.}, \bibinfo{author}{{Aikawa}, Y.},
  \bibinfo{author}{{Yamamoto}, S.}, \bibinfo{year}{2014}b.
\newblock \bibinfo{title}{{Change in the chemical composition of infalling gas
  forming a disk around a protostar}}.
\newblock \bibinfo{journal}{\nat} \bibinfo{volume}{507},
  \bibinfo{pages}{78--80}.
\newblock \DOIprefix\doi{10.1038/nature13000}.
%Type = Article
\bibitem[{{Sakai} et~al.(2008){Sakai}, {Sakai}, {Hirota} and
  {Yamamoto}}]{sakai08}
\bibinfo{author}{{Sakai}, N.}, \bibinfo{author}{{Sakai}, T.},
  \bibinfo{author}{{Hirota}, T.}, \bibinfo{author}{{Yamamoto}, S.},
  \bibinfo{year}{2008}.
\newblock \bibinfo{title}{{Abundant Carbon-Chain Molecules toward the Low-Mass
  Protostar IRAS 04368+2557 in L1527}}.
\newblock \bibinfo{journal}{\apj} \bibinfo{volume}{672},
  \bibinfo{pages}{371--381}.
\newblock \DOIprefix\doi{10.1086/523635}.
%Type = Article
\bibitem[{{Sakai} and {Yamamoto}(2013)}]{sakai13}
\bibinfo{author}{{Sakai}, N.}, \bibinfo{author}{{Yamamoto}, S.},
  \bibinfo{year}{2013}.
\newblock \bibinfo{title}{{Warm Carbon-Chain Chemistry}}.
\newblock \bibinfo{journal}{Chemical Reviews} \bibinfo{volume}{113},
  \bibinfo{pages}{8981--9015}.
\newblock \DOIprefix\doi{10.1021/cr4001308}.
%Type = Article
\bibitem[{{S{\'a}nchez-Monge} et~al.(2013){S{\'a}nchez-Monge},
  {L{\'o}pez-Sepulcre}, {Cesaroni}, {Walmsley}, {Codella}, {Beltr{\'a}n},
  {Pestalozzi} and {Molinari}}]{Sanchez13}
\bibinfo{author}{{S{\'a}nchez-Monge}, {\'A}.},
  \bibinfo{author}{{L{\'o}pez-Sepulcre}, A.}, \bibinfo{author}{{Cesaroni}, R.},
  \bibinfo{author}{{Walmsley}, C.M.}, \bibinfo{author}{{Codella}, C.},
  \bibinfo{author}{{Beltr{\'a}n}, M.T.}, \bibinfo{author}{{Pestalozzi}, M.},
  \bibinfo{author}{{Molinari}, S.}, \bibinfo{year}{2013}.
\newblock \bibinfo{title}{{Evolution and excitation conditions of outflows in
  high-mass star-forming regions}}.
\newblock \bibinfo{journal}{\aap} \bibinfo{volume}{557}, \bibinfo{pages}{A94}.
\newblock \DOIprefix\doi{10.1051/0004-6361/201321589},
  \href{http://arxiv.org/abs/1305.3471}{{\tt arXiv:1305.3471}}.
%Type = Article
\bibitem[{{Santos} et~al.(2023){Santos}, {Chuang}, {Schrauwen}, {Traspas
  Mui{\~n}a}, {Zhang}, {Cuppen}, {Redlich}, {Linnartz} and
  {Ioppolo}}]{santos23}
\bibinfo{author}{{Santos}, J.C.}, \bibinfo{author}{{Chuang}, K.J.},
  \bibinfo{author}{{Schrauwen}, J.G.M.}, \bibinfo{author}{{Traspas Mui{\~n}a},
  A.}, \bibinfo{author}{{Zhang}, J.}, \bibinfo{author}{{Cuppen}, H.M.},
  \bibinfo{author}{{Redlich}, B.}, \bibinfo{author}{{Linnartz}, H.},
  \bibinfo{author}{{Ioppolo}, S.}, \bibinfo{year}{2023}.
\newblock \bibinfo{title}{{Resonant infrared irradiation of CO and CH$_{3}$OH
  interstellar ices}}.
\newblock \bibinfo{journal}{\aap} \bibinfo{volume}{672}, \bibinfo{pages}{A112}.
\newblock \DOIprefix\doi{10.1051/0004-6361/202245704},
  \href{http://arxiv.org/abs/2302.11591}{{\tt arXiv:2302.11591}}.
%Type = Article
\bibitem[{{Sanz-Novo} et~al.(2020){Sanz-Novo}, {Belloche}, {Alonso},
  {Kolesnikov{\'a}}, {Garrod}, {Mata}, {M{\"u}ller}, {Menten} and
  {Gong}}]{sanz-novo20}
\bibinfo{author}{{Sanz-Novo}, M.}, \bibinfo{author}{{Belloche}, A.},
  \bibinfo{author}{{Alonso}, J.L.}, \bibinfo{author}{{Kolesnikov{\'a}}, L.},
  \bibinfo{author}{{Garrod}, R.T.}, \bibinfo{author}{{Mata}, S.},
  \bibinfo{author}{{M{\"u}ller}, H.S.P.}, \bibinfo{author}{{Menten}, K.M.},
  \bibinfo{author}{{Gong}, Y.}, \bibinfo{year}{2020}.
\newblock \bibinfo{title}{{Interstellar glycolamide: A comprehensive rotational
  study and an astronomical search in Sgr B2(N)}}.
\newblock \bibinfo{journal}{\aap} \bibinfo{volume}{639}, \bibinfo{pages}{A135}.
\newblock \DOIprefix\doi{10.1051/0004-6361/202038149},
  \href{http://arxiv.org/abs/2006.13634}{{\tt arXiv:2006.13634}}.
%Type = Article
\bibitem[{{Sanz-Novo} et~al.(2022){Sanz-Novo}, {Belloche}, {Rivilla}, {Garrod},
  {Alonso}, {Redondo}, {Barrientos}, {Kolesnikov{\'a}}, {Valle},
  {Rodr{\'\i}guez-Almeida}, {Jimenez-Serra}, {Mart{\'\i}n-Pintado},
  {M{\"u}ller} and {Menten}}]{sanz-novo22}
\bibinfo{author}{{Sanz-Novo}, M.}, \bibinfo{author}{{Belloche}, A.},
  \bibinfo{author}{{Rivilla}, V.M.}, \bibinfo{author}{{Garrod}, R.T.},
  \bibinfo{author}{{Alonso}, J.L.}, \bibinfo{author}{{Redondo}, P.},
  \bibinfo{author}{{Barrientos}, C.}, \bibinfo{author}{{Kolesnikov{\'a}}, L.},
  \bibinfo{author}{{Valle}, J.C.}, \bibinfo{author}{{Rodr{\'\i}guez-Almeida},
  L.}, \bibinfo{author}{{Jimenez-Serra}, I.},
  \bibinfo{author}{{Mart{\'\i}n-Pintado}, J.}, \bibinfo{author}{{M{\"u}ller},
  H.S.P.}, \bibinfo{author}{{Menten}, K.M.}, \bibinfo{year}{2022}.
\newblock \bibinfo{title}{{Toward the limits of complexity of interstellar
  chemistry: Rotational spectroscopy and astronomical search for n- and
  i-butanal}}.
\newblock \bibinfo{journal}{\aap} \bibinfo{volume}{666}, \bibinfo{pages}{A114}.
\newblock \DOIprefix\doi{10.1051/0004-6361/202142848},
  \href{http://arxiv.org/abs/2203.07334}{{\tt arXiv:2203.07334}}.
%Type = Article
\bibitem[{{Schilke} et~al.(1997){Schilke}, {Walmsley}, {Pineau des Forets} and
  {Flower}}]{Schilke97}
\bibinfo{author}{{Schilke}, P.}, \bibinfo{author}{{Walmsley}, C.M.},
  \bibinfo{author}{{Pineau des Forets}, G.}, \bibinfo{author}{{Flower}, D.R.},
  \bibinfo{year}{1997}.
\newblock \bibinfo{title}{{SiO production in interstellar shocks.}}
\newblock \bibinfo{journal}{\aap} \bibinfo{volume}{321},
  \bibinfo{pages}{293--304}.
%Type = Article
\bibitem[{{Scibelli} and {Shirley}(2020)}]{Scibelli20}
\bibinfo{author}{{Scibelli}, S.}, \bibinfo{author}{{Shirley}, Y.},
  \bibinfo{year}{2020}.
\newblock \bibinfo{title}{{Prevalence of Complex Organic Molecules in Starless
  and Prestellar Cores within the Taurus Molecular Cloud}}.
\newblock \bibinfo{journal}{\apj} \bibinfo{volume}{891}, \bibinfo{pages}{73}.
\newblock \DOIprefix\doi{10.3847/1538-4357/ab7375},
  \href{http://arxiv.org/abs/2002.02469}{{\tt arXiv:2002.02469}}.
%Type = Article
\bibitem[{{Scibelli} et~al.(2021){Scibelli}, {Shirley}, {Vasyunin} and
  {Launhardt}}]{scibelli21}
\bibinfo{author}{{Scibelli}, S.}, \bibinfo{author}{{Shirley}, Y.},
  \bibinfo{author}{{Vasyunin}, A.}, \bibinfo{author}{{Launhardt}, R.},
  \bibinfo{year}{2021}.
\newblock \bibinfo{title}{{Detection of complex organic molecules in young
  starless core L1521E}}.
\newblock \bibinfo{journal}{\mnras} \bibinfo{volume}{504},
  \bibinfo{pages}{5754--5767}.
\newblock \DOIprefix\doi{10.1093/mnras/stab1151},
  \href{http://arxiv.org/abs/2104.07683}{{\tt arXiv:2104.07683}}.
%Type = Article
\bibitem[{{Sewi{\l}o} et~al.(2022){Sewi{\l}o}, {Cordiner}, {Charnley},
  {Oliveira}, {Garcia-Berrios}, {Schilke}, {Ward}, {Wiseman}, {Indebetouw},
  {Tokuda}, {van Loon}, {S{\'a}nchez-Monge}, {Allen}, {Chen}, {Hamedani
  Golshan}, {Karska}, {Kristensen}, {Kurtz}, {M{\"o}ller}, {Onishi} and
  {Zahorecz}}]{Sewilo22}
\bibinfo{author}{{Sewi{\l}o}, M.}, \bibinfo{author}{{Cordiner}, M.},
  \bibinfo{author}{{Charnley}, S.B.}, \bibinfo{author}{{Oliveira}, J.M.},
  \bibinfo{author}{{Garcia-Berrios}, E.}, \bibinfo{author}{{Schilke}, P.},
  \bibinfo{author}{{Ward}, J.L.}, \bibinfo{author}{{Wiseman}, J.},
  \bibinfo{author}{{Indebetouw}, R.}, \bibinfo{author}{{Tokuda}, K.},
  \bibinfo{author}{{van Loon}, J.T.}, \bibinfo{author}{{S{\'a}nchez-Monge},
  {\'A}.}, \bibinfo{author}{{Allen}, V.}, \bibinfo{author}{{Chen}, C.H.R.},
  \bibinfo{author}{{Hamedani Golshan}, R.}, \bibinfo{author}{{Karska}, A.},
  \bibinfo{author}{{Kristensen}, L.E.}, \bibinfo{author}{{Kurtz}, S.E.},
  \bibinfo{author}{{M{\"o}ller}, T.}, \bibinfo{author}{{Onishi}, T.},
  \bibinfo{author}{{Zahorecz}, S.}, \bibinfo{year}{2022}.
\newblock \bibinfo{title}{{ALMA Observations of Molecular Complexity in the
  Large Magellanic Cloud: The N 105 Star-forming Region}}.
\newblock \bibinfo{journal}{\apj} \bibinfo{volume}{931}, \bibinfo{pages}{102}.
\newblock \DOIprefix\doi{10.3847/1538-4357/ac4e8f},
  \href{http://arxiv.org/abs/2201.09945}{{\tt arXiv:2201.09945}}.
%Type = Article
\bibitem[{{Sewi{\l}o} et~al.(2018){Sewi{\l}o}, {Indebetouw}, {Charnley},
  {Zahorecz}, {Oliveira}, {van Loon}, {Ward}, {Chen}, {Wiseman}, {Fukui},
  {Kawamura}, {Meixner}, {Onishi} and {Schilke}}]{Sewilo18}
\bibinfo{author}{{Sewi{\l}o}, M.}, \bibinfo{author}{{Indebetouw}, R.},
  \bibinfo{author}{{Charnley}, S.B.}, \bibinfo{author}{{Zahorecz}, S.},
  \bibinfo{author}{{Oliveira}, J.M.}, \bibinfo{author}{{van Loon}, J.T.},
  \bibinfo{author}{{Ward}, J.L.}, \bibinfo{author}{{Chen}, C.H.R.},
  \bibinfo{author}{{Wiseman}, J.}, \bibinfo{author}{{Fukui}, Y.},
  \bibinfo{author}{{Kawamura}, A.}, \bibinfo{author}{{Meixner}, M.},
  \bibinfo{author}{{Onishi}, T.}, \bibinfo{author}{{Schilke}, P.},
  \bibinfo{year}{2018}.
\newblock \bibinfo{title}{{The Detection of Hot Cores and Complex Organic
  Molecules in the Large Magellanic Cloud}}.
\newblock \bibinfo{journal}{\apjl} \bibinfo{volume}{853}, \bibinfo{pages}{L19}.
\newblock \DOIprefix\doi{10.3847/2041-8213/aaa079},
  \href{http://arxiv.org/abs/1801.10275}{{\tt arXiv:1801.10275}}.
%Type = Article
\bibitem[{{Shara} et~al.(1985){Shara}, {Moffat} and {Webbink}}]{Shara85}
\bibinfo{author}{{Shara}, M.M.}, \bibinfo{author}{{Moffat}, A.F.J.},
  \bibinfo{author}{{Webbink}, R.F.}, \bibinfo{year}{1985}.
\newblock \bibinfo{title}{{Unraveling the oldest and faintest recovered nova :
  CK Vulpeculae (1670).}}
\newblock \bibinfo{journal}{\apj} \bibinfo{volume}{294},
  \bibinfo{pages}{271--285}.
\newblock \DOIprefix\doi{10.1086/163296}.
%Type = Article
\bibitem[{{Shimonishi} et~al.(2016a){Shimonishi}, {Dartois}, {Onaka} and
  {Boulanger}}]{Shimonishi16a}
\bibinfo{author}{{Shimonishi}, T.}, \bibinfo{author}{{Dartois}, E.},
  \bibinfo{author}{{Onaka}, T.}, \bibinfo{author}{{Boulanger}, F.},
  \bibinfo{year}{2016a}.
\newblock \bibinfo{title}{{VLT/ISAAC infrared spectroscopy of embedded
  high-mass YSOs in the Large Magellanic Cloud: Methanol and the 3.47
  {\ensuremath{\mu}}m band}}.
\newblock \bibinfo{journal}{\aap} \bibinfo{volume}{585}, \bibinfo{pages}{A107}.
\newblock \DOIprefix\doi{10.1051/0004-6361/201526559},
  \href{http://arxiv.org/abs/1511.04141}{{\tt arXiv:1511.04141}}.
%Type = Article
\bibitem[{{Shimonishi} et~al.(2020){Shimonishi}, {Das}, {Sakai}, {Tanaka},
  {Aikawa}, {Onaka}, {Watanabe} and {Nishimura}}]{Shimonishi20}
\bibinfo{author}{{Shimonishi}, T.}, \bibinfo{author}{{Das}, A.},
  \bibinfo{author}{{Sakai}, N.}, \bibinfo{author}{{Tanaka}, K.E.I.},
  \bibinfo{author}{{Aikawa}, Y.}, \bibinfo{author}{{Onaka}, T.},
  \bibinfo{author}{{Watanabe}, Y.}, \bibinfo{author}{{Nishimura}, Y.},
  \bibinfo{year}{2020}.
\newblock \bibinfo{title}{{Chemistry and Physics of a Low-metallicity Hot Core
  in the Large Magellanic Cloud}}.
\newblock \bibinfo{journal}{\apj} \bibinfo{volume}{891}, \bibinfo{pages}{164}.
\newblock \DOIprefix\doi{10.3847/1538-4357/ab6e6b},
  \href{http://arxiv.org/abs/2001.06982}{{\tt arXiv:2001.06982}}.
%Type = Article
\bibitem[{{Shimonishi} et~al.(2021){Shimonishi}, {Izumi}, {Furuya} and
  {Yasui}}]{Shimonishi21}
\bibinfo{author}{{Shimonishi}, T.}, \bibinfo{author}{{Izumi}, N.},
  \bibinfo{author}{{Furuya}, K.}, \bibinfo{author}{{Yasui}, C.},
  \bibinfo{year}{2021}.
\newblock \bibinfo{title}{{The Detection of a Hot Molecular Core in the Extreme
  Outer Galaxy}}.
\newblock \bibinfo{journal}{\apj} \bibinfo{volume}{922}, \bibinfo{pages}{206}.
\newblock \DOIprefix\doi{10.3847/1538-4357/ac289b},
  \href{http://arxiv.org/abs/2109.11123}{{\tt arXiv:2109.11123}}.
%Type = Article
\bibitem[{{Shimonishi} et~al.(2016b){Shimonishi}, {Onaka}, {Kawamura} and
  {Aikawa}}]{Shimonishi16b}
\bibinfo{author}{{Shimonishi}, T.}, \bibinfo{author}{{Onaka}, T.},
  \bibinfo{author}{{Kawamura}, A.}, \bibinfo{author}{{Aikawa}, Y.},
  \bibinfo{year}{2016b}.
\newblock \bibinfo{title}{{The Detection of a Hot Molecular Core in the Large
  Magellanic Cloud with ALMA}}.
\newblock \bibinfo{journal}{\apj} \bibinfo{volume}{827}, \bibinfo{pages}{72}.
\newblock \DOIprefix\doi{10.3847/0004-637X/827/1/72},
  \href{http://arxiv.org/abs/1606.02823}{{\tt arXiv:1606.02823}}.
%Type = Article
\bibitem[{{Shimonishi} et~al.(2018){Shimonishi}, {Watanabe}, {Nishimura},
  {Aikawa}, {Yamamoto}, {Onaka}, {Sakai} and {Kawamura}}]{Shimonishi18}
\bibinfo{author}{{Shimonishi}, T.}, \bibinfo{author}{{Watanabe}, Y.},
  \bibinfo{author}{{Nishimura}, Y.}, \bibinfo{author}{{Aikawa}, Y.},
  \bibinfo{author}{{Yamamoto}, S.}, \bibinfo{author}{{Onaka}, T.},
  \bibinfo{author}{{Sakai}, N.}, \bibinfo{author}{{Kawamura}, A.},
  \bibinfo{year}{2018}.
\newblock \bibinfo{title}{{A Multiline Study of a High-mass Young Stellar
  Object in the Small Magellanic Cloud with ALMA: The Detection of Methanol Gas
  at 0.2 Solar Metallicity}}.
\newblock \bibinfo{journal}{\apj} \bibinfo{volume}{862}, \bibinfo{pages}{102}.
\newblock \DOIprefix\doi{10.3847/1538-4357/aacd0c},
  \href{http://arxiv.org/abs/1806.07120}{{\tt arXiv:1806.07120}}.
%Type = Article
\bibitem[{{Sinclair} et~al.(1992){Sinclair}, {Carrad}, {Caswell}, {Norris} and
  {Whiteoak}}]{Sinclair92}
\bibinfo{author}{{Sinclair}, M.W.}, \bibinfo{author}{{Carrad}, G.J.},
  \bibinfo{author}{{Caswell}, J.L.}, \bibinfo{author}{{Norris}, R.P.},
  \bibinfo{author}{{Whiteoak}, J.B.}, \bibinfo{year}{1992}.
\newblock \bibinfo{title}{{A methanol maser in the Large Magellanic Cloud.}}
\newblock \bibinfo{journal}{\mnras} \bibinfo{volume}{256},
  \bibinfo{pages}{33P--34}.
\newblock \DOIprefix\doi{10.1093/mnras/256.1.33P}.
%Type = Article
\bibitem[{{Skouteris} et~al.(2017){Skouteris}, {Vazart}, {Ceccarelli},
  {Balucani}, {Puzzarini} and {Barone}}]{Skouteris17}
\bibinfo{author}{{Skouteris}, D.}, \bibinfo{author}{{Vazart}, F.},
  \bibinfo{author}{{Ceccarelli}, C.}, \bibinfo{author}{{Balucani}, N.},
  \bibinfo{author}{{Puzzarini}, C.}, \bibinfo{author}{{Barone}, V.},
  \bibinfo{year}{2017}.
\newblock \bibinfo{title}{{New quantum chemical computations of formamide
  deuteration support gas-phase formation of this prebiotic molecule}}.
\newblock \bibinfo{journal}{\mnras} \bibinfo{volume}{468},
  \bibinfo{pages}{L1--L5}.
\newblock \DOIprefix\doi{10.1093/mnrasl/slx012},
  \href{http://arxiv.org/abs/1701.06138}{{\tt arXiv:1701.06138}}.
%Type = Article
\bibitem[{{Snow} and {McCall}(2006)}]{Snow06}
\bibinfo{author}{{Snow}, T.P.}, \bibinfo{author}{{McCall}, B.J.},
  \bibinfo{year}{2006}.
\newblock \bibinfo{title}{{Diffuse Atomic and Molecular Clouds}}.
\newblock \bibinfo{journal}{\araa} \bibinfo{volume}{44},
  \bibinfo{pages}{367--414}.
\newblock \DOIprefix\doi{10.1146/annurev.astro.43.072103.150624}.
%Type = Article
\bibitem[{{Snyder} et~al.(1974){Snyder}, {Buhl}, {Schwartz}, {Clark},
  {Johnson}, {Lovas} and {Giguere}}]{Snyder74}
\bibinfo{author}{{Snyder}, L.E.}, \bibinfo{author}{{Buhl}, D.},
  \bibinfo{author}{{Schwartz}, P.R.}, \bibinfo{author}{{Clark}, F.O.},
  \bibinfo{author}{{Johnson}, D.R.}, \bibinfo{author}{{Lovas}, F.J.},
  \bibinfo{author}{{Giguere}, P.T.}, \bibinfo{year}{1974}.
\newblock \bibinfo{title}{{Radio Detection of Interstellar Dimethyl Ether}}.
\newblock \bibinfo{journal}{\apjl} \bibinfo{volume}{191}, \bibinfo{pages}{L79}.
\newblock \DOIprefix\doi{10.1086/181554}.
%Type = Article
\bibitem[{{Soma} et~al.(2018){Soma}, {Sakai}, {Watanabe} and
  {Yamamoto}}]{soma18}
\bibinfo{author}{{Soma}, T.}, \bibinfo{author}{{Sakai}, N.},
  \bibinfo{author}{{Watanabe}, Y.}, \bibinfo{author}{{Yamamoto}, S.},
  \bibinfo{year}{2018}.
\newblock \bibinfo{title}{{Complex Organic Molecules in Taurus Molecular
  Cloud-1}}.
\newblock \bibinfo{journal}{\apj} \bibinfo{volume}{854}, \bibinfo{pages}{116}.
\newblock \DOIprefix\doi{10.3847/1538-4357/aaa70c}.
%Type = Article
\bibitem[{{Spezzano} et~al.(2016){Spezzano}, {Bizzocchi}, {Caselli}, {Harju}
  and {Br{\"u}nken}}]{spezzano16}
\bibinfo{author}{{Spezzano}, S.}, \bibinfo{author}{{Bizzocchi}, L.},
  \bibinfo{author}{{Caselli}, P.}, \bibinfo{author}{{Harju}, J.},
  \bibinfo{author}{{Br{\"u}nken}, S.}, \bibinfo{year}{2016}.
\newblock \bibinfo{title}{{Chemical differentiation in a prestellar core traces
  non-uniform illumination}}.
\newblock \bibinfo{journal}{\aap} \bibinfo{volume}{592}, \bibinfo{pages}{L11}.
\newblock \DOIprefix\doi{10.1051/0004-6361/201628652},
  \href{http://arxiv.org/abs/1607.03242}{{\tt arXiv:1607.03242}}.
%Type = Article
\bibitem[{{Spezzano} et~al.(2020){Spezzano}, {Caselli}, {Pineda}, {Bizzocchi},
  {Prudenzano} and {Nagy}}]{spezzano20}
\bibinfo{author}{{Spezzano}, S.}, \bibinfo{author}{{Caselli}, P.},
  \bibinfo{author}{{Pineda}, J.E.}, \bibinfo{author}{{Bizzocchi}, L.},
  \bibinfo{author}{{Prudenzano}, D.}, \bibinfo{author}{{Nagy}, Z.},
  \bibinfo{year}{2020}.
\newblock \bibinfo{title}{{Distribution of methanol and cyclopropenylidene
  around starless cores}}.
\newblock \bibinfo{journal}{\aap} \bibinfo{volume}{643}, \bibinfo{pages}{A60}.
\newblock \DOIprefix\doi{10.1051/0004-6361/201936598},
  \href{http://arxiv.org/abs/2009.04768}{{\tt arXiv:2009.04768}}.
%Type = Article
\bibitem[{{Stahler} et~al.(1994){Stahler}, {Korycansky}, {Brothers} and
  {Touma}}]{Stahler94}
\bibinfo{author}{{Stahler}, S.W.}, \bibinfo{author}{{Korycansky}, D.G.},
  \bibinfo{author}{{Brothers}, M.J.}, \bibinfo{author}{{Touma}, J.},
  \bibinfo{year}{1994}.
\newblock \bibinfo{title}{{The Early Evolution of Protostellar Disks}}.
\newblock \bibinfo{journal}{\apj} \bibinfo{volume}{431}, \bibinfo{pages}{341}.
\newblock \DOIprefix\doi{10.1086/174489}.
%Type = Article
\bibitem[{{Sugimura} et~al.(2011){Sugimura}, {Yamaguchi}, {Sakai}, {Umemoto},
  {Sakai}, {Takano}, {Aikawa}, {Hirano}, {Liu}, {Millar}, {Nomura}, {Su},
  {Takakuwa} and {Yamamoto}}]{Sugimura11}
\bibinfo{author}{{Sugimura}, M.}, \bibinfo{author}{{Yamaguchi}, T.},
  \bibinfo{author}{{Sakai}, T.}, \bibinfo{author}{{Umemoto}, T.},
  \bibinfo{author}{{Sakai}, N.}, \bibinfo{author}{{Takano}, S.},
  \bibinfo{author}{{Aikawa}, Y.}, \bibinfo{author}{{Hirano}, N.},
  \bibinfo{author}{{Liu}, S.Y.}, \bibinfo{author}{{Millar}, T.J.},
  \bibinfo{author}{{Nomura}, H.}, \bibinfo{author}{{Su}, Y.N.},
  \bibinfo{author}{{Takakuwa}, S.}, \bibinfo{author}{{Yamamoto}, S.},
  \bibinfo{year}{2011}.
\newblock \bibinfo{title}{{Early Results of the 3mm Spectral Line Survey toward
  the Lynds 1157 B1 Shocked Region}}.
\newblock \bibinfo{journal}{\pasj} \bibinfo{volume}{63},
  \bibinfo{pages}{459--472}.
\newblock \DOIprefix\doi{10.1093/pasj/63.2.459}.
%Type = Article
\bibitem[{{Tabone} et~al.(2020){Tabone}, {Cabrit}, {Pineau des For{\^e}ts},
  {Ferreira}, {Gusdorf}, {Podio}, {Bianchi}, {Chapillon}, {Codella} and
  {Gueth}}]{Tabone20}
\bibinfo{author}{{Tabone}, B.}, \bibinfo{author}{{Cabrit}, S.},
  \bibinfo{author}{{Pineau des For{\^e}ts}, G.}, \bibinfo{author}{{Ferreira},
  J.}, \bibinfo{author}{{Gusdorf}, A.}, \bibinfo{author}{{Podio}, L.},
  \bibinfo{author}{{Bianchi}, E.}, \bibinfo{author}{{Chapillon}, E.},
  \bibinfo{author}{{Codella}, C.}, \bibinfo{author}{{Gueth}, F.},
  \bibinfo{year}{2020}.
\newblock \bibinfo{title}{{Constraining MHD disk winds with ALMA. Apparent
  rotation signatures and application to HH212}}.
\newblock \bibinfo{journal}{\aap} \bibinfo{volume}{640}, \bibinfo{pages}{A82}.
\newblock \DOIprefix\doi{10.1051/0004-6361/201834377},
  \href{http://arxiv.org/abs/2004.08804}{{\tt arXiv:2004.08804}}.
%Type = Article
\bibitem[{{Tafalla} et~al.(2004){Tafalla}, {Myers}, {Caselli} and
  {Walmsley}}]{tafalla04}
\bibinfo{author}{{Tafalla}, M.}, \bibinfo{author}{{Myers}, P.C.},
  \bibinfo{author}{{Caselli}, P.}, \bibinfo{author}{{Walmsley}, C.M.},
  \bibinfo{year}{2004}.
\newblock \bibinfo{title}{{On The Internal Structure Of Starless Cores.
  Physical and Chemical Properties of L1498 and L1517B}}.
\newblock \bibinfo{journal}{\apss} \bibinfo{volume}{292},
  \bibinfo{pages}{347--354}.
\newblock \DOIprefix\doi{10.1023/B:ASTR.0000045036.76044.bd},
  \href{http://arxiv.org/abs/astro-ph/0401148}{{\tt arXiv:astro-ph/0401148}}.
%Type = Article
\bibitem[{{Taillard} et~al.(2023){Taillard}, {Wakelam}, {Gratier}, {Dartois},
  {Chabot}, {Noble}, {Keane}, {Boogert} and {Harsono}}]{taillard23}
\bibinfo{author}{{Taillard}, A.}, \bibinfo{author}{{Wakelam}, V.},
  \bibinfo{author}{{Gratier}, P.}, \bibinfo{author}{{Dartois}, E.},
  \bibinfo{author}{{Chabot}, M.}, \bibinfo{author}{{Noble}, J.A.},
  \bibinfo{author}{{Keane}, J.V.}, \bibinfo{author}{{Boogert}, A.C.A.},
  \bibinfo{author}{{Harsono}, D.}, \bibinfo{year}{2023}.
\newblock \bibinfo{title}{{Constraints on the non-thermal desorption of
  methanol in the cold core LDN 429-C}}.
\newblock \bibinfo{journal}{\aap} \bibinfo{volume}{670}, \bibinfo{pages}{A141}.
\newblock \DOIprefix\doi{10.1051/0004-6361/202245157},
  \href{http://arxiv.org/abs/2301.01288}{{\tt arXiv:2301.01288}}.
%Type = Article
\bibitem[{{Taquet} et~al.(2015){Taquet}, {L{\'o}pez-Sepulcre}, {Ceccarelli},
  {Neri}, {Kahane} and {Charnley}}]{Taquet15}
\bibinfo{author}{{Taquet}, V.}, \bibinfo{author}{{L{\'o}pez-Sepulcre}, A.},
  \bibinfo{author}{{Ceccarelli}, C.}, \bibinfo{author}{{Neri}, R.},
  \bibinfo{author}{{Kahane}, C.}, \bibinfo{author}{{Charnley}, S.B.},
  \bibinfo{year}{2015}.
\newblock \bibinfo{title}{{Constraining the Abundances of Complex Organics in
  the Inner Regions of Solar-type Protostars}}.
\newblock \bibinfo{journal}{\apj} \bibinfo{volume}{804}, \bibinfo{pages}{81}.
\newblock \DOIprefix\doi{10.1088/0004-637X/804/2/81},
  \href{http://arxiv.org/abs/1502.06427}{{\tt arXiv:1502.06427}}.
%Type = Article
\bibitem[{{Taquet} et~al.(2017){Taquet}, {Wirstr{\"o}m}, {Charnley}, {Faure},
  {L{\'o}pez-Sepulcre} and {Persson}}]{taquet17}
\bibinfo{author}{{Taquet}, V.}, \bibinfo{author}{{Wirstr{\"o}m}, E.S.},
  \bibinfo{author}{{Charnley}, S.B.}, \bibinfo{author}{{Faure}, A.},
  \bibinfo{author}{{L{\'o}pez-Sepulcre}, A.}, \bibinfo{author}{{Persson},
  C.M.}, \bibinfo{year}{2017}.
\newblock \bibinfo{title}{{Chemical complexity induced by efficient ice
  evaporation in the Barnard 5 molecular cloud}}.
\newblock \bibinfo{journal}{\aap} \bibinfo{volume}{607}, \bibinfo{pages}{A20}.
\newblock \DOIprefix\doi{10.1051/0004-6361/201630023},
  \href{http://arxiv.org/abs/1706.01368}{{\tt arXiv:1706.01368}}.
%Type = Article
\bibitem[{{Tercero} et~al.(2020){Tercero}, {Cernicharo}, {Cuadrado}, {de
  Vicente} and {Gu{\'e}lin}}]{tercero20}
\bibinfo{author}{{Tercero}, B.}, \bibinfo{author}{{Cernicharo}, J.},
  \bibinfo{author}{{Cuadrado}, S.}, \bibinfo{author}{{de Vicente}, P.},
  \bibinfo{author}{{Gu{\'e}lin}, M.}, \bibinfo{year}{2020}.
\newblock \bibinfo{title}{{New molecular species at redshift z = 0.89}}.
\newblock \bibinfo{journal}{\aap} \bibinfo{volume}{636}, \bibinfo{pages}{L7}.
\newblock \DOIprefix\doi{10.1051/0004-6361/202037837},
  \href{http://arxiv.org/abs/2004.02486}{{\tt arXiv:2004.02486}}.
%Type = Article
\bibitem[{{Tercero} et~al.(2010){Tercero}, {Cernicharo}, {Pardo} and
  {Goicoechea}}]{tercero10}
\bibinfo{author}{{Tercero}, B.}, \bibinfo{author}{{Cernicharo}, J.},
  \bibinfo{author}{{Pardo}, J.R.}, \bibinfo{author}{{Goicoechea}, J.R.},
  \bibinfo{year}{2010}.
\newblock \bibinfo{title}{{A line confusion limited millimeter survey of Orion
  KL . I. Sulfur carbon chains}}.
\newblock \bibinfo{journal}{\aap} \bibinfo{volume}{517}, \bibinfo{pages}{A96}.
\newblock \DOIprefix\doi{10.1051/0004-6361/200913501},
  \href{http://arxiv.org/abs/1004.2711}{{\tt arXiv:1004.2711}}.
%Type = Article
\bibitem[{{Tercero} et~al.(2018){Tercero}, {Cuadrado}, {L{\'o}pez},
  {Brouillet}, {Despois} and {Cernicharo}}]{Tercero18}
\bibinfo{author}{{Tercero}, B.}, \bibinfo{author}{{Cuadrado}, S.},
  \bibinfo{author}{{L{\'o}pez}, A.}, \bibinfo{author}{{Brouillet}, N.},
  \bibinfo{author}{{Despois}, D.}, \bibinfo{author}{{Cernicharo}, J.},
  \bibinfo{year}{2018}.
\newblock \bibinfo{title}{{Chemical segregation of complex organic O-bearing
  species in Orion KL}}.
\newblock \bibinfo{journal}{\aap} \bibinfo{volume}{620}, \bibinfo{pages}{L6}.
\newblock \DOIprefix\doi{10.1051/0004-6361/201834417},
  \href{http://arxiv.org/abs/1811.08765}{{\tt arXiv:1811.08765}}.
%Type = Article
\bibitem[{{Tercero} et~al.(2012){Tercero}, {Margul{\`e}s}, {Carvajal},
  {Motiyenko}, {Huet}, {Alekseev}, {Kleiner}, {Guillemin}, {M{\o}llendal} and
  {Cernicharo}}]{tercero12}
\bibinfo{author}{{Tercero}, B.}, \bibinfo{author}{{Margul{\`e}s}, L.},
  \bibinfo{author}{{Carvajal}, M.}, \bibinfo{author}{{Motiyenko}, R.A.},
  \bibinfo{author}{{Huet}, T.R.}, \bibinfo{author}{{Alekseev}, E.A.},
  \bibinfo{author}{{Kleiner}, I.}, \bibinfo{author}{{Guillemin}, J.C.},
  \bibinfo{author}{{M{\o}llendal}, H.}, \bibinfo{author}{{Cernicharo}, J.},
  \bibinfo{year}{2012}.
\newblock \bibinfo{title}{{Microwave and submillimeter spectroscopy and first
  ISM detection of $^{18}$O-methyl formate}}.
\newblock \bibinfo{journal}{\aap} \bibinfo{volume}{538}, \bibinfo{pages}{A119}.
\newblock \DOIprefix\doi{10.1051/0004-6361/201117072}.
%Type = Article
\bibitem[{{Thiel} et~al.(2017){Thiel}, {Belloche}, {Menten}, {Garrod} and
  {M{\"u}ller}}]{Thiel17}
\bibinfo{author}{{Thiel}, V.}, \bibinfo{author}{{Belloche}, A.},
  \bibinfo{author}{{Menten}, K.M.}, \bibinfo{author}{{Garrod}, R.T.},
  \bibinfo{author}{{M{\"u}ller}, H.S.P.}, \bibinfo{year}{2017}.
\newblock \bibinfo{title}{{Complex organic molecules in diffuse clouds along
  the line of sight to Sagittarius B2}}.
\newblock \bibinfo{journal}{\aap} \bibinfo{volume}{605}, \bibinfo{pages}{L6}.
\newblock \DOIprefix\doi{10.1051/0004-6361/201731495},
  \href{http://arxiv.org/abs/1708.07292}{{\tt arXiv:1708.07292}}.
%Type = Article
\bibitem[{{Thiel} et~al.(2019){Thiel}, {Belloche}, {Menten}, {Giannetti},
  {Wiesemeyer}, {Winkel}, {Gratier}, {M{\"u}ller}, {Colombo} and
  {Garrod}}]{Thiel19}
\bibinfo{author}{{Thiel}, V.}, \bibinfo{author}{{Belloche}, A.},
  \bibinfo{author}{{Menten}, K.M.}, \bibinfo{author}{{Giannetti}, A.},
  \bibinfo{author}{{Wiesemeyer}, H.}, \bibinfo{author}{{Winkel}, B.},
  \bibinfo{author}{{Gratier}, P.}, \bibinfo{author}{{M{\"u}ller}, H.S.P.},
  \bibinfo{author}{{Colombo}, D.}, \bibinfo{author}{{Garrod}, R.T.},
  \bibinfo{year}{2019}.
\newblock \bibinfo{title}{{Small-scale physical and chemical structure of
  diffuse and translucent molecular clouds along the line of sight to Sgr B2}}.
\newblock \bibinfo{journal}{\aap} \bibinfo{volume}{623}, \bibinfo{pages}{A68}.
\newblock \DOIprefix\doi{10.1051/0004-6361/201834467},
  \href{http://arxiv.org/abs/1901.03231}{{\tt arXiv:1901.03231}}.
%Type = Article
\bibitem[{{Tielens} et~al.(1993){Tielens}, {Meixner}, {van der Werf},
  {Bregman}, {Tauber}, {Stutzki} and {Rank}}]{Tielens93}
\bibinfo{author}{{Tielens}, A.G.G.M.}, \bibinfo{author}{{Meixner}, M.M.},
  \bibinfo{author}{{van der Werf}, P.P.}, \bibinfo{author}{{Bregman}, J.},
  \bibinfo{author}{{Tauber}, J.A.}, \bibinfo{author}{{Stutzki}, J.},
  \bibinfo{author}{{Rank}, D.}, \bibinfo{year}{1993}.
\newblock \bibinfo{title}{{Anatomy of the Photodissociation Region in the Orion
  Bar}}.
\newblock \bibinfo{journal}{Science} \bibinfo{volume}{262},
  \bibinfo{pages}{86--89}.
\newblock \DOIprefix\doi{10.1126/science.262.5130.86}.
%Type = Article
\bibitem[{{Tosaki} et~al.(2017){Tosaki}, {Kohno}, {Harada}, {Tanaka}, {Egusa},
  {Izumi}, {Takano}, {Nakajima}, {Taniguchi} and {Tamura}}]{Tosaki17}
\bibinfo{author}{{Tosaki}, T.}, \bibinfo{author}{{Kohno}, K.},
  \bibinfo{author}{{Harada}, N.}, \bibinfo{author}{{Tanaka}, K.},
  \bibinfo{author}{{Egusa}, F.}, \bibinfo{author}{{Izumi}, T.},
  \bibinfo{author}{{Takano}, S.}, \bibinfo{author}{{Nakajima}, T.},
  \bibinfo{author}{{Taniguchi}, A.}, \bibinfo{author}{{Tamura}, Y.},
  \bibinfo{year}{2017}.
\newblock \bibinfo{title}{{A statistical study of giant molecular clouds traced
  by $^{13}$CO, C$^{18}$O, CS, and CH$_{3}$OH in the disk of NGC 1068 based on
  ALMA observations}}.
\newblock \bibinfo{journal}{\pasj} \bibinfo{volume}{69}, \bibinfo{pages}{18}.
\newblock \DOIprefix\doi{10.1093/pasj/psw122},
  \href{http://arxiv.org/abs/1612.00948}{{\tt arXiv:1612.00948}}.
%Type = Article
\bibitem[{{Turner}(1989)}]{Turner89}
\bibinfo{author}{{Turner}, B.E.}, \bibinfo{year}{1989}.
\newblock \bibinfo{title}{{A Molecular Line Survey of Sagittarius B2 and
  Orion-KL from 70 to 115 GHz. I. The Observational Data}}.
\newblock \bibinfo{journal}{\apjs} \bibinfo{volume}{70}, \bibinfo{pages}{539}.
\newblock \DOIprefix\doi{10.1086/191348}.
%Type = Article
\bibitem[{{Turner}(1991)}]{Turner91}
\bibinfo{author}{{Turner}, B.E.}, \bibinfo{year}{1991}.
\newblock \bibinfo{title}{{A Molecular Line Survey of Sagittarius B2 and
  Orion--KL from 70 to 115 GHz. II. Analysis of the Data}}.
\newblock \bibinfo{journal}{\apjs} \bibinfo{volume}{76}, \bibinfo{pages}{617}.
\newblock \DOIprefix\doi{10.1086/191577}.
%Type = Article
\bibitem[{{van Dishoeck} et~al.(1995){van Dishoeck}, {Blake}, {Jansen} and
  {Groesbeck}}]{vanDishoeck95}
\bibinfo{author}{{van Dishoeck}, E.F.}, \bibinfo{author}{{Blake}, G.A.},
  \bibinfo{author}{{Jansen}, D.J.}, \bibinfo{author}{{Groesbeck}, T.D.},
  \bibinfo{year}{1995}.
\newblock \bibinfo{title}{{Molecular Abundances and Low-Mass Star Formation.
  II. Organic and Deuterated Species toward IRAS 16293-2422}}.
\newblock \bibinfo{journal}{\apj} \bibinfo{volume}{447}, \bibinfo{pages}{760}.
\newblock \DOIprefix\doi{10.1086/175915}.
%Type = Article
\bibitem[{{van Gelder} et~al.(2022b){van Gelder}, {Jaspers}, {Nazari},
  {Ahmadi}, {van Dishoeck}, {Beltr{\'a}n}, {Fuller}, {S{\'a}nchez-Monge} and
  {Schilke}}]{vanGelder22b}
\bibinfo{author}{{van Gelder}, M.L.}, \bibinfo{author}{{Jaspers}, J.},
  \bibinfo{author}{{Nazari}, P.}, \bibinfo{author}{{Ahmadi}, A.},
  \bibinfo{author}{{van Dishoeck}, E.F.}, \bibinfo{author}{{Beltr{\'a}n},
  M.T.}, \bibinfo{author}{{Fuller}, G.A.},
  \bibinfo{author}{{S{\'a}nchez-Monge}, {\'A}.}, \bibinfo{author}{{Schilke},
  P.}, \bibinfo{year}{2022b}.
\newblock \bibinfo{title}{{Methanol deuteration in high-mass protostars}}.
\newblock \bibinfo{journal}{\aap} \bibinfo{volume}{667}, \bibinfo{pages}{A136}.
\newblock \DOIprefix\doi{10.1051/0004-6361/202244471},
  \href{http://arxiv.org/abs/2208.06515}{{\tt arXiv:2208.06515}}.
%Type = Article
\bibitem[{{van Gelder} et~al.(2022a){van Gelder}, {Nazari}, {Tabone}, {Ahmadi},
  {van Dishoeck}, {Beltr{\'a}n}, {Fuller}, {Sakai}, {S{\'a}nchez-Monge},
  {Schilke}, {Yang} and {Zhang}}]{vanGelder22a}
\bibinfo{author}{{van Gelder}, M.L.}, \bibinfo{author}{{Nazari}, P.},
  \bibinfo{author}{{Tabone}, B.}, \bibinfo{author}{{Ahmadi}, A.},
  \bibinfo{author}{{van Dishoeck}, E.F.}, \bibinfo{author}{{Beltr{\'a}n},
  M.T.}, \bibinfo{author}{{Fuller}, G.A.}, \bibinfo{author}{{Sakai}, N.},
  \bibinfo{author}{{S{\'a}nchez-Monge}, {\'A}.}, \bibinfo{author}{{Schilke},
  P.}, \bibinfo{author}{{Yang}, Y.L.}, \bibinfo{author}{{Zhang}, Y.},
  \bibinfo{year}{2022a}.
\newblock \bibinfo{title}{{Importance of source structure on complex organics
  emission. I. Observations of CH$_{3}$OH from low-mass to high-mass
  protostars}}.
\newblock \bibinfo{journal}{\aap} \bibinfo{volume}{662}, \bibinfo{pages}{A67}.
\newblock \DOIprefix\doi{10.1051/0004-6361/202142769},
  \href{http://arxiv.org/abs/2202.04723}{{\tt arXiv:2202.04723}}.
%Type = Article
\bibitem[{{van 't Hoff} et~al.(2018){van 't Hoff}, {Tobin}, {Trapman},
  {Harsono}, {Sheehan}, {Fischer}, {Megeath} and {van Dishoeck}}]{vantHoff18}
\bibinfo{author}{{van 't Hoff}, M.L.R.}, \bibinfo{author}{{Tobin}, J.J.},
  \bibinfo{author}{{Trapman}, L.}, \bibinfo{author}{{Harsono}, D.},
  \bibinfo{author}{{Sheehan}, P.D.}, \bibinfo{author}{{Fischer}, W.J.},
  \bibinfo{author}{{Megeath}, S.T.}, \bibinfo{author}{{van Dishoeck}, E.F.},
  \bibinfo{year}{2018}.
\newblock \bibinfo{title}{{Methanol and its Relation to the Water Snowline in
  the Disk around the Young Outbursting Star V883 Ori}}.
\newblock \bibinfo{journal}{\apjl} \bibinfo{volume}{864}, \bibinfo{pages}{L23}.
\newblock \DOIprefix\doi{10.3847/2041-8213/aadb8a},
  \href{http://arxiv.org/abs/1808.08258}{{\tt arXiv:1808.08258}}.
%Type = Article
\bibitem[{{Vastel} et~al.(2014){Vastel}, {Ceccarelli}, {Lefloch} and
  {Bachiller}}]{vastel14}
\bibinfo{author}{{Vastel}, C.}, \bibinfo{author}{{Ceccarelli}, C.},
  \bibinfo{author}{{Lefloch}, B.}, \bibinfo{author}{{Bachiller}, R.},
  \bibinfo{year}{2014}.
\newblock \bibinfo{title}{{The Origin of Complex Organic Molecules in
  Prestellar Cores}}.
\newblock \bibinfo{journal}{\apjl} \bibinfo{volume}{795}, \bibinfo{pages}{L2}.
\newblock \DOIprefix\doi{10.1088/2041-8205/795/1/L2},
  \href{http://arxiv.org/abs/1409.6565}{{\tt arXiv:1409.6565}}.
%Type = Article
\bibitem[{{Vasyunin} et~al.(2017){Vasyunin}, {Caselli}, {Dulieu} and
  {Jim{\'e}nez-Serra}}]{vasyunin17}
\bibinfo{author}{{Vasyunin}, A.I.}, \bibinfo{author}{{Caselli}, P.},
  \bibinfo{author}{{Dulieu}, F.}, \bibinfo{author}{{Jim{\'e}nez-Serra}, I.},
  \bibinfo{year}{2017}.
\newblock \bibinfo{title}{{Formation of Complex Molecules in Prestellar Cores:
  A Multilayer Approach}}.
\newblock \bibinfo{journal}{\apj} \bibinfo{volume}{842}, \bibinfo{pages}{33}.
\newblock \DOIprefix\doi{10.3847/1538-4357/aa72ec},
  \href{http://arxiv.org/abs/1705.04747}{{\tt arXiv:1705.04747}}.
%Type = Article
\bibitem[{{Vasyunin} and {Herbst}(2013)}]{vasyunin13}
\bibinfo{author}{{Vasyunin}, A.I.}, \bibinfo{author}{{Herbst}, E.},
  \bibinfo{year}{2013}.
\newblock \bibinfo{title}{{Reactive Desorption and Radiative Association as
  Possible Drivers of Complex Molecule Formation in the Cold Interstellar
  Medium}}.
\newblock \bibinfo{journal}{\apj} \bibinfo{volume}{769}, \bibinfo{pages}{34}.
\newblock \DOIprefix\doi{10.1088/0004-637X/769/1/34},
  \href{http://arxiv.org/abs/1303.7266}{{\tt arXiv:1303.7266}}.
%Type = Article
\bibitem[{{Viti} et~al.(2004){Viti}, {Collings}, {Dever}, {McCoustra} and
  {Williams}}]{Viti04}
\bibinfo{author}{{Viti}, S.}, \bibinfo{author}{{Collings}, M.P.},
  \bibinfo{author}{{Dever}, J.W.}, \bibinfo{author}{{McCoustra}, M.R.S.},
  \bibinfo{author}{{Williams}, D.A.}, \bibinfo{year}{2004}.
\newblock \bibinfo{title}{{Evaporation of ices near massive stars: models based
  on laboratory temperature programmed desorption data}}.
\newblock \bibinfo{journal}{\mnras} \bibinfo{volume}{354},
  \bibinfo{pages}{1141--1145}.
\newblock \DOIprefix\doi{10.1111/j.1365-2966.2004.08273.x},
  \href{http://arxiv.org/abs/astro-ph/0406054}{{\tt arXiv:astro-ph/0406054}}.
%Type = Article
\bibitem[{{Wakelam} et~al.(2021){Wakelam}, {Dartois}, {Chabot}, {Spezzano},
  {Navarro-Almaida}, {Loison} and {Fuente}}]{wakelam21}
\bibinfo{author}{{Wakelam}, V.}, \bibinfo{author}{{Dartois}, E.},
  \bibinfo{author}{{Chabot}, M.}, \bibinfo{author}{{Spezzano}, S.},
  \bibinfo{author}{{Navarro-Almaida}, D.}, \bibinfo{author}{{Loison}, J.C.},
  \bibinfo{author}{{Fuente}, A.}, \bibinfo{year}{2021}.
\newblock \bibinfo{title}{{Efficiency of non-thermal desorptions in cold-core
  conditions. Testing the sputtering of grain mantles induced by cosmic rays}}.
\newblock \bibinfo{journal}{\aap} \bibinfo{volume}{652}, \bibinfo{pages}{A63}.
\newblock \DOIprefix\doi{10.1051/0004-6361/202039855},
  \href{http://arxiv.org/abs/2106.08621}{{\tt arXiv:2106.08621}}.
%Type = Article
\bibitem[{{Walsh} et~al.(2016){Walsh}, {Loomis}, {{\"O}berg}, {Kama}, {van 't
  Hoff}, {Millar}, {Aikawa}, {Herbst}, {Widicus Weaver} and {Nomura}}]{Walsh16}
\bibinfo{author}{{Walsh}, C.}, \bibinfo{author}{{Loomis}, R.A.},
  \bibinfo{author}{{{\"O}berg}, K.I.}, \bibinfo{author}{{Kama}, M.},
  \bibinfo{author}{{van 't Hoff}, M.L.R.}, \bibinfo{author}{{Millar}, T.J.},
  \bibinfo{author}{{Aikawa}, Y.}, \bibinfo{author}{{Herbst}, E.},
  \bibinfo{author}{{Widicus Weaver}, S.L.}, \bibinfo{author}{{Nomura}, H.},
  \bibinfo{year}{2016}.
\newblock \bibinfo{title}{{First Detection of Gas-phase Methanol in a
  Protoplanetary Disk}}.
\newblock \bibinfo{journal}{\apjl} \bibinfo{volume}{823}, \bibinfo{pages}{L10}.
\newblock \DOIprefix\doi{10.3847/2041-8205/823/1/L10},
  \href{http://arxiv.org/abs/1606.06492}{{\tt arXiv:1606.06492}}.
%Type = Article
\bibitem[{{Walsh} et~al.(2014){Walsh}, {Millar}, {Nomura}, {Herbst}, {Widicus
  Weaver}, {Aikawa}, {Laas} and {Vasyunin}}]{Walsh14}
\bibinfo{author}{{Walsh}, C.}, \bibinfo{author}{{Millar}, T.J.},
  \bibinfo{author}{{Nomura}, H.}, \bibinfo{author}{{Herbst}, E.},
  \bibinfo{author}{{Widicus Weaver}, S.}, \bibinfo{author}{{Aikawa}, Y.},
  \bibinfo{author}{{Laas}, J.C.}, \bibinfo{author}{{Vasyunin}, A.I.},
  \bibinfo{year}{2014}.
\newblock \bibinfo{title}{{Complex organic molecules in protoplanetary disks}}.
\newblock \bibinfo{journal}{\aap} \bibinfo{volume}{563}, \bibinfo{pages}{A33}.
\newblock \DOIprefix\doi{10.1051/0004-6361/201322446},
  \href{http://arxiv.org/abs/1403.0390}{{\tt arXiv:1403.0390}}.
%Type = Article
\bibitem[{{Wang} et~al.(2009){Wang}, {Chin}, {Henkel}, {Whiteoak} and
  {Cunningham}}]{Wang09}
\bibinfo{author}{{Wang}, M.}, \bibinfo{author}{{Chin}, Y.N.},
  \bibinfo{author}{{Henkel}, C.}, \bibinfo{author}{{Whiteoak}, J.B.},
  \bibinfo{author}{{Cunningham}, M.}, \bibinfo{year}{2009}.
\newblock \bibinfo{title}{{Abundances and Isotope Ratios in the Magellanic
  Clouds: The Star-Forming Environment of N 113}}.
\newblock \bibinfo{journal}{\apj} \bibinfo{volume}{690},
  \bibinfo{pages}{580--597}.
\newblock \DOIprefix\doi{10.1088/0004-637X/690/1/580},
  \href{http://arxiv.org/abs/0809.1706}{{\tt arXiv:0809.1706}}.
%Type = Article
\bibitem[{{Watanabe} et~al.(2014){Watanabe}, {Sakai}, {Sorai} and
  {Yamamoto}}]{Watanabe14}
\bibinfo{author}{{Watanabe}, Y.}, \bibinfo{author}{{Sakai}, N.},
  \bibinfo{author}{{Sorai}, K.}, \bibinfo{author}{{Yamamoto}, S.},
  \bibinfo{year}{2014}.
\newblock \bibinfo{title}{{Spectral Line Survey toward the Spiral Arm of M51 in
  the 3 and 2 mm Bands}}.
\newblock \bibinfo{journal}{\apj} \bibinfo{volume}{788}, \bibinfo{pages}{4}.
\newblock \DOIprefix\doi{10.1088/0004-637X/788/1/4},
  \href{http://arxiv.org/abs/1404.1202}{{\tt arXiv:1404.1202}}.
%Type = Article
\bibitem[{{Wenzel} et~al.(2024a){Wenzel}, {Cooke}, {Changala}, {Bergin},
  {Zhang}, {Burkhardt}, {Byrne}, {Charnley}, {Cordiner}, {Duffy}, {Fried},
  {Gupta}, {Holdren}, {Lipnicky}, {Loomis}, {Shay}, {Shingledecker}, {Siebert},
  {Stewart}, {Willis}, {Xue}, {Remijan}, {Wendlandt}, {McCarthy} and
  {McGuire}}]{wenzel2024a}
\bibinfo{author}{{Wenzel}, G.}, \bibinfo{author}{{Cooke}, I.R.},
  \bibinfo{author}{{Changala}, P.B.}, \bibinfo{author}{{Bergin}, E.A.},
  \bibinfo{author}{{Zhang}, S.}, \bibinfo{author}{{Burkhardt}, A.M.},
  \bibinfo{author}{{Byrne}, A.N.}, \bibinfo{author}{{Charnley}, S.B.},
  \bibinfo{author}{{Cordiner}, M.A.}, \bibinfo{author}{{Duffy}, M.},
  \bibinfo{author}{{Fried}, Z.T.P.}, \bibinfo{author}{{Gupta}, H.},
  \bibinfo{author}{{Holdren}, M.S.}, \bibinfo{author}{{Lipnicky}, A.},
  \bibinfo{author}{{Loomis}, R.A.}, \bibinfo{author}{{Shay}, H.T.},
  \bibinfo{author}{{Shingledecker}, C.N.}, \bibinfo{author}{{Siebert}, M.A.},
  \bibinfo{author}{{Stewart}, D.A.}, \bibinfo{author}{{Willis}, R.H.J.},
  \bibinfo{author}{{Xue}, C.}, \bibinfo{author}{{Remijan}, A.J.},
  \bibinfo{author}{{Wendlandt}, A.E.}, \bibinfo{author}{{McCarthy}, M.C.},
  \bibinfo{author}{{McGuire}, B.A.}, \bibinfo{year}{2024a}.
\newblock \bibinfo{title}{{Detection of interstellar 1-cyanopyrene: A four-ring
  polycyclic aromatic hydrocarbon}}.
\newblock \bibinfo{journal}{Science} \bibinfo{volume}{386},
  \bibinfo{pages}{810--813}.
\newblock \DOIprefix\doi{10.1126/science.adq6391}.
%Type = Article
\bibitem[{{Wenzel} et~al.(2024b){Wenzel}, {Speak}, {Changala}, {Willis},
  {Burkhardt}, {Zhang}, {Bergin}, {Byrne}, {Charnley}, {Fried}, {Gupta},
  {Herbst}, {Holdren}, {Lipnicky}, {Loomis}, {Shingledecker}, {Xue}, {Remijan},
  {Wendlandt}, {McCarthy}, {Cooke} and {McGuire}}]{wenzel2024b}
\bibinfo{author}{{Wenzel}, G.}, \bibinfo{author}{{Speak}, T.H.},
  \bibinfo{author}{{Changala}, P.B.}, \bibinfo{author}{{Willis}, R.H.J.},
  \bibinfo{author}{{Burkhardt}, A.M.}, \bibinfo{author}{{Zhang}, S.},
  \bibinfo{author}{{Bergin}, E.A.}, \bibinfo{author}{{Byrne}, A.N.},
  \bibinfo{author}{{Charnley}, S.B.}, \bibinfo{author}{{Fried}, Z.T.P.},
  \bibinfo{author}{{Gupta}, H.}, \bibinfo{author}{{Herbst}, E.},
  \bibinfo{author}{{Holdren}, M.S.}, \bibinfo{author}{{Lipnicky}, A.},
  \bibinfo{author}{{Loomis}, R.A.}, \bibinfo{author}{{Shingledecker}, C.N.},
  \bibinfo{author}{{Xue}, C.}, \bibinfo{author}{{Remijan}, A.J.},
  \bibinfo{author}{{Wendlandt}, A.E.}, \bibinfo{author}{{McCarthy}, M.C.},
  \bibinfo{author}{{Cooke}, I.R.}, \bibinfo{author}{{McGuire}, B.A.},
  \bibinfo{year}{2024b}.
\newblock \bibinfo{title}{{Detections of interstellar aromatic nitriles
  2-cyanopyrene and 4-cyanopyrene in TMC-1}}.
\newblock \bibinfo{journal}{Nature Astronomy}
  \DOIprefix\doi{10.1038/s41550-024-02410-9},
  \href{http://arxiv.org/abs/2410.00670}{{\tt arXiv:2410.00670}}.
%Type = Article
\bibitem[{{Wiklind} and {Combes}(1998)}]{wiklind98}
\bibinfo{author}{{Wiklind}, T.}, \bibinfo{author}{{Combes}, F.},
  \bibinfo{year}{1998}.
\newblock \bibinfo{title}{{The Complex Molecular Absorption Line System at Z =
  0.886 toward PKS 1830-211}}.
\newblock \bibinfo{journal}{\apj} \bibinfo{volume}{500},
  \bibinfo{pages}{129--137}.
\newblock \DOIprefix\doi{10.1086/305701},
  \href{http://arxiv.org/abs/astro-ph/9709141}{{\tt arXiv:astro-ph/9709141}}.
%Type = Article
\bibitem[{{Williams} et~al.(2022){Williams}, {Cyganowski}, {Brogan}, {Hunter},
  {Ilee}, {Nazari}, {Kruijssen}, {Smith} and {Bonnell}}]{williams22}
\bibinfo{author}{{Williams}, G.M.}, \bibinfo{author}{{Cyganowski}, C.J.},
  \bibinfo{author}{{Brogan}, C.L.}, \bibinfo{author}{{Hunter}, T.R.},
  \bibinfo{author}{{Ilee}, J.D.}, \bibinfo{author}{{Nazari}, P.},
  \bibinfo{author}{{Kruijssen}, J.M.D.}, \bibinfo{author}{{Smith}, R.J.},
  \bibinfo{author}{{Bonnell}, I.A.}, \bibinfo{year}{2022}.
\newblock \bibinfo{title}{{ALMA observations of the Extended Green Object
  G19.01-0.03 - I. A Keplerian disc in a massive protostellar system}}.
\newblock \bibinfo{journal}{\mnras} \bibinfo{volume}{509},
  \bibinfo{pages}{748--762}.
\newblock \DOIprefix\doi{10.1093/mnras/stab2973},
  \href{http://arxiv.org/abs/2110.06262}{{\tt arXiv:2110.06262}}.
%Type = Article
\bibitem[{{Willis} et~al.(2020){Willis}, {Garrod}, {Belloche}, {M{\"u}ller},
  {Barger}, {Bonfand} and {Menten}}]{willis20}
\bibinfo{author}{{Willis}, E.R.}, \bibinfo{author}{{Garrod}, R.T.},
  \bibinfo{author}{{Belloche}, A.}, \bibinfo{author}{{M{\"u}ller}, H.S.P.},
  \bibinfo{author}{{Barger}, C.J.}, \bibinfo{author}{{Bonfand}, M.},
  \bibinfo{author}{{Menten}, K.M.}, \bibinfo{year}{2020}.
\newblock \bibinfo{title}{{Exploring molecular complexity with ALMA (EMoCA):
  complex isocyanides in Sgr B2(N)}}.
\newblock \bibinfo{journal}{\aap} \bibinfo{volume}{636}, \bibinfo{pages}{A29}.
\newblock \DOIprefix\doi{10.1051/0004-6361/201936489},
  \href{http://arxiv.org/abs/2003.07423}{{\tt arXiv:2003.07423}}.
%Type = Article
\bibitem[{{Winn} et~al.(2002){Winn}, {Kochanek}, {McLeod}, {Falco}, {Impey} and
  {Rix}}]{winn02}
\bibinfo{author}{{Winn}, J.N.}, \bibinfo{author}{{Kochanek}, C.S.},
  \bibinfo{author}{{McLeod}, B.A.}, \bibinfo{author}{{Falco}, E.E.},
  \bibinfo{author}{{Impey}, C.D.}, \bibinfo{author}{{Rix}, H.W.},
  \bibinfo{year}{2002}.
\newblock \bibinfo{title}{{PKS 1830-211: A Face-on Spiral Galaxy Lens}}.
\newblock \bibinfo{journal}{\apj} \bibinfo{volume}{575},
  \bibinfo{pages}{103--110}.
\newblock \DOIprefix\doi{10.1086/341265},
  \href{http://arxiv.org/abs/astro-ph/0201551}{{\tt arXiv:astro-ph/0201551}}.
%Type = Article
\bibitem[{{Yamaguchi} et~al.(2012){Yamaguchi}, {Takano}, {Watanabe}, {Sakai},
  {Sakai}, {Liu}, {Su}, {Hirano}, {Takakuwa}, {Aikawa}, {Nomura} and
  {Yamamoto}}]{Yamaguchi12}
\bibinfo{author}{{Yamaguchi}, T.}, \bibinfo{author}{{Takano}, S.},
  \bibinfo{author}{{Watanabe}, Y.}, \bibinfo{author}{{Sakai}, N.},
  \bibinfo{author}{{Sakai}, T.}, \bibinfo{author}{{Liu}, S.Y.},
  \bibinfo{author}{{Su}, Y.N.}, \bibinfo{author}{{Hirano}, N.},
  \bibinfo{author}{{Takakuwa}, S.}, \bibinfo{author}{{Aikawa}, Y.},
  \bibinfo{author}{{Nomura}, H.}, \bibinfo{author}{{Yamamoto}, S.},
  \bibinfo{year}{2012}.
\newblock \bibinfo{title}{{The 3 mm Spectral Line Survey toward the Lynds 1157
  B1 Shocked Region. I. Data}}.
\newblock \bibinfo{journal}{\pasj} \bibinfo{volume}{64}, \bibinfo{pages}{105}.
\newblock \DOIprefix\doi{10.1093/pasj/64.5.105}.
%Type = Article
\bibitem[{{Yang} et~al.(2021){Yang}, {Sakai}, {Zhang}, {Murillo}, {Zhang},
  {Higuchi}, {Zeng}, {L{\'o}pez-Sepulcre}, {Yamamoto}, {Lefloch}, {Bouvier},
  {Ceccarelli}, {Hirota}, {Imai}, {Oya}, {Sakai} and {Watanabe}}]{Yang21}
\bibinfo{author}{{Yang}, Y.L.}, \bibinfo{author}{{Sakai}, N.},
  \bibinfo{author}{{Zhang}, Y.}, \bibinfo{author}{{Murillo}, N.M.},
  \bibinfo{author}{{Zhang}, Z.E.}, \bibinfo{author}{{Higuchi}, A.E.},
  \bibinfo{author}{{Zeng}, S.}, \bibinfo{author}{{L{\'o}pez-Sepulcre}, A.},
  \bibinfo{author}{{Yamamoto}, S.}, \bibinfo{author}{{Lefloch}, B.},
  \bibinfo{author}{{Bouvier}, M.}, \bibinfo{author}{{Ceccarelli}, C.},
  \bibinfo{author}{{Hirota}, T.}, \bibinfo{author}{{Imai}, M.},
  \bibinfo{author}{{Oya}, Y.}, \bibinfo{author}{{Sakai}, T.},
  \bibinfo{author}{{Watanabe}, Y.}, \bibinfo{year}{2021}.
\newblock \bibinfo{title}{{The Perseus ALMA Chemistry Survey (PEACHES). I. The
  Complex Organic Molecules in Perseus Embedded Protostars}}.
\newblock \bibinfo{journal}{\apj} \bibinfo{volume}{910}, \bibinfo{pages}{20}.
\newblock \DOIprefix\doi{10.3847/1538-4357/abdfd6},
  \href{http://arxiv.org/abs/2101.11009}{{\tt arXiv:2101.11009}}.
%Type = Article
\bibitem[{{Yen} et~al.(2019){Yen}, {Gu}, {Hirano}, {Koch}, {Lee}, {Liu} and
  {Takakuwa}}]{yen19}
\bibinfo{author}{{Yen}, H.W.}, \bibinfo{author}{{Gu}, P.G.},
  \bibinfo{author}{{Hirano}, N.}, \bibinfo{author}{{Koch}, P.M.},
  \bibinfo{author}{{Lee}, C.F.}, \bibinfo{author}{{Liu}, H.B.},
  \bibinfo{author}{{Takakuwa}, S.}, \bibinfo{year}{2019}.
\newblock \bibinfo{title}{{HL Tau Disk in HCO$^{+}$ (3-2) and (1-0) with ALMA:
  Gas Density, Temperature, Gap, and One-arm Spiral}}.
\newblock \bibinfo{journal}{\apj} \bibinfo{volume}{880}, \bibinfo{pages}{69}.
\newblock \DOIprefix\doi{10.3847/1538-4357/ab29f8},
  \href{http://arxiv.org/abs/1906.05535}{{\tt arXiv:1906.05535}}.
%Type = Article
\bibitem[{{Zeng} et~al.(2018){Zeng}, {Jim{\'e}nez-Serra}, {Rivilla},
  {Mart{\'\i}n}, {Mart{\'\i}n-Pintado}, {Requena-Torres},
  {Armijos-Abenda{\~n}o}, {Riquelme} and {Aladro}}]{zeng18}
\bibinfo{author}{{Zeng}, S.}, \bibinfo{author}{{Jim{\'e}nez-Serra}, I.},
  \bibinfo{author}{{Rivilla}, V.M.}, \bibinfo{author}{{Mart{\'\i}n}, S.},
  \bibinfo{author}{{Mart{\'\i}n-Pintado}, J.},
  \bibinfo{author}{{Requena-Torres}, M.A.},
  \bibinfo{author}{{Armijos-Abenda{\~n}o}, J.}, \bibinfo{author}{{Riquelme},
  D.}, \bibinfo{author}{{Aladro}, R.}, \bibinfo{year}{2018}.
\newblock \bibinfo{title}{{Complex organic molecules in the Galactic Centre:
  the N-bearing family}}.
\newblock \bibinfo{journal}{\mnras} \bibinfo{volume}{478},
  \bibinfo{pages}{2962--2975}.
\newblock \DOIprefix\doi{10.1093/mnras/sty1174},
  \href{http://arxiv.org/abs/1804.11321}{{\tt arXiv:1804.11321}}.
%Type = Article
\bibitem[{{Zeng} et~al.(2021){Zeng}, {Jim{\'e}nez-Serra}, {Rivilla},
  {Mart{\'\i}n-Pintado}, {Rodr{\'\i}guez-Almeida}, {Tercero}, {de Vicente},
  {Rico-Villas}, {Colzi}, {Mart{\'\i}n} and {Requena-Torres}}]{zeng21}
\bibinfo{author}{{Zeng}, S.}, \bibinfo{author}{{Jim{\'e}nez-Serra}, I.},
  \bibinfo{author}{{Rivilla}, V.M.}, \bibinfo{author}{{Mart{\'\i}n-Pintado},
  J.}, \bibinfo{author}{{Rodr{\'\i}guez-Almeida}, L.F.},
  \bibinfo{author}{{Tercero}, B.}, \bibinfo{author}{{de Vicente}, P.},
  \bibinfo{author}{{Rico-Villas}, F.}, \bibinfo{author}{{Colzi}, L.},
  \bibinfo{author}{{Mart{\'\i}n}, S.}, \bibinfo{author}{{Requena-Torres},
  M.A.}, \bibinfo{year}{2021}.
\newblock \bibinfo{title}{{Probing the Chemical Complexity of Amines in the
  ISM: Detection of Vinylamine (C$_{2}$H$_{3}$NH$_{2}$) and Tentative Detection
  of Ethylamine (C$_{2}$H$_{5}$NH$_{2}$)}}.
\newblock \bibinfo{journal}{\apjl} \bibinfo{volume}{920}, \bibinfo{pages}{L27}.
\newblock \DOIprefix\doi{10.3847/2041-8213/ac2c7e},
  \href{http://arxiv.org/abs/2110.01791}{{\tt arXiv:2110.01791}}.
%Type = Article
\bibitem[{{Zeng} et~al.(2019){Zeng}, {Qu{\'e}nard}, {Jim{\'e}nez-Serra},
  {Mart{\'\i}n-Pintado}, {Rivilla}, {Testi} and
  {Mart{\'\i}n-Dom{\'e}nech}}]{Zeng19}
\bibinfo{author}{{Zeng}, S.}, \bibinfo{author}{{Qu{\'e}nard}, D.},
  \bibinfo{author}{{Jim{\'e}nez-Serra}, I.},
  \bibinfo{author}{{Mart{\'\i}n-Pintado}, J.}, \bibinfo{author}{{Rivilla},
  V.M.}, \bibinfo{author}{{Testi}, L.},
  \bibinfo{author}{{Mart{\'\i}n-Dom{\'e}nech}, R.}, \bibinfo{year}{2019}.
\newblock \bibinfo{title}{{First detection of the pre-biotic molecule
  glycolonitrile (HOCH$_{2}$CN) in the interstellar medium}}.
\newblock \bibinfo{journal}{\mnras} \bibinfo{volume}{484},
  \bibinfo{pages}{L43--L48}.
\newblock \DOIprefix\doi{10.1093/mnrasl/slz002},
  \href{http://arxiv.org/abs/1901.02576}{{\tt arXiv:1901.02576}}.
%Type = Article
\bibitem[{{Zeng} et~al.(2023){Zeng}, {Rivilla}, {Jim{\'e}nez-Serra}, {Colzi},
  {Mart{\'\i}n-Pintado}, {Tercero}, {de Vicente}, {Mart{\'\i}n} and
  {Requena-Torres}}]{zeng23}
\bibinfo{author}{{Zeng}, S.}, \bibinfo{author}{{Rivilla}, V.M.},
  \bibinfo{author}{{Jim{\'e}nez-Serra}, I.}, \bibinfo{author}{{Colzi}, L.},
  \bibinfo{author}{{Mart{\'\i}n-Pintado}, J.}, \bibinfo{author}{{Tercero}, B.},
  \bibinfo{author}{{de Vicente}, P.}, \bibinfo{author}{{Mart{\'\i}n}, S.},
  \bibinfo{author}{{Requena-Torres}, M.A.}, \bibinfo{year}{2023}.
\newblock \bibinfo{title}{{Amides inventory towards the G+0.693-0.027 molecular
  cloud}}.
\newblock \bibinfo{journal}{\mnras} \bibinfo{volume}{523},
  \bibinfo{pages}{1448--1463}.
\newblock \DOIprefix\doi{10.1093/mnras/stad1478},
  \href{http://arxiv.org/abs/2305.18715}{{\tt arXiv:2305.18715}}.
%Type = Article
\bibitem[{{Zeng} et~al.(2020){Zeng}, {Zhang}, {Jim{\'e}nez-Serra}, {Tercero},
  {Lu}, {Mart{\'\i}n-Pintado}, {de Vicente}, {Rivilla} and {Li}}]{zeng20}
\bibinfo{author}{{Zeng}, S.}, \bibinfo{author}{{Zhang}, Q.},
  \bibinfo{author}{{Jim{\'e}nez-Serra}, I.}, \bibinfo{author}{{Tercero}, B.},
  \bibinfo{author}{{Lu}, X.}, \bibinfo{author}{{Mart{\'\i}n-Pintado}, J.},
  \bibinfo{author}{{de Vicente}, P.}, \bibinfo{author}{{Rivilla}, V.M.},
  \bibinfo{author}{{Li}, S.}, \bibinfo{year}{2020}.
\newblock \bibinfo{title}{{Cloud-cloud collision as drivers of the chemical
  complexity in Galactic Centre molecular clouds}}.
\newblock \bibinfo{journal}{\mnras} \bibinfo{volume}{497},
  \bibinfo{pages}{4896--4909}.
\newblock \DOIprefix\doi{10.1093/mnras/staa2187},
  \href{http://arxiv.org/abs/2007.14362}{{\tt arXiv:2007.14362}}.
%Type = Article
\bibitem[{{Zuckerman} et~al.(1975){Zuckerman}, {Turner}, {Johnson}, {Clark},
  {Lovas}, {Fourikis}, {Palmer}, {Morris}, {Lilley}, {Ball}, {Gottlieb},
  {Litvak} and {Penfield}}]{Zuckerman75}
\bibinfo{author}{{Zuckerman}, B.}, \bibinfo{author}{{Turner}, B.E.},
  \bibinfo{author}{{Johnson}, D.R.}, \bibinfo{author}{{Clark}, F.O.},
  \bibinfo{author}{{Lovas}, F.J.}, \bibinfo{author}{{Fourikis}, N.},
  \bibinfo{author}{{Palmer}, P.}, \bibinfo{author}{{Morris}, M.},
  \bibinfo{author}{{Lilley}, A.E.}, \bibinfo{author}{{Ball}, J.A.},
  \bibinfo{author}{{Gottlieb}, C.A.}, \bibinfo{author}{{Litvak}, M.M.},
  \bibinfo{author}{{Penfield}, H.}, \bibinfo{year}{1975}.
\newblock \bibinfo{title}{{Detection of interstellar trans-ethyl alcohol.}}
\newblock \bibinfo{journal}{\apjl} \bibinfo{volume}{196},
  \bibinfo{pages}{L99--L102}.
\newblock \DOIprefix\doi{10.1086/181753}.

\end{thebibliography}





% \end{thebibliography}
\end{document}